\documentclass[10pt,a4paper,twocolumn,journal]{IEEEtran}
\usepackage[T1]{fontenc}
\usepackage{amsmath,amsfonts,amssymb,amsxtra,bm}
\usepackage{setspace}
\usepackage[latin9]{inputenc}
\usepackage{graphicx}
\usepackage{tikz}
\usetikzlibrary{matrix}
\usepackage{amsthm}
\usepackage{color}
\usepackage{cite}              
\usepackage{epsfig,psfrag}
\usepackage{caption}
\usepackage{pgfplots}
\usepackage{algorithm}
\usepackage{algpseudocode}
\usepackage{enumitem}

\allowdisplaybreaks
\newcommand{\rd}{\textcolor{black}}
\newcommand{\rdd}{\textcolor{black}}
\newcommand{\red}{\textcolor{black}}

\newcommand{\be}{\begin{equation}}
\newcommand{\ee}{\end{equation}}
\newcommand{\ist}{\hspace*{.3mm}}
\newcommand{\rmv}{\hspace*{-.3mm}}

\newcommand{\rrmv}{\hspace*{-1mm}}
\newcommand{\bd}[1]{\mathbf{#1}}

\newcommand{\nn}{\nonumber}
\newcommand{\T}{\text{T}}

\newcommand{\va}[1]{#1}  
\newcommand{\rv}[1]{\mathsf{#1}}  
\newcommand{\V}[1]{\bm{#1}}
\newcommand{\RV}[1]{\boldsymbol{\mathsf{#1}}}
\newcommand{\RS}[1]{\mathsf{\uppercase{#1}}}
\newcommand{\Se}[1]{\uppercase{#1}}

\newcommand{\st}{\V{x}} 								
\newcommand{\stR}{\RV{x}}							  
\newcommand{\RFSst}{\Se{X}}							
\newcommand{\RFSstR}{\RS{X}}					  
\newcommand{\lRFSst}{\tilde{\RFSst}}		
\newcommand{\lRFSstR}{\RS{\lRFSst}}			
\newcommand{\ca}{\rho}									
\newcommand{\ex}{r}											
\newcommand{\sd}{s}	 										
\newcommand{\intf}{\lambda}							
\newcommand{\la}{l}											
\newcommand{\laR}{\rv{l}}								
\newcommand{\wei}{w}										
\newcommand{\me}{\V{z}} 								
\newcommand{\meR}{\RV{z}} 							
\newcommand{\RFSme}{\Se{Z}}							
\newcommand{\RFSmeR}{\RS{Z}}						
\newcommand{\su}{p_{\text{S}}}
\newcommand{\de}{p_{\text{D}}}

\newcommand{\ass}{\va{a}}								
\newcommand{\assv}{\V{a}}								
\newcommand{\assR}{\rv{a}}							
\newcommand{\assvR}{\RV{a}}							

\newcommand{\TO}{}
\newcommand{\assw}{\beta}

\pgfplotsset{compat=1.17}

\begin{document}

\title{An Efficient Labeled/Unlabeled Random Finite Set\\ Algorithm for Multiobject Tracking \vspace{2mm}}

\author{Thomas Kropfreiter, Florian Meyer, and Franz Hlawatsch\thanks{T.\ Kropfrei\-ter 
and F.\ Hlawatsch are with the Institute of Telecommunications, TU Wien, 1040 Vienna, Austria
(e-mail: \{thomas.kropfreiter,\,franz. hlawatsch\}@tuwien.ac.at). 
F.~Meyer is with the Scripps Institution of Oceanography and the Department of Electrical and Computer Engineering, University of California San Diego, La Jolla, CA 92037, USA (e-mail: flmeyer@ucsd.edu).
This work was supported by the Austrian Science Fund (FWF) under grant P 32055-N31, by the Czech Science Foundation (GA\v{C}R) under grant 17-19638S,
and by the Office of Naval Research (ONR) under grant N00014-21-1-2267. Parts of this paper were previously presented at SDF 2018, Bonn, Germany, October 2018.}}

\maketitle

\begin{abstract}
We propose an efficient random finite set (RFS) based algorithm for multiobject tracking in which the object states are modeled by a combination of a labeled multi-Bernoulli (LMB) RFS and a Poisson RFS. The less computationally demanding Poisson part of the algorithm is used to track potential objects whose existence is unlikely. Only if a quantity characterizing the plausibility of object existence is above a threshold, a new labeled Bernoulli component is created and the object is tracked by the more accurate but more computationally demanding LMB part of the algorithm. Conversely, a labeled Bernoulli component is transferred back to the Poisson RFS if the corresponding existence probability falls below another threshold. Contrary to existing hybrid algorithms based on multi-Bernoulli and Poisson RFSs, the proposed method  facilitates track continuity and implements complexity-reducing features. Simulation results demonstrate a large complexity reduction relative to other RFS-based algorithms with comparable performance.
\vspace{-1mm}
\end{abstract}

\begin{IEEEkeywords}
Filtering,
multiobject tracking, 
multitarget tracking,
object detection,
point processes,
random finite sets,
sequential estimation.
\end{IEEEkeywords}

\section{Introduction}
\label{sec:intro}

Multiobject tracking aims to estimate the time-dependent states of an unknown, time-dependent number of objects from a sequence of measurements \cite{Bar11,Mah07,Mah14,Cha11,Bla99}. \red{This task is complicated by a measurement-origin uncertainty, i.e., the fact that it is unknown which measurement was generated by which object. Most established multiobject tracking algorithms address measurement-origin uncertainty by solving a data association problem \cite{Bar11}.} Here, we propose a multiobject tracking algorithm that uses random finite sets (RFSs) and the framework of finite set statistics (FISST) \cite{Mah07,Mah14} to model the object states and measurements. 

\vspace{-6mm}

\rd{
\subsection{State of the Art}
\label{sec:intro_stateoftheart}
}

Existing RFS-based multiobject tracking methods include the probability hypothesis density (PHD) filter \cite{Mah03,Mah07}, the cardinalized PHD (CPHD) filter \cite{Mah07b,Mah07}, and multi-Bernoulli (MB) filters \cite{Mah07,Mah14,Vo09}. \rd{These filters do not require a data association step. 
They have a low or moderate computational complexity but can exhibit poor accuracy in more challenging scenarios. They do not maintain track continuity, in that they do not estimate entire trajectories of consecutive object 
\pagebreak 
states.}

In many applications, 
track continuity is required.
A widely used approach to achieving track continuity is to model the multiobject state by a labeled RFS \cite{Vo13,Vo14,Reu14,Reu17LMBGibbs,Kro19LMB,VoVoHoa:J17,BeaVoVo:J20,VoVo:J19}. Related tracking filters include the generalized labeled multi-Bernoulli (GLMB) filter \cite{Vo13,Vo14,VoVoHoa:J17}, which is based on the GLMB RFS, and the labeled multi-Bernoulli (LMB) filter \cite{Reu14,Reu17LMBGibbs,Kro19LMB}, which is based on the LMB RFS. Compared to the GLMB filter, the LMB filter incorporates certain approximations resulting in a much lower complexity. 
Recently, (G)LMB methods that are suitable for large-scale tracking scenarios \cite{Kro19LMB,VoVoHoa:J17,BeaVoVo:J20,Reu17LMBGibbs} and that consider information from multiple consecutive measurements at each filtering step \cite{VoVo:J19} have been proposed. On the other hand, the track-oriented marginal multi-Bernoulli/Poisson (TOMB/P) filter \cite{Wil15} is based on the union of two unlabeled RFSs, namely, a Poisson RFS and an MB RFS. 
The TOMB/P filter creates a new Bernoulli component for each measurement and prunes Bernoulli components with low existence probability. 
A modification of the TOMB/P filter \cite{Wil12} transfers Bernoulli components with low existence probability to the Poisson RFS instead of pruning them; this transfer is referred to as \emph{recycling} in \cite{Wil12}. A ``label-augmented'' version of the TOMB/P filter that maintains track continuity was obtained in \cite{Mey18Proc} by heuristically introducing labels in the formulation of the TOMB/P filter.

An alternative approach to multiobject tracking with track continuity is the paradigm of partially distinguishable populations \cite{HouCla:J18}. 
This approach can lead to methods with a computational complexity that is linear in the number of tracks and the number of 
measurements.\rd{\footnote{\rd{Although 
relying on a different theoretical paradigm, the concept of distinguishable versus indistinguishable objects is partly similar to our concept of labeled versus unlabeled objects.
However, whereas indistinguishable/unlabeled objects are considered in both approaches as an entity (modeled by a Poisson RFS in our method), in our approach
also unlabeled objects are tracked within the Poisson part, i.e., the Poisson part is also updated by measurements. 
Furthermore, in \cite{HouCla:J18} track continuity is achieved by using partially distinguishable populations, whereas in our method it is achieved 
by using a labeled RFS. Finally, the two methods employ different approximations to reduce computational 
complexity.}}} 
Finally, track continuity can be achieved by modeling the multiobject state as an RFS of trajectories \cite{Gar20SoT,Gra18SoTPMBM,Gar19SoTPHD,Xia19TPMBM}, where each trajectory is characterized by its initial time, its length, and the sequence of object states it contains. Algorithms based on this approach comprise the trajectory PHD and CPHD filters \cite{Gar19SoTPHD}, the trajectory multi-Bernoulli mixture filter \cite{Gar20SoT}, and the trajectory Poisson multi-Bernoulli mixture filter \cite{Gra18SoTPMBM}. These methods can have performance advantages over the methods proposed in \cite{Vo13,Vo14,Reu14,Reu17LMBGibbs,Wil15,Wil12,Mey18Proc,Kro19LMB,VoVoHoa:J17,BeaVoVo:J20,HouCla:J18}, but also a significantly increased 
\pagebreak 
computational complexity.

\subsection{Contribution}
\label{sec:intro_contr}

Here, we propose a multiobject tracking algorithm with track continuity, termed LMB/P filter, 
that combines the strengths of the LMB filter and the PHD filter and is inspired by the label-augmented TOMB/P filter. We model the multiobject state as a combination of an LMB RFS (i.e., a labeled RFS) and a Poisson RFS (i.e., an unlabeled RFS). Whereas in the TOMB/P filter the Poisson RFS facilitates the creation of new Bernoulli components, the proposed LMB/P filter extends the use of the Poisson RFS to the tracking of ``unlikely'' objects. Only if a quantity characterizing the plausibility of object existence is above a threshold, the LMB/P filter creates a new labeled Bernoulli component, and the corresponding object is tracked within the more accurate but less efficient LMB part. Conversely, the LMB/P filter transfers labeled Bernoulli components to the Poisson RFS if the probability of object existence falls below another threshold. The fact that unlikely objects are tracked within the more efficient Poisson part results in a large reduction of computational complexity.

\rd{Our derivation of the proposed LMB/P filter is based on a new system model for labeled/unlabeled objects in which the multiobject state is modeled by a tuple of 
a labeled RFS and an unlabeled RFS. This system model is interesting in its own right as a basis for deriving further new labeled/unlabeled multiobject tracking filters.}

\rd{The proposed LMB/P filter is rooted in the framework of Bayes-optimal multiobject tracking and employs several
approximations to achieve computational feasibility and efficiency.
Since an exact implementation of the Bayes-optimal multiobject tracking filter is computationally infeasible, certain
approximations are 
employed by all practical multiobject tracking algorithms.
For example, in the 
popular PHD filter, the posterior multiobject pdf is approximated by a Poisson pdf.
While this is a rather strong approximation, it can be motivated and justified by the fact that the PHD filter has a very low computational complexity while 
still achieving good performance in multiobject tracking scenarios of low to moderate difficulty.}

\rd{The proposed LMB/P filter employs a sequence of approximations that are considerably less severe and more sophisticated.
Our goal is to combine the strengths of the PHD and LMB filters. In fact, the LMB/P filter can be interpreted as a combination of an LMB filter and a PHD filter 
that run in parallel but not independently of each other, even though the update relations of the PHD part are different from the update relations of the original PHD filter. The derivation of our filter is based on approximating the posterior multiobject pdf by a combined LMB--Poisson pdf. To further decrease the computational complexity, we introduce certain additional approximations and modifications. More specifically, we propose a clustering scheme based on a new 
criterion in order to reduce the complexity of data association, and we employ a flexible transfer between labeled and unlabeled objects
in order to track ``unlikely'' objects with low complexity and ``likely'' objects with high accuracy. These approximations can be justified by the fact that they result in a low complexity and an excellent performance even in challenging multiobject tracking scenarios.}

This paper differs from our conference publication \cite{Kro18LMBPConf} in that it proposes an improved label and measurement partitioning scheme, which results in a lower complexity; it presents a detailed derivation of the approximations used in the update step; it provides a detailed step-by-step statement of the proposed algorithm; and it presents an improved experimental performance evaluation. Furthermore, the proposed method differs from the TOMB/P filter with recycling \cite{Wil12} in that 
it uses a labeled RFS in order to facilitate track continuity, 
it incorporates a label and measurement partitioning scheme resulting in a complexity reduction, and 
it updates the Poisson RFS based on measurements that are unlikely to originate from a labeled object.

\vspace{-1mm}

\rd{
\subsection{Paper Organization and Notation}
\label{sec:intro_org}
}

The remainder of this paper is organized as follows. After a brief review of RFSs in Section \ref{sec:fund_RFS}, \rd{Section \ref{sec:sysmod-1} presents a system model for labeled/unlabeled objects.}
The prediction step and (exact) update step are presented in Sections \ref{sec:pred_SM1} and \ref{sec:upda_SM1}, respectively. In Sections \ref{sec:first_appr_CO} and \ref{sec:second_appr_CO}, we describe the complexity-reducing approximations used in the update step of the LMB/P filter. Section \ref{sec:LMB/P} summarizes the LMB/P filter algorithm. Simulation results are presented in Section \ref{sec:Sim}.

We will use the following notation. Vectors are denoted by small boldface letters (e.g., $\st$), unlabeled finite sets by capital letters (e.g., $\RFSst$), and labeled finite sets by capital letters with a tilde (e.g., $\lRFSst$). Labeled states are denoted as $(\st,\la)$, where $\st$ is a state vector and $\la$ is a label. Randomness is indicated by a sans serif font, such as in $\bm{{\sf x}}$ or ${\sf X}$. We write probability density functions (pdfs) as $f(\cdot)$ or $s(\cdot)$ and probability mass functions (pmfs) as $p(\cdot)$. The expectation operator is denoted by $\mathrm{E}\{ \cdot \}$ and the probability by $\mathrm{Pr}\{ \cdot \}$. Integrals are over the entire space of the integration variable unless noted otherwise. The superscript ${}^{\T}$ indicates transposition, and $\mathbf{I}_N$ denotes the $N \!\times\rmv N$ identity matrix.

\vspace{3mm}

\section{Fundamentals of RFSs}
\label{sec:fund_RFS}

\subsection{Unlabeled RFSs}
\label{sec:fund_unlabeled}

An (unlabeled) RFS $\RFSstR = \{\stR^{(1)}\rmv,\ldots,\stR^{(\rv{n})}\}$ is a random variable whose realizations $X$ are finite sets $\{\st^{(1)}\rmv,\ldots,\st^{(n)}\}$ of vectors $\st^{(i)} \!\rmv\in\rmv \mathbb{R}^{n_x}\rmv$. Both the vectors $\stR^{(i)}$ and their number $\rv{n} \rmv=\rmv |\RFSstR|$ (the cardinality of $\RFSstR$) are random, and the elements $\stR^{(i)}$ are unordered.
We define $\ca(n) \triangleq \mathrm{Pr}\{|{\sf\RFSst}| \rmv = \rmv n\}$ as the pmf of the cardinality $\rv{n} \rmv=\rmv |\RFSstR|$. The \emph{set integral} $\int \!g(\RFSst) \ist \delta \RFSst$ of a real-valued set function $g(\RFSst)$ is defined as described in \cite{Mah07}.

The statistics of an RFS $\RFSstR$ can be described by the \emph{multiobject pdf} $f_{\RFSstR}(\RFSst)$, briefly denoted $f(\RFSst)$, or equivalently by the \emph{probability generating functional} 
(pgfl) \cite{Mah07}
\[
G_{\RFSstR}[h] \triangleq\rmv \int \rmv h^{\RFSst} f(\RFSst) \ist \delta\RFSst .
\vspace{-.8mm}
\]
Here, $h^{\RFSst} \triangleq \prod_{\st \in \RFSst} h(\st)$, where $h \!: \mathbb{R}^{n_x} \!\rmv\to\! [0,\infty)$ is any nonnegative vector function. 
The pgfl of the union $\RFSstR = \bigcup_{j=1}^J \rmv \RFSstR^{(j)}$ of statistically independent RFSs $\RFSstR^{(j)}\rmv$, $j \rmv\in\rmv \mathcal{J} \triangleq \{1,\ldots,J\}$ is the product of the individual pgfls $G_{\RFSstR^{(j)}}[h]$, 
i.e,
\vspace{0mm}
\begin{equation}
\label{eq:fund_pgfl_prod}
G_{\RFSstR}[h] = \prod_{j \in \mathcal{J}} G_{\RFSstR^{(j)}}[h] \ist.
\end{equation} 

The \emph{PHD} or \emph{intensity function} $\lambda_{\RFSstR}(\st) \!: \mathbb{R}^{n_x} \!\rmv\to\! [0,\infty)$ of an RFS $\RFSstR$, briefly denoted $\lambda(\st)$, 
\rd{is a first-order moment of $\RFSstR$ with the property that} for any region $\mathcal{S} \subseteq \mathbb{R}^{n_x}\rmv$, the integral $\int_{\mathcal{S}} \lambda(\st) \ist \text{d}\st$ yields the expected number of objects whose states are located in that region, i.e., $\mathrm{E} \{ |\RFSstR \cap S| \} \rmv=\! \int \rmv |\RFSst \cap S| \ist f(\RFSst)\ist \delta\RFSst$. 
The PHD can be obtained from the pgfl according to
\begin{equation}
\label{eq:PHD_PGFL}
\lambda(\st) = \frac{\delta}{\delta\st} \ist G_{\RFSstR}[h] \bigg|_{h = 1} \rmv,
\vspace{-.5mm}
\end{equation}
where $\frac{\delta}{\delta\st} \ist G_{\RFSstR}[h]$ denotes the functional derivative of $G_{\RFSstR}[h]$ \cite{Mah07}.

For a \emph{Poisson RFS}, the cardinality $\rv{n}$ is Poisson distributed with mean $\mu$, i.e., $\ca(n) \rmv= e^{-\mu}\mu^{n}/n! \ist$, $n \rmv\in\rmv \mathbb{N}_0$. \red{For each cardinality $\sf{n} \rmv=\rmv |\RFSstR|$}, the individual elements $\stR$ are independent and identically distributed (iid) with some ``spatial pdf'' $f(\st)$. The pgfl is \cite{Mah07}
\begin{equation}
\label{eq:fund1_5}
G_{\RFSstR}[h] = \rd{P[h;\lambda] \triangleq} \; e^{\lambda[h-1]},
\end{equation}
where $\lambda[h \rmv-\! 1] = \int (h(\st) \rmv-\rmv 1) \ist \intf(\st) \ist \text{d}\st$, and the PHD (intensity function) is $\intf(\st) \rmv=\rmv \mu f(\st)$.

A \emph{Bernoulli RFS} is parametrized by a probability of existence $\ex$ and a spatial pdf $\sd(\st)$. It is either empty with probability $1 \!-\rmv \ex$ or it contains one element $\stR \sim \sd(\st)$ with probability $r$. The pgfl 
\vspace{-.5mm}
is \cite{Mah07}
\begin{equation}
\label{eq:fund2_2}
G_{\RFSstR}[h] = \rd{B[h;\ex,\sd] \ist\triangleq} \; 1 \rmv-\rmv \ex + \ex \ist \sd[h] \ist,
\end{equation}
with $\sd[h] = \int\rmv h(\st) \ist \sd(\st) \ist \text{d}\st$. 
\red{A linear combination of Bernoulli pgfls is again a Bernoulli pgfl: more specifically,}
for weights $\gamma_i$ satisfying $\gamma_i \!\ge\! 0$ and $\sum_{i} \rmv \gamma_i \!=\! 1$, we have
\begin{equation}
\label{eq:fund2_lincom-Bernoulli}
\sum_i \rmv \gamma_i \ist \rd{B\big[h;\ex^{(i)}\rmv,\sd^{(i)}\big]} = \rd{B[h;\ex,\sd]} \ist,
 \vspace{-2.5mm}
 \end{equation}
where
 \vspace{-1mm}
 \begin{equation}
\label{eq:fund2_lincom-Bernoulli_rs}
r = \sum_i \rmv \gamma_i \ist r^{(i)} , \quad \sd(\st) = \frac{1}{r} \sum_i \rmv \gamma_i \ist r^{(i)} \sd^{(i)}(\st) \ist.
 \vspace{-.3mm}
 \end{equation}

An \emph{MB RFS} is the union of a fixed number $J$ of statistically independent Bernoulli RFSs $\RFSstR^{(j)}\rmv$, 
$j \!\in\! \mathcal{J}$ parametrized by possibly different probabilities of existence $\ex^{(j)}$ and spatial pdfs $\sd^{(j)}(\st)$. The pgfl is (cf.\ \eqref{eq:fund_pgfl_prod} and \eqref{eq:fund2_2}) 
\begin{equation}
\label{eq:fund2_7}
G_{\RFSstR}[h]=\rmv \rd{M_{\mathcal{J}}\big[h;\ex^{(\cdot)}\rmv,\sd^{(\cdot)}\big] \triangleq} \prod_{j\in\mathcal{J}} \! \rd{B\big[h;\ex^{(j)}\rmv,\sd^{(j)}\big]} ,
 \vspace{-1mm}
\end{equation}
where $\sd^{(j)}[h] = \int \rmv h(\st) \ist \sd^{(j)}(\st) \ist \text{d}\st$.
\rd{Here, the superscript $^{(\cdot)}$ used in $M_{\mathcal{J}}\big[h;\ex^{(\cdot)}\rmv,\sd^{(\cdot)}\big]$ indicates that $M_{\mathcal{J}}\big[h;\ex^{(\cdot)}\rmv,\sd^{(\cdot)}\big]$
involves the set of existence probabilities ${\{ \ex^{(j)} \}}_{j \in \mathcal{J}}$ and the set of spatial pdfs ${\{ \sd^{(j)}(\st) \}}_{j \in \mathcal{J}}$.}

\vspace{-1mm}

\subsection{Labeled RFSs}
\label{sec:fund_labeled}

In a labeled RFS $\lRFSstR$, each element is a tuple of the form $(\stR,\laR) \rmv \in \mathbb{R}^{n_x} \!\rmv\times\rmv \mathbb{L}$, 
where the label space $\mathbb{L}$ is a countable set.
\red{Thus, a realization of $\lRFSstR$ has the form} $\lRFSst \rmv= \big\{(\st^{(1)}\rmv,\la^{(1)}),\ldots,\linebreak 
(\st^{(n)}\rmv,\la^{(n)})\big\}$.
The set integral $\int\! g(\lRFSst) \ist \delta \lRFSst$ of a real-valued function $g(\lRFSst)$ can be defined as described in \cite{Vo13,Mah14}. Analogously to an unlabeled RFS, the statistics of a labeled RFS can be described by the multiobject pdf $f(\lRFSst)$ \cite{Mah14,Vo13,Vo14} or by the pgfl $G_{\lRFSstR}[\tilde{h}] \triangleq \int\rmv \tilde{h}^{\lRFSst} f(\lRFSst) \ist \delta\lRFSst$ \cite[p.\ 449]{Mah14}, where $\tilde{h}^{\lRFSst} \triangleq \prod_{(\st,\la) \in \lRFSst} \tilde{h}(\st,\la)$ with $\tilde{h}\!: \mathbb{R}^{n_x} \!\rmv\times\rmv \mathbb{L} \rmv\to\! [0,\infty)$.

An \emph{LMB RFS} $\lRFSstR$ is an MB RFS where for any realization $\lRFSst$ each single-vector set $\{\st\}$ corresponding to a Bernoulli component $\RFSstR^{(j)}$ is augmented by a distinct label $\la \rmv\in\rmv \mathbb{L}^{*}\rmv$. 
Here, adopting the labeling procedure of \cite{Mah14}, the \emph{same} label $\la$ is assigned to each state realization $\st$ of a given Bernoulli component $\RFSstR^{(j)}\rmv$, and $\mathbb{L}^{*} \!\rmv\subseteq\rmv \mathbb{L}$ denotes the finite set of assigned labels. To simplify the notation, we index the Bernoulli RFSs directly by their labels $\la$, i.e., they are denoted ${\sf \RFSst}^{(\la)}\rmv$, $l \!\in\! \mathbb{L}^{*}$ with corresponding existence probabilities $\ex^{(\la)}$ and spatial distributions $\sd^{(\la)}(\st)$ \cite{Reu14}. The LMB RFS $\lRFSstR$ is completely specified by the parameter set $\big\{ \big( \ex^{(\la)}\rmv,\sd^{(\la)}(\st) \big) \big\}_{\la\in\mathbb{L}^{*}}$. The pgfl is given 
\vspace{-.5mm}
by \cite{Mah14}
\begin{equation}
\label{eq:fund4_pgflLMB}
G_{\lRFSstR}[\tilde{h}] = \rd{L_{\mathbb{L^{*}}}\big[\tilde{h},\ex^{(\cdot)}\rmv,\sd^{(\cdot)}\big] \triangleq} \prod_{\la \in \mathbb{L}^{*}} \!\rmv \rd{B\big[\tilde{h};\ex^{(\la)}\rmv,\sd^{(\la)}\big]} ,
\vspace{-.7mm}
\end{equation}
with $\sd^{(\la)}[\tilde{h}] \rmv=\! \int\rmv \tilde{h}(\st,\la) \ist \sd^{(\la)}(\st) \ist \text{d}\st$ (cf.\ \eqref{eq:fund2_2}). 

An \emph{LMB mixture (LMBM) RFS} generalizes the LMB RFS in that its pgfl is a mixture of a finite number of LMB pgfls with identical label set $\mathbb{L}^{*}\rmv$, i.e.,
\begin{align*}
G_{\lRFSstR}[\tilde{h}] &= \sum_{i} \rmv \wei_i\ist \rd{L_{\mathbb{L}^{*}}\big[\tilde{h},\ex^{(\cdot,i)}\rmv,\sd^{(\cdot,i)}\big]} \nn \\[.5mm]
&= \sum_{i} \rmv \wei_i \rmv\prod_{\la \in \mathbb{L}^{*}} \!\rmv \rd{B\big[\tilde{h};\ex^{(\la,i)}\rmv,\sd^{(\la,i)}\big]} . 
\\[-6mm]
\end{align*}
Here, the weights satisfy $\wei_i \!\ge\! 0$ and $\sum_{i} \rmv \wei_i \!=\! 1$, and $\sd^{(\la,i)}[\tilde{h}] \rmv=\! \int\rmv \tilde{h}(\st,\la) \ist \sd^{(\la,i)}(\st) \ist \text{d}\st$.

\vspace{-1mm}

\section{System Model}
\label{sec:sysmod-1}

\rd{In this section, we present a new labeled/unlabeled RFS-based system model that provides statistical descriptions
of the state evolution process and the measurement process. The proposed model is valid for all types of labeled/unlabeled multiobject state RFSs;
the specific RFS type used for the multiobject state in our LMB/P filter will be described in Section \ref{sec:pred_SM1}.
The multiobject state is composed of a labeled RFS part and an unlabeled RFS part. The labeled RFS part encodes the identities of the modeled objects and thus allows these objects to be distinguished. By contrast, the objects modeled by the unlabeled RFS part are indistinguishable.} 

More specifically, the
multiobject state at time $k\!-\!1$ is constituted by the tuple $(\lRFSstR_{k-1},\RFSstR_{k-1})$ of a labeled RFS $\lRFSstR_{k-1}$  and an unlabeled RFS $\RFSstR_{k-1}$. The elements of $\lRFSstR_{k-1}$ are random tuples $(\stR_{k-1},\laR) \in \mathbb{R}^{n_x} \!\times \mathbb{L}_{k-1}^{*}$, while the elements of $\RFSstR_{k-1}^{\TO}$ are random vectors $\stR_{k-1} \!\in\! \mathbb{R}^{n_x}\rmv$. Here, $\stR_{k-1}$ typically consists of the object's position and possibly further parameters, and $\mathbb{L}_{k-1}^{*}$ is the set of labels corresponding to $\lRFSstR_{k-1}$, which is a subset of the label space $\mathbb{L}_{k-1} \rmv=\rmv \{1,\ldots,k\!-\!1\} \times \mathbb{N}$. Each label $\la \!\in\! \mathbb{L}_{k-1}$ is a tuple of the form $l \rmv=\rmv (k'\!,\nu)$, where $k' \!\rmv\in\! \{1,\ldots,k \!-\! 1\}$ represents the object's time of birth and $\nu \!\in\! \mathbb{N}$ distinguishes objects born at the same time.

\vspace{-1mm}

\subsection{State-Evolution Model}
\label{sec:SM_ST} 

\red{The state-evolution model describes the statistics of the multiobject state at time $k$, $(\lRFSstR_{k},\RFSstR_{k})$, 
for a given multiobject state at time $k-1$, $(\lRFSst_{k-1},\RFSst_{k-1})$, as detailed in what follows.}
At time $k \rmv-\! 1$, an object with labeled state $(\st_{k-1},\la) \!\in \lRFSst_{k-1}$ either survives with probability $\su(\st_{k-1},\la)$ or dies with probability $1 \!-\rmv \su(\st_{k-1},\la)$. If it survives, its new state $\stR_{k}$ (without the label $\la$) is distributed according to the 
\linebreak 
transition pdf $f(\st_k | \st_{k-1},\la)$, and the label is preserved by the state transition. 
This means that the labels of surviving objects do not change, and thus we denote them as $\la$ rather than $\la_{k}$. The states of different objects evolve independently,
i.e.,\footnote{\red{We note 
that $(\stR_k,\la)$ is short for $(\stR_k,\laR \!=\! \la)$, which denotes the state of an object with a specific (thus, deterministic) label $\la$, 
whereas $(\stR_k,\laR)$ denotes the state of an object with an arbitrary (thus, random) label 
$\laR$.}} 
$(\stR_k,\la)$ is conditionally independent, given $(\st_{k-1},\la)$, of all $(\stR'_k,\la')$ with $\la' \!\not=\rmv \la$ and also of all states $\stR_k'' \!\in\! \RFSstR_k$. Due to these assumptions, the multiobject state of the labeled objects at time $k$, given $(\lRFSst_{k-1},\RFSst_{k-1})$, is described by an LMB RFS (see\ Section \ref{sec:fund_labeled})
\[
\lRFSstR_k =\rmv \bigcup_{\la \in \mathbb{L}_{k-1}^{*}} \!\!\rmv\tilde{\RS{S}}_k(\st_{k-1},\la) \ist,
\vspace{-1mm}
\]
where $\tilde{\RS{S}}_k(\st_{k-1},\la)$ is a labeled Bernoulli RFS with existence probability $r_k^{(\la)} \rmv=\rmv \su(\st_{k-1},\la)$ and spatial pdf $\sd_k^{(\la)}(\st_{k}) \rmv=$\linebreak 
$f(\st_{k} | \st_{k-1},\la)$. Thus, $\lRFSstR_k$ is characterized by the Bernoulli parameter set $\big\{ \big( \ist\su(\st_{k-1},\la), f(\st_{k} | \st_{k-1},\la) \big) \big\}_{\la \in \mathbb{L}_{k-1}^{*}}\rmv$.

Furthermore, at time $k \rmv-\! 1$, an object with unlabeled state $\st_{k-1} \!\in\rmv \RFSst_{k-1}$ either survives with 
probability\footnote{With 
an abuse of notation, $\su(\cdot)$ is used to denote both the survival probability of labeled objects (with argument $(\st_{k-1},\la)$) and of unlabeled objects 
(with argument $\st_{k-1}$). A similar remark applies to the detection probability $\de(\cdot)$ considered in 
Section \ref{sec:MM}.} 
$\su(\st_{k-1})$ or dies with probability $1 \!-\rmv \su(\st_{k-1})$. If it survives, its new state $\stR_k$ is distributed according to the transition pdf $f(\st_k|\st_{k-1})$. The states of different unlabeled objects evolve independently, i.e., $\stR_k$ is conditionally independent, given $\st_{k-1}$, of all the other $\stR_k'$ and also of the states $(\stR_k'',\laR) \in \lRFSstR_k$. 
Accordingly, the multiobject state of the survived unlabeled objects at time $k$, given $(\lRFSst_{k-1},\RFSst_{k-1})$, is modeled as an MB RFS (see Section \ref{sec:fund_unlabeled}) $\RFSstR_k^{\text{S}} \rmv=\rmv \bigcup_{\st_{k-1} \in \RFSst_{k-1}} \!\RS{S}_k(\st_{k-1})$, where $\RS{S}_k(\st_{k-1})$ is a Bernoulli RFS with parameters $\ex_k \rmv=\rmv \su(\st_{k-1})$ and $\sd_k(\st_k) \rmv= f(\st_k|\st_{k-1})$. Thus, $\RFSstR_k^{\text{S}}$ is characterized by the Bernoulli parameter set $\big\{ \big( \ist\su(\st_{k-1}),f(\st_{k}|\st_{k-1}) \big) \big\}_{\st_{k-1} \in \RFSst_{k-1}}\rmv$.

\red{Object birth is modeled by an (unlabeled) Poisson RFS $\RFSstR_{k}^{\text{B}}\rmv$ with mean parameter $\mu_{\text{B}}$ and spatial pdf 
$f_{\text{B}}(\st_k)$ and, hence, PHD $\lambda^{\text{B}}_k(\st_k) \rmv= \mu_{\text{B}} \ist f_{\text{B}}(\st_k)$.}\footnote{In 
our system model, newborn objects may not be labeled objects. As we will explain in Section \ref{sec:upda_SM1_det}, there do exist
``new'' labeled objects, which are previously unlabeled objects that are augmented by a new distinct label and thereby are transferred from the unlabeled RFS to the labeled RFS.
Thus, this creation of new labeled objects is not modeled by a birth process as in the LMB filter \cite{Reu14}; it is considered as a part of the tracking algorithm, rather than of the system 
model.}
Thus, the entirety of unlabeled objects at time $k$, given $(\lRFSst_{k-1},\RFSst_{k-1})$, is described by the 
\vspace{-.5mm}
RFS
\[
\RFSstR_k = \RFSstR_k^{\text{S}} \cup \RFSstR_k^{\text{B}} 
  = \Bigg( \bigcup_{\st_{k-1} \in \RFSst_{k-1}} \!\!\!\RS{S}_k\big(\st_{k-1}\big) \rmv \Bigg) \cup \RFSstR_k^{\text{B}}\ist . 
\vspace{-.5mm}
\]
\rd{We assume that all newborn unlabeled object states $\stR_k \!\in\! \RFSstR_{k}^{\text{B}}$ are independent of all $\stR_k' \!\in\! \RFSstR_k$, 
all $(\stR_k'',\la) \rmv\in\rmv \lRFSstR_k$, and all measurements (see below) $\meR_k \!\in\rmv \RFSmeR_k$. Due to our above
independence assumptions,} the RFSs $\RFSstR_k^{\text{S}}$ and $\RFSstR_k^{\text{B}}$ are conditionally independent given $\big(\lRFSst_{k-1},\RFSst_{k-1}\big)$.

\vspace{-1mm}

\subsection{Measurement Model}
\label{sec:MM}

At time $k$, a sensor produces ${\sf M}_k$ measurements $\meR_k^{(1)}\rmv,\ldots,$\linebreak 
$\meR_k^{({\sf M}_k)}\rmv$, which are modeled as an (unlabeled) RFS 
$\RFSmeR_k \rmv\triangleq\rmv \big\{ \meR_k^{(1)}\rmv,$\linebreak 
$\ldots,\meR_k^{({\sf M}_k)} \big\}$.\rd{\footnote{\rd{The 
measurement model describes the statistical dependence of the random (unobserved) measurements on the 
multiobject state. Accordingly, at this point, the measurements are considered random and thus denoted as $\RFSmeR_k \rmv=\linebreak 
\big\{ \meR_k^{(1)}\rmv,\ldots,\meR_k^{({\sf M}_k)} \big\}$.
However, in the context of our tracking algorithm (see Sections \ref{sec:upda_SM1}--\ref{sec:LMB/P}), the measurements will be considered as 
deterministic (observed) and will thus be denoted as 
$\RFSme_k \rmv=\rmv \big\{ \me_k^{(1)}\rmv,\ldots,\me_k^{(M_k)} \big\}$.}}} 
The measurements may originate from a labeled object, an unlabeled object, or clutter.

A labeled object with state $(\st_{k},\la) \!\in\! \lRFSst_k$ is detected (i.e., it generates a measurement) with probability $\de(\st_k,\la)$ or is missed (i.e., it does not generate a measurement) with probability $1 \!-\rmv \de(\st_k,\la)$. In the first case, the object generates exactly one measurement $\meR_k$, which is distributed according to the likelihood function $f(\me_k | \st_k,\la)$. 
We assume that $\meR_k$ is conditionally independent, given $(\st_{k},\la)$, of all the other $\meR_k'$, all the other $(\stR_{k}',\laR') \!\in\! \lRFSstR_k$, and all the $\stR_{k}'' \!\in\! \RFSstR_k$. 
Accordingly, the measurements originating from labeled objects, given $(\lRFSst_k,\RFSst_k)$, are modeled by an MB RFS 
\rd{$\RFSmeR_k^{\text{L}} = \bigcup_{\la \in \mathbb{L}_{k-1}^{*}} \!\rmv \RS{\Theta}^{\text{L}}_k(\st_k,\la)$, 
\vspace{-0.5mm}
where $\RS{\Theta}^{\text{L}}_k(\st_k,\la)$ is a Bernoulli RFS with parameters $\ex^{(\la)}_k = \de(\st_k,\la)$ and $\sd^{(\la)}(\st_k) = f(\me_k | \st_k,\la)$. Thus, $\RFSmeR_k^{\text{L}}$ is characterized by the} Bernoulli parameter set $\big\{ \big( \ist\de(\st_k,\la),f(\me_k | \st_k,\la) \big) \big\}_{\la \in \mathbb{L}_{k-1}^{*}}\rmv$. 

An unlabeled object with state $\st_k \!\in\rmv \RFSst_k$ is detected with probability $\de(\st_k)$ or is missed with probability $1 \!-\rmv \de(\st_k)$. In the first case, it generates exactly one measurement $\meR_k$, which is distributed according to the likelihood function $f(\me_k | \st_k)$. We assume that $\meR_k$ is conditionally independent, given $\st_k$, of all the other $\meR_k'$, all the other $\stR_{k}' \!\in\! \RFSstR_k$, and all the $(\stR_{k}'',\laR) \!\in\! \lRFSstR_k$. Hence, the measurements originating from unlabeled objects, given $(\lRFSst_k,\RFSst_k)$, are modeled by an MB RFS \rd{$\RFSmeR_k^{\text{U}} \!=\rmv \bigcup_{\st_k \in \RFSst_k} \rmv \RS{\Theta}^{\text{U}}_k(\st_k)$, where $\RS{\Theta}^{\text{U}}_k(\st_k)$ is a Bernoulli RFS with parameters $\ex_k = \de(\st_k)$ and $\sd(\st_k) = f(\me_k |\st_k)$. Thus, $\RFSmeR_k^{\text{U}}$ is characterized by the} Bernoulli parameter set $\big\{ \rmv\big( \ist\de(\st_k),f(\me_k | \st_k) \big) \rmv\big\}_{\st_k \in \RFSst_k}$. 

Finally, the clutter-originated measurements are modeled by a Poisson RFS $\RFSmeR_k^{\text{C}}$ with mean parameter $\mu_\text{C}$ and spatial pdf $f_\text{C}(\me_k)$ and, hence, PHD $\lambda^{\text{C}}_k(\me_k) \rmv=\rmv \mu_\text{C} \ist f_\text{C}(\me_k)$. 
It thus follows that the overall measurement RFS at time $k$, given the multiobject state $(\lRFSst_k,\RFSst_k)$, is 
\begin{align*}
\RFSmeR_k &= \RFSmeR_k^{\text{L}} \cup \RFSmeR_k^{\text{U}} \cup \RFSmeR_k^{\text{C}} \nn \\
&= \Bigg( \bigcup_{\la \in \mathbb{L}_{k-1}^{*}} \!\!\rmv\RS{\Theta}_k^{\text{L}}(\st_k,\la) \rmv \Bigg) 
  \cup \Bigg( \bigcup_{\st_k \in \RFSst_k} \!\!\rmv\RS{\Theta}_k^{\text{U}}(\st_k) \rmv \Bigg)
  \cup \RFSmeR_k^{\text{C}} \ist. \\[-6mm]
\end{align*} 
\rd{We assume that all clutter-originated measurements $\meR_k \!\in\rmv \RFSmeR_k^{\text{C}}$ are independent of all $\meR'_k \!\in\rmv \RFSmeR_k^{\text{U}}$ 
and $\meR''_k \!\in\rmv \RFSmeR_k^{\text{L}}$ and all  $(\stR_{k},\laR) \!\in\! \lRFSstR_k$ and $\stR_{k}' \!\in\! \RFSstR_k$.} 
\rd{Due to our above independence assumptions,} the RFSs $\RFSmeR_k^{\text{L}}\ist$, $\RFSmeR_k^{\text{U}}\ist$, and $\RFSmeR_k^{\text{C}}$ 
are conditionally independent given $(\lRFSst_k,\RFSst_k)$. 
\rd{We note that equivalent independence assumptions, although possibly formulated in a different manner, underlie
many established RFS-based \cite{Mah07,Mah14} and 
other \cite{Bar11,Cha11} tracking algorithms.}

\vspace{-1mm}

\section{Prediction Step}
\label{sec:pred_SM1}

Adopting a Bayesian sequential inference framework, the fundamental quantity to be calculated recursively is the joint posterior multiobject pdf of $\lRFSstR_{k}$ and $\RFSstR_{k}$, $f(\lRFSst_k,\RFSst_k|\RFSme_{1:k})$ with $\RFSme_{1:k} \rmv\triangleq\rmv (\RFSme_1, \ldots, \RFSme_k)$, or equivalently the joint posterior pgfl $G_{\lRFSstR_{k},\RFSstR_{k}}[\tilde{h},h|\RFSme_{1:k}] \triangleq 
\linebreak 
\int\!\int \tilde{h}^{\lRFSst_{k}} h^{\RFSst_{k}} f(\lRFSst_{k},\RFSst_{k}|\RFSme_{1:k}) \, \delta\lRFSst_{k} \ist \delta\RFSst_{k}$.
\red{We make the simplifying approximation that, at the previous time $k\rmv-\!1$,} $\lRFSstR_{k-1}$ and $\RFSstR_{k-1}$ are conditionally independent given $\RFSme_{1:k-1}$, so that
\be
G_{\lRFSstR_{k-1},\RFSstR_{k-1}}[\tilde{h},h|\RFSme_{1:k-1}] = G_{\lRFSstR_{k-1}}[\tilde{h}] \, G_{\RFSstR_{k-1}}[h] \ist.
\label{eq:SM1-indep_pgfl}
\ee
(Note that in all pgfl factors and approximating pgfls, we suppress the conditions $\RFSme_{1:k-1}$ and $\RFSme_{1:k}$ for notational simplicity.) 
\red{The above factorization will be preserved automatically over time. That is, using the proposed algorithm---in particular, the approximations in the update step
described in Sections \ref{sec:first_appr_CO} and \ref{sec:second_appr_CO}---, the joint posterior pgfl will factor into a labeled part 
and an unlabeled part also at time $k$ and at all future times.}

The pgfl factors $G_{\lRFSstR_{k-1}}[\tilde{h}]$ and $G_{\RFSstR_{k-1}}[h]$ in \eqref{eq:SM1-indep_pgfl} are given as follows. We model $\lRFSstR_{k-1}$ as an LMB RFS consisting of $|\mathbb{L}_{k-1}^{*}|$ labeled Bernoulli RFSs with existence probabilities $\ex_{k-1}^{(\la)}$ and spatial pdfs $\sd_{k-1}^{(\la)}(\st_{k-1})$, $\la \!\in\rmv \mathbb{L}_{k-1}^{*}$. Here, $\mathbb{L}_{k-1}^{*} \!\subseteq\rmv \mathbb{L}_{k-1}$ is the set of labels underlying $\lRFSstR_{k-1}$. Thus, according to \eqref{eq:fund4_pgflLMB}, 
\begin{equation}
\label{eq:SM1-detobj}
G_{\lRFSstR_{k-1}}[\tilde{h}] \ist =\rmv  \prod_{\la \in \mathbb{L}_{k-1}^{*}} \!\!\! \rd{B\big[\tilde{h};\ex_{k-1}^{(\la)},\sd_{k-1}^{(\la)}\big]} ,
\vspace{-1.5mm}
\end{equation}
where $\sd_{k-1}^{(\la)}[\tilde{h}] =\rmv \int \rmv \tilde{h}(\st_{k-1},\la) \ist\sd_{k-1}^{(\la)}(\st_{k-1}) \ist\text{d}\st_{k-1}$. Furthermore, we model $\RFSstR_{k-1}$ as a Poisson RFS with PHD $\lambda_{k-1}(\st_{k-1})$. Thus, according to \eqref{eq:fund1_5}, 
\begin{equation}
\label{eq:SM1-undetobj}
G_{\RFSstR_{k-1}}[h] \ist = \rd{P[h;\lambda_{k-1}]} \ist.
\end{equation} 
Taken together, Eqs.\ \eqref{eq:SM1-indep_pgfl}--\eqref{eq:SM1-undetobj} express the fact that \rdd{all the object states}---both the labeled states, 
$(\stR_{k-1},\laR)\rmv \in \lRFSstR_{k-1}$, and the unlabeled states, $\stR_{k-1}\rmv \in\rmv \RFSstR_{k-1}$---are conditionally independent given $\RFSme_{1:k-1}$. 
A similar approximation, though formulated in a different manner, is used by many established RFS-based \cite{Mah07,Mah14} and 
other \cite{Bar11,Cha11} 
tracking algorithms.

\rd{The joint pgfl $G_{\lRFSstR_{k-1},\RFSstR_{k-1}}[\tilde{h},h|\RFSme_{1:k-1}]$ in \eqref{eq:SM1-indep_pgfl} represents the joint RFS $(\lRFSstR_{k-1},\RFSstR_{k-1})$. 
Since the elements of the labeled RFS $\lRFSstR_{k-1}$ are defined on the space $\mathbb{R}^{n_x} \!\times \mathbb{L}_{k-1}^{*}$ and the elements 
of the unlabeled RFS $\RFSstR_{k-1}$ 
on the space $\mathbb{R}^{n_x}\rmv$, the elements of $(\lRFSstR_{k-1},\RFSstR_{k-1})$ are defined on the space 
$\mathbb{R}^{n_x} \!\times \mathbb{L}_{k-1}^{*} \rmv\times \mathbb{R}^{n_x}\rmv$. 
Accordingly, in \eqref{eq:SM1-indep_pgfl}, the LMB pgfl  $G_{\lRFSstR_{k-1}}[\tilde{h}]$ (cf.\ \eqref{eq:SM1-detobj}) describes labeled object states that are defined on the space $\mathbb{R}^{n_x} \!\times \mathbb{L}_{k-1}^{*}$, and the Poisson pgfl $G_{\RFSstR_{k-1}}[h]$ (cf. \eqref{eq:SM1-undetobj}) describes unlabeled object states that are defined on the space $\mathbb{R}^{n_x}\rmv$.}

As previously stated in Section \ref{sec:sysmod-1}, the labeled state RFS, i.e, the LMB RFS $\lRFSstR_{k-1}$, allows the corresponding objects to be distinguished,
whereas the objects modeled by the unlabeled state RFS, i.e., the Poisson RFS $\RFSstR_{k-1}$, are indistinguishable.
On the other hand, the Poisson RFS is parametrized by a single function,
i.e., its PHD, and it enables a much more efficient representation and processing of a large number of potentially existing objects. 
Therefore, we will model objects that are likely to exist by the computationally more demanding LMB part and objects that are unlikely to exist by the computationally less demanding Poisson part. 
The LMB part guarantees track continuity and thereby allows the consistent tracking of distinguishable objects over consecutive time 
\linebreak 
steps.

The proposed LMB/P filter propagates the posterior pgfl $G_{\lRFSstR_k,\RFSstR_k}[\tilde{h},h|\RFSme_{1:k}]$ from one time step to the next. This consists of a prediction step and an update step. In the prediction step, the previous posterior pgfl $G_{\lRFSstR_{k-1},\RFSstR_{k-1}}[\tilde{h},h|\RFSme_{1:k-1}]$ given by \eqref{eq:SM1-indep_pgfl}--\eqref{eq:SM1-undetobj} is converted into a predicted posterior pgfl $G_{\lRFSstR_{k}\rmv,\RFSstR_{k}}[\tilde{h},h|\RFSme_{1:k-1}] \rmv\triangleq\! \int \rmv \tilde{h}^{\lRFSst_k} \rmv h^{\RFSst_k} f(\lRFSst_k\rmv,\RFSst_k|\RFSme_{1:k-1}) \ist \delta \lRFSst_k \ist \delta \RFSst_k$, 
\red{where $f(\lRFSst_k\rmv,\RFSst_k|\RFSme_{1:k-1})$ is the predicted posterior multiobject pdf.} 
This conversion involves the state-transition parameters $\su(\st_{k-1},\la)$, $f(\st_k | \st_{k-1},\la)$, $\su(\st_{k-1})$, $f(\st_k|\st_{k-1})$, and $\lambda^{\text{B}}_k(\st_k) = \mu_{\text{B}}\ist f_{\text{B}}(\st_k)$ introduced in Section \ref{sec:SM_ST}.

The derivation of the prediction step is analogous to that in \cite{Wil15} but extends it from an unlabeled to a partly labeled multiobject state. Following \cite{Wil15}, one obtains that 
the predicted posterior pgfl factors analogously to \eqref{eq:SM1-indep_pgfl}, i.e.,
\be	
G_{\lRFSstR_k,\RFSstR_k}[\tilde{h},h|\RFSme_{1:k-1}] = G_{\lRFSstR_{k}}^{\text{P}}[\tilde{h}] \, G_{\RFSstR_{k}}^{\text{P}}[h] \ist.  
   \label{eq:pred3}
\ee
Here, the factor $G_{\lRFSstR_{k}}^{\text{P}}[\tilde{h}]$ is of LMB form, i.e.,
\[
G_{\lRFSstR_{k}}^{\text{P}}[\tilde{h}] = \prod_{\la \in \mathbb{L}_{k-1}^{*}} \!\!\! \rd{B\big[\tilde{h};\ex_{k|k-1}^{(\la)},\sd_{k|k-1}^{(\la)}\big]} ,
\vspace{-2.5mm}
\]
where 
\begin{align}
\hspace{-.5mm}\ex_{k|k-1}^{(\la)} &= \ex_{k-1}^{(\la)} \int \rmv\su(\st_{k-1},\la) \ist \sd_{k-1}^{(\la)}(\st_{k-1}) \ist \text{d}\st_{k-1} \ist , \label{eq:pred7}\\[.5mm]
\hspace{-.5mm}\sd_{k|k-1}^{(\la)}(\st_k) &= \frac{\int f(\st_k | \st_{k-1},\la) \ist \su(\st_{k-1},\la) \ist \sd_{k-1}^{(\la)}(\st_{k-1}) \ist \text{d}\st_{k-1}
  }{\int \su(\st_{k-1}',\la ) \ist \sd_{k-1}^{(\la)}(\st_{k-1}') \ist \text{d} \st'_{k-1} } \ist , \nn\\[-2.5mm]
\label{eq:pred8} \\[-7.5mm]
\nn
\end{align}
for $\la \!\in\rmv \mathbb{L}_{k-1}^{*}$. We recall that $\ex_{k-1}^{(\la)}$ and $\sd_{k-1}^{(\la)}(\st_{k-1})$ are the parameters of $G_{\lRFSstR_{k-1}}[\tilde{h}]$ in \eqref{eq:SM1-detobj}. Relations \eqref{eq:pred7} and \eqref{eq:pred8} equal the prediction relations of the LMB filter \cite{Reu14}.

The other factor in \eqref{eq:pred3}, $G_{\RFSstR_{k}}^{\text{P}}[h]$, is not a Poisson pgfl anymore but a weighted Poisson pgfl \cite{Wil15}. 
Still following \cite{Wil15}, we approximate it
by the pgfl of the Poisson RFS whose PHD equals the PHD corresponding to $G_{\RFSstR_{k}}^{\text{P}}[h]$. This yields
\be
\label{eq:G_D_pred}
G_{\RFSstR_{k}}^{\text{P}}[h] \ist\approx \rd{P[h;\lambda_{k|k-1}]} \ist,
\vspace{-2.5mm}
\ee
with
\begin{align}
&\hspace{-2.3mm}\lambda_{k|k-1}(\st_{k}) \nn \\[-1mm]
&\hspace{-3.8mm}\, = \lambda^{\text{B}}_k(\st_k) +\! \int \!\rmv f(\st_k|\st_{k-1}) \ist \su(\st_{k-1}) \ist \lambda_{k-1}(\st_{k-1}) \ist \text{d}\st_{k-1} \ist. \!\! \label{eq:pred6}
\end{align}
Here, we recall that $\lambda_{k-1}(\st_{k-1})$ is the PHD corresponding to $G_{\RFSstR_{k-1}}[h]$ in \eqref{eq:SM1-undetobj} \red{ and $\lambda^{\text{B}}_k(\st_k)$ is the birth PHD modeling the birth of objects as explained in Section \ref{sec:SM_ST}.}
\rd{We note that the above Poisson pgfl approximation is also used in the prediction step of the PHD filter \cite{Mah03}, and in fact}
relation \eqref{eq:pred6} equals the prediction relation of the PHD filter \cite{Mah03}. \rd{We furthermore note that the approximation can be interpreted as the minimization of a Kullback-Leibler divergence \cite{Gar15PHDKul}.}

We conclude that when the approximation \eqref{eq:G_D_pred} is used, the prediction step preserves the LMB--Poisson form of the previous posterior pgfl $G_{\lRFSstR_{k-1},\RFSstR_{k-1}}[\tilde{h},h|\RFSme_{1:k-1}]$.

\section{Exact Update Step}
\label{sec:upda_SM1}

In the update step, the predicted posterior pgfl $G_{\lRFSstR_{k}\rmv,\RFSstR_{k}}[\tilde{h},h|\RFSme_{1:k-1}]$ is converted into the new posterior pgfl at time $k$, $G_{\lRFSstR_k,\RFSstR_k}[\tilde{h},h|\RFSme_{1:k}]$. This conversion involves the current measurement set $\RFSme_k$ as well as the measurement parameters $\de(\st_k,\la)$, $f(\me_k | \st_k,\la)$, $\de(\st_k)$, $f(\me_k | \st_k)$, 
and $\lambda^{\text{C}}_k(\me_k) = \mu_\text{C} \ist f_\text{C}(\me_k)$ introduced in Section \ref{sec:MM}. The derivation of the update step is again analogous to that in \cite{Wil15}. 
It turns out that $G_{\lRFSstR_k,\RFSstR_k}[\tilde{h},h|\RFSme_{1:k}]$ factors according to
\begin{equation}
\label{eq:up_post_pgfl_3}
G_{\lRFSstR_k,\RFSstR_k}[\tilde{h},h|\RFSme_{1:k}] \ist = \ist G'_{\lRFSstR_k,\RFSstR_k}[\tilde{h},h]\, \bar{G}_{\RFSstR_k}[h] \ist,
\end{equation}
\rdd{where the factor $G'_{\lRFSstR_k,\RFSstR_k}[\tilde{h},h]$ represents \emph{detected objects} and the factor $\bar{G}_{\RFSstR_k}[h]$ \emph{undetected objects}. Detected objects are labeled or unlabeled objects---either likely to exist or not---that generated a measurement in the current or a previous update step, while undetected objects are unlabeled objects that are unlikely to exist and did not generate a measurement in the current update step.}
Expressions of $G'_{\lRFSstR_k,\RFSstR_k}[\tilde{h},h]$ and $\bar{G}_{\RFSstR_k}[h]$ will be provided in the next two subsections.

The ``exact'' update step discussed in this section has a high complexity. We emphasize that the update step of the proposed LMB/P filter is different in that it involves several complexity-reducing modifications and approximations, to be described in Sections \ref{sec:first_appr_CO} and \ref{sec:second_appr_CO}. 

\vspace{-1mm}

\subsection{\rd{Expression of the pgfl of Detected Objects}}
\label{sec:upda_SM1_det}

\rd{Next, we will provide an expression of the pgfl of detected objects, $G'_{\lRFSstR_k,\RFSstR_k}[\tilde{h},h]$.}
Let $\mathcal{M}_k \rmv\triangleq\rmv \{1,\dots,M_k\}$ denote the set of measurement indices (cf.\ Section \ref{sec:MM}). We introduce the \rd{random}
\emph{association vector} $\assvR_k \rmv\in\rmv (\{0\} \cup \mathcal{M}_k)^{|\mathbb{L}_{k-1}^{*}|}$, whose entries $\assR_k^{(\la)}\rmv$, $\la \!\in\! \mathbb{L}_{k-1}^{*}$ are given as $\assR_{k}^{(\la)} \!\triangleq\rmv m \rmv\in\rmv \mathcal{M}_k$ if 
the labeled object with state $(\stR_k,\la)$ generates measurement $\me_k^{(m)}$ and $\assR_{k}^{(\la)} \!\triangleq\rmv 0$ 
if it does not generate a measurement. Note that in the first case, the labeled object with state $(\stR_k,\la)$ is detected, and in the second case, it is missed. We call each possible value $\assv_k$ of the association vector $\assvR_k$ an \emph{association hypothesis}, and we call $\assv_k$ \emph{admissible} if all the nonzero entries $\ass_k^{(\la)}$ are different, which implies that at most one measurement is assigned to a labeled object and no measurement is assigned to more than one labeled object. The \emph{association alphabet} $\mathcal{A}_k$ is defined as the set of all admissible $\assv_k$.

Using $\assvR_k$, a derivation analogous to \cite{Wil15} shows that $G'_{\lRFSstR_k,\RFSstR_k}[\tilde{h},h]$ is a mixture of pgfls, where each pgfl
\vspace{-1mm}
is the product of an LMB pgfl 
\rd{$L_{\mathbb{L}_{k-1}^{*}}\big[\tilde{h};\ex_k^{(\cdot,\ass_k^{(\cdot)})}\!,\sd_k^{(\cdot,\ass_k^{(\cdot)})}\big]$ (see \eqref{eq:fund4_pgflLMB})} and 
\vspace{0.2mm}
an MB pgfl 
$M_{\mathcal{M}_{\assv_k}}\big[h;\bar{\ex}_k^{(\cdot)}\!,\bar{\sd}_k^{(\cdot)}\big]$ (see \eqref{eq:fund2_7}), i.e.,
\begin{align}
G'_{\lRFSstR_k,\RFSstR_k}[\tilde{h},h] &=\! \sum_{\assv_k \in \mathcal{A}_{k}} \!\!\rmv\wei_{\assv_k} 
  \ist\rd{L_{\mathbb{L}_{k-1}^{*}}\big[\tilde{h};\ex_k^{(\cdot,\ass_k^{(\cdot)})}\!,\sd_k^{(\cdot,\ass_k^{(\cdot)})}\big]} \nn\\[-2.5mm]
  &\hspace{22mm} \rd{\times\ist M_{\mathcal{M}_{\assv_k}}\big[h;\bar{\ex}_k^{(\cdot)}\!,\bar{\sd}_k^{(\cdot)}\big]} \label{eq:upd_LMB2} \\[1.2mm] 
&=\! \rd{\sum_{\assv_k \in \mathcal{A}_{k}} \!\!\rmv\wei_{\assv_k}
  \bigg( \prod_{\la \in \mathbb{L}_{k-1}^{*}}\rrmv\rrmv B\big[\tilde{h};\ex_k^{(\la,\ass_k^{(\la)})}\!,\sd_k^{(\la,\ass_k^{(\la)})}\big] \rmv\bigg)} \nn\\[0mm]
  &\hspace{22mm} \rd{\times\!\rmv\prod_{m \in \mathcal{M}_{\assv_k}}\rrmv\rrmv B\big[h;\bar{\ex}_k^{(m)}\!,\bar{\sd}_k^{(m)}\big]} . \label{eq:upd_LMB}\\[-6mm]
\nn
\end{align}
Here, $\mathcal{M}_{\assv_k} \!\subseteq \rmv \mathcal{M}_k$ is the index set of all measurements that are not associated with any labeled object via $\assv_k \!\in\! \mathcal{A}_k$; note in particular that $\mathcal{M}_{\assv_k} \!\rmv=\rmv \emptyset$ indicates that all measurements are associated with labeled objects. 
Expressions of $\ex_k^{(\la,\ass_k^{(\la)})}\rmv$, $\sd_k^{(\la,\ass_k^{(\la)})}(\st_k)$ and $\bar{\ex}_k^{(m)}\rmv$, $\bar{\sd}_k^{(m)}(\st_k)$ will be presented shortly. 
Furthermore,
the weights $\wei_{\assv_k}$ in \eqref{eq:upd_LMB2} and \eqref{eq:upd_LMB} are given \rd{up to a normalization constant} 
\vspace{-1mm}
by
\begin{equation}
\label{eq:upd_LMB_2}
\wei_{\assv_k} \rmv\propto  \Bigg( \prod_{\la \in \mathbb{L}_{k-1}^{*}} \!\!\rmv \assw_k^{(\la,\ass_k^{(\la)})} \Bigg) \!\rmv\prod_{m \in \mathcal{M}_{\assv_k}} \!\!\rmv \assw_k^{(m)},
\vspace{-1mm}
\end{equation} 
where $\assw_k^{(\la,\ass_k^{(\la)})}$ and $\assw_k^{(m)}$  are referred to as \emph{association weights} \cite{Wil15}. 
Note that in \eqref{eq:upd_LMB}, each mixture component corresponds to one of the admissible association hypotheses $\assv_k \!\in\! \mathcal{A}_{k}$. 
\rd{The LMB pgfl $L_{\mathbb{L}_{k-1}^{*}}\big[\tilde{h};\ex_k^{(\cdot,\ass_k^{(\cdot)})}\!,\sd_k^{(\cdot,\ass_k^{(\cdot)})}\big]$ represents objects that are 
likely to exist and are either detected or undetected in the current update step, and the MB pgfl $M_{\mathcal{M}_{\assv_k}}\big[h;\bar{\ex}_k^{(\cdot)}\!,\bar{\sd}_k^{(\cdot)}\big]$
represents objects that are unlikely to exist but, nevertheless, are detected in the current update 
\vspace{-1.5mm}
step.}
 
Next, we present expressions of $\assw_k^{(\la,\ass_k^{(\la)})}\rmv$, $\ex_k^{(\la,\ass_k^{(\la)})}\rmv$, and $\sd_k^{(\la,\ass_k^{(\la)})}(\st_k)$ for $\la \!\in\! \mathbb{L}_{k-1}^{*}$ \cite{Wil15}. 
For $\ass_k^{(\la)} \!\rmv=\rmv m \rmv\in\rmv \mathcal{M}_k$, 
we have
\vspace{1mm}
\begin{align}
\assw_k^{(\la,m)} \rmv&= \ex_{k|k-1}^{(l)} \ist b_k^{(l,m)} , \label{eq:up4_1} \\[1mm]
\ex_k^{(\la,m)}\rmv &= 1 \ist, \label{eq:up4_2} \\[0mm]
\sd_k^{(\la,m)}(\st_k) & =\ist \frac{\de(\st_k,\la) \ist f\big(\me_k^{(m)} \big| \st_k,\la\big) \ist \sd_{k|k-1}^{(\la)}(\st_k)}{b_k^{(l,m)}} \ist, \label{eq:up4_3}
\end{align}
with $b_k^{(l,m)} \rmv\triangleq\rmv \int \rmv \de(\st_k,\la) \ist f\big(\me_k^{(m)} \big| \st_k,\la\big) \ist \sd_{k|k-1}^{(\la)}(\st_k) \ist\text{d}\st_k\ist$. 
Here, $\ex_{k|k-1}^{(\la)}$ and $\sd_{k|k-1}^{(\la)}(\st_k)$ were calculated in the prediction step, see \eqref{eq:pred7} and \eqref{eq:pred8}. 
Note that \eqref{eq:up4_2} indicates that the object with label $\la$ exists; its state $(\stR_k,\la)$ is distributed according to $\sd_k^{(\la,m)}(\st_k)$ in \eqref{eq:up4_3}. 
\rd{The plausibility of this event (i.e., that the object with state $(\stR_k,\la)$ exists and generates measurement $\me_k^{(m)}\rmv$) is quantified by $\assw_k^{(\la,m)}$ in \eqref{eq:up4_1}.}
On the other hand, for $\ass_k^{(\la)} \rmv\!=\rmv 0$, we 
\vspace{-1.5mm}
have
\begin{align}
\assw_k^{(\la,0)} \rmv&= 1 - \ex_{k|k-1}^{(\la)} + \ex_{k|k-1}^{(\la)} c_k^{(l)} , \label{eq:up5_1} \\[1.2mm]
\ex_k^{(\la,0)} \rmv & = \frac{\ex_{k|k-1}^{(l)} c_k^{(l)} }{\assw_k^{(\la,0)}} \ist, \label{eq:up5_2} \\[.4mm]
\sd_k^{(\la,0)}(\st_k) & =\ist \frac{\big(1\!-\rmv \de(\st_k,\la) \big) \ist \sd_{k|k-1}^{(\la)}(\st_k) }{c_k^{(l)} } \ist,\label{eq:up5_3} \\[-6mm]
\nn
\end{align}
with $c_k^{(l)} \rmv\triangleq\rmv \int \rmv\big(1\!-\rmv \de(\st_k,\la) \big) \ist \sd_{k|k-1}^{(\la)}(\st_k) \ist\text{d}\st_k\ist$. Thus, the existence of the object with label $\la$ is uncertain (as described by the existence probability $\ex_k^{(\la,0)}$ in \eqref{eq:up5_2}). 
Note that $\ex_k^{(\la,0)} \!=\rmv 0$ would indicate that the labeled object with state $(\stR_k,\la)$ does not exist and $\ex_k^{(\la,0)} \!=\! 1$ would indicate that the object exists but 
does not generate a measurement. 
If the object exists, its state $(\stR_k,\la)$ is distributed according to $\sd_k^{(\la,0)}(\st_k)$ in \eqref{eq:up5_3}. 
The plausibility of these events (i.e., that the labeled object with state $(\stR_k,\la)$ does not exist or it exists but does not generate a measurement) 
is quantified by $\assw_k^{(\la,0)}$ in \eqref{eq:up5_1}.
\rdd{Note that in the latter case, the labeled object with state $(\stR_k,\la)$ does not generate a measurement in the current update step, 
but it did generate a measurement in a previous update step.}

Finally, expressions of $\assw_k^{(m)}\rmv$, $\bar{\ex}_k^{(m)}\rmv$, and $\bar{\sd}_k^{(m)}(\st_k)$ for $m \!\in\! \mathcal{M}_k$ are given by \cite{Wil15}  
\begin{align}
\beta_{k}^{(m)} \rmv&= \lambda^{\text{C}}_k(\me_k^{(m)}) + d_k^{(m)} , \label{eq:undet3_1} \\[1mm]
\bar{\ex}_{k}^{(m)} \rmv&= \frac{d_k^{(m)}}{\beta_k^{(m)} } \ist, \label{eq:undet3_2} \\[0mm]
\bar{\sd}_{k}^{(m)}(\st_k) & =\ist \frac{\de(\st_k) \ist f\big(\me_k^{(m)} \big| \st_k\big) \ist \lambda_{k|k-1}(\st_k) }{d_k^{(m)}} \ist, \label{eq:undet3_3}
\end{align}
with $d_k^{(m)} \rmv\triangleq\rmv \int \rmv \de(\st_k) \ist f\big(\me_k^{(m)} \big| \st_k\big) \ist \lambda_{k|k-1}(\st_k) \ist\text{d}\st_k\ist$. Here, $\lambda_{k|k-1}(\st_{k})$ was calculated in the prediction step, see \eqref{eq:pred6}, 
\red{and $\lambda^{\text{C}}_k(\me_k^{(m)})$ is the clutter PHD introduced in Section \ref{sec:MM}.}
 Note that $\bar{\ex}_{k}^{(m)} \!=\! 1$ would indicate that measurement $\me_k^{(m)}\rmv$ originates from an unlabeled object; the state $\stR_k$ of that object is distributed according to  $\bar{\sd}_{k}^{(m)}(\st_k)$ in \eqref{eq:undet3_3}. On the other hand, $\bar{\ex}_{k}^{(m)} \!=\rmv 0$ would indicate that $\me_k^{(m)}$ originates from clutter.
The plausibility of this event (i.e., that measurement $\me_k^{(m)}\rmv$ originates from an unlabeled object or from clutter) is quantified by $\beta_{k}^{(m)}$ in \eqref{eq:undet3_1}.

\vspace{-1mm}

\subsection{\rd{Expression of the pgfl of Undetected Objects}} 
\label{sec:upda_SM1_undet}

It remains to provide an expression of the 
pgfl of undetected objects, $\bar{G}_{\RFSstR_k}[h]$
in \eqref{eq:up_post_pgfl_3}. 
\rdd{(Recall that an undetected object is an unlabeled object that is unlikely to exist and did not generate a measurement in the current update step.)}
A derivation analogous to \cite{Wil15} yields the Poisson pgfl (see \eqref{eq:fund1_5})
\begin{equation}
\label{eq:G_U_update}
\bar{G}_{\RFSstR_k}[h] = \rd{P[h;\lambda_k]} \ist,
\vspace{-2mm}
\end{equation}
with
\vspace{1mm}
\begin{equation}
\label{eq:upd_PHD}
\lambda_k(\st_k) \ist=\ist (1 \!-\rmv \de(\st_k)) \ist \lambda_{k|k-1}(\st_k) \ist.
\end{equation}
\rd{We note that $\bar{G}_{\RFSstR_k}[h]$ represents objects that are unlikely to exist and are also undetected.}

In summary, the exact update step transforms the predicted posterior pgfl $G_{\lRFSstR_k,\RFSstR_k}[\tilde{h},h|\RFSme_{1:k-1}]$ in \eqref{eq:pred3}, which is approximately the product of an LMB pgfl 
and a Poisson pgfl, into the new posterior pgfl $G_{\lRFSstR_k,\RFSstR_k}[\tilde{h},h|\RFSme_{1:k}]$, which, according to \eqref{eq:up_post_pgfl_3} and our discussion above, is the product of the LMB--MB mixture pgfl $G'_{\lRFSstR_k,\RFSstR_k}[\tilde{h},h]$ in \eqref{eq:upd_LMB2}, \eqref{eq:upd_LMB} and the Poisson pgfl $\bar{G}_{\RFSstR_k}[h]$ in \eqref{eq:G_U_update}. 
The exact update step also takes into account the detection of objects that are unlikely to exist. 
This is achieved by the MB pgfl $M_{\mathcal{M}_{\assv_k}}\big[h;\bar{\ex}_k^{(\cdot)}\!,\bar{\sd}_k^{(\cdot)}\big]$ involved in \eqref{eq:upd_LMB2}, which comprises one Bernoulli component for each observed measurement.

\section{Update Step of the LMB/P Filter:\\[-.6mm]First Approximation Stage}
\label{sec:first_appr_CO} 

\vspace{.7mm}

The proposed LMB/P filter is now obtained by two successive approximations of the exact update step discussed above, which result in a significant reduction of complexity. 
\rd{The first approximation stage results in a transformation of certain unlabeled objects into labeled objects. 
More concretely, to reduce the complexity of data association,
we first cluster the LMB--MB mixture pgfl $G'_{\lRFSstR_k,\RFSstR_k}[\tilde{h},h]$ in \eqref{eq:upd_LMB} into $C$
LMB--MB mixture pgfls. Then we transfer unlabeled objects that were previously unlikely to exist but satisfy a suitable threshold criterion 
to the labeled object part, which means that they are now considered as objects that are likely to exist.} 

\vspace{-1mm}

\subsection{Partitioning of Label and Measurement Sets}
\label{sec:first_appr:part}
\rd{The clustering of 
$G'_{\lRFSstR_k,\RFSstR_k}[\tilde{h},h]$ is based on a partitioning of the label set $\mathbb{L}_{k-1}^{*}$ and of the measurement 
index set $\mathcal{M}_k \rmv= \{1,\dots,M_k\}$.}
We partition the label set $\mathbb{L}_{k-1}^{*}$ into $C \rmv\in\rmv \mathbb{N}$ \rd{disjoint} subsets, i.e.,
\begin{equation}
\label{eq:upda_Ldecom}
\mathbb{L}_{k-1}^{*} = \bigcup_{c \in \mathcal{C}} \mathbb{L}_{k-1}^{(c)}\ist, 
\vspace{-.5mm}
\end{equation}
where $\mathcal{C}\! \triangleq\! \{1,\ldots,C\}$, and we partition the measurement index set $\mathcal{M}_k$ into $C + 1$ \rd{disjoint} subsets, i.e., 
\begin{equation}
\label{eq:upda_Zdecom2}
\mathcal{M}_k = \Bigg(\bigcup_{c \in \mathcal{C}} \mathcal{M}_k^{(c)} \rmv\Bigg) \cup \mathcal{M}^{\text{res}}_{k} . 
\vspace{-.5mm}
\end{equation}
Each measurement index subset $\mathcal{M}_k^{(c)}\!\rmv \subseteq\rmv \mathcal{M}_k$ is associated with a corresponding label subset $\mathbb{L}_{k-1}^{(c)}\!\rmv \subseteq\rmv \mathbb{L}_{k-1}^{*}$, whereas the residual measurement index subset $\mathcal{M}^{\text{res}}_{k} = \mathcal{M}_k \setminus\ist \bigcup_{c\in\mathcal{C}} \mathcal{M}_k^{(c)}$ is not associated with any label set.  
More specifically, the partitionings \eqref{eq:upda_Ldecom} and \eqref{eq:upda_Zdecom2} are chosen such that for any $c \!\in\! \mathcal{C}$, the association (described by $\ass_k^{(\la)}$) of an object with state $(\stR_k,\la)$, $\la \rmv\in\rmv \mathbb{L}_{k-1}^{(c)}$ with a measurement with index $m$ is plausible for $m \rmv\in\rmv \mathcal{M}_k^{(c)}$ and implausible for $m \rmv\in\rmv \mathcal{M}_k^{(c')}$ with $c' \!\neq\rmv c$. Here, the plausibility of an association is quantified by the association weight $\assw_k^{(\la,m)}$ in \eqref{eq:up4_1}. An algorithm for constructing the partitionings \eqref{eq:upda_Ldecom} and \eqref{eq:upda_Zdecom2} is presented in Appendix \ref{sec:app}. This algorithm uses a nonnegative threshold $\gamma_{\text{C}}$ that determines
$\mathbb{L}_{k-1}^{(c)}$, $\mathcal{M}_k^{(c)}\rmv$, and $\mathcal{M}_k^{\text{res}}\rmv$.

\begin{figure}[t!]
\hspace{-3mm}
\includegraphics[scale=0.97]{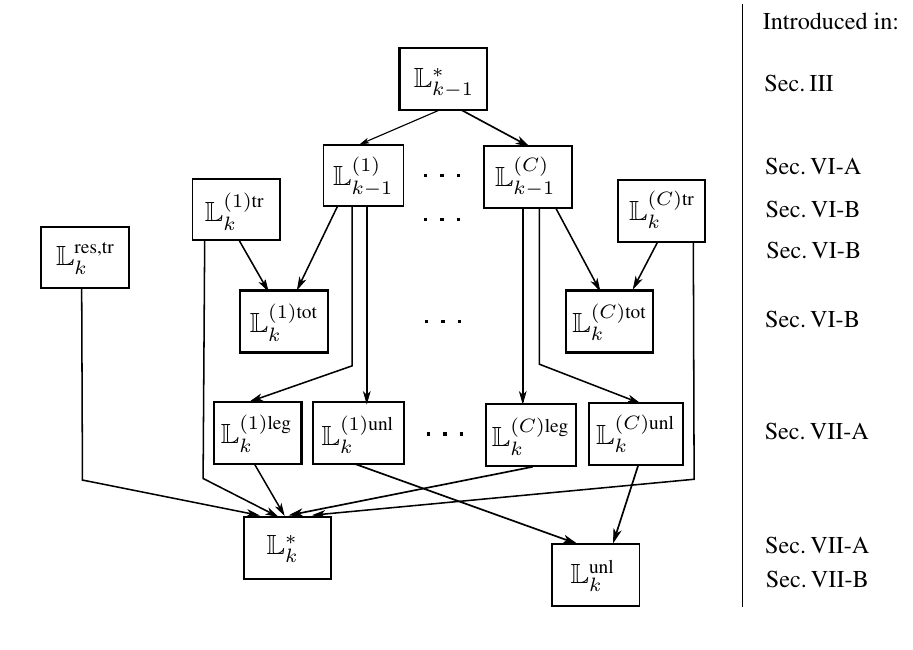}
\vspace{-7mm}
  \caption{\rdd{Overview of the} label sets 
  involved in the approximations described in Sections \ref{sec:first_appr_CO} and \ref{sec:second_appr_CO}.}
	\label{fig:Labels}
\vspace{-2.5mm}
\end{figure}

\begin{figure}[t!]
\hspace{9.5mm}
\includegraphics[scale=1]{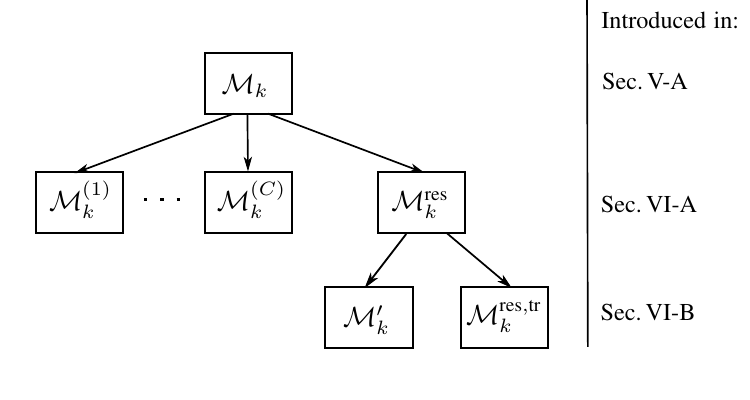}
\vspace{-3mm}
  \caption{\rdd{Overview of the} measurement index sets involved in the approximations described in Sections \ref{sec:first_appr_CO} and \ref{sec:second_appr_CO}.}
	\label{fig:Meas}
\vspace{-1.5mm}
\end{figure}

The partitionings of $\mathbb{L}_{k-1}^{*}$ and $\mathcal{M}_k$ are illustrated in Fig.~\ref{fig:Labels} and Fig.~\ref{fig:Meas}, respectively. The overall partitioning scheme is similar in spirit to the classical gating procedure used, e.g., in the joint probabilistic data association filter \cite{Bar11}. However, it is different in that it considers also the (non)existence of objects, it uses the association weights $\assw_k^{(l,m)}$ as plausibility measures, 
and it collects all the residual measurement indices in $\mathcal{M}^{\text{res}}_k\rmv$.

\subsection{\rd{Approximation of the pgfls of Detected and Undetected Objects}}
\label{sec:Stage1_approx}

\rd{Based on the label and measurement partitionings described above, we approximate the posterior pgfl $G_{\lRFSstR_k,\RFSstR_k}[\tilde{h},h|\RFSme_{1:k}]$ in \eqref{eq:up_post_pgfl_3} according to}  
\begin{equation}
\label{eq:up_post_pgfl_6}
G_{\lRFSstR_k,\RFSstR_k}[\tilde{h},h|\RFSme_{1:k}] \ist \approx \ist G'_{\lRFSstR_k}[\tilde{h}]\ist G'_{\RFSstR_k}[h] \ist,
\vspace{.5mm}
\end{equation}
\rd{where expressions of the factors $G'_{\lRFSstR_k}[\tilde{h}]$ and $G'_{\RFSstR_k}[h]$ will be provided presently. As mentioned earlier,
this approximation involves the clustering of the LMB-MB mixture pgfl $G_{\lRFSstR_k,\RFSstR_k}[\tilde{h},h|\RFSme_{1:k}]$
into $C$ LMB-MB mixture pgfls and the transfer of certain unlabeled objects to labeled objects. 
The clustering step \rdd{combined with} the pruning of implausible association hypotheses 
significantly reduces the complexity of data association. 
The transfer step implicates
that unlabeled objects that are likely to exist are now modeled by the labeled object part.
A detailed description of the clustering and transfer steps is provided in Appendix \ref{sec:appB}.}
Most of the pgfls involved in the approximations described in Sections \ref{sec:first_appr_CO} and \ref{sec:second_appr_CO} and in Appendix \ref{sec:appB} are illustrated in Fig.\ \ref{fig:PGFLs}.
\\[2mm]
\rd{
\emph{Labeled pgfl factor:\,}
The labeled pgfl factor $G'_{\lRFSstR_k}[\tilde{h}]$ in \eqref{eq:up_post_pgfl_6} represents objects that are likely to exist; it is given by} 
\begin{equation}
\label{eq:final_appr_stage1_1}
G'_{\lRFSstR_k}[\tilde{h}] \ist\triangleq L_{\mathbb{L}_k^{\text{res,tr}}}\big[\tilde{h};\bar{\ex}_k^{(\cdot)}\rmv,\bar{\sd}_k^{(\cdot)}\big] \prod_{c\ist\in\mathcal{C}} \rmv G^{(c)}[\tilde{h}] \ist.
\vspace{-.5mm}
\end{equation}
\rd{Here, according to the derivation described in Appendix \ref{sec:first_appr_part2}, 
the labeled objects represented by the LMB pgfl $L_{\mathbb{L}_k^{\text{res,tr}}}\big[\tilde{h};\bar{\ex}_k^{(\cdot)}\rmv,\bar{\sd}_k^{(\cdot)}\big]$ 
include objects that were transferred from the set of unlabeled objects.}
The label set $\mathbb{L}_k^{\text{res,tr}}$ consists of all labels $\la \rmv=\rmv (k,m)$ with $m \rmv\in\!\mathcal{M}^{\text{res,tr}}_k\rmv$, 
where $\mathcal{M}^{\text{res,tr}}_k \subseteq \mathcal{M}_k^{\text{res}}$ comprises all $m \rmv\in\rmv \mathcal{M}_k^{\text{res}}$ for which
$\bar{\ex}_k^{(m)} \!\geq\rmv \gamma_{\text{tr}}\ist$, with $\gamma_{\text{tr}}$ being a positive threshold. 
Furthermore, $\bar{\ex}_k^{(m)}$ and $\bar{\sd}^{(m)}(\st_k)$ 
are given by \eqref{eq:undet3_2} and \eqref{eq:undet3_3}, respectively.

\begin{figure}
\includegraphics[scale=1]{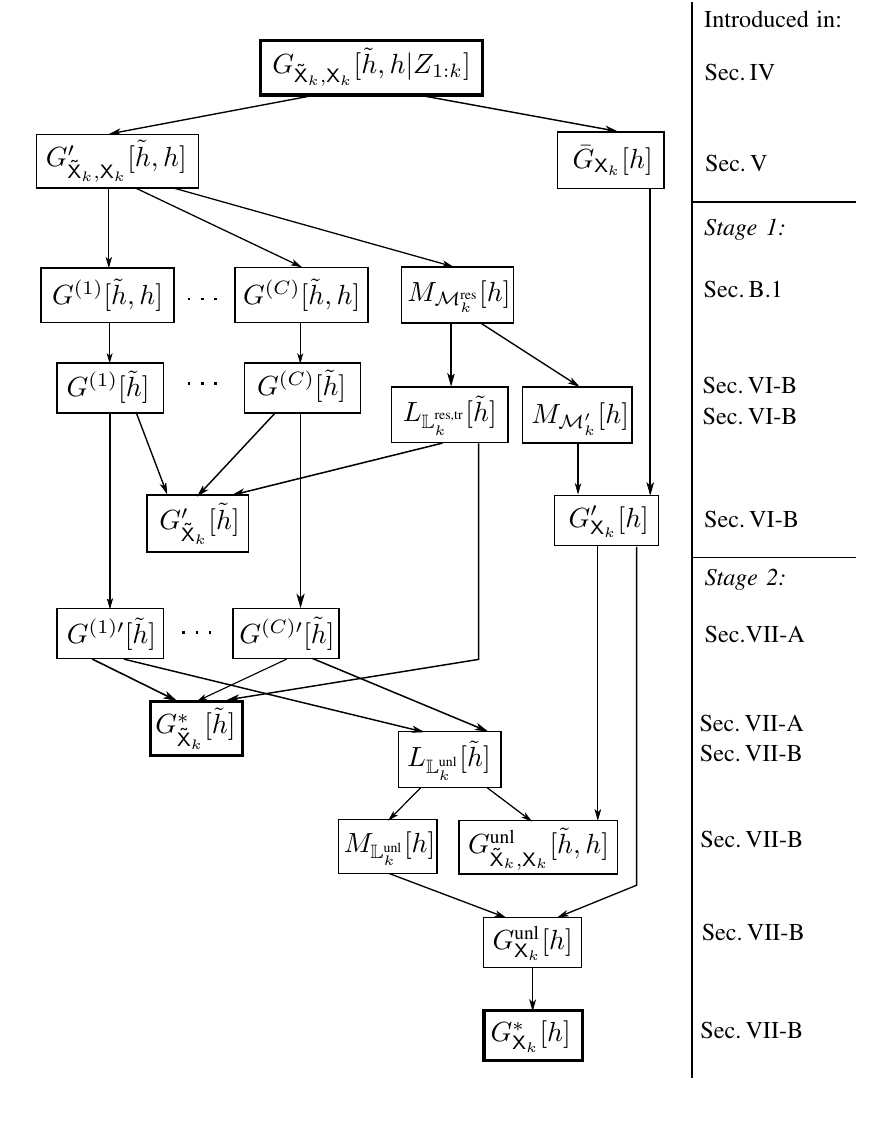}
\vspace{-6.5mm}
	\caption{\rdd{Overview of} some of the pgfls involved in the approximations described in Sections \ref{sec:first_appr_CO} and \ref{sec:second_appr_CO} and in Appendix \ref{sec:appB}. 
	\red{For simplicity of notation, we omit the existence probabilities and spatial pdfs in the pgfls; e.g., we write $M_{\mathcal{M}^{\text{res}}_k}[h]$ instead of
	$M_{\mathcal{M}^{\text{res}}_k}\big[h;\bar{\ex}_k^{(\cdot)}\!,\bar{\sd}_k^{(\cdot)}\big]$.}} 
	\label{fig:PGFLs}
\vspace{-3mm}
\end{figure}

The factors $G^{(c)}[\tilde{h}]$ \rd{in \eqref{eq:final_appr_stage1_1}, just as the factor $L_{\mathbb{L}_k^{\text{res,tr}}}\big[\tilde{h};\bar{\ex}_k^{(\cdot)}\rmv,\bar{\sd}_k^{(\cdot)}\big]$,
represent labeled objects that are likely to exist. As described in Appendix \ref{sec:approx3}, some of these objects were transferred from the set of unlabeled objects 
within the respective cluster $c$. 
The underlying clustering step, described in Appendix \ref{sec:first_appr:prun-1}, significantly reduces the complexity of data association.}
For an expression of the factors $G^{(c)}[\tilde{h}]$, we first introduce the random association vectors  
$\assvR_k^{(c)}\rmv\! \in\rmv \tilde{\mathcal{A}}_{k}^{(c)} \triangleq \big(\{0\} \cup \mathcal{M}_k^{(c)}\big)^{|\mathbb{L}_{k-1}^{(c)}|} \rmv\times \{0,1\}^{|\mathbb{L}_k^{(c)\text{tr}}|}$, where the entries $\ass_{k}^{(c,\la)}$ of a realization $\assv_k^{(c)}$ are as follows. 
For $\la \!\in\! \mathbb{L}_{k-1}^{(c)}$, $\ass_{k}^{(c,\la)}$ is defined similarly to $\ass_k^{(\la)}$ in Section \ref{sec:upda_SM1_det}
as $\ass_{k}^{(c,\la)} \!\triangleq m \rmv\in\rmv \mathcal{M}_k^{(c)}$ if the labeled object with state $(\stR_k,\la)$ generates measurement $\me_k^{(m)}$ 
and $\ass_{k}^{(c,\la)} \!\triangleq\rmv 0$ if it does not generate a measurement.
\vspace{0.2mm}
For $\la \!\in\! \mathbb{L}_k^{(c)\text{tr}}\rmv$, $\ass_{k}^{(c,\la)}$ is $1$ if 
the labeled object with state $(\stR_k,\la)$ \rd{with $\la \!=\! (k,m)$, $m \!\in\! \mathcal{M}_k^{(c)}$} generates measurement $\me_k^{(m)}$  
and $0$ if it does not generate a measurement. 
Similarly to Section \ref{sec:upda_SM1_det}, we call $\assv_k^{(c)}$ admissible if at most one measurement is assigned to a labeled object and no measurement is assigned to more than one labeled object. 
The set $\mathcal{A}_{k}^{(c)}\! \subseteq\rmv \tilde{\mathcal{A}}_{k}^{(c)}$ collects all admissible association vectors $\assv_k^{(c)}\rmv$.

The factors $G^{(c)}[\tilde{h}]$ in \eqref{eq:final_appr_stage1_1} are LMBM pgfls given by 
\begin{equation}
\label{eq:cluster_pgfl2_2}
G^{(c)}[\tilde{h}] \ist\triangleq\! \sum_{\assv_k^{(c)} \in \mathcal{A}_{k}^{(c)}} \!\!\wei_{\assv_k^{(c)}} \ist 
  \rd{L_{\mathbb{L}_k^{(c)\text{tot}}}\big[\tilde{h};\ex_k^{(\cdot,\ass_k^{(c,\cdot)})} \!, \sd_k^{(\cdot,\ass_k^{(c,\cdot)})}\big]} .
\end{equation}
Here, the label set $\mathbb{L}_{k}^{(c)\text{tot}}$ is given as 
(see Fig.~\ref{fig:Labels})
\be
\mathbb{L}_{k}^{(c)\text{tot}} \triangleq\ist \mathbb{L}_{k-1}^{(c)}\rmv \cup\ist \mathbb{L}_{k}^{(c)\text{tr}}\rmv, 
  \quad\! \text{with}\;\, \mathbb{L}_{k-1}^{(c)}\rmv \cap \mathbb{L}_{k}^{(c)\text{tr}} \!\rmv=\rmv \emptyset \ist,
\label{eq:L-decomp}
\ee
where the label set $\mathbb{L}_{k}^{(c)\text{tr}}$ consists of all labels $\la \rmv=\rmv (k,m)$ with $m \rmv\in\rmv \mathcal{M}_k^{(c)}$ such that 
$\bar{\ex}_k^{(m)} \!\geq\rmv \gamma_{\text{tr}}$\ist.
Furthermore, $\ex_k^{(\la,m)}$ and $\sd_k^{(\la,m)}(\st_k)$ are as follows. 
For $\la \!\in\! \mathbb{L}_{k-1}^{(c)}$, they are given for $m \rmv\in\rmv \mathcal{M}_k^{(c)}$ by \eqref{eq:up4_2} and \eqref{eq:up4_3}, respectively 
and for $m \!=\! 0$ by \eqref{eq:up5_2} and \eqref{eq:up5_3}, respectively.
For $\la \!\in\! \mathbb{L}_k^{(c)\text{tr}}\rmv$, $\ex_k^{(\la,1)}$ and $\sd_k^{(\la,1)}(\st_k)$ \rd{with $\la \!=\! (k,m)$, $m \!\in\! \mathcal{M}_k^{(c)}$} 
are given by \eqref{eq:undet3_2} and \eqref{eq:undet3_3}, respectively;
furthermore, $\ex_k^{(\la,0)} \!=\! 0$ whereas $\sd_k^{(\la,0)}(\st_k)$ is not defined since the corresponding object does not exist.   
Finally, the weights $\wei_{\assv_k^{(c)}}$
are given up to a normalization constant 
\vspace{-.5mm}
as 
\begin{equation}
\label{eq:upd_LMB1_approx_2}
\wei_{\assv_k^{(c)}} \rmv\propto \ist\bigg( \prod_{\la \in \mathbb{L}_{k}^{(c)\text{tot}}} \!\rmv \assw_k^{(\la,\ass_k^{(c,\la)})} \bigg) 
   \!\prod_{m \in \mathcal{M}_{\!\assv_k^{(c)}}} \!\!\rmv \assw_k^{(m)},
\vspace{-1mm}
\end{equation}
where $\mathcal{M}_{\rmv\assv_k^{(c)}} \!\subseteq\rmv \mathcal{M}_{k}^{(c)}$ comprises all $m \!\in\! \mathcal{M}_{k}^{(c)}$ 
\vspace{-1mm}
that are not associated with any object label $\la \!\in\! \mathbb{L}_k^{(c)\text{tot}}\rmv$.
For $\la \!\in\! \mathbb{L}_{k-1}^{(c)}$, the association weights $\assw_k^{(\la,m)}$ are given for $m \in \mathcal{M}_k^{(c)}$ by \eqref{eq:up4_1} and for $m = 0$ by \eqref{eq:up5_1}, 
and for $\la \!\in\! \mathbb{L}_{k}^{(c)\text{tr}}\rmv$, the $\assw_k^{(\la,m)}$ are given for $m = 1$ by \eqref{eq:undet3_1} and for $m = 0$ by $1$.
Furthermore, the $\assw_k^{(m)}$ are given by \eqref{eq:undet3_1}. \\[2mm]
\emph{\rd{Unlabeled pgfl factor:}\,}
\rd{The unlabeled pgfl factor $G'_{\RFSstR_k}[h]$ in \eqref{eq:up_post_pgfl_6} represents unlabeled objects that are unlikely to exist; it is given 
\vspace{0mm}
by}
\begin{equation}
\label{eq:final_appr_stage1_2}
G'_{\RFSstR_k}[h] \ist\triangleq M_{\mathcal{M}'_k}\big[h;\bar{\ex}_k^{(\cdot)}\!,\bar{\sd}_k^{(\cdot)}\big] \,\bar{G}_{\RFSstR_k}[h] \ist. 
 \vspace{0.5mm}
\end{equation}
\rd{Here, $\mathcal{M}'_k \triangleq \mathcal{M}^{\text{res}}_k \setminus \mathcal{M}^{\text{res,tr}}_k\rmv$, and
$\bar{\ex}_k^{(m)}$ and $\bar{\sd}_k^{(m)}(\st_k)$ are given by \eqref{eq:undet3_2} and \eqref{eq:undet3_3}, respectively.} 
Furthermore, $\bar{G}_{\RFSstR_k}[h]$ is the Poisson pgfl given by \eqref{eq:G_U_update} and \eqref{eq:upd_PHD}. Thus, $G'_{\RFSstR_k}[h]$ is an MB--Poisson pgfl.\\[2mm]
\emph{Summary of the first approximation stage:\,}
In summary, in the first approximation stage, the exact posterior pgfl \red{$G_{\lRFSstR_k,\RFSstR_k}[\tilde{h},h|\RFSme_{1:k}]$ in \eqref{eq:up_post_pgfl_3}, which is the product of the labeled/unlabeled pgfl $G'_{\lRFSstR_k,\RFSstR_k}[\tilde{h},h]$ and the unlabeled pgfl $\bar{G}_{\RFSstR_k}[h]$,} is approximated by $G'_{\lRFSstR_k}[\tilde{h}] \ist G'_{\RFSstR_k}[h]$ in \eqref{eq:up_post_pgfl_6}. 
Here, the factor $G'_{\lRFSstR_k}[\tilde{h}]$ is the pgfl of a labeled RFS \rd{representing objects that are likely to exist}. 
More specifically, it is the product of the LMB pgfl \rd{$L_{\mathbb{L}_k^{\text{res,tr}}}\big[\tilde{h};\bar{\ex}_k^{(\cdot)}\rmv,\bar{\sd}_k^{(\cdot)}\big]$} 
and the LMBM pgfls $G^{(c)}[\tilde{h}]$, $c = 1,\ldots,C$. 
The other factor, $G'_{\RFSstR_k}[h]$, is the pgfl of an unlabeled RFS representing objects that are unlikely to exist. More specifically, it is 
the product of the MB pgfl \rd{$M_{\mathcal{M}'_k}\big[h;\bar{\ex}_k^{(\cdot)}\!,\bar{\sd}_k^{(\cdot)}\big]$} and the Poisson pgfl $\bar{G}_{\RFSstR_k}[h]$. 
The effect of the first approximation stage is to reduce the overall complexity (based on the clustering described in Section \ref{sec:first_appr:part}) 
and to transfer the part of the unlabeled RFS representing likely unlabeled objects to the labeled RFS (as described 
in Appendices \ref{sec:approx3} and \ref{sec:first_appr_part2}).
Note that 
the resulting creation of new labeled objects 
is an inherent part of our tracking algorithm, 
and not due to a birth process in our system model (cf.\ Section \ref{sec:SM_ST}).

\section{Update Step of the LMB/P Filter:\\[-.3mm]Second Approximation Stage}
\label{sec:second_appr_CO}

In the second approximation stage, we approximate $G'_{\lRFSstR_k}[\tilde{h}]$ in \eqref{eq:up_post_pgfl_6} and \eqref{eq:final_appr_stage1_1}, which is the product of an LMB pgfl and $C$ LMBM pgfls, 
by an LMB pgfl. Furthermore, we modify $G'_{\RFSstR_k}[h]$ in \eqref{eq:up_post_pgfl_6} and \eqref{eq:final_appr_stage1_2}, which is the product of an MB pgfl and a Poisson pgfl. This modification consists of first combining $G'_{\RFSstR_k}[h]$ with the ``unlikely'' legacy Bernoulli components of the LMB pgfl approximating $G'_{\lRFSstR_k}[\tilde{h}]$ and then approximating the resulting pgfl by a Poisson pgfl.

\vspace{-2mm}

\subsection{Labeled Objects}
\label{sec:appr_CO} 

\rd{We first approximate the pgfl of labeled objects, $G'_{\lRFSstR_k}[\tilde{h}]$, by an LMB pgfl, and then we 
transfer labeled objects that are unlikely to exist to the unlabeled RFS part. This transfer is known as recycling \cite{Wil12}.}
 
According to \eqref{eq:final_appr_stage1_1}, \red{the pgfl of labeled objects $G'_{\lRFSstR_k}[\tilde{h}]$ is the product of the pgfl representing objects transferred from the set of unlabeled nonclustered objects, $L_{\mathbb{L}_k^{\text{res,tr}}}\big[\tilde{h};\bar{\ex}_k^{(\cdot)}\rmv,\bar{\sd}_k^{(\cdot)}\big]$, and the product of all $C$ pgfls $G^{(c)}[\tilde{h}]$ representing labeled clustered objects.}
To approximate $G'_{\lRFSstR_k}[\tilde{h}]$ by an LMB pgfl, we first note that the product of LMB pgfls is again an LMB pgfl, and that $L_{\mathbb{L}_k^{\text{res,tr}}}\big[\tilde{h};\bar{\ex}_k^{(\cdot)}\rmv,\bar{\sd}_k^{(\cdot)}\big]$ is already an LMB pgfl. Therefore, we will approximate the LMBM pgfls $G^{(c)}[\tilde{h}]$, $c \!\in\! \mathcal{C}$ by LMB pgfls. For this, we start from expression \eqref{eq:cluster_pgfl2_2} and exploit the fact that the weights $\wei_{\assv_k^{(c)}}$, $\assv^{(c)}_k \!\!\in\! \mathcal{A}_k^{(c)}\rmv$ in \eqref{eq:upd_LMB1_approx_2} satisfy $\sum_{\assv^{(c)}_k \rmv\in \mathcal{A}^{(c)}_k} \rmv\wei_{\assv^{(c)}_k} \!=\! 1$. Thus,
we are able to formally interpret these weights as the pmf of the joint association vector $\assvR^{(c)}_k\rmv$, i.e., we set
\vspace{-1mm}
\be
p\big(\assv^{(c)}_k\big) \triangleq \begin{cases} 
   \wei_{\assv^{(c)}_k} \ist, &\! \assv^{(c)}_k \!\rmv\in\! \mathcal{A}^{(c)}_k \rmv,\\[-.50mm] 
   0 \ist, &\! \text{otherwise}.
  \end{cases}
\label{eq:p-a_w-a}
\vspace{-.5mm}
\ee
Expression \eqref{eq:cluster_pgfl2_2} can then be rewritten 
\vspace{-.5mm}
as
\be
\hspace{.6mm}G^{(c)}[\tilde{h}] 
\ist=\! \sum_{\assv^{(c)}_k \in \tilde{\mathcal{A}}^{(c)}_k} \!\! p\big(\assv^{(c)}_k\big) 
  \ist \rd{L_{\mathbb{L}_k^{(c)\text{tot}}}\big[\tilde{h};\ex_k^{(\cdot,\ass_k^{(c,\cdot)})} \!, \sd_k^{(\cdot,\ass_k^{(c,\cdot)})}\big]} . \!\!
\label{eq:appr_marg1}
\vspace{-1.5mm}
\ee
Note that the summation over the larger set $\tilde{\mathcal{A}}^{(c)}_k
\!=\rmv \big(\{0\} \cup \mathcal{M}_k^{(c)}\big)^{|\mathbb{L}_{k-1}^{(c)}|} \rmv\times \{0,1\}^{|\mathbb{L}_k^{(c)\text{tr}}|}$ 
(i.e., larger than $\mathcal{A}^{(c)}_k$ in \eqref{eq:cluster_pgfl2_2}) is possible because $p\big(\assv^{(c)}_k\big) \rmv=\rmv 0$ for 
$\assv^{(c)}_k \!\rmv\in\rmv \tilde{\mathcal{A}}^{(c)}_k \!\setminus\rmv \mathcal{A}^{(c)}_k\rmv$.\\
\indent 
Following \cite{Wil15}, we now approximate $p\big(\assv^{(c)}_k\big)$ by the product of the marginal pmfs $p\big(\ass_k^{(c,\la)}\big)$, i.e., 
\[
p\big(\assv^{(c)}_k\big) \ist\approx\ist p'\big(\assv^{(c)}_k\big) \triangleq\! \prod_{\la \in \mathbb{L}^{(c)\text{tot}}_k} \! p\big(\ass_k^{(c,\la)}\big) \ist, 
  \quad \assv^{(c)}_k \!\rmv\in\! \tilde{\mathcal{A}}^{(c)}_k \rmv.
\vspace{-2.5mm}
\] 
Here,
\vspace{-1mm}
\begin{equation}
p\big(\ass_k^{(c,\la)}\big) \triangleq 
 \begin{cases} 
   \sum_{\assv_k^{(c)\sim l} \rmv\in \tilde{\mathcal{A}}_k^{(c)\text{leg}}}\ist p\big(\assv^{(c)}_k\big) \ist, &\! \la \rmv\in\rmv \mathbb{L}_{k-1}^{(c)} \ist,\\[1.5mm] 
   \sum_{\assv_k^{(c)\sim l} \rmv\in \tilde{\mathcal{A}}_k^{(c)\text{tr}}} \ist p\big(\assv^{(c)}_k\big) \ist, &\! \la \rmv\in\rmv \mathbb{L}_k^{(c)\text{tr}}
  \end{cases} \!\!\!
\label{eq:appr_marg2}
\vspace{.5mm}
\end{equation} 
(recall from \eqref{eq:L-decomp} that $\mathbb{L}^{(c)\text{tot}}_k \!= \mathbb{L}_{k-1}^{(c)} \rmv\cup\ist \mathbb{L}_{k}^{(c)\text{tr}}$), where $\assv_k^{(c)\sim l}$ denotes $\assv^{(c)}_k$ without entry $\ass_k^{(c,\la)}\rmv$, $\tilde{\mathcal{A}}_k^{(c)\text{leg}} \!\triangleq\rmv \big(\{0\} \cup \mathcal{M}^{(c)}_k\big)^{|\mathbb{L}_{k-1}^{(c)}|-1}$\linebreak 
$\times \{0,1\}^{|\mathbb{L}_{k}^{(c)\text{tr}}|}\rmv$, and $\tilde{\mathcal{A}}_k^{(c)\text{tr}} \!\triangleq\! \big(\{0\} \rmv\cup\rmv \mathcal{M}^{(c)}_k\big)^{|\mathbb{L}_{k-1}^{(c)}|} \rmv\times
\{0,1\}^{|\mathbb{L}_{k}^{(c)\text{tr}}|-1}\rmv$. 
\red{We note that an efficient and scalable approximate implementation of the marginalization in \eqref{eq:appr_marg2} is provided by the belief propagation 
algorithm proposed in \cite{Wil15}.}
Substituting $p'\big(\assv^{(c)}_k\big)$ for $p\big(\assv^{(c)}_k\big)$ in \eqref{eq:appr_marg1} and using \red{the fact that the LMB pgfl $L_{\mathbb{L}_k^{(c)\text{tot}}}\big[\tilde{h};\ex_k^{(\cdot,\ass_k^{(c,\cdot)})} \!, \sd_k^{(\cdot,\ass_k^{(c,\cdot)})}\big]$ representing all (labeled) objects within cluster $c$ is the product of all corresponding labeled Bernoulli pgfls $B\big[\tilde{h};\ex_k^{(\la,\ass_k^{(c,\la)})} \!, \sd_k^{(\la,\ass_k^{(c,\la)})}\big]$}  
(see \eqref{eq:fund4_pgflLMB}), we obtain the following approximation of 
\vspace{-.5mm}
$G^{(c)}[\tilde{h}]$:
\[
G^{(c)\prime}[\tilde{h}] \rmv\triangleq\!\sum_{\assv^{(c)}_k \in \tilde{\mathcal{A}}^{(c)}_k} \prod_{\la \in \mathbb{L}^{(c)\text{tot}}_k} \! p\big(\ass_k^{(c,\la)}\big) \ist 
  \rd{B\big[\tilde{h};\ex_k^{(\la,\ass_k^{(c,\la)})} \!, \sd_k^{(\la,\ass_k^{(c,\la)})}\big]} .
\]
Using the identities
$\prod_{\la \in \mathbb{L}^{(c)\text{tot}}_k} = \Big(\prod_{\la \in \mathbb{L}_{k-1}^{(c)}}\Big) \prod_{\la \in \mathbb{L}_{k}^{(c)\text{tr}}}$
and 
\rdd{
\begin{align*}
\sum_{\assv^{(c)}_k \in \tilde{\mathcal{A}}^{(c)}_k} &\!=\! \sum_{\ass^{(c,1)}_k \in\ist \{0\} \cup \mathcal{M}^{(c)}_k} \! \cdots\rmv \sum_{\ass^{(c,|\mathbb{L}_{k-1}^{(c)}|)}_k \rmv\in\ist \{0\} \cup \mathcal{M}^{(c)}_k} \nn \\
&\hspace{5mm}\times\!\! \sum_{\ass^{(c,|\mathbb{L}_{k-1}^{(c)}|+1)}_k \rmv\in \{0,1\}} \!\cdots\rmv \sum_{\ass^{(c,|\mathbb{L}_{k-1}^{(c)}|+|\mathbb{L}_{k}^{(c)\text{tr}}|)}_k \rmv\in \{0,1\}}\!\rmv, \\[-7mm]
\end{align*}
}
this becomes 
\begin{align*}
&G^{(c)\prime}[\tilde{h}] \nn\\[.5mm]
&\;=\rmv\Bigg(\prod_{\la \in \mathbb{L}_{k-1}^{(c)}} \sum_{\ass_k^{(c,\la)} \rmv\in \{0\} \cup \mathcal{M}^{(c)}_k} \!\!\!\rmv p\big(\ass_k^{(c,\la)}\big) \ist 
  \rd{B\big[\tilde{h};\ex_k^{(\la,\ass_k^{(c,\la)})} \!, \sd_k^{(\la,\ass_k^{(c,\la)})}\big]} 
  \!\Bigg) \nn \\[-.5mm]
&\quad\;\; \times\!\! \prod_{\la \in \mathbb{L}_{k}^{(c)\text{tr}}} \sum_{\ass_k^{(c,\la)} \rmv\in \{0,1\}} \!\!\rmv p\big(\ass_k^{(c,\la)}\big) \ist 
  \rd{B\big[\tilde{h};\ex_k^{(\la,\ass_k^{(c,\la)})} \!, \sd_k^{(\la,\ass_k^{(c,\la)})}\big]} .
\end{align*}
Using \eqref{eq:fund2_lincom-Bernoulli}, this can be written as the LMB pgfl
\be
G^{(c)\prime}[\tilde{h}] = L_{\mathbb{L}^{(c)\text{tot}}_k}\big[\tilde{h};\ex_k^{(\cdot)}\rmv,\sd_k^{(\cdot)}\big] ,
\label{eq:appr_marg3_1}
\ee
where, according to \eqref{eq:fund2_lincom-Bernoulli_rs}, $\ex_{k}^{(\la)}$ and $\sd_{k}^{(\la)}(\st_k)$ are given for $\la \!\in\! \mathbb{L}_{k-1}^{(c)}$ 
\vspace{-.5mm}
by
\begin{align}
\ex_{k}^{(\la)} &=\rmv \sum_{\ass_k^{(c,\la)}\! \in \{0\} \cup \mathcal{M}_k^{(c)}} \!\!\rmv p\big(\ass_k^{(c,\la)}\big) \ist \ex_k^{(\la,\ass_k^{(c,\la)})} , \label{eq:appr_marg4} \\[1mm]
\sd_{k}^{(\la)}(\st_k) &=\ist \frac{1}{\ex_{k}^{(\la)}} \!\rmv \sum_{\ass_k^{(c,\la)}\rmv \in \{0\} \cup \mathcal{M}_k^{(c)}} \!\!\rmv p\big(\ass_k^{(c,\la)}\big) \ist \ex_k^{(\la,\ass_k^{(c,\la)})} \sd_k^{(\la,\ass_k^{(c,\la)})}(\st_k) \ist, \nn\\[-3mm]
\label{eq:appr_marg5} \\[-7.5mm]
\nn
\end{align}
and for $\la \!\in\rmv \mathbb{L}_{k}^{(c)\text{tr}}$ by  
\vspace{-1mm}
\begin{align}
\ex_{k}^{(\la)}\rmv &= p\big(\ass_k^{(c,\la)} \!\rmv=\! 1\big) \ist \ex_k^{(\la,\ass_k^{(c,\la)} \rmv= 1)} , \label{eq:appr_marg6} \\[.5mm]
\sd_{k}^{(\la)}(\st_k) &=   \sd_k^{(\la,\ass_k^{(c,\la)} \rmv= 1)}(\st_k) \ist.
\label{eq:appr_marg7} \\[-6mm]
\nn
\end{align}
(To obtain \eqref{eq:appr_marg6} and \eqref{eq:appr_marg7}, we used the fact that $\ex_k^{(\la,\ass_k^{(c,\la)} \rmv= 0)} \!\rmv=\rmv 0$\linebreak 
for $\la \!\in\rmv \mathbb{L}_{k}^{(c)\text{tr}}\rmv$, as mentioned in Section \ref{sec:Stage1_approx}.)
Note that \eqref{eq:appr_marg4}--\eqref{eq:appr_marg7} are update equations for the labeled objects; more specifically, \eqref{eq:appr_marg4} and \eqref{eq:appr_marg5} for the legacy Bernoulli components and \eqref{eq:appr_marg6} and \eqref{eq:appr_marg7} for the transferred Bernoulli components. It can be shown that our LMB approximation of the LMBM pgfls---which is based on interpreting the weights $\wei_{\assv^{(c)}_k}$ 
\vspace{-1mm}
as the joint association pmf $p\big(\assv_k^{(c)}\big)$ and approximating that pmf by the product of its marginals---is equivalent to the LMB approximation of the LMBM pgfls that is obtained by matching the PHD of each LMB pgfl to that of the corresponding LMBM pgfl (similarly to \cite{Reu14}).

Let $\mathbb{L}_{k}^{(c)\text{leg}} \!\subseteq\rmv \mathbb{L}_{k-1}^{(c)}$ collect the labels $\la \!\in\rmv \mathbb{L}_{k-1}^{(c)}$ of those legacy Bernoulli components that are ``likely'' in the sense that  their existence probability $\ex_k^{(\la)}$ in \eqref{eq:appr_marg4} satisfies $\ex_k^{(\la)} \!\geq\rmv \gamma_{\text{leg}}$, where $\gamma_{\text{leg}}$ is another positive threshold.
The total label set of all ``likely'' legacy Bernoulli components and transferred Bernoulli components is then given by (see 
\vspace{-.5mm}
Fig.~\ref{fig:Labels})
\begin{equation}
\label{eq:labobj_final_lab}
\mathbb{L}_k^{*} \triangleq \Bigg(\bigcup_{c\in\mathcal{C}}\big(\mathbb{L}_{k}^{(c)\text{leg}} \cup\ist \mathbb{L}_{k}^{(c)\text{tr} }\ist\big) \rmv\Bigg) \rmv\cup \mathbb{L}_k^{\text{res,tr}},  
\vspace{-1mm}
\end{equation}
where $\mathbb{L}_k^{\text{res,tr}}$ was introduced in Section \ref{sec:Stage1_approx}. 
The LMB pgfl corresponding to $\mathbb{L}_k^{*}$ is now given by
\begin{align}
G_{\lRFSstR_k}^{*}[\tilde{h}] &\triangleq L_{\mathbb{L}_k^{*}}\big[\tilde{h};\ex_k^{(\cdot)}\rmv,\sd_k^{(\cdot)}\big] \nn \\[1mm]
& = L_{\mathbb{L}_k^{\text{res,tr}}}\big[\tilde{h};\bar{\ex}_k^{(\cdot)}\rmv,\bar{\sd}_k^{(\cdot)}\big] \prod_{c\ist\in\mathcal{C}} \rmv L_{\mathbb{L}_{k}^{(c)\text{leg}}\cup\ist \mathbb{L}_{k}^{(c)\text{tr}}}\big[\tilde{h};\ex_k^{(\cdot)}\rmv,\sd_k^{(\cdot)}\big] \nn \\[-3mm]
\label{eq:labobj_final2} \\[-5.5mm] \nn
\end{align}
(see Fig.~\ref{fig:PGFLs}).
According to \eqref{eq:labobj_final2}, $G_{\lRFSstR_k}^{*}[\tilde{h}]$
\vspace{-0.4mm}
equals the product of the LMB pgfl $L_{\mathbb{L}_k^{\text{res,tr}}}\big[\tilde{h};\bar{\ex}_k^{(\cdot)}\rmv,\bar{\sd}_k^{(\cdot)}\big]$ involved in \eqref{eq:final_appr_stage1_1} and the $C$ LMB pgfls obtained by restricting the LMB pgfls in \eqref{eq:appr_marg3_1} to the label sets
$\mathbb{L}_{k}^{(c)\text{leg}} \cup\ist \mathbb{L}_{k}^{(c)\text{tr}}\rmv$, for all $c \in \mathcal{C}$. 
This is our final approximation of the labeled object part, i.e., of the pgfl $G'_{\lRFSstR_k}[\tilde{h}]$ in \eqref{eq:final_appr_stage1_1}. That is, we 
\vspace{-.5mm}
have
\[
G'_{\lRFSstR_k}[\tilde{h}] \ist\approx\ist G_{\lRFSstR_k}^{*}[\tilde{h}] \ist.
\]

The ``unlikely'' legacy Bernoulli components correspond to the labels $\la \!\in\rmv \mathbb{L}_{k-1}^{(c)}$ with $\ex_k^{(\la)} \!\rmv<\rmv \gamma_{\text{leg}}$, or equivalently
$\la \!\in\rmv \mathbb{L}_{k}^{(c)\text{unl}}$\linebreak 
$\triangleq \mathbb{L}_{k-1}^{(c)} \rmv\setminus\rmv \mathbb{L}_{k}^{(c)\text{leg}}$. Instead of discarding them, as is done, e.g., in the LMB filter \cite{Reu14}, we use recycling \cite{Wil12}, i.e., we transfer them to the unlabeled RFS part. As a consequence, these unlikely objects are still being tracked but with a smaller computational cost.  
A higher threshold $\gamma_{\text{leg}}$ tends to imply that fewer Bernoulli components remain in the labeled RFS part and more are transferred to the unlabeled RFS part. 
\red{In particular, when many measurements are missing 
(due to, e.g., object death or object occlusion), then $\ex_k^{(\la)}$ is decreased, and if $\ex_k^{(\la)} \!<\rmv \gamma_{\text{leg}}$,
then the corresponding labeled Bernoulli component will be transferred to the unlabeled RFS part.}
We note that the Bernoulli components transferred to the unlabeled RFS part comprise only \emph{legacy} Bernoulli components and do not include Bernoulli components that were transferred from the unlabeled RFS part to the labeled RFS part in the current time step.
This is due to the fact that the corresponding label sets $\mathbb{L}_{k-1}^{(c)}$ and $\mathbb{L}_k^{(c)\text{tr}}$ are disjoint (cf.\ \eqref{eq:L-decomp}) and, thus, Bernoulli components that were transferred from the unlabeled RFS part to the labeled RFS part are not transferred back in the current time step.

\vspace{-1mm}

\subsection{Unlabeled Objects}
\label{sec:fappr_undobj}

We proceed by representing unlabeled and currently labeled objects that are unlikely to exist by a Poisson RFS. 
Compared to our previous use of an LMB RFS to represent objects that are likely to exist, using
a Poisson RFS reduces the computational complexity at the expense of a decreased tracking accuracy and the loss of track continuity for the respective
objects.

Consider the unlikely legacy objects
defined by the label set (see 
\vspace{-.5mm}
Fig.~\ref{fig:Labels}) 
\begin{equation}
\label{eq:L_bar}
\mathbb{L}_k^{\text{unl}} \triangleq \bigcup_{c \in \mathcal{C}} \mathbb{L}_{k}^{(c)\text{unl}}.
\vspace{-.5mm}
\end{equation} 
The labeled pgfl comprising the corresponding Bernoulli components is given by \rd{$L_{\mathbb{L}_k^{\text{unl}}}\big[\tilde{h};\ex_k^{(\cdot)},\sd_k^{(\cdot)}\big]$} 
(see Fig.~\ref{fig:PGFLs}).
We now combine this labeled pgfl with the unlabeled pgfl $G'_{\RFSstR_k}[h]$ in \eqref{eq:final_appr_stage1_2} by defining 
\be
\vspace{1mm}
G^{\text{unl}}_{\lRFSstR_k,\RFSstR_k}[\tilde{h},h] \ist \triangleq\ist \rd{L_{\mathbb{L}_k^{\text{unl}}}\big[\tilde{h};\ex_k^{(\cdot)},\sd_k^{(\cdot)}\big]}
  \ist G'_{\RFSstR_k}[h] \ist.
\label{eq:up_post_pgfl_6a} 
\vspace{-.5mm}
\ee 
We recall that $G'_{\RFSstR_k}[h]$ is the product of an MB pgfl and a Poisson pgfl (see \eqref{eq:final_appr_stage1_2}), and it represents unlabeled objects that are unlikely. Thus, the LMB--MB--Poisson pgfl $G^{\text{unl}}_{\lRFSstR_k,\RFSstR_k}[\tilde{h},h]$ represents the labeled and unlabeled objects that are unlikely.

To further reduce the complexity of the update step, we next approximate $G^{\text{unl}}_{\lRFSstR_k,\RFSstR_k}[\tilde{h},h]$ by a Poisson pgfl, i.e.\  
\vspace{-.5mm}
(see Fig.~\ref{fig:PGFLs})
\begin{equation}
\label{eq:appr_Poisson}
G^{\text{unl}}_{\lRFSstR_k,\RFSstR_k}[\tilde{h},h] \ist\approx\ist G_{\RFSstR_k}^{*}[h] \ist\triangleq \rd{P[h;\lambda^{*}_{k}]} \ist.
\end{equation}
To find the PHD $\lambda^{*}_k(\st_k)$, we first ``unlabel'' the LMB pgfl $\rd{L_{\mathbb{L}_k^{\text{unl}}}\big[\tilde{h};\ex_k^{(\cdot)},\sd_k^{(\cdot)}\big]}$. 
This results in the MB pgfl $\rd{M_{\mathbb{L}_k^{\text{unl}}}\big[h;\ex_k^{(\cdot)}\rmv,\sd_k^{(\cdot)}\big]}$, wherein $\la \!\in\rmv \mathbb{L}^{\text{unl}}_k$ is used solely to index the Bernoulli components, and not as the label of a labeled state $(\stR_k,\la)$. Through this unlabeling, the mixed labeled/unlabeled (LMB--MB--Poisson) pgfl $G^{\text{unl}}_{\lRFSstR_k,\RFSstR_k}[\tilde{h},h]$ in \eqref{eq:up_post_pgfl_6a} is converted into the unlabeled (MB--Poisson) 
\vspace{-1mm}
pgfl 
\begin{align*}
\vspace{1mm}
G^{\text{unl}}_{\RFSstR_k}[h] &\triangleq \rd{M_{\mathbb{L}_k^{\text{unl}}}\big[h;\ex_k^{(\cdot)}\rmv,\sd_k^{(\cdot)}\big]} \ist G'_{\RFSstR_k}[h] \ist. 
\end{align*}
The PHD $\lambda^{*}_k(\st_k)$ in \eqref{eq:appr_Poisson} is now chosen as the PHD corresponding to $G^{\text{unl}}_{\RFSstR_k}[h]$. That is, invoking \eqref{eq:PHD_PGFL}, we set $\lambda^{*}_k(\st_k) = \delta G^{\text{unl}}_{\RFSstR_k}[h]/\delta \st_k\big|_{h=1}$. 
Using \eqref{eq:final_appr_stage1_2}, \eqref{eq:G_U_update}, and \eqref{eq:upd_PHD}, this can be shown to 
yield
\begin{align}
\lambda^{*}_k(\st_k) 
&=\rmv \sum_{\la \in \mathbb{L}^{\text{unl}}_{k}} \! \ex_k^{(\la)} \rmv\sd_k^{(\la)}(\st_k) 
   +\! \sum_{m \in \mathcal{M}_{k}'} \!\!\rmv \bar{\ex}_k^{(m)} \bar{\sd}_k^{(m)}(\st_k)\nn \\[0mm] 
&\hspace{20mm} + \big(1 \!-\rmv \de(\st_k)\big) \lambda_{k|k-1}(\st_k) \ist,  
\label{eq:appr_TO}
\end{align}  
where $\ex_{k}^{(\la)}$ and $\sd_{k}^{(\la)}(\st_k)$ are given by \eqref{eq:appr_marg4} and \eqref{eq:appr_marg5}, respectively, $\bar{\ex}_k^{(m)}$ and $\bar{\sd}_k^{(m)}(\st_k)$  are given by \eqref{eq:undet3_2} and \eqref{eq:undet3_3}, respectively, and $\lambda_{k|k-1}(\st_{k})$ is given by \eqref{eq:pred6}. The first term in \eqref{eq:appr_TO}, $\sum_{\la \in \mathbb{L}^{\text{unl}}_{k}} \ex_k^{(\la)} \rmv\sd_k^{(\la)}(\st_k)$, corresponds to originally labeled objects that are unlikely---either because the objects already disappeared or because no measurement was associated with them for some time. The second term, $\sum_{m \in \mathcal{M}_k'} \!\bar{\ex}_k^{(m)} \bar{\sd}_k^{(m)}(\st_k)$, corresponds to measurements that are not likely to originate from any labeled objects. The third term, $\big(1 \!-\rmv \de(\st_k)\big)\lambda_{k|k-1}(\st_k)$, corresponds to unlabeled objects that are undetected. The Poisson pgfl $G^{*}_{\RFSstR_k}[h]$ defined in \eqref{eq:appr_Poisson} is our final approximation of the unlabeled object part.

\section{The Proposed LMB/P Filter}
\label{sec:LMB/P}

The core of the proposed LMB/P filter algorithm is the approximate update step developed in Sections \ref{sec:first_appr_CO} and \ref{sec:second_appr_CO}. 
\red{We recall that this approximate update step transforms the predicted posterior pgfl $G_{\lRFSstR_k,\RFSstR_k}[\tilde{h},h|\RFSme_{1:k-1}]$, 
which according to \eqref{eq:pred3} is the product of the labeled pgfl $G_{\lRFSstR_{k}}^{\text{P}}[\tilde{h}]$ and the unlabeled pgfl $G_{\RFSstR_{k}}^{\text{P}}[h]$}, 
into the following approximation of the new posterior pgfl $G_{\lRFSstR_k,\RFSstR_k}[\tilde{h},h|\RFSme_{1:k}]$ in \eqref{eq:up_post_pgfl_3}:
\[
G_{\lRFSstR_k,\RFSstR_k}[\tilde{h},h|\RFSme_{1:k}] \approx \, G^{*}_{\lRFSstR_k}[\tilde{h}] \ist G^{*}_{\RFSstR_k}[h] \ist.
\]
This is the product of the LMB pgfl $G^{*}_{\lRFSstR_k}[\tilde{h}]$, which is given by \eqref{eq:labobj_final2} and \eqref{eq:appr_marg4}--\eqref{eq:appr_marg7}, and the Poisson pgfl $G^{*}_{\RFSstR_k}[h]$, which is given by \eqref{eq:appr_Poisson} and \eqref{eq:appr_TO}. The update relations are \eqref{eq:appr_marg4}--\eqref{eq:appr_marg7} for the LMB parameters (existence probabilities and spatial pdfs) and \eqref{eq:appr_TO} for the Poisson parameter (PHD). 

\rd{These update relations can be viewed as those of an LMB filter and a PHD filter that run in parallel but not independently of each other. 
The LMB part models objects that are likely to exist and uses in the update step measurements that are likely (plausible) to originate from these objects. 
It maintains track continuity of the modeled objects and offers a better
tracking accuracy than
the Poisson part. 
The Poisson part, on the other hand, models objects that are unlikely to exist, and it uses in the update step all those measurements that are unlikely (implausible) 
to originate from a labeled object and thus
likely to originate from an unlabeled object or from clutter. 
Each measurement is used only once in the update step, either by the LMB part or by the Poisson part.
The overall approximate update step includes transfers between the labeled and unlabeled RFS parts. 
That is, based on newly observed measurements, some objects that were previously considered unlikely to exist are considered likely to exist and vice versa.}
These transfers are controlled by the thresholds $\gamma_{\text{C}}$, $\gamma_{\text{tr}}$, and $\gamma_{\text{leg}}$.

The proposed LMB/P filter algorithm is finally obtained by cascading the prediction step (Section \ref{sec:pred_SM1}) and the approximate update step (Sections \ref{sec:first_appr_CO} and \ref{sec:second_appr_CO}), and by adding a detection-estimation step. Since the unlabeled RFS part represents objects that are unlikely to exist, object detection and state estimation are based solely on the labeled RFS part. An object with label $\la \!\in\rmv \mathbb{L}_k^{*}$ is detected---i.e., declared to exist---if its existence probability $\ex_k^{(\la)}$ 
is larger than a positive detection threshold $\gamma_{\text{D}}$; the label $\la$ is then included in the ``detected label set'' $\mathbb{L}_k^{\text{D}} \rmv\subseteq\rmv \mathbb{L}_k^{*}$. Subsequently, for each detected object $\la \rmv\in\rmv \mathbb{L}_k^{\text{D}}$, a state estimate is calculated according to 
\begin{equation}
\label{eq:state_est}
\hat{\st}_k^{(\la)} =\rmv \int \rmv \st_k \ist \sd_k^{(\la)}(\st_k) \ist \text{d}\st_k \ist.
\end{equation} 
Table \ref{tab:alg} summarizes the proposed LMB/P filter algorithm.

  \begin{table}[t!]
\vspace{2.5mm}
      \small    
      {\hrule height .5pt} 
\vspace{1.1mm}
      \caption{{\ist}Proposed LMB/P filter algorithm---recursion at time $k \rmv\ge\! 1$} 
	  \label{tab:alg}
      \vspace{-1.5mm}

      {\hrule height .5pt} 
  
  
      \vspace{1.5mm}

      \textbf{Input:}\, Previous existence probabilities $\ex_{k-1}^{(\la)}$ and previous spatial pdfs $\sd_{k-1}^{(\la)}(\st_{k-1})$ for $\la \!\in\rmv \mathbb{L}_{k-1}^{*}$;
      previous PHD $\lambda_{k-1}(\st_{k-1})$ (in practice, this is replaced by the previously calculated approximation $\lambda^{*}_{k-1}(\st_{k-1})$);
measurements $\me_k^{(m)}$ for $m \!\in\! \mathcal{M}_k$.

      \vspace{1.5mm}

      \textbf{Output:}\, Existence probabilities $\ex_{k}^{(\la)}$ and spatial pdfs $\sd_{k}^{(\la)}(\st_{k})$ for $\la \!\in\rmv \mathbb{L}_{k}^{*}$;
      approximate PHD $\lambda^{*}_{k}(\st_k)$;
object state estimates $\hat{\st}_k^{(\la)}$ for $\la \!\in\! \mathbb{L}_k^{\text{D}}$. 
 
      \vspace{1.5mm}

      \textbf{Operations:}\,

      \vspace{1.5mm}
      
      \emph{Step 1 -- Prediction}: 
      
	  \begin{enumerate} 
	    \vspace{.8mm}
	    
	    \item[1.1)] 
		For $\la \!\in\rmv \mathbb{L}_{k-1}^{*}$, calculate the predicted existence probabilities $\ex_{k|k-1}^{(\la)}$ and the predicted spatial pdfs 
		$\sd_{k|k-1}^{(\la)}(\st_k)$ according to \eqref{eq:pred7} and \eqref{eq:pred8}, respectively.

		\vspace{1mm}
 
 	    \item[1.2)] 
		Calculate the predicted posterior PHD $\lambda_{k|k-1}(\st_k)$ according to \eqref{eq:pred6}.
		
	  \vspace{1.2mm}

	  \end{enumerate} 
      
      \emph{Step 2 -- Preparations for Update}:\, 

	  \begin{enumerate} 
	    \vspace{.5mm}
			
	    \item[2.1)] 
			For $\la \!\in\rmv \mathbb{L}_{k-1}^{*}$, calculate the association weights $\assw_k^{(\la,m)}\rmv$, existence probabilities $\ex_k^{(\la,m)}\rmv$, and spatial
			pdfs $\sd_k^{(\la,m)}(\st_k)$ according to \eqref{eq:up4_1}--\eqref{eq:up4_3} (for $m \!\in\! \mathcal{M}_k$)
			or \eqref{eq:up5_1}--\eqref{eq:up5_3} (for $m \!=\! 0$). 

	    \vspace{1mm}

			\item[2.2)] For $m \in \mathcal{M}_k$, calculate $\assw_k^{(m)}\rmv$, $\bar{\ex}_k^{(m)}\rmv$, and $\bar{\sd}_k^{(m)}(\st_k)$ 
			according to \eqref{eq:undet3_1}--\eqref{eq:undet3_3}.

	    \vspace{1mm}
	    
	    \item[2.3)] 
			Partition the label set $\mathbb{L}_{k-1}^{*}$ and the measurement index set $\mathcal{M}_k$ as described 
			in Section \ref{sec:first_appr:part}. This yields $\mathbb{L}_{k-1}^{(c)}$ and $\mathcal{M}_k^{(c)}$ for $c \rmv\in\rmv \mathcal{C}$ as well as $\mathcal{M}^{\text{res}}_k$.

			\vspace{1mm}
	    
	    \item[2.4)]
			Determine $\mathbb{L}_k^{(c)\text{tr}}$ for $c \rmv\in\rmv \mathcal{C}$,
			$\mathbb{L}_k^{\text{res,tr}}$ (corresponding to $\mathcal{M}^{\text{res},\text{tr}}_k$), and $\mathcal{M}'_k$ as described in 
			Section \ref{sec:Stage1_approx}.

	    \vspace{1.2mm}
	\end{enumerate}
	
      \emph{Step 3 -- Update for Labeled Objects}:\, 
			
			 \begin{enumerate} 
	    \vspace{.5mm}			
	   
	    \item[3.1)]
	    For $c \rmv\in\rmv \mathcal{C}$, calculate the weights $\wei_{\assv_k^{(c)}}$ 
	    \vspace{-.7mm}
	    according to \eqref{eq:upd_LMB1_approx_2} and the joint association pmf $p\big(\assv^{(c)}_k\big)$ according to \eqref{eq:p-a_w-a}.

 	    \vspace{1mm}

	    \item[3.2)] 
			For $c \!\in\! \mathcal{C}$ and $\la \!\in\rmv \mathbb{L}_k^{(c)\text{tot}} \!\rmv=\rmv \mathbb{L}_{k-1}^{(c)} \rmv\cup \mathbb{L}_k^{(c)\text{tr}}\rmv$, 
			calculate the marginal\linebreak 
			association pmf $p\big(\ass_k^{(c,l)}\big)$ according to \eqref{eq:appr_marg2}. 
			\red{(An efficient and scalable belief propagation algorithm for computing the $p\big(\ass_k^{(c,l)}\big)$ is presented in \cite{Wil15}.)}
	    
	    \vspace{1mm}
	    
	    \item[3.3)] 
			For $c \rmv\in\rmv \mathcal{C}$, calculate the updated existence probabilities $\ex_k^{(\la)}$ and spatial pdfs $\sd_k^{(\la)}(\st_k)$ according to \eqref{eq:appr_marg4} and \eqref{eq:appr_marg5}
			(for $\la \!\in\rmv \mathbb{L}_{k-1}^{(c)}$) or \eqref{eq:appr_marg6} and \eqref{eq:appr_marg7} (for $\la \!\in\rmv \mathbb{L}_{k}^{(c)\text{tr}}$).
	    
	    \vspace{1mm}
	    
	    \item[3.4)] 
			For $c \rmv\in\rmv \mathcal{C}$, determine $\mathbb{L}_k^{(c)\text{leg}}$ and $\mathbb{L}_k^{(c)\text{unl}}$ as described in Section \ref{sec:appr_CO}.

			\vspace{1mm}
	    
	    \item[3.5)]
			Determine $\mathbb{L}_k^{*}$ according to \eqref{eq:labobj_final_lab} and $\mathbb{L}^{\text{unl}}_k$ according to \eqref{eq:L_bar}.
		    
	    \vspace{1.2mm}
		\end{enumerate}
		
      \emph{Step 4 -- Update for Unlabeled Objects}:\, Calculate the approximate updated posterior PHD $\lambda^{*}_k(\st_k)$ according to \eqref{eq:appr_TO}.

	    \vspace{1.8mm}

      \emph{Step 5 -- Object Detection and State Estimation}: 
		
			 \begin{enumerate} 

	    \vspace{.5mm}			
	   
	    \item[5.1)]
			Determine $\mathbb{L}_k^{\text{D}}$ as described in Section \ref{sec:LMB/P}.

			\vspace{1mm}

	    \item[5.2)] 
			For $\la \rmv\in\rmv \mathbb{L}_k^{\text{D}}$, calculate an object state estimate $\hat{\st}_k^{(\la)}$
			according to \eqref{eq:state_est}.

	         \vspace{1mm}
	         
	         	\end{enumerate}

    \textbf{Initialization at time $k \!=\! 0$:}\,
	$\mathbb{L}_0^{*} \!=\! \emptyset$, $\lambda_0(\st_0)$.
					
\vspace{1.5mm}
{\hrule height .5pt}
\label{tab:summary}
\vspace{-3mm}
  \end{table}

\section{Simulation Study}
\label{sec:Sim}

\subsection{Simulation Setup}
\label{sec:sim_setup}

\begin{figure}[t!]
\vspace*{-1mm}
\hspace{-7mm}
\centering
\footnotesize
\begin{minipage}[H!]{0.25\textwidth}
\vspace{0mm}
 \includegraphics[scale=0.7]{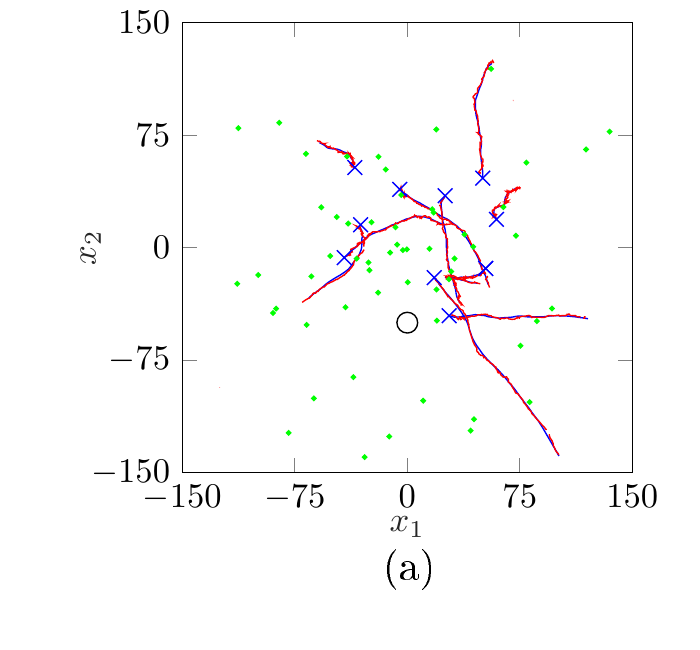}
\end{minipage}
\begin{minipage}[H!]{0.25\textwidth}
\vspace{0mm}
 \includegraphics[scale=0.7]{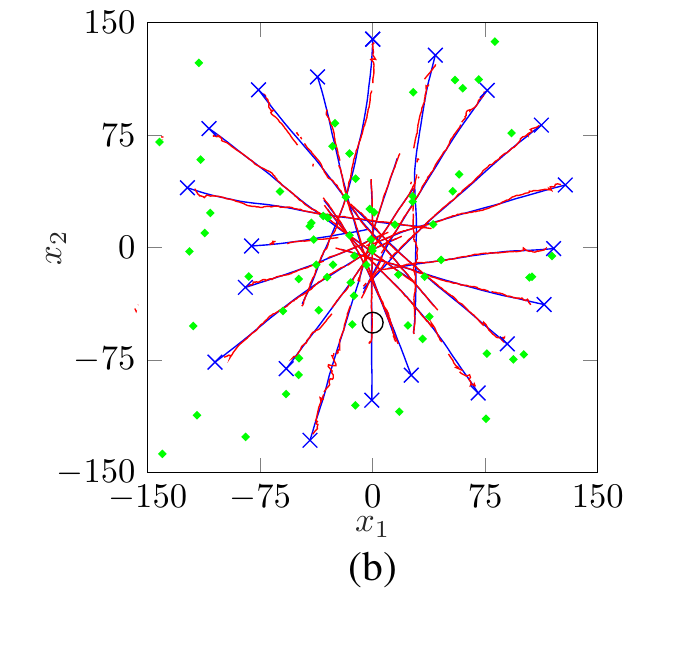}
\end{minipage}
\vspace{-4mm}
\caption{Examples of the true object trajectories (represented by blue lines, starting positions indicated by blue crosses)
and the corresponding estimates obtained with the proposed BP-LMB/P filter (represented by red lines) for (a) TS1 and (b) TS2.
The position of the sensor is indicated by a black circle. The green circles show the measurements of the sensor at time $k \!=\! 100$ within the region 
$[-150,150]\rmv\times\rmv[-150,150]$.}
\label{fig:setup} 
\vspace{-4mm}
\end{figure}

We evaluate the performance of the proposed LMB/P filter in two two-dimensional (2D) tracking scenarios, termed TS1 and TS2. In TS1, ten objects appear at randomly chosen positions in the region of interest (ROI) before time $k \!=\! 40$ and disappear after $k \!=\! 150$. In TS2, 20 objects appear before $k \!=\! 100$ and disappear after $k \!=\! 140$; they conform to the object generation scheme of \cite{Mey17MSBP}, according to which all objects move toward the point $(0,0)$ and simultaneously come in close proximity around that point at $k \!=\! 120$. The object states consist of 2D position and velocity, i.e, $\stR_k \!= [\rv{x}_{1,k} \,\ist \rv{x}_{2,k} \,\ist \dot{\rv{x}}_{1,k} \,\ist \dot{\rv{x}}_{2,k} ]^{\T}\rmv$. They evolve according to the nearly constant velocity motion model, i.e., $\stR_k \rmv=\rmv \bd{A} \stR_{k-1} + \bd{W} \RV{u}_k$, where $\bd{A} \!\in\! \mathbb{R}^{4\times 4}$ and $\bd{W} \!\in\! \mathbb{R}^{4\times 2}$ are chosen as in \cite[Sec. 6.3.2]{Bar02} and $\RV{u}_k$
is an iid sequence of 2D zero-mean Gaussian random vectors with independent components and component variance $\sigma^2_u \!=\rmv 10^{-4}\rmv$. The sensor is located at position $\V{p} \rmv=\rmv [p_1\;p_2]^{\text{T}} \!=\rmv [0\;\,{-50}]^{\bd{\text{T}}}$ and has a measurement range of 300. 
The ROI is equal to the disk determined by the sensor's measurement range. 
\red{Realizations of the object trajectories for TS1 and TS2 are shown in Fig.~\ref{fig:setup}.}

The object-originated measurements conform to the nonlinear range-bearing model $\meR_{k} \rmv=\rmv \big[ \rho(\stR_k) \;\ist \theta(\stR_k) \big]^{\T} \!+ \RV{v}_{k}$. Here, $\rho(\stR_k) \rmv\triangleq\rmv \| \RV{x}'_k \!-\rmv\V{p} \ist\|$, where $\RV{x}'_k \!\triangleq\rmv [ \rv{x}_{1,k} \,\, \rv{x}_{2,k} ]^{\text{T}}$ is the object position, and $\theta(\stR_k) \rmv\triangleq\rmv \tan^{-1}\!\big(\frac{\rv{x}_{2,k}-\ist p_2}{\rv{x}_{1,k}-\ist p_1}\big)$. 
Furthermore, $\RV{v}_{k}$ is 2D zero-mean white Gaussian measurement noise with independent components and component standard deviations $\sigma_{\rho} \rmv=\rmv 2$ and $\sigma_{\theta} \rmv=\rmv 1^{\circ}\rmv$. The detection probability of the sensor is modeled as $\de(\st_k) \rmv=\rmv p_{\text{D,max}} \exp(-\|\st'_k\|^2/450^{2})$ \cite{Reu14} with $p_{\text{D,max}} \!=\rmv 0.7$ for TS1 and $p_{\text{D,max}} \!=\rmv 0.5$ for TS2. Thus, the detection probability has its maximum of $0.7$ for TS1 and $0.5$ for TS2 at the ROI center and decreases towards the ROI border, where it is $0.45$ for TS1 and $0.32$ for TS2. The clutter pdf $f_{\text{C}}(\me_k)$ is uniform (in polar coordinates) on the ROI with mean parameter $\mu_{\text{C}} \!=\! 100$ for TS1 and $\mu_{\text{C}} \!=\! 150$ for TS2.

We compare the performance of particle implementations of the proposed LMB/P filter, the LMB filter \cite{Reu17LMBGibbs}, the fast LMB filter presented in \cite{Kro19LMB}, and a version of the TOMB/P filter \cite{Wil15,Kro16} that performs recycling of Bernoulli components as proposed in \cite{Wil12}. We remark that our performance comparison does not consider algorithms with a significantly higher complexity, such as the GLMB filter \cite{Vo13,Vo14,VoVoHoa:J17} or the trajectory-based filters proposed in \cite{Gar20SoT,Gra18SoTPMBM,Xia19TPMBM,Gar19SoTPHD}. Note also that the latter filters use Gaussian representations of spatial distributions and thus presuppose a linear-Gaussian system model, 
and moreover they assume a spatially constant detection probability,
both of which are incompatible with the considered measurement model.
\rd{Our performance comparison uses 1,000 Monte Carlo runs for each experiment. The object trajectories 
are randomly generated for each run according to the state-evolution model described above.}

The proposed LMB/P filter and the TOMB/P filter use the belief propagation (BP) algorithm of \cite{Wil15} to calculate approximations of the marginal association probabilities (cf.\ Eq.\ \eqref{eq:appr_marg2} and Step 3.2 in Table \ref{tab:summary}), and the fast LMB filter uses for this task the modified BP algorithm described in \cite{Kro19LMB}. We will therefore refer to these filters as BP-LMB/P, BP-TOMB/P, and BP-LMB, respectively. The LMB filter of \cite{Reu17LMBGibbs} is based on the Gibbs sampler and will be referred to as Gibbs-LMB. BP-LMB/P and BP-TOMB/P use 5,000 particles to represent, respectively, the posterior PHD of unlabeled objects and the posterior PHD of undetected objects. 
Another 5,000 particles are used by BP-LMB/P and BP-TOMB/P to represent the PHD of newborn unlabeled objects and the PHD of newborn undetected objects, respectively, but the resulting 10,000 particles are reduced to 5,000 particles after the update step. All filters represent the spatial pdf of each Bernoulli component by 1,000 particles. BP-LMB/P, BP-LMB, and BP-TOMB/P use 20 BP iterations to calculate the approximate marginal probabilities. The Gibbs sampler in Gibbs-LMB uses 100 samples for TS1 and 1,000 samples for TS2. All filters declare an object as detected if the existence probability of the corresponding Bernoulli component exceeds $\gamma_{\text{D}} \!=\! 0.5$, and when this is the case, they calculate a sample mean approximation of \eqref{eq:state_est} from the particle representation of the corresponding spatial pdf.

The birth statistics of all filters are established using the previous measurements $\me_{k-1}^{(m)}$, $m \!\in\! \mathcal{M}_{k-1}$. More precisely, BP-LMB/P and BP-TOMB/P choose their birth pdf as a mixture of the pdfs
\begin{align*}
&\tilde{f}_{\text{B}}^{(m)}(\st_k) \propto\! \int\! f(\st_k|\st_{k-1}) \ist f\big(\me_{k-1}^{(m)} \big| x_{1,k-1},x_{2,k-1}\big) \\[0mm]
&\hspace{35mm} \times f_{\text{v}}(\dot{x}_{1,k-1} ,\ist \dot{x}_{2,k-1}) \ist \text{d} \st_{k-1} \ist,
\end{align*}
for $m \!\in\! \mathcal{M}_{k-1}$. 
Here, $f\big(\me_{k-1}^{(m)} \big| x_{1,k-1},x_{2,k-1}\big)$ is the likelihood function corresponding to our measurement model and $f_{\text{v}}(\dot{x}_{1,k-1} ,\ist \dot{x}_{2,k-1})$ is the pdf of independent, zero-mean, Gaussian random variables $\dot{x}_{1,k-1}$, $\dot{x}_{2,k-1}$ with variance 0.25.
BP-LMB and Gibbs-LMB create a new Ber\-noul\-li component for each measurement $\me_{k-1}^{(m)}$, $m \!\in\! \mathcal{M}_{k-1}$, with spatial pdf $\sd_{\text{B}}^{(\la = (k,m))}(\st_k) \rmv=\rmv \tilde{f}_{\text{B}}^{(m)}(\st_k)$. The mean number of newborn objects is $\mu_{\text{B}} \!=\rmv 0.1$ for all filters. In BP-LMB/P and BP-TOMB/P, the mean number of, respectively, unlabeled objects and undetected objects is initialized as 0.01.

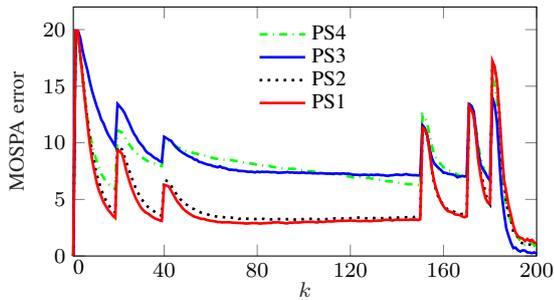
\begin{figure}[t!]
\vspace*{.3mm}
\centering
\footnotesize
\hspace*{-2mm}
   
\begin{tikzpicture}

\begin{axis}[%
width=2.4in,
height=1.3in,
at={(0.758in,0.481in)},
scale only axis,
clip=false,
xmin=1,
xmax=200,
xtick={  0,  40,  80, 120, 160, 200},
xlabel style={at={(0.5,-2.5mm)},font=\color{white!15!black}},
xlabel={$k$},
ymin=0,
ymax=22,
ytick={ 0,  5, 10, 15, 20},
ylabel style={at={(-5.5mm,0.5)},font=\color{white!15!black}},
ylabel={MOSPA error},
axis background/.style={fill=white},
legend style={at={(0.5,0.97)}, anchor=north,row sep=-0.7mm, legend cell align=left, align=left, fill=none, draw=none}
]

\addplot [color=green, dashdotted, line width=1.0pt]
  table[row sep=crcr]{%
1	0\\
2	19.917698665505\\
3	19.8860936408148\\
4	17.6393902383375\\
5	15.603711602307\\
6	13.9841808454954\\
7	12.3181518918598\\
8	10.9225367075334\\
9	9.90787468610982\\
10	9.08909226793323\\
11	8.39101352793302\\
12	7.7770356362627\\
13	7.27590755550766\\
14	6.92767628784186\\
15	6.57881500079555\\
16	6.39629339132769\\
17	6.18272338995197\\
18	6.06550162374055\\
19	5.88145769448947\\
20	11.0893030905511\\
21	11.0216970820363\\
22	10.8848410730499\\
23	10.5416800308421\\
24	10.1378881373419\\
25	9.86320731923529\\
26	9.58073672003608\\
27	9.3105688401783\\
28	9.12010848382288\\
29	8.91682471462397\\
30	8.78731553995274\\
31	8.64216034400367\\
32	8.49909661769162\\
33	8.3782933511153\\
34	8.31059432220364\\
35	8.23191045046892\\
36	8.15398739999691\\
37	8.0443279049255\\
38	7.96967150160406\\
39	7.93848979301995\\
40	10.2664172622866\\
41	10.208637941941\\
42	10.1601590861815\\
43	10.0308605092187\\
44	9.84655677065154\\
45	9.71058355241134\\
46	9.5624554079077\\
47	9.51432645538098\\
48	9.38707287650541\\
49	9.29182579139501\\
50	9.24664170236087\\
51	9.20069696213358\\
52	9.14200511324158\\
53	9.07669637724095\\
54	9.04554215191456\\
55	8.95388515728825\\
56	8.94494288931991\\
57	8.89368345835107\\
58	8.83369448629962\\
59	8.78951568287397\\
60	8.7068530166066\\
61	8.66605622449763\\
62	8.60385639005729\\
63	8.58797582289329\\
64	8.54627238093858\\
65	8.51798312017624\\
66	8.50042813711341\\
67	8.52167115852771\\
68	8.46420697114278\\
69	8.47621750457887\\
70	8.41601005858986\\
71	8.39596305720122\\
72	8.37806728122345\\
73	8.32915546926467\\
74	8.27327229470512\\
75	8.2348362749335\\
76	8.22031759909517\\
77	8.18665457299437\\
78	8.13729883582443\\
79	8.11704568975129\\
80	8.04526603805643\\
81	8.02712353590006\\
82	8.01593410350797\\
83	7.95753592634491\\
84	7.94850255717615\\
85	7.97181875884182\\
86	7.89102959853301\\
87	7.87004319955595\\
88	7.83575487904215\\
89	7.8225510810054\\
90	7.75288439510064\\
91	7.75602592653276\\
92	7.76283141768462\\
93	7.7335119554664\\
94	7.70041048485848\\
95	7.70380857653589\\
96	7.69382346590582\\
97	7.62413555362429\\
98	7.60919879011265\\
99	7.57454775159888\\
100	7.57244195991848\\
101	7.56152270770514\\
102	7.49650775150162\\
103	7.48387074713168\\
104	7.46279385753374\\
105	7.43590622585094\\
106	7.42845720606531\\
107	7.38980579257527\\
108	7.31533458462815\\
109	7.33597687844784\\
110	7.26586123279429\\
111	7.26200278139785\\
112	7.25838632067169\\
113	7.2315629178536\\
114	7.21705022676376\\
115	7.21457976472363\\
116	7.17746549645351\\
117	7.10995465834189\\
118	7.07332049813004\\
119	7.03210886841872\\
120	7.00708084810725\\
121	6.93992848135247\\
122	6.99801143971703\\
123	6.94019496916967\\
124	6.96890545436655\\
125	6.94578653817833\\
126	6.91182853676963\\
127	6.88755170988407\\
128	6.89893007778671\\
129	6.83971304989525\\
130	6.82766294270396\\
131	6.79088965858949\\
132	6.78272836628258\\
133	6.710797770437\\
134	6.6933728805495\\
135	6.63382563805874\\
136	6.60313723622389\\
137	6.60052534041245\\
138	6.57535998512602\\
139	6.53625078738219\\
140	6.47665369149249\\
141	6.43950504151259\\
142	6.43059789478595\\
143	6.40402456179851\\
144	6.36038218052765\\
145	6.32510821985673\\
146	6.34329769443115\\
147	6.3238042363667\\
148	6.27233442001858\\
149	6.27126616934414\\
150	6.23735226613124\\
151	12.4657102933269\\
152	12.2274383415046\\
153	11.7315994839803\\
154	10.6471519418753\\
155	9.71249704157166\\
156	8.95743008534349\\
157	8.58155680751365\\
158	8.32497931533359\\
159	8.07297943379946\\
160	7.82912376008207\\
161	7.72592015023006\\
162	7.55639555719175\\
163	7.50283747872273\\
164	7.33580161685466\\
165	7.31108191086081\\
166	7.20615591973743\\
167	7.05862773326116\\
168	6.88036588111886\\
169	6.83079148227091\\
170	6.71006432964567\\
171	13.5523490809695\\
172	13.2506815601214\\
173	12.6453498281988\\
174	11.2556142925992\\
175	9.65407786198518\\
176	8.59361306793666\\
177	7.89734210043261\\
178	7.36182430616999\\
179	6.89580882360425\\
180	6.74699311095644\\
181	15.6851110200449\\
182	15.1012331626175\\
183	13.746714805476\\
184	11.2009026845857\\
185	7.80523669635297\\
186	5.81083990895678\\
187	4.52893263059809\\
188	3.63790968159655\\
189	3.01823469012612\\
190	2.48229657656484\\
191	2.11814517248455\\
192	1.62117073095276\\
193	1.35449817235688\\
194	1.1826617482256\\
195	1.0207262076164\\
196	1.12758414676943\\
197	1.12242480456818\\
198	1.19632349417959\\
199	0.931786272646866\\
200	0.904876704112974\\
};
\addlegendentry{PS4}

\addplot [color=blue, line width=1.0pt]
  table[row sep=crcr]{%
1	0\\
2	19.9027133221704\\
3	19.9024039393453\\
4	19.2478420787737\\
5	18.1229365443789\\
6	17.2272055564633\\
7	16.4605331159467\\
8	15.4764874079259\\
9	14.6166300693578\\
10	13.9255933057413\\
11	13.2548168462827\\
12	12.7003424828099\\
13	12.0584447256604\\
14	11.7094793281667\\
15	11.1622871205236\\
16	10.7253090258041\\
17	10.3866970852969\\
18	10.05387471498\\
19	9.70211123534801\\
20	13.4235706430518\\
21	13.1952131890829\\
22	12.957801767915\\
23	12.6403402235534\\
24	12.1761533373945\\
25	11.7440306329651\\
26	11.2611859036515\\
27	10.8170220658433\\
28	10.5253262009172\\
29	10.1175054094404\\
30	9.87273568731231\\
31	9.64580411687315\\
32	9.29815051602822\\
33	9.09047301111447\\
34	8.99506766898258\\
35	8.81611665986145\\
36	8.64752515524736\\
37	8.56626411416387\\
38	8.38557031902069\\
39	8.31053849570387\\
40	10.5298874848692\\
41	10.4111381080979\\
42	10.2775334165471\\
43	10.1774119224314\\
44	9.88309216198835\\
45	9.64101211306584\\
46	9.42976058519618\\
47	9.26383235472412\\
48	9.11473413761803\\
49	9.00095550588661\\
50	8.86767374102696\\
51	8.76922116513495\\
52	8.68632734904339\\
53	8.52172733535606\\
54	8.49753200970063\\
55	8.31834672214455\\
56	8.28029925687512\\
57	8.19324805822636\\
58	8.12765987071702\\
59	8.06066872664494\\
60	8.03424468190665\\
61	7.9365349157202\\
62	7.90096983926893\\
63	7.83169979109397\\
64	7.92891986353633\\
65	7.79989842152957\\
66	7.77620673393903\\
67	7.70170668022147\\
68	7.61675356377566\\
69	7.62519960937781\\
70	7.59362873300477\\
71	7.59625432520706\\
72	7.51360068058358\\
73	7.51325247752302\\
74	7.50099187326855\\
75	7.47898496665444\\
76	7.43254071837643\\
77	7.44830691469999\\
78	7.42943284694729\\
79	7.38545627417536\\
80	7.44730354491633\\
81	7.40816274157779\\
82	7.43713569772804\\
83	7.42209334091216\\
84	7.43925254071033\\
85	7.42342569422529\\
86	7.39707094098029\\
87	7.4113368667604\\
88	7.3938769587753\\
89	7.35256493179867\\
90	7.36161297010666\\
91	7.37783408519632\\
92	7.38186980364397\\
93	7.4181136926051\\
94	7.36855772085439\\
95	7.39088846781879\\
96	7.35579397100533\\
97	7.36539656860003\\
98	7.34002798484343\\
99	7.36903694533618\\
100	7.36997594417182\\
101	7.32748026683051\\
102	7.33598536566883\\
103	7.3366585406115\\
104	7.30272497037981\\
105	7.31483832092274\\
106	7.29403162105253\\
107	7.3119876378047\\
108	7.30870996830282\\
109	7.28660853274485\\
110	7.27648158242258\\
111	7.2857395634668\\
112	7.26491046597536\\
113	7.27229673705219\\
114	7.29668858363654\\
115	7.26837779722136\\
116	7.27554825250226\\
117	7.29699022826646\\
118	7.26216701574653\\
119	7.27446290254662\\
120	7.24583286611995\\
121	7.18021452225455\\
122	7.18184632785747\\
123	7.17482195357814\\
124	7.15376461615664\\
125	7.16498226603457\\
126	7.18536742687641\\
127	7.1702135634965\\
128	7.14023459633586\\
129	7.18220149164955\\
130	7.15322198626498\\
131	7.18124573626977\\
132	7.18766871664369\\
133	7.13922485279564\\
134	7.14745497282943\\
135	7.11119270109084\\
136	7.07568602305668\\
137	7.12075729288937\\
138	7.13124383437638\\
139	7.1663540116522\\
140	7.19959632493641\\
141	7.17039652638809\\
142	7.14767588442358\\
143	7.19551200729739\\
144	7.23454673599519\\
145	7.17599657282228\\
146	7.15731580706362\\
147	7.13586174326654\\
148	7.14266745439971\\
149	7.11792308867556\\
150	7.12264382806894\\
151	11.5499396487723\\
152	11.2751735778934\\
153	10.6605364584357\\
154	9.4849996717446\\
155	8.64947953582692\\
156	8.18806506145452\\
157	7.95200938965514\\
158	7.7525318396238\\
159	7.58296070174573\\
160	7.34919101358621\\
161	7.34579388466152\\
162	7.25426865965035\\
163	7.2290137092278\\
164	7.19846519878846\\
165	7.12024130615997\\
166	6.9538502166151\\
167	7.03129813320473\\
168	7.07326372220958\\
169	7.02734616493084\\
170	7.04560865208317\\
171	13.409655897793\\
172	13.2749783688538\\
173	12.7545727655138\\
174	10.8649524299785\\
175	9.1413393711173\\
176	8.06066779735966\\
177	7.63322581978554\\
178	7.31920148297309\\
179	6.95127167601712\\
180	6.67642773700727\\
181	13.9682533508982\\
182	13.3489453673941\\
183	11.7962531458001\\
184	8.47561321227138\\
185	5.49183574976825\\
186	3.8\\
187	2.42\\
188	1.76\\
189	1.15895346371802\\
190	0.7\\
191	0.74\\
192	0.5\\
193	0.4\\
194	0.5\\
195	0.48\\
196	0.32\\
197	0.3\\
198	0.254674876337339\\
199	0.315124039666251\\
200	0.215211063162824\\
};
\addlegendentry{PS3}

\addplot [color=black, dotted, line width=1.0pt]
  table[row sep=crcr]{%
1	0\\
2	19.9050146293559\\
3	19.8761806701332\\
4	17.8345674726841\\
5	15.8804291543047\\
6	14.0736793786576\\
7	12.1945395018934\\
8	10.5640540125917\\
9	9.31386114895408\\
10	8.24369900333653\\
11	7.23867190009178\\
12	6.54100825131464\\
13	5.87305571156973\\
14	5.35071183666533\\
15	4.92216673977004\\
16	4.55047730411438\\
17	4.24961078642753\\
18	4.05460511516495\\
19	3.83855129216903\\
20	9.6800826785062\\
21	9.54529181629787\\
22	9.38798373268759\\
23	8.74496657142471\\
24	8.1389292218757\\
25	7.56505773492262\\
26	6.97647594745337\\
27	6.39309231808171\\
28	5.89148948938389\\
29	5.49971065251281\\
30	5.13507326673493\\
31	4.83564004366979\\
32	4.5608399365909\\
33	4.33853532254114\\
34	4.18872077738017\\
35	4.04218027968846\\
36	3.89892201061129\\
37	3.74189336382872\\
38	3.64063626017317\\
39	3.54196091997545\\
40	6.7382300479177\\
41	6.64978307690217\\
42	6.53170103432707\\
43	6.20220795594432\\
44	5.87686680248978\\
45	5.55584771049497\\
46	5.29299578694551\\
47	4.99918655230144\\
48	4.83312620178835\\
49	4.62087164089696\\
50	4.48084318638559\\
51	4.29299704111135\\
52	4.19449083360665\\
53	4.13628225772256\\
54	3.98589022486142\\
55	3.91212342907578\\
56	3.82371663208121\\
57	3.75707660638349\\
58	3.73063130162387\\
59	3.65231287951121\\
60	3.55496418020669\\
61	3.5495181997128\\
62	3.50594046778295\\
63	3.4684168261089\\
64	3.47609543115821\\
65	3.44422850981038\\
66	3.41780990388316\\
67	3.39661082422136\\
68	3.41160334873142\\
69	3.39236173975804\\
70	3.34777641469646\\
71	3.37610019510623\\
72	3.27864286098858\\
73	3.32260914560266\\
74	3.25514847483005\\
75	3.30161369629117\\
76	3.29123895035011\\
77	3.26505444383932\\
78	3.27177973930202\\
79	3.25753633083399\\
80	3.28185969181884\\
81	3.24889708491694\\
82	3.29373725786056\\
83	3.27208956123233\\
84	3.20507336387829\\
85	3.23161773067099\\
86	3.23249776607146\\
87	3.25727170720761\\
88	3.27059386107412\\
89	3.27406358024233\\
90	3.24697619856931\\
91	3.27278307730677\\
92	3.26865331628798\\
93	3.24936593065421\\
94	3.29024026685929\\
95	3.19904524725257\\
96	3.22240879490852\\
97	3.19767349087597\\
98	3.25098358772432\\
99	3.25403866139591\\
100	3.25841288997406\\
101	3.30840980102485\\
102	3.28713816500725\\
103	3.28092612788373\\
104	3.24812031278548\\
105	3.28566677416297\\
106	3.23175437899519\\
107	3.26278350766931\\
108	3.28403069776379\\
109	3.24315246073295\\
110	3.26799250903134\\
111	3.28986798562842\\
112	3.27766269410834\\
113	3.31646462036131\\
114	3.3248552297459\\
115	3.32586505208554\\
116	3.30623678986261\\
117	3.3179959068812\\
118	3.31372357370284\\
119	3.34074645962565\\
120	3.33646377038405\\
121	3.36139476102812\\
122	3.39645843549166\\
123	3.39339689143374\\
124	3.39622369897239\\
125	3.33004844742503\\
126	3.35219232249724\\
127	3.32538856708168\\
128	3.31099799824003\\
129	3.36202932067352\\
130	3.35244330583156\\
131	3.34884020574847\\
132	3.3713415379083\\
133	3.33434300410307\\
134	3.34282782682899\\
135	3.35802261738967\\
136	3.33235930802852\\
137	3.36810473846843\\
138	3.40077449304315\\
139	3.38485746109352\\
140	3.39273725343638\\
141	3.3976742647011\\
142	3.41823193628348\\
143	3.4179902913795\\
144	3.42182122892922\\
145	3.42166922347639\\
146	3.47226652273809\\
147	3.42089945250428\\
148	3.48716985458384\\
149	3.495538629157\\
150	3.53292480447314\\
151	11.4030072469357\\
152	11.1777170624487\\
153	10.6408462640762\\
154	9.26221062199562\\
155	7.56447446822343\\
156	6.37924599066356\\
157	5.70319342673506\\
158	5.06718953711662\\
159	4.78480275414219\\
160	4.51143076575314\\
161	4.31227917137155\\
162	4.02893378843224\\
163	3.96844394318639\\
164	3.94049455903721\\
165	3.87683396161327\\
166	3.77993995099979\\
167	3.68523861054513\\
168	3.72053066619006\\
169	3.69268179212117\\
170	3.68578656760577\\
171	13.2814494184538\\
172	13.0011503781745\\
173	12.4202956362944\\
174	10.8239134437822\\
175	8.64249767521297\\
176	7.30318281994241\\
177	6.54812890431945\\
178	5.6288965731661\\
179	4.82766217825691\\
180	4.54729462603091\\
181	16.9492764555266\\
182	16.4249921861985\\
183	15.3547039854807\\
184	12.5217675132115\\
185	8.6694235362754\\
186	6.46726480255543\\
187	4.71756013968348\\
188	3.55097002110259\\
189	3.25756796701023\\
190	2.11526477241605\\
191	1.86\\
192	1.65984189683384\\
193	1.2087261886088\\
194	1.12303710472007\\
195	1.0428958531492\\
196	1.00209592084034\\
197	1.00646793832903\\
198	0.94678627849814\\
199	1.05617104658844\\
200	1.02979693851733\\
};
\addlegendentry{PS2}

\addplot [color=red, line width=1.0pt]
  table[row sep=crcr]{%
1	0\\
2	19.9015883756484\\
3	19.8686071188776\\
4	17.7891616830116\\
5	15.8210240378719\\
6	13.7194365331378\\
7	11.7965436386873\\
8	10.0135040221014\\
9	8.61743569180283\\
10	7.54471675257273\\
11	6.50998381772692\\
12	5.78589318465485\\
13	5.27993041509833\\
14	4.7407769396576\\
15	4.39953668780961\\
16	4.06267408459329\\
17	3.82593987075512\\
18	3.55944160920541\\
19	3.36987850446933\\
20	9.33457631773609\\
21	9.23640800352188\\
22	9.00802834778193\\
23	8.19990205998332\\
24	7.34544762088702\\
25	6.66532841856176\\
26	6.03484138143556\\
27	5.51388208585305\\
28	5.02606773358902\\
29	4.64823430540837\\
30	4.28774851319538\\
31	4.06058749755123\\
32	3.85619285856973\\
33	3.75409739379624\\
34	3.59852763760425\\
35	3.45075962016838\\
36	3.32821961423617\\
37	3.24269175468472\\
38	3.15679309179682\\
39	3.10593236241245\\
40	6.28162973736052\\
41	6.22951047029713\\
42	6.13872169791397\\
43	5.76856591708493\\
44	5.41309023889552\\
45	5.03978092561911\\
46	4.71327800240106\\
47	4.48466680214249\\
48	4.20110774973125\\
49	4.01819017324961\\
50	3.85835658358425\\
51	3.72457171411176\\
52	3.58914268071115\\
53	3.47757552689705\\
54	3.42390653037342\\
55	3.34461780347941\\
56	3.27661026811707\\
57	3.28324534129639\\
58	3.20610917324852\\
59	3.15540361423925\\
60	3.11189728793272\\
61	3.08708132886348\\
62	3.04864079718567\\
63	2.99888379963063\\
64	2.99588404290468\\
65	2.96778974458836\\
66	2.97709592814118\\
67	2.97928813029947\\
68	2.97523142171215\\
69	2.95982629584287\\
70	2.93740334204226\\
71	2.93003958681935\\
72	2.92320052188569\\
73	2.91396707410541\\
74	2.90112746762116\\
75	2.86718033599794\\
76	2.87468447201578\\
77	2.89935364540849\\
78	2.90974226715921\\
79	2.90573701733247\\
80	2.90597226695858\\
81	2.86211215582034\\
82	2.88136300933411\\
83	2.89766484803332\\
84	2.90453079083786\\
85	2.92850910774315\\
86	2.92259142472394\\
87	2.97225866351629\\
88	2.96242292252201\\
89	2.9847889697557\\
90	2.94236611376308\\
91	2.94140188071931\\
92	2.89509168724509\\
93	2.88307334519952\\
94	2.95208249165222\\
95	2.9299223389365\\
96	2.96800540186823\\
97	2.9584289278595\\
98	2.95342138517376\\
99	2.98795712002061\\
100	3.00355887473928\\
101	3.01400821116641\\
102	3.01166954844886\\
103	3.00532888029692\\
104	3.02412228681437\\
105	2.96898786881293\\
106	2.9712407702246\\
107	3.04142362017167\\
108	3.00270395084223\\
109	3.01569497932622\\
110	3.06366237489046\\
111	3.05125883194019\\
112	3.04786115108745\\
113	3.06322336743072\\
114	3.08479144785812\\
115	3.10092665682602\\
116	3.10710631078881\\
117	3.10703986289786\\
118	3.107915012345\\
119	3.11828414840008\\
120	3.13892791834549\\
121	3.11297589468341\\
122	3.12241009837066\\
123	3.08171878826013\\
124	3.08471775268607\\
125	3.08838078002202\\
126	3.10734365355748\\
127	3.09157619659596\\
128	3.07628472668565\\
129	3.07844580618847\\
130	3.13762197470692\\
131	3.10417161379521\\
132	3.13577737174597\\
133	3.1211503046463\\
134	3.16209909562858\\
135	3.19258217387769\\
136	3.20101343605245\\
137	3.18556946305371\\
138	3.13128132766474\\
139	3.18574413865614\\
140	3.22412098984205\\
141	3.1869746954062\\
142	3.19154302653796\\
143	3.19864148855542\\
144	3.1992976712801\\
145	3.18751756565038\\
146	3.22478650352414\\
147	3.19915750959608\\
148	3.22667138058353\\
149	3.17652522701752\\
150	3.2408220487166\\
151	11.3392457259919\\
152	11.1701558136898\\
153	10.6607704139923\\
154	9.27570199640098\\
155	7.66002605633302\\
156	6.59478982123188\\
157	5.84071272074355\\
158	5.24812601054016\\
159	4.75441437361\\
160	4.40167115719986\\
161	4.13384161057072\\
162	3.94782669704665\\
163	3.9146950959539\\
164	3.80738070759517\\
165	3.66887850172365\\
166	3.64818746330124\\
167	3.61437405062628\\
168	3.49885090058161\\
169	3.41834818784697\\
170	3.50750537363395\\
171	13.218041946404\\
172	12.8902377136113\\
173	12.3489114231035\\
174	10.4743254878292\\
175	8.3889154992592\\
176	7.03962981097475\\
177	5.97821312491153\\
178	5.1385058690685\\
179	4.7358485099607\\
180	4.68310320838173\\
181	17.224717262341\\
182	16.724261397615\\
183	15.4289936803749\\
184	12.6810787739933\\
185	9.41463576036639\\
186	6.717539262941\\
187	5.62379462437402\\
188	4.44937810595806\\
189	3.60683560376633\\
190	2.68745057622322\\
191	2.06952190849893\\
192	1.75173189312927\\
193	1.60941519804964\\
194	1.54301946531536\\
195	1.32338895945513\\
196	1.50681279997888\\
197	1.3186503169905\\
198	1.42253980550851\\
199	1.18233677161807\\
200	1.12205500685042\\
};
\addlegendentry{PS1}

\node[right, align=left]
at (axis cs:-0.8mm,-0.35mm) {0};

\end{axis}
\end{tikzpicture}%
\caption{MOSPA error of BP-LMB/P versus time for TS1 using parameter settings PS1 through PS4.} 
\label{fig:thresh} 
\vspace{-1mm}
\end{figure}

\begin{table}
\centering
\vspace*{2.5mm}
\rd{\small
\hspace*{-.5mm}	\begin{tabular}{|llll|}	
	\hline
	\rule[0mm]{0mm}{2mm}&$\gamma_{\text{tr}}$& $\gamma_{\text{C}}$& $\gamma_{\text{leg}}$\\[.15mm] \hline
	\rule{0mm}{3.5mm}\!\!PS1 &  $10^{-2}$&  $10^{-10}$&  $10^{-2}$    \\[.2mm]
	\!\!PS2  &  $10^{-2}$&  $10^{-10}$&  $10^{-1}$  \\[.2mm]
	\!\!PS3  &  $10^{-2}$&  $10^{-3}$&  $10^{-2}$  \\[.2mm]
	\!\!PS4  &  $10^{-1}$&  $10^{-10}$&  $10^{-2}$  \\[.2mm]
	\hline
	\end{tabular}
}
\caption{\rd{Threshold parameter settings (PSs) used for TS1.}}
\label{tab:PS} 
\vspace{-1.5mm}
\end{table}

\vspace{-1mm}

\subsection{Simulation Results}
\label{sec:res_TS1}  

In Fig.\ \ref{fig:thresh}, we study the performance of BP-LMB/P for TS1, using four different choices of the thresholds $\gamma_{\text{tr}}$, $\gamma_{\text{C}}$, and $\gamma_{\text{leg}}$. The figure displays the Euclidean distance based mean optimal subpattern assignment (MOSPA) metric with cutoff parameter $c \!=\!  20$ and order $p \!=\! 2$ \cite{Sch08} versus time $k$. Each curve is based on a specific threshold parameter setting (PS) and was obtained by averaging over 
1,000 Monte Carlo runs. 
\rd{The PSs are defined by the values of $\gamma_{\text{tr}}$, $\gamma_{\text{C}}$, and $\gamma_{\text{leg}}$ specified in Table~\ref{tab:PS}; in particular, PS2 uses a higher value of $\gamma_{\text{leg}}$, PS3 a higher value of $\gamma_{\text{C}}$, and PS4 a higher value of $\gamma_{\text{tr}}$.}

One can see in Fig.\ \ref{fig:thresh} that the lowest MOSPA curve is achieved for PS1, i.e., for the lowest threshold values. However, a further reduction of the thresholds would not decrease the MOSPA curves further but would result in a higher filter runtime. If $\gamma_{\text{leg}}$ is increased (as in PS2), then according to Section \ref{sec:appr_CO}, there tend to be more Bernoulli components $\la$ such that $\ex_k^{(\la)}$ falls below $\gamma_{\text{leg}}$, and which are hence transferred from the LMB part to the Poisson part. In challenging scenarios, such as low $\de(\st_k)$ and/or high clutter, it is then possible that Bernoulli components are transferred to the Poisson part even though the corresponding objects exist, and this will generally reduce the tracking performance. If $\gamma_{\text{C}}$ is increased (as in PS3), then according to Section \ref{sec:first_appr:part} and \nolinebreak 
Appendix~\ref{sec:app}, this generally results in a larger number of subsets $\mathbb{L}_{k-1}^{(c)}$, which may imply that some labeled objects are no longer correctly associated with the measurements and thus the tracking performance is again reduced. Finally, if $\gamma_{\text{tr}}$ is increased (as in PS4), then according to Appendices \ref{sec:approx3} and \ref{sec:first_appr_part2}, fewer Bernoulli components are transferred to the labeled RFS part, which may again result in a poorer tracking performance.

Therefore, for TS1, we will hereafter use the thresholds of PS1. 
\rd{These thresholds are shown again in Table \ref{tab:thresh}, along with the thresholds used in TS2. 
In fact, for the more challenging TS2, we observed that the thresholds in Table \ref{tab:thresh}
resulted in a better MOSPA performance; 
in particular, we use smaller values of $\gamma_{\text{tr}}$ and $\gamma_{\text{leg}}$.
Table \ref{tab:thresh} furthermore shows the threshold $\gamma_{\text{P}}$ used by BP-LMB and Gibbs-LMB for pruning Bernoulli components and the threshold $\gamma_{\text{T}}$ 
used by BP-TOMB/P for transferring Bernoulli components of the MB part of the posterior state RFS to the Poisson part.}

\rd{Fig.~\ref{fig:setup} shows an example of the estimated object trajectories obtained with BP-LMB/P for TS1 and for TS2,
along with the true trajectories. One can see that the estimated trajectories closely match the true trajectories in both 
scenarios.}

\begin{table}[t!]
\centering
\vspace*{2.5mm}
\rd{\small
\hspace*{-.5mm}	\begin{tabular}{|llllll|}	
	\hline
	\rule[0mm]{0mm}{2mm}&$\gamma_{\text{tr}}$& $\gamma_{\text{C}}$& $\gamma_{\text{leg}}$&$\gamma_{\text{P}}$&$\gamma_{\text{T}}$\\[.15mm] \hline
	\rule{0mm}{3.5mm}\!\!TS1 &  $10^{-2}$&  $10^{-10}$&  $10^{-2}$&  $10^{-3}$&  $10^{-3}$  \\[.2mm]
	\!\!TS2  &  $10^{-3}$&  $10^{-10}$&  $10^{-3}$&  $10^{-4}$&  $10^{-4}$  \\[.2mm]
	\hline
	\end{tabular}
}
\vspace{.5mm}
\caption{\rd{Thresholds $\gamma_{\text{tr}}$, $\gamma_{\text{C}}$, and $\gamma_{\text{leg}}$ used by BP-LMB/P, $\gamma_{\text{P}}$ used by BP-LMB and Gibbs-LMB, 
and $\gamma_{\text{T}}$ used by BP-TOMB/P.}}
\label{tab:thresh} 
\vspace{-2mm}
\end{table}

\begin{figure*}[t!]
\centering
\footnotesize
\begin{minipage}[H!]{0.45\textwidth}
\hspace*{-2mm}
    \begin{tikzpicture}

\begin{axis}[%
width=2.4in,
height=1.3in,
at={(0.758in,0.481in)},
scale only axis,
clip=false,
xmin=1,
xmax=200,
xtick={  0,  40,  80, 120, 160, 200},
xlabel style={at={(0.5,-2.5mm)},font=\color{white!15!black}},
xlabel={$k$},
ymin=0,
ymax=22,
ytick={ 0,  5, 10, 15, 20},
ylabel style={at={(-5.5mm,0.5)},font=\color{white!15!black}},
ylabel={MOSPA error},
axis background/.style={fill=white},
legend style={at={(0.5,0.97)}, anchor=north,row sep=-0.7mm, legend cell align=left, align=left, fill=none, draw=none}
]
\addplot [color=blue, line width=1.0pt]
  table[row sep=crcr]{%
1	0\\
2	19.9678478723568\\
3	19.966674527372\\
4	18.5384964228267\\
5	17.6821262838667\\
6	16.8083521087681\\
7	15.5961375309588\\
8	14.5430744929233\\
9	13.6548121647247\\
10	12.7333891842656\\
11	11.9642405360236\\
12	10.7692423404918\\
13	10.0075653016148\\
14	9.24386847927408\\
15	8.62926417373875\\
16	8.04502886514262\\
17	7.71291318291973\\
18	7.31217906988555\\
19	7.01335194702473\\
20	11.6734146743453\\
21	11.4229007492161\\
22	11.148680506289\\
23	10.618891780782\\
24	10.2246047576939\\
25	9.78640574634772\\
26	9.45582037125777\\
27	9.02879735456764\\
28	8.53152510693754\\
29	8.16191132996242\\
30	7.82640195400045\\
31	7.51511034130089\\
32	7.17914286330438\\
33	6.83082808125959\\
34	6.58239821338095\\
35	6.27108041219747\\
36	6.15000487022961\\
37	5.86916299818117\\
38	5.58684049575642\\
39	5.4460680653534\\
40	8.27899945631159\\
41	8.1876328242226\\
42	8.07199579527717\\
43	7.76870881019455\\
44	7.55169576320506\\
45	7.24868098373653\\
46	7.0768921046353\\
47	6.76164814639269\\
48	6.58500542490136\\
49	6.33165728064369\\
50	6.11306931206483\\
51	6.10778931304191\\
52	5.94889399065015\\
53	5.70076786875358\\
54	5.58139818586595\\
55	5.45597604155231\\
56	5.39734388828567\\
57	5.19838205554685\\
58	5.10905062737937\\
59	4.91713376591013\\
60	4.81046568543999\\
61	4.79328263169588\\
62	4.60130989956702\\
63	4.55634573452232\\
64	4.53022614350125\\
65	4.49422033079902\\
66	4.45732272064597\\
67	4.39894961885106\\
68	4.38707472572697\\
69	4.3491077162352\\
70	4.23788382067707\\
71	4.19116892205311\\
72	4.14120039493911\\
73	4.12193801230593\\
74	4.10058578914989\\
75	3.96226833166819\\
76	3.90794927424598\\
77	3.89382279392047\\
78	3.8576482821929\\
79	3.88795408607668\\
80	3.90976783565877\\
81	3.94510642186433\\
82	3.89600513214772\\
83	3.93805229926719\\
84	3.83546901831797\\
85	3.84497557619475\\
86	3.86034640711678\\
87	3.80718165707103\\
88	3.77719105253035\\
89	3.77583620889527\\
90	3.72073979900411\\
91	3.74483817211253\\
92	3.67914551184918\\
93	3.63126039179024\\
94	3.65860251055889\\
95	3.52668205007059\\
96	3.51101586797552\\
97	3.53694656526699\\
98	3.45614538784942\\
99	3.48261499413543\\
100	3.50306764745844\\
101	3.44580569674038\\
102	3.44790539068988\\
103	3.43450436014328\\
104	3.44007070460872\\
105	3.40475998443634\\
106	3.46395859323494\\
107	3.46333566920268\\
108	3.56134196789371\\
109	3.52423264759316\\
110	3.48903035409209\\
111	3.49714813639759\\
112	3.39036325901791\\
113	3.35213105021321\\
114	3.40441482993442\\
115	3.49068557061279\\
116	3.4498653519874\\
117	3.35739319587651\\
118	3.49443093635683\\
119	3.49805680782832\\
120	3.52334025214441\\
121	3.37771436220058\\
122	3.45419346055058\\
123	3.60149297164808\\
124	3.45017920684248\\
125	3.43698919026992\\
126	3.44744472254039\\
127	3.47572463391659\\
128	3.397048181941\\
129	3.32197854360839\\
130	3.35366547252655\\
131	3.44121995813329\\
132	3.4186196226615\\
133	3.41499291401424\\
134	3.30091244033827\\
135	3.41681421700633\\
136	3.4491367580784\\
137	3.474798789067\\
138	3.49544004476176\\
139	3.42871143582305\\
140	3.55795055573176\\
141	3.56124426160348\\
142	3.64043459690321\\
143	3.61770196037023\\
144	3.67666991760673\\
145	3.60585893551705\\
146	3.58835900212221\\
147	3.64713421193924\\
148	3.58442871629016\\
149	3.67430834539069\\
150	3.66526379667966\\
151	11.6022123759347\\
152	11.4831004630369\\
153	11.3399311268796\\
154	11.0731891215833\\
155	10.2200276442757\\
156	9.46620228019614\\
157	8.50653418915414\\
158	7.94891666913025\\
159	7.50801264645972\\
160	7.36937464860594\\
161	6.93879995563199\\
162	6.63920953844785\\
163	6.0814734420523\\
164	5.70901694189674\\
165	5.38170081821117\\
166	4.87441005745843\\
167	5.00041371096262\\
168	4.91667486384851\\
169	4.65744830520653\\
170	4.50953216833493\\
171	13.3970752694554\\
172	13.1863208442625\\
173	12.7250994456637\\
174	11.9214704927977\\
175	10.953193887857\\
176	9.65113024321991\\
177	8.72595094014631\\
178	7.91183446110547\\
179	7.47644650218388\\
180	6.19595867973293\\
181	17.9621373014975\\
182	17.563789656298\\
183	16.3896960603605\\
184	14.7864307842777\\
185	13.3910778943452\\
186	10.789576346908\\
187	7.77989348486963\\
188	6.17740321096914\\
189	5.17497113669443\\
190	4.2\\
191	3.6\\
192	3.2\\
193	2.6\\
194	2\\
195	1.4\\
196	0.8\\
197	0.2\\
198	0.6\\
199	0.4\\
200	0.4\\
};
\addlegendentry{Gibbs-LMB}

\addplot [color=green, dashdotted, line width=1.0pt]
  table[row sep=crcr]{%
1	0\\
2	19.9678478723568\\
3	19.966674527372\\
4	18.7614966593227\\
5	17.3791553503692\\
6	15.5031492443367\\
7	13.8180641576973\\
8	11.9884553768822\\
9	10.4442995180877\\
10	9.56178102794076\\
11	8.47151642179052\\
12	7.4353427613449\\
13	6.3480852926562\\
14	5.52634366048367\\
15	5.00920805543775\\
16	4.62963085618225\\
17	4.1748778804779\\
18	3.65757813268837\\
19	3.47554576253651\\
20	9.33200781339794\\
21	9.44939394278457\\
22	9.09211016607889\\
23	8.73275474172944\\
24	7.92998869622974\\
25	7.44045670159125\\
26	6.69810555165694\\
27	6.15674797355664\\
28	5.577964482893\\
29	4.89697058129538\\
30	4.71629995214612\\
31	4.47711082052641\\
32	4.30934856214712\\
33	3.85438623541983\\
34	3.92640044713334\\
35	3.50052621531499\\
36	3.54483080374596\\
37	3.32555157610407\\
38	3.22599943581768\\
39	3.21116359250852\\
40	6.47910187813341\\
41	6.3057737109283\\
42	6.30911506370431\\
43	6.08276191033347\\
44	5.69735498910012\\
45	5.37272712429284\\
46	4.80934382291737\\
47	4.53611986880803\\
48	4.24525871882334\\
49	4.06577062472699\\
50	3.91947430890385\\
51	3.84660995882404\\
52	3.77142509662021\\
53	3.46846393794433\\
54	3.35128083887084\\
55	3.23949863455745\\
56	3.2118943123898\\
57	3.16064552455981\\
58	3.08790875091846\\
59	2.94337229390008\\
60	2.91441518205387\\
61	2.92083347361345\\
62	3.02486857933494\\
63	2.99487050192577\\
64	2.9031205811789\\
65	2.86574623317961\\
66	2.92244732780815\\
67	2.92955240306963\\
68	2.99528696414075\\
69	2.87304139734234\\
70	2.81032552433899\\
71	2.81161240901353\\
72	2.7317694687609\\
73	2.81296388397342\\
74	2.87738439914505\\
75	2.77752476883968\\
76	2.79539841442306\\
77	2.70615537045793\\
78	2.73028072279158\\
79	2.68528375659972\\
80	2.71678810395584\\
81	2.74330042592621\\
82	2.75813204620172\\
83	2.81615656888812\\
84	2.79886671943537\\
85	2.85656733907372\\
86	2.77766039518225\\
87	2.87509005698853\\
88	2.80593449834263\\
89	2.7531002841324\\
90	2.870016133103\\
91	2.81340881312534\\
92	2.72539506642627\\
93	2.66514346360907\\
94	2.66476352361936\\
95	2.56356253211117\\
96	2.61665879475639\\
97	2.70893129471124\\
98	2.54867333937999\\
99	2.74245930899936\\
100	2.73954663490892\\
101	2.72689217566257\\
102	2.75831864605088\\
103	2.91519902154554\\
104	2.73962559155153\\
105	2.84630537701172\\
106	2.82770865572098\\
107	2.81111491495676\\
108	2.82227782976778\\
109	2.85023981967548\\
110	2.79991700081469\\
111	2.68248927238581\\
112	2.91686852589387\\
113	2.80940612044927\\
114	2.84404640909855\\
115	2.93254260297178\\
116	2.92532588013118\\
117	2.92448794474896\\
118	2.96131315666551\\
119	2.87745957757789\\
120	3.07150077275292\\
121	3.07160828106357\\
122	3.07718652170915\\
123	3.16833841886401\\
124	3.10844311147259\\
125	3.18623466742403\\
126	3.0063422989305\\
127	3.05575648915352\\
128	2.97742617322448\\
129	2.89967873732874\\
130	2.94849318324046\\
131	2.98687715168024\\
132	2.95009362497029\\
133	2.89807906137086\\
134	2.84681182163825\\
135	2.88310353987143\\
136	2.82402188395555\\
137	2.86524640411659\\
138	2.97652463803211\\
139	3.0557158999157\\
140	3.14866968715704\\
141	3.16220857500009\\
142	3.14203938143482\\
143	3.23429757864228\\
144	3.14215525859923\\
145	3.14263622720328\\
146	3.19116116290427\\
147	3.25540069684695\\
148	3.04819029307807\\
149	2.96936863047559\\
150	3.03582201611414\\
151	11.1318403753911\\
152	10.9165456562732\\
153	10.4091263730727\\
154	8.90721198444775\\
155	7.60396822864384\\
156	6.09426045566601\\
157	5.18742996639769\\
158	4.35301828398127\\
159	4.36400335385995\\
160	4.18838663449232\\
161	3.98743016491321\\
162	3.88100756072395\\
163	3.71041740301578\\
164	3.58742881434628\\
165	3.21626745963697\\
166	2.99576933478115\\
167	3.13455408708796\\
168	3.369298658181\\
169	3.21742262925571\\
170	3.38835409288558\\
171	12.8753232771428\\
172	12.5399358148273\\
173	11.6817349602988\\
174	10.0334709754042\\
175	8.61957392847304\\
176	7.5831144087627\\
177	7.131189287465\\
178	6.04957359071154\\
179	5.86526624488031\\
180	4.55716629352288\\
181	16.7641765266095\\
182	16.5543399112459\\
183	15.5927278650551\\
184	12.5008123017642\\
185	9.4139701926036\\
186	6.59333092738701\\
187	4.059578988479\\
188	3.38113081815127\\
189	3.81537015797201\\
190	1.8\\
191	2.8\\
192	2.4\\
193	1.25038384847686\\
194	0.857183542912087\\
195	0.650532668879531\\
196	0.305441368828176\\
197	0.291438586358054\\
198	0.509500515609691\\
199	0.289829728348789\\
200	0.633993867947185\\
};
\addlegendentry{BP-TOMB/P}

\addplot [color=black, dotted, line width=1.0pt]
  table[row sep=crcr]{%
1	0\\
2	19.9678478723568\\
3	19.966674527372\\
4	18.0399034680553\\
5	16.5273905146073\\
6	14.9214560523205\\
7	12.9987185869498\\
8	11.3452904153376\\
9	9.82484667639533\\
10	8.67366862909557\\
11	7.94803201131013\\
12	6.98818520275415\\
13	6.18705890066311\\
14	5.10803921080556\\
15	4.7719128492704\\
16	4.15889550544377\\
17	3.74613691839676\\
18	3.47816169406773\\
19	3.1972855577027\\
20	9.29465348806573\\
21	9.30571937624135\\
22	9.04045136894258\\
23	8.42507200151364\\
24	7.75976218329135\\
25	7.26979817306225\\
26	6.6131085692078\\
27	5.93494925158531\\
28	5.41000423106757\\
29	4.83805099487999\\
30	4.48356617363746\\
31	4.3379329536031\\
32	4.1145206191555\\
33	3.76696970513288\\
34	3.90910378042341\\
35	3.53422013438849\\
36	3.54386240499675\\
37	3.29084048744319\\
38	3.11578457393808\\
39	3.23060692810843\\
40	6.37398887597362\\
41	6.24849898636544\\
42	6.12464259768676\\
43	5.83107902852822\\
44	5.5387013842401\\
45	5.13737165199984\\
46	4.48313551677622\\
47	4.29811035399178\\
48	4.15360806733895\\
49	3.95047928140351\\
50	3.77543114863844\\
51	3.57161077516077\\
52	3.63123242587512\\
53	3.33926220728818\\
54	3.15486231712004\\
55	3.07468741847265\\
56	3.04017941617386\\
57	2.98853375225071\\
58	3.03299068166377\\
59	2.82522833379784\\
60	2.88287419599064\\
61	2.78771540125405\\
62	2.9579909381789\\
63	2.9224344159103\\
64	2.84926361346148\\
65	2.77478516377085\\
66	2.89380606195808\\
67	2.90581012452267\\
68	2.88960363226441\\
69	2.77650341942973\\
70	2.68148498135408\\
71	2.79155822756851\\
72	2.71818333421058\\
73	2.74807183959137\\
74	2.75596879616186\\
75	2.70824220869505\\
76	2.64472233183618\\
77	2.67827388006348\\
78	2.72544551175711\\
79	2.66590064700031\\
80	2.71062738981135\\
81	2.7709658675984\\
82	2.71302503193391\\
83	2.75733591357309\\
84	2.66102510746031\\
85	2.67153979228922\\
86	2.63251949397268\\
87	2.76542507887491\\
88	2.77010686796611\\
89	2.68645550478488\\
90	2.72009656941509\\
91	2.71042003432197\\
92	2.63498205256498\\
93	2.62254862558638\\
94	2.614791638066\\
95	2.57556576215212\\
96	2.64289397866425\\
97	2.70722582255767\\
98	2.5746018552127\\
99	2.72851323382064\\
100	2.82079870767912\\
101	2.851256836655\\
102	2.82248876083843\\
103	2.90037942607972\\
104	2.81013856062727\\
105	2.85706716591709\\
106	2.84567005186216\\
107	2.84221688697495\\
108	2.90477539446239\\
109	2.96398769829343\\
110	2.89474088958201\\
111	2.75949798939131\\
112	2.94827824984439\\
113	2.91948115137082\\
114	3.01754802089794\\
115	2.95558275099868\\
116	2.90027000988133\\
117	2.93826435071246\\
118	2.92488713208245\\
119	2.88902615826704\\
120	3.04019986774887\\
121	3.09384400491628\\
122	3.02299525962724\\
123	3.09544914587497\\
124	3.06531952520689\\
125	3.05690091284441\\
126	2.92866795944613\\
127	2.97358070675251\\
128	2.92136441996817\\
129	2.83720554648867\\
130	2.90864617789186\\
131	2.97273306665579\\
132	2.88960562912959\\
133	2.82376063932904\\
134	2.84492382154024\\
135	2.90123863541942\\
136	2.82038269441223\\
137	2.84299381033369\\
138	3.08391606484528\\
139	3.06428458114511\\
140	3.14009694537501\\
141	3.1576106032168\\
142	3.1498710656459\\
143	3.21818361145691\\
144	3.1545206365255\\
145	3.17926073609784\\
146	3.23252397966253\\
147	3.22453530210933\\
148	3.02281753882831\\
149	3.01130255432579\\
150	3.08276418541845\\
151	11.1552577469965\\
152	10.8508101769076\\
153	10.4435994947733\\
154	8.87743846673124\\
155	7.50554918658144\\
156	6.00137184786783\\
157	5.23113183340063\\
158	4.45175046901651\\
159	4.63960514116039\\
160	4.30157787608727\\
161	4.03477257660847\\
162	3.93060529558712\\
163	3.8496542636045\\
164	3.68287310547414\\
165	3.23610139112308\\
166	3.06643639362686\\
167	3.03657486789654\\
168	3.39733057306293\\
169	3.20601489003048\\
170	3.22038947650895\\
171	12.9266012272989\\
172	12.65635377835\\
173	11.6635147382483\\
174	10.0295544135516\\
175	8.77745613957544\\
176	7.63249509571639\\
177	7.12046237026276\\
178	5.84573900804207\\
179	5.73982733750378\\
180	4.74233223476965\\
181	16.9586926178642\\
182	16.7603432996894\\
183	15.3854827411881\\
184	12.1827964788412\\
185	9.78984061367644\\
186	6.3884980786673\\
187	4.77753336314933\\
188	3.85422663164548\\
189	3.6\\
190	2.6\\
191	3\\
192	2.8\\
193	2\\
194	1.25482010038467\\
195	1.6\\
196	0.644894119761257\\
197	0.2\\
198	0.8\\
199	0.4\\
200	0.639571243634631\\
};
\addlegendentry{BP-LMB}

\addplot [color=red, line width=1.0pt]
  table[row sep=crcr]{%
1	0\\
2	19.9232847412521\\
3	19.8900741443434\\
4	17.6999961405869\\
5	15.5970498826029\\
6	13.5821527477889\\
7	11.4112626105659\\
8	9.6720745217781\\
9	8.34669591517727\\
10	7.31280759470601\\
11	6.35671567704546\\
12	5.64824997828062\\
13	5.04120120031701\\
14	4.56968450179536\\
15	4.22124441931757\\
16	3.91327637833297\\
17	3.70040244483994\\
18	3.45836144853659\\
19	3.27002739589367\\
20	9.24469861902966\\
21	9.20099330898486\\
22	8.95911139158855\\
23	8.186079610927\\
24	7.34951156903688\\
25	6.75866929017343\\
26	6.00647360109145\\
27	5.50510333360764\\
28	5.09339400292893\\
29	4.66759632984536\\
30	4.39261766986881\\
31	4.12973345253379\\
32	3.88772010744521\\
33	3.76371438874208\\
34	3.58831145731627\\
35	3.41881585517074\\
36	3.37231832734115\\
37	3.2378001279239\\
38	3.15815875192793\\
39	3.0558183086901\\
40	6.27551243439371\\
41	6.25958854945134\\
42	6.12530859869793\\
43	5.78223684662177\\
44	5.37243077383114\\
45	5.03855601532182\\
46	4.70976556608888\\
47	4.4375389640042\\
48	4.17984172195865\\
49	4.04227104485069\\
50	3.86393528951933\\
51	3.71157498390139\\
52	3.61810558142865\\
53	3.5154534393889\\
54	3.49780777961439\\
55	3.41167086333608\\
56	3.32784269598986\\
57	3.24126046410547\\
58	3.21792925286985\\
59	3.18019602819338\\
60	3.12653674041658\\
61	3.07133938003563\\
62	3.05938320163838\\
63	3.03200153904926\\
64	3.02527007975379\\
65	3.00580542560176\\
66	3.01135362411943\\
67	2.94679391135532\\
68	2.96997567210365\\
69	2.99840086487969\\
70	2.93721782188254\\
71	2.95829908648054\\
72	2.96342921081216\\
73	2.9558890335077\\
74	2.93233017601303\\
75	2.90688535192052\\
76	2.93601322079099\\
77	2.92106853325994\\
78	2.89388072190593\\
79	2.88601543568382\\
80	2.90593564009382\\
81	2.89048792342052\\
82	2.8591315522364\\
83	2.88352849646049\\
84	2.90751482705069\\
85	2.90413928562007\\
86	2.94272582459908\\
87	2.89789306734317\\
88	2.88679291814595\\
89	2.85772638975179\\
90	2.87927121984699\\
91	2.86529275165447\\
92	2.86779962266854\\
93	2.90688089404439\\
94	2.93235989117103\\
95	2.88327241835856\\
96	2.91315557068341\\
97	2.9488737612379\\
98	2.9032187034647\\
99	2.89761597833475\\
100	2.90692129346406\\
101	2.91736745896973\\
102	2.90044289894005\\
103	2.933857064982\\
104	2.97875180505885\\
105	2.97016241185556\\
106	2.95468202862032\\
107	2.9677969560048\\
108	2.98959268840057\\
109	3.01274879374297\\
110	2.99919725239229\\
111	3.02145038928322\\
112	3.0276950964419\\
113	3.02019587157404\\
114	3.0017741328975\\
115	3.04456166730946\\
116	3.08282169468366\\
117	3.03847900876941\\
118	3.09071960646129\\
119	3.07228400287158\\
120	3.07990751816286\\
121	3.12906230793636\\
122	3.14015349863738\\
123	3.18370626565941\\
124	3.15191504519145\\
125	3.14506888235628\\
126	3.16439222695801\\
127	3.11643022681337\\
128	3.16638055257943\\
129	3.12052610589842\\
130	3.12345603808124\\
131	3.05087783686844\\
132	3.07893864738081\\
133	3.1508197749162\\
134	3.13540439484697\\
135	3.14687869103216\\
136	3.18122684219413\\
137	3.2323127602187\\
138	3.2499315859729\\
139	3.27808890978327\\
140	3.28802244040013\\
141	3.28309642041965\\
142	3.30683264531632\\
143	3.22920048233782\\
144	3.28441487024205\\
145	3.25941384588248\\
146	3.25327734196359\\
147	3.21794097872538\\
148	3.21876200815591\\
149	3.17759942199932\\
150	3.22019028447208\\
151	11.3876290861896\\
152	11.1596962468511\\
153	10.5880197368366\\
154	9.21182095710934\\
155	7.5060879941119\\
156	6.3273364315933\\
157	5.53191163207463\\
158	5.02861236631407\\
159	4.67207591854589\\
160	4.26794864822911\\
161	4.04524661190766\\
162	3.94241584579417\\
163	3.76660465513679\\
164	3.80148682084687\\
165	3.69336773032133\\
166	3.5822902918655\\
167	3.55248028733966\\
168	3.57031433471449\\
169	3.44232767195957\\
170	3.4367567639959\\
171	12.9781998157544\\
172	12.7106942510055\\
173	12.0996999087503\\
174	10.4692700918485\\
175	8.69259172649154\\
176	7.3109732065729\\
177	6.2192944576036\\
178	5.65052856053439\\
179	5.14386509872357\\
180	4.63124745387591\\
181	17.580198076037\\
182	17.0846263389535\\
183	15.8133268082904\\
184	13.0904747061629\\
185	9.730299020856\\
186	7.32920617564646\\
187	5.29448554014816\\
188	4.52060558801281\\
189	3.25031466494459\\
190	2.87542447199937\\
191	2.39363238781828\\
192	2.18985175561408\\
193	2.06949206498464\\
194	1.820576149421\\
195	1.81294892287161\\
196	1.56456927641286\\
197	1.37080274099402\\
198	1.22026940204712\\
199	1.67669989556711\\
200	1.72625534061357\\
};
\addlegendentry{BP-LMB/P (proposed)}

\node[right, align=left]
at (axis cs:-0.8mm,-0.35mm) {0};
\node[right, align=left,scale=1.1]
at (axis cs:32mm,-2mm) {(a)};
\end{axis}
\end{tikzpicture}%
\end{minipage}
\begin{minipage}[H!]{0.4\textwidth}
\hspace*{1mm}
    \begin{tikzpicture}

\begin{axis}[%
width=2.4in,
height=1.3in,
at={(0.758in,0.481in)},
scale only axis,
clip=false,
xmin=1,
xmax=200,
xtick={  0,  40,  80, 120, 160, 200},
xlabel style={at={(0.5,-2.5mm)},font=\color{white!15!black}},
xlabel={$k$},
ymin=0,
ymax=22,
ytick={ 0,  5, 10, 15, 20},
ylabel style={at={(-5.5mm,0.5)},font=\color{white!15!black}},
ylabel={MOSPA error},
axis background/.style={fill=white},
legend style={row sep=-0.7mm, legend cell align=left, align=left, fill=none, draw=none}
]
\addplot [color=blue, line width=1.0pt]
  table[row sep=crcr]{%
1	0\\
2	20\\
3	20\\
4	20\\
5	20\\
6	19.8373001331381\\
7	19.718423262189\\
8	19.5468066245586\\
9	19.2585264410565\\
10	19.0414525752382\\
11	18.6956489856584\\
12	18.5148990434376\\
13	17.8976656557618\\
14	17.3419738888714\\
15	16.9979335862512\\
16	16.5058270059252\\
17	15.939516197247\\
18	15.7749653639639\\
19	15.5793093154424\\
20	15.0926358990932\\
21	14.79268631376\\
22	14.5764389182304\\
23	14.0033914525257\\
24	13.4217532833542\\
25	15.0276516543991\\
26	14.850973535864\\
27	15.8577956287861\\
28	15.6592968473531\\
29	15.4557925410516\\
30	15.8458172301244\\
31	15.6986405725768\\
32	15.5014981984227\\
33	15.490128953791\\
34	15.2068533920857\\
35	15.0299103287685\\
36	14.7220344562512\\
37	14.3425392581692\\
38	14.3390746440918\\
39	13.9558002283364\\
40	13.8219099798489\\
41	13.5290772907712\\
42	13.2574514116634\\
43	13.2022021723636\\
44	12.9164170737567\\
45	12.6238659789664\\
46	12.5137412604492\\
47	12.1800786015123\\
48	12.0875934339804\\
49	11.7786224146468\\
50	12.3316897804177\\
51	12.1903927664421\\
52	12.1202423653245\\
53	12.7045419222356\\
54	12.6066805311396\\
55	13.1037437799451\\
56	13.029795122238\\
57	12.890804685199\\
58	12.7314134171664\\
59	12.5829234655296\\
60	13.0167341739461\\
61	12.8071098244791\\
62	12.7275283500674\\
63	13.0453837801798\\
64	13.0157410550711\\
65	12.6927370629053\\
66	12.6152441797751\\
67	12.6461968196105\\
68	12.4642320565118\\
69	12.3713374383614\\
70	12.6736746605081\\
71	12.5482260709939\\
72	12.5540148878823\\
73	12.5000230386786\\
74	12.3754205766668\\
75	12.6382719347156\\
76	12.5758854010276\\
77	12.4840330321609\\
78	12.2583308930818\\
79	12.15524485774\\
80	12.8344303238193\\
81	12.8105579361567\\
82	12.7071657115575\\
83	12.7212277684349\\
84	12.6311545014181\\
85	12.4817449325126\\
86	12.3632266691495\\
87	12.3034132860281\\
88	12.2260911796679\\
89	12.0901229428732\\
90	12.036260311867\\
91	11.989431504287\\
92	11.9491639683537\\
93	11.9149202197392\\
94	11.7857432293683\\
95	11.7882475327232\\
96	11.6017259131363\\
97	11.5891185442321\\
98	11.4913558331972\\
99	11.4478599106305\\
100	11.7890036218909\\
101	11.7495679346247\\
102	11.7576757728861\\
103	11.7355325478096\\
104	11.6796959276259\\
105	11.5773943893903\\
106	11.520292795151\\
107	11.4632546778883\\
108	11.4444989987996\\
109	11.4163109244948\\
110	11.3547130431571\\
111	11.3655669107818\\
112	11.2673396485818\\
113	11.212572423742\\
114	11.1826487205248\\
115	11.0646790714407\\
116	11.0917407802628\\
117	11.0990161281284\\
118	11.0763887732115\\
119	11.0231275839459\\
120	10.9262600032642\\
121	10.8718376592432\\
122	10.8087485952272\\
123	10.7656038825205\\
124	10.6890040702887\\
125	10.6085163720494\\
126	10.5821551982765\\
127	10.5729754444006\\
128	10.4969330366353\\
129	10.4478044634273\\
130	10.3583095126281\\
131	10.3337988768899\\
132	10.2516003943119\\
133	10.2303613694046\\
134	10.233490502965\\
135	10.2062606663245\\
136	10.1728463444485\\
137	10.0436163901656\\
138	10.0177614962508\\
139	9.95274442884649\\
140	9.87454281247268\\
141	9.23901204549863\\
142	9.19331888048356\\
143	9.16582766053054\\
144	9.09523156762614\\
145	9.07892170842726\\
146	9.09109679790992\\
147	9.1184638044931\\
148	9.11713316508279\\
149	9.12378000614715\\
150	9.14297073125143\\
151	8.37820001672932\\
152	8.32394425137484\\
153	8.38286280228719\\
154	8.42711118953732\\
155	8.43671662812597\\
156	8.51464867116543\\
157	8.59026394365476\\
158	8.67175257595693\\
159	8.73478013063322\\
160	8.81080799685497\\
161	7.65135896877766\\
162	7.7772971142204\\
163	7.81390774860059\\
164	7.9906055914324\\
165	8.07181331871995\\
166	8.2444490108232\\
167	8.45133664443196\\
168	8.68706744392966\\
169	8.72626558825489\\
170	8.89059330081847\\
171	9.1001362096343\\
172	9.08325346860775\\
173	9.16949159092025\\
174	9.18854680201439\\
175	9.24005352449921\\
176	9.24399398718194\\
177	9.1583798051622\\
178	9.15413022965566\\
179	9.14059684185291\\
180	9.24301678140074\\
181	9.244005014937\\
182	9.05366830856989\\
183	8.97453217031699\\
184	8.84802115637004\\
185	8.63992845899965\\
186	8.48562277376968\\
187	8.43844194242197\\
188	8.37238100567964\\
189	8.21089320107359\\
190	8.22831525645211\\
191	10.7280214659002\\
192	10.4498180758751\\
193	10.1680046763391\\
194	9.97875263309118\\
195	9.71308799365032\\
196	9.34244430177311\\
197	9.22164916115554\\
198	9.05571515440679\\
199	8.81275807528262\\
200	8.54795444239837\\
};
\addlegendentry{Gibbs-LMB}

\addplot [color=green, dashdotted, line width=1.0pt]
  table[row sep=crcr]{%
1	0\\
2	20\\
3	20\\
4	20\\
5	20\\
6	19.9468849226766\\
7	19.7914570817725\\
8	19.5852847243044\\
9	19.2762342039391\\
10	19.1027926814737\\
11	18.8462022439639\\
12	18.336874672929\\
13	17.8872691525067\\
14	17.3006909236156\\
15	16.9609457741902\\
16	16.0712175041587\\
17	15.3705372727071\\
18	15.2658064915053\\
19	14.6745533863217\\
20	14.0676794458835\\
21	13.4023977785943\\
22	12.6407446644772\\
23	12.1797796084961\\
24	11.6736785894129\\
25	13.8255009527314\\
26	13.2517586268795\\
27	14.7476401321266\\
28	14.3044921702519\\
29	14.1415931351981\\
30	14.4788887272604\\
31	14.3411338160558\\
32	13.8866613604157\\
33	13.7334149569506\\
34	13.3603431223598\\
35	12.9102619142954\\
36	12.1938936870748\\
37	11.8118207394278\\
38	11.4575805358868\\
39	10.9034102310037\\
40	10.4827946162568\\
41	10.0923801059088\\
42	9.43909492783897\\
43	9.08846266803976\\
44	8.90650951715972\\
45	8.26671366821857\\
46	8.07965824112869\\
47	7.68744219099455\\
48	7.50646283128145\\
49	7.06639192473431\\
50	7.9429907319623\\
51	7.62189687689585\\
52	7.36812674748483\\
53	8.46357034506311\\
54	8.10674648516082\\
55	9.05896911909212\\
56	8.99002514977335\\
57	8.56923105739006\\
58	8.24393554806615\\
59	7.99611225176488\\
60	8.82667071364462\\
61	8.64656821762294\\
62	8.33992785169388\\
63	9.08756561830028\\
64	8.70970138709689\\
65	8.38987251562075\\
66	8.28280186145094\\
67	7.88507025368582\\
68	7.6814077394176\\
69	7.38389170446328\\
70	7.78659907916085\\
71	7.54117336335144\\
72	7.32595033238877\\
73	7.06519261744096\\
74	6.90165940659004\\
75	7.37320816828239\\
76	7.00742700444297\\
77	6.9524065766817\\
78	6.61507562847762\\
79	6.48619571133876\\
80	7.8132793593295\\
81	7.48925966783569\\
82	7.38122027182088\\
83	7.00737089216062\\
84	6.84211798562029\\
85	6.80823220100645\\
86	6.51776419271105\\
87	6.29223804633383\\
88	6.06849575800246\\
89	5.8781335454908\\
90	5.72648331631709\\
91	5.5313772967908\\
92	5.29808261553342\\
93	5.22735924687787\\
94	4.87879607040426\\
95	4.74375481468811\\
96	4.71466286540734\\
97	4.53834469923978\\
98	4.40162162913739\\
99	4.22381741608626\\
100	4.84577835878121\\
101	4.66789999069124\\
102	4.54838890463609\\
103	4.48852505842645\\
104	4.31090404644772\\
105	4.17673221138185\\
106	4.04328278162021\\
107	3.99022138317577\\
108	3.86069369606575\\
109	3.70270172918457\\
110	3.61063164637988\\
111	3.5204693713731\\
112	3.41750080888829\\
113	3.33646323911332\\
114	3.19223685260031\\
115	3.17295989564055\\
116	3.08148322674193\\
117	2.93674452596427\\
118	2.91170482619475\\
119	2.86662964666097\\
120	2.76536521403591\\
121	2.74418956153904\\
122	2.75278820232784\\
123	2.66948978035386\\
124	2.62378244028257\\
125	2.68160883326659\\
126	2.63760106739954\\
127	2.51536735943484\\
128	2.48871944328444\\
129	2.49831838322782\\
130	2.31228035577626\\
131	2.34155232577176\\
132	2.18790327990928\\
133	2.24778814577249\\
134	2.23947780787107\\
135	2.28691212565583\\
136	2.14425669171844\\
137	2.03048115704361\\
138	2.03822249119226\\
139	2.0683135142469\\
140	2.09887297723833\\
141	3.02956955085087\\
142	2.94341397286567\\
143	2.91671723904548\\
144	2.78260982538033\\
145	2.63071781264774\\
146	2.40445735488531\\
147	2.32081386063163\\
148	2.29134176606906\\
149	2.27168657836083\\
150	2.16764198362588\\
151	3.36761400869258\\
152	3.24008365510771\\
153	3.13577156524993\\
154	3.00242345301661\\
155	2.80426301555011\\
156	2.5488239860172\\
157	2.49089514193161\\
158	2.32238374800425\\
159	2.28159818270667\\
160	2.23400525960098\\
161	5.82168451457398\\
162	5.4859105870932\\
163	5.25739309136346\\
164	4.85888993522501\\
165	4.30211782592148\\
166	3.43887139846298\\
167	2.84444755691121\\
168	2.68483299139359\\
169	2.57282530717827\\
170	2.29774699467013\\
171	7.49070941100808\\
172	7.24900640228549\\
173	6.88273582620067\\
174	6.30173601686191\\
175	5.52304140713119\\
176	4.27851148048655\\
177	3.57668748330863\\
178	3.34703664576274\\
179	2.97474629461033\\
180	2.60941798345332\\
181	8.59026952658997\\
182	8.22215984573251\\
183	7.90762305749984\\
184	7.50471645482798\\
185	6.36536645105192\\
186	4.69674529027111\\
187	4.03815785490919\\
188	3.72029325742577\\
189	3.17744445043505\\
190	2.79767847434246\\
191	9.03254959848509\\
192	8.75274130899689\\
193	8.17438019279377\\
194	7.64848434710755\\
195	6.48620636405371\\
196	5.15138966584977\\
197	4.57864392282998\\
198	4.23270427080692\\
199	3.56640466264284\\
200	3.03296854385075\\
};
\addlegendentry{BP-TOMB/P}

\addplot [color=black, dotted, line width=1.0pt]
  table[row sep=crcr]{%
1	0\\
2	20\\
3	20\\
4	20\\
5	19.9449558511927\\
6	19.7212872555427\\
7	19.5464158837355\\
8	19.4659167683576\\
9	18.8842975949281\\
10	18.7194199815636\\
11	18.2276220786207\\
12	17.890540633847\\
13	17.2909230635476\\
14	16.5699190906751\\
15	16.0764988165593\\
16	15.5933147846175\\
17	14.6782280938243\\
18	14.3140527688109\\
19	13.6018317864079\\
20	13.1005406284789\\
21	12.247573254671\\
22	11.5474600758876\\
23	11.0982393452495\\
24	10.226501612146\\
25	12.6643090461734\\
26	12.3699312579881\\
27	14.0347903391681\\
28	13.6404905141013\\
29	13.4036989052438\\
30	13.9344478197395\\
31	13.5813198252776\\
32	13.1646795087375\\
33	12.8617692134664\\
34	12.3941121029531\\
35	11.8900521670581\\
36	11.4334683682038\\
37	11.0030669067177\\
38	10.6505373567295\\
39	10.3212058866795\\
40	9.69582059411653\\
41	9.41375788047336\\
42	8.79810230206535\\
43	8.38946510082899\\
44	8.28121739314991\\
45	7.71371547636277\\
46	7.48256081515073\\
47	7.22014432381699\\
48	6.93333234737204\\
49	6.75538326505949\\
50	7.72092743359122\\
51	7.45839212861693\\
52	7.48920548928163\\
53	8.19711735519698\\
54	8.04662538083647\\
55	8.67725228214552\\
56	8.73396941266117\\
57	8.4746812491798\\
58	8.22404543092887\\
59	7.91794406449131\\
60	8.62162974549333\\
61	8.49042493411202\\
62	8.26553777038531\\
63	8.73429156092881\\
64	8.31739917559866\\
65	8.07110364303819\\
66	7.73410585187539\\
67	7.53917227576804\\
68	7.35105397908925\\
69	6.88759785185383\\
70	7.52951886546083\\
71	7.2162832865191\\
72	7.07733836376165\\
73	6.70376273981269\\
74	6.56544328721515\\
75	7.07973752771215\\
76	6.77094055681272\\
77	6.65633392617021\\
78	6.43488722289474\\
79	6.19428829997525\\
80	7.49731214320504\\
81	7.35367123211135\\
82	7.11130283590321\\
83	6.78938332816867\\
84	6.60792098674993\\
85	6.32979031163506\\
86	6.25146512373496\\
87	6.02092114637886\\
88	5.75948564350959\\
89	5.53463464914787\\
90	5.43025951785349\\
91	5.25406741670651\\
92	4.98638705239892\\
93	4.89172583450692\\
94	4.59323396395584\\
95	4.58056198490093\\
96	4.30167468154956\\
97	4.2052131075122\\
98	4.09581763019102\\
99	3.99308200648868\\
100	4.67404870316047\\
101	4.47726511915249\\
102	4.40707249099228\\
103	4.34744013010494\\
104	4.15451639373441\\
105	4.00563630113643\\
106	3.87537943048111\\
107	3.79498572595217\\
108	3.69231399902652\\
109	3.66787858494395\\
110	3.50759943428034\\
111	3.41070387770215\\
112	3.25096758395793\\
113	3.05468612135855\\
114	3.07677956773294\\
115	3.04306078882349\\
116	2.90623966997166\\
117	2.84507950254206\\
118	2.81817779709049\\
119	2.73089746884805\\
120	2.70643729470368\\
121	2.70558892252269\\
122	2.57412286574087\\
123	2.53049355367927\\
124	2.42432665390336\\
125	2.4213499532576\\
126	2.39045266335257\\
127	2.37681293559421\\
128	2.36026158200696\\
129	2.31183697034991\\
130	2.24144313187547\\
131	2.2338740231486\\
132	2.15822610355512\\
133	2.14967783971081\\
134	2.17106560643603\\
135	2.13261356629283\\
136	2.20089646955676\\
137	2.01950120889223\\
138	2.06663755529582\\
139	1.99846448222239\\
140	2.07640050921478\\
141	3.0325362919443\\
142	2.90906888264781\\
143	2.86396757904818\\
144	2.75920758161615\\
145	2.54085480353593\\
146	2.35870888821285\\
147	2.20944070019115\\
148	2.16739336965108\\
149	2.10155829581288\\
150	2.13326952257707\\
151	3.36337639352704\\
152	3.3322175143388\\
153	3.1604294982297\\
154	2.96506636510594\\
155	2.78437670824257\\
156	2.49938706498604\\
157	2.43388830074112\\
158	2.31987735832225\\
159	2.2941756404323\\
160	2.07152047442877\\
161	5.86183211901843\\
162	5.59789134015947\\
163	5.31353438497804\\
164	4.91255214190967\\
165	4.34498360951582\\
166	3.45736249599699\\
167	2.94243775473362\\
168	2.63954977315664\\
169	2.61062696260973\\
170	2.42683451976525\\
171	7.63074955379514\\
172	7.2834634024325\\
173	6.97924980759509\\
174	6.41092306678794\\
175	5.60507263264157\\
176	4.40123842738382\\
177	3.66610143293638\\
178	3.38060919750375\\
179	3.09396162693908\\
180	2.61700144302743\\
181	8.54682543489711\\
182	8.15564135437616\\
183	7.96073102329465\\
184	7.47514660909119\\
185	6.44959978217634\\
186	4.66842247315278\\
187	4.05923791551733\\
188	3.68608022941289\\
189	3.37613215615657\\
190	2.99965380081651\\
191	9.19117101589022\\
192	8.7148225961664\\
193	8.17477438520964\\
194	7.70556976528618\\
195	6.42997181943287\\
196	5.15798294748597\\
197	4.65803264327725\\
198	4.12641800525711\\
199	3.70831903809962\\
200	3.10765301084462\\
};
\addlegendentry{BP-LMB}

\addplot [color=red, line width=1.0pt]
  table[row sep=crcr]{%
1	0\\
2	20\\
3	20\\
4	20\\
5	19.7484353892104\\
6	19.5926727948199\\
7	19.2438221127184\\
8	19.036818180467\\
9	18.6178117644974\\
10	18.7186145100428\\
11	18.1849191622271\\
12	17.6886521557738\\
13	16.9881129053686\\
14	16.5153145412621\\
15	15.9505165685789\\
16	15.2812757196045\\
17	14.6033522785854\\
18	13.8983091660137\\
19	13.4467601243085\\
20	12.9387384248876\\
21	11.8294564524885\\
22	11.2949087240255\\
23	10.7078756198555\\
24	10.1498859118929\\
25	12.5727316690463\\
26	12.181912576497\\
27	13.8015612035251\\
28	13.5586040760513\\
29	13.382609282814\\
30	13.7694445115235\\
31	13.3673139072629\\
32	12.8819135182881\\
33	12.6383092536381\\
34	12.0472298044334\\
35	11.5620258868155\\
36	11.2551519173055\\
37	10.9095185340079\\
38	10.5911315935257\\
39	9.967077623471\\
40	9.52868524529937\\
41	9.03927597824819\\
42	8.74464888521182\\
43	8.43174794664032\\
44	8.17147012790952\\
45	7.72742983224748\\
46	7.26661653003945\\
47	7.11946373475091\\
48	6.75262482468744\\
49	6.41755079058371\\
50	7.35966666621805\\
51	7.23029312564809\\
52	6.86193860960941\\
53	7.7686914966884\\
54	7.55142953529062\\
55	8.39700841175204\\
56	8.29222710384896\\
57	7.98534458434418\\
58	7.90840954763728\\
59	7.72953876117339\\
60	8.24682296590026\\
61	8.09339244112876\\
62	7.82715774514404\\
63	8.27534325183448\\
64	8.06760945427257\\
65	7.81286131154235\\
66	7.53250393629371\\
67	7.3031549126483\\
68	7.06923063190632\\
69	6.95259212903222\\
70	7.49252159180541\\
71	7.18131945003622\\
72	6.98814082941289\\
73	6.79270504795247\\
74	6.63167777332843\\
75	7.10489360216291\\
76	6.88408375988947\\
77	6.61636907904541\\
78	6.42641102306273\\
79	6.30399594820102\\
80	7.5098143006328\\
81	7.35410586329898\\
82	7.09122566432386\\
83	6.83366182370507\\
84	6.59869974745962\\
85	6.43751933740108\\
86	6.27672401489926\\
87	6.05954211548546\\
88	5.7951299384752\\
89	5.65168386818567\\
90	5.45990567835862\\
91	5.38710484257674\\
92	5.17693012019624\\
93	5.02493016432476\\
94	4.88437447542122\\
95	4.6664653254059\\
96	4.44291748139806\\
97	4.30569101660387\\
98	4.13158187641118\\
99	4.01851847673067\\
100	4.48074284447434\\
101	4.37799268934355\\
102	4.33342508455266\\
103	4.28127491818614\\
104	4.17452033773031\\
105	4.09241037528374\\
106	4.03728060130204\\
107	4.02465227338138\\
108	3.91957803819212\\
109	3.86829643981297\\
110	3.83835587048681\\
111	3.72702382458763\\
112	3.65409003094994\\
113	3.46989154782227\\
114	3.4152173994915\\
115	3.31480002827814\\
116	3.24511343461217\\
117	3.09588597547557\\
118	3.06460953573979\\
119	3.07079382761005\\
120	2.94980442003\\
121	2.88500466446016\\
122	2.9325357572871\\
123	2.84988151734733\\
124	2.84939508304406\\
125	2.79680762325077\\
126	2.78151094306806\\
127	2.68177948149724\\
128	2.69212562311083\\
129	2.67632227083352\\
130	2.53963442225653\\
131	2.53841319894794\\
132	2.51327252150776\\
133	2.46474273519749\\
134	2.40527165054002\\
135	2.39958831795974\\
136	2.29686225057247\\
137	2.27101827099343\\
138	2.2599112181044\\
139	2.27204970873405\\
140	2.22192035996072\\
141	3.07493030997507\\
142	2.9902511950648\\
143	2.92776376248821\\
144	2.89592599682151\\
145	2.77638015630035\\
146	2.60513206789219\\
147	2.55717095241653\\
148	2.41133454567927\\
149	2.4402189817069\\
150	2.4168083050717\\
151	3.36734366036227\\
152	3.30590636622527\\
153	3.24888119564555\\
154	3.11368457441685\\
155	2.93122307265685\\
156	2.69557840637413\\
157	2.59357632837149\\
158	2.51842678199746\\
159	2.44391088506477\\
160	2.35033672219729\\
161	5.8608203769042\\
162	5.65351433559406\\
163	5.36630306437923\\
164	5.03471436733232\\
165	4.36353933022993\\
166	3.55114110965326\\
167	3.14266159585732\\
168	2.90357094928992\\
169	2.7523718177309\\
170	2.49818633869252\\
171	7.54046840743207\\
172	7.28513417826384\\
173	7.00878431678979\\
174	6.32975107162679\\
175	5.7104011261175\\
176	4.35217353525154\\
177	3.76741921315014\\
178	3.57963332904925\\
179	3.14053846939555\\
180	2.72291787821449\\
181	8.65986552826993\\
182	8.43587290873425\\
183	7.99055810388653\\
184	7.51341936911128\\
185	6.34542709739738\\
186	4.67045709236888\\
187	4.05080701953082\\
188	3.65942703463226\\
189	3.39425864775292\\
190	3.15230965434208\\
191	9.37099047555758\\
192	9.0698464669189\\
193	8.69552482849342\\
194	8.09197039357859\\
195	7.40597452377876\\
196	5.43418212176685\\
197	4.47473449089804\\
198	4.20364592872646\\
199	3.86165393370746\\
200	3.31812358198734\\
};
\addlegendentry{BP-LMB/P (proposed)}

\node[right, align=left]
at (axis cs:-0.8mm,-0.35mm) {0};
\node[right, align=left,scale=1.1]
at (axis cs:32mm,-2mm) {(b)};
\end{axis}
\end{tikzpicture}%
\end{minipage}
\caption{MOSPA error of the four filters versus time for (a) TS1 and (b) TS2.}
\label{fig:results} 
\vspace{-3mm}
\end{figure*}
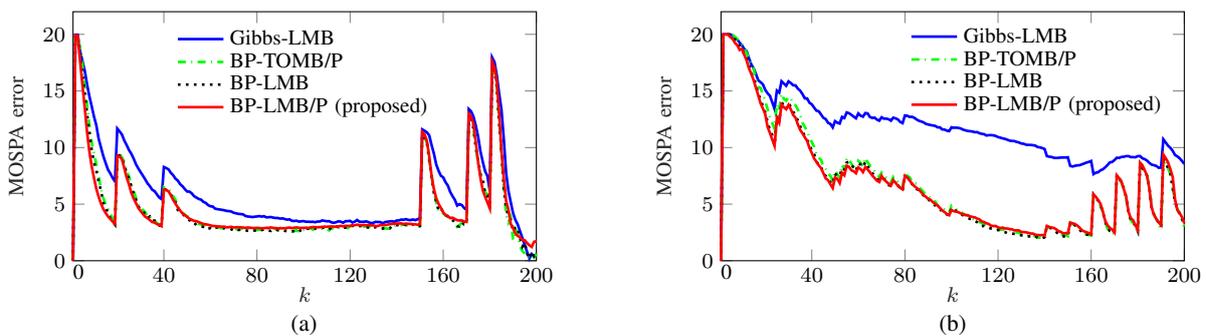

Fig.\ \ref{fig:results} compares the MOSPA performance of BP-LMB/P, Gibbs-LMB, BP-LMB, and BP-TOMB/P for TS1 and TS2. It is seen that for TS1, the performance of BP-LMB/P is almost identical to that of BP-LMB and BP-TOMB/P whereas the performance of Gibbs-LMB is noticeably poorer. For TS2, the results are similar except that the performance gap of Gibbs-LMB is much larger. This performance gap is due to the fact that Gibbs-LMB tends to ignore relevant association information in challenging scenarios. The amount of relevant association information taken into account by Gibbs-LMB grows with the number of samples used in the Gibbs sampler, but this comes at the cost of a higher computational complexity. In challenging scenarios such as TS2, more association information is required to obtain good results; this explains the larger performance gap of Gibbs-LMB in that case (even though for TS2, our Gibbs-LMB implementation used ten times more samples than for TS1). Overall, these results also demonstrate the excellent performance of the BP algorithm used by BP-LMB/P, BP-LMB, and BP-TOMB/P to compute the marginal association probabilities.

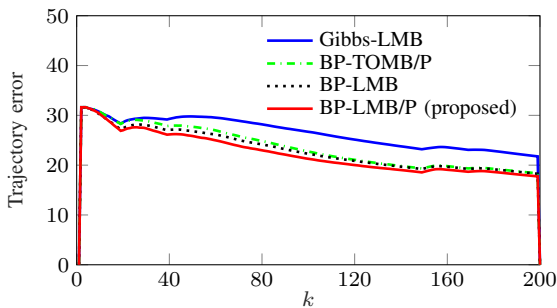
\begin{figure}[t!]
\vspace*{.3mm}
\centering
\footnotesize
\hspace*{-2mm}
   \begin{tikzpicture}

\begin{axis}[%
width=2.4in,
height=1.3in,
at={(0.758in,0.481in)},
scale only axis,
clip=false,
xmin=0,
xmax=200,
xtick={0,40,80,120,160,200},
xlabel style={at={(0.5,-2.5mm)},font=\color{white!15!black}},
xlabel={$k$},
ymin=0,
ymax=50,
ytick={0,10,20,30,40,50},
ylabel style={font=\color{white!15!black}},
ylabel style={at={(-5.5mm,0.5)},font=\color{white!15!black}},
ylabel={Trajectory error},
axis background/.style={fill=white},
legend style={row sep=-0.7mm, legend cell align=left, align=left, fill=none, draw=none}
]

\addplot [color=blue, line width=1.0pt]
  table[row sep=crcr]{%
1	0\\
2	31.6227766016838\\
3	31.6227766016838\\
4	31.6227766016838\\
5	31.4323651707604\\
6	31.3019828086737\\
7	31.2070187073209\\
8	31.0566621618502\\
9	30.8692706797539\\
10	30.649776075637\\
11	30.4092947253263\\
12	30.1555266971308\\
13	29.9609453395304\\
14	29.7742012162486\\
15	29.4556659484151\\
16	29.149601826875\\
17	28.8726095002201\\
18	28.6219533304522\\
19	28.3415742973656\\
20	28.6013465824129\\
21	28.8511706220408\\
22	29.0195935540329\\
23	29.1334646012684\\
24	29.2145604876991\\
25	29.3012732941525\\
26	29.3445506105954\\
27	29.3980317440464\\
28	29.4482201071408\\
29	29.4936805451597\\
30	29.4911986029498\\
31	29.4814239797764\\
32	29.4515544624802\\
33	29.4250332398006\\
34	29.3880859642874\\
35	29.3441623252073\\
36	29.2972103139576\\
37	29.2614797815125\\
38	29.2111876641962\\
39	29.164872957878\\
40	29.2875899744163\\
41	29.3908342499675\\
42	29.4866439083167\\
43	29.592875803854\\
44	29.6430901326353\\
45	29.7044462947485\\
46	29.7438227136525\\
47	29.7626279454205\\
48	29.7761281104043\\
49	29.7828571994393\\
50	29.7768573211019\\
51	29.7602432440375\\
52	29.7391008217619\\
53	29.723565039956\\
54	29.712717531214\\
55	29.6905635648054\\
56	29.673942521085\\
57	29.6457056538848\\
58	29.632766737577\\
59	29.5830435305975\\
60	29.5342477468839\\
61	29.4912440653732\\
62	29.444042547478\\
63	29.3883799168034\\
64	29.3184408441202\\
65	29.2519570902693\\
66	29.1820668060352\\
67	29.1093566338326\\
68	29.0435162855635\\
69	28.9746315694452\\
70	28.9024051436813\\
71	28.8272372202398\\
72	28.7584220933114\\
73	28.6908866414427\\
74	28.6161619567\\
75	28.5391611445718\\
76	28.4598691209671\\
77	28.3955928517256\\
78	28.3375342208931\\
79	28.2637991242326\\
80	28.1923271102535\\
81	28.1098415031285\\
82	28.0324943114606\\
83	27.9519747635628\\
84	27.8642635629952\\
85	27.7786037769225\\
86	27.6983204567901\\
87	27.6240465815233\\
88	27.551090507401\\
89	27.4757785918476\\
90	27.3968204226651\\
91	27.3277386299432\\
92	27.2683843207009\\
93	27.1934152871357\\
94	27.1240715386513\\
95	27.0518897633058\\
96	26.984336777973\\
97	26.9072157769506\\
98	26.8372401372649\\
99	26.7646357344457\\
100	26.693422124181\\
101	26.6330731769792\\
102	26.5657668567877\\
103	26.4857500860492\\
104	26.4026771177085\\
105	26.3185877132975\\
106	26.2323624239096\\
107	26.1465347774554\\
108	26.0614498835833\\
109	25.9815363994405\\
110	25.900481567862\\
111	25.8165694969067\\
112	25.7337197034757\\
113	25.6559606703821\\
114	25.5781471849931\\
115	25.5011632952104\\
116	25.4255937224344\\
117	25.3471792105823\\
118	25.2730208811373\\
119	25.2030820575356\\
120	25.1313594655496\\
121	25.0633705819856\\
122	24.9871278765699\\
123	24.914603186458\\
124	24.8399505658064\\
125	24.7749742793887\\
126	24.7079808155888\\
127	24.6387916571466\\
128	24.5647277415088\\
129	24.4951085725084\\
130	24.4326112164053\\
131	24.3651633970881\\
132	24.2983066510621\\
133	24.2353970631302\\
134	24.1728190419919\\
135	24.1082584769156\\
136	24.0375678511677\\
137	23.9675226245049\\
138	23.9017747786626\\
139	23.8366702038083\\
140	23.7754410518009\\
141	23.7155230019354\\
142	23.6535000301641\\
143	23.5883514107302\\
144	23.5213661452138\\
145	23.4587187649247\\
146	23.3934474508765\\
147	23.3251763630624\\
148	23.257258482725\\
149	23.1900905186211\\
150	23.2618978368258\\
151	23.3325551487243\\
152	23.3988145436237\\
153	23.4591837602127\\
154	23.5158582874585\\
155	23.568766549033\\
156	23.6212201710837\\
157	23.6467877144431\\
158	23.6417520854413\\
159	23.6195625118041\\
160	23.576816932411\\
161	23.5261182729407\\
162	23.4707344524929\\
163	23.4159646779398\\
164	23.3643544638854\\
165	23.3133000636459\\
166	23.2545251326055\\
167	23.1941062464903\\
168	23.1341562779847\\
169	23.0720223392205\\
170	23.0847682302267\\
171	23.0948423564421\\
172	23.0971847478027\\
173	23.092073944831\\
174	23.0870875480078\\
175	23.0773161248838\\
176	23.0549831599182\\
177	23.0079131367241\\
178	22.9536994384155\\
179	22.902749797317\\
180	22.844862886617\\
181	22.7901751341695\\
182	22.7359462615559\\
183	22.6794038856062\\
184	22.6210447713295\\
185	22.560411597007\\
186	22.5002091360165\\
187	22.4405373244125\\
188	22.3813347030758\\
189	22.3225828218173\\
190	22.2642878199138\\
191	22.2064994722751\\
192	22.1492083901783\\
193	22.0922555545842\\
194	22.0357688519041\\
195	21.9797741037318\\
196	21.9242194281872\\
197	21.8690940585884\\
198	21.8144170997459\\
199	21.760105139165\\
200	0\\
};
\addlegendentry{Gibbs-LMB}

\addplot [color=green, dashdotted, line width=1.0pt]
  table[row sep=crcr]{%
1	0\\
2	31.6227766016838\\
3	31.6227766016838\\
4	31.6227766016838\\
5	31.4321899231674\\
6	31.4096500262908\\
7	31.3012202868213\\
8	31.1783876227443\\
9	30.9425403151373\\
10	30.6838032942468\\
11	30.4081381982264\\
12	30.1232550971824\\
13	29.8766668092746\\
14	29.6658497015147\\
15	29.4195458333439\\
16	29.0978505093174\\
17	28.7622998470586\\
18	28.5425861948374\\
19	28.2734001673945\\
20	28.5301426484972\\
21	28.7211372771477\\
22	28.8823220673249\\
23	28.9922624578528\\
24	29.0497776423448\\
25	29.1006532944867\\
26	29.0916028677558\\
27	29.0528950402746\\
28	28.9940016320149\\
29	28.9597546871717\\
30	28.8504088196484\\
31	28.7868003750575\\
32	28.6829113925491\\
33	28.5822856487295\\
34	28.4672447127994\\
35	28.3693050179123\\
36	28.2638292689646\\
37	28.1481118214665\\
38	27.996468555983\\
39	27.8535760557127\\
40	27.9018649579506\\
41	27.9302494198883\\
42	27.9408115206\\
43	27.9477006032048\\
44	27.9343348936747\\
45	27.943348599146\\
46	27.9002810803238\\
47	27.8424897311407\\
48	27.8007944468329\\
49	27.7612225552931\\
50	27.7087800222623\\
51	27.6504664391276\\
52	27.5949686140578\\
53	27.528714232139\\
54	27.4518919813393\\
55	27.3816184891707\\
56	27.3302962538017\\
57	27.2598656948563\\
58	27.1913906809184\\
59	27.0847019086628\\
60	26.9738449560042\\
61	26.8642344286234\\
62	26.7322334519539\\
63	26.6113206764578\\
64	26.4948994665371\\
65	26.3720555277147\\
66	26.2528860901576\\
67	26.1595664789813\\
68	26.055097365027\\
69	25.9437017311705\\
70	25.8230954076424\\
71	25.7002337484662\\
72	25.5840999227238\\
73	25.4938450611791\\
74	25.4204813892264\\
75	25.3220016177365\\
76	25.2361563874127\\
77	25.1425497593131\\
78	25.0343534012296\\
79	24.935717806288\\
80	24.8545665910073\\
81	24.7525231092095\\
82	24.657003947813\\
83	24.5531621257523\\
84	24.4374582252816\\
85	24.3197849101092\\
86	24.2090637786776\\
87	24.0964334401203\\
88	23.9803468104133\\
89	23.8787527350058\\
90	23.7757182615922\\
91	23.6734479988861\\
92	23.5649339864439\\
93	23.4579038224317\\
94	23.3525084055226\\
95	23.2471928392469\\
96	23.1433439716455\\
97	23.0365450170873\\
98	22.9360434658446\\
99	22.8293261391282\\
100	22.7340363626093\\
101	22.6432580267834\\
102	22.5460432091098\\
103	22.4459803819594\\
104	22.3434015331505\\
105	22.2469087430787\\
106	22.1515881246417\\
107	22.0659307667914\\
108	21.9821095136319\\
109	21.8990902809676\\
110	21.81713200361\\
111	21.7332171197239\\
112	21.6579306142879\\
113	21.5872896620991\\
114	21.5059435072764\\
115	21.4298867658502\\
116	21.3543190402273\\
117	21.2793842904092\\
118	21.2092355697692\\
119	21.1402059449758\\
120	21.0691219751372\\
121	21.0056948218617\\
122	20.9326464041528\\
123	20.8520717512006\\
124	20.7839866732303\\
125	20.7276019627146\\
126	20.6607027312712\\
127	20.5916510999111\\
128	20.5282099604898\\
129	20.4642636499388\\
130	20.4013164380932\\
131	20.3358204918304\\
132	20.2780578071249\\
133	20.2254690541194\\
134	20.1731031439883\\
135	20.1251243026542\\
136	20.0592851859078\\
137	19.9975593221459\\
138	19.940032985229\\
139	19.8832599487711\\
140	19.830330437157\\
141	19.7744525685831\\
142	19.7190155169293\\
143	19.6643538647892\\
144	19.6100933608531\\
145	19.5597674656149\\
146	19.5059463475987\\
147	19.4527331605598\\
148	19.3943150747638\\
149	19.3325138805567\\
150	19.4426361382185\\
151	19.5435708588591\\
152	19.646299830174\\
153	19.7510392690788\\
154	19.8436400387944\\
155	19.9123880839854\\
156	19.9253961239644\\
157	19.915483293638\\
158	19.8925430508335\\
159	19.8467045628141\\
160	19.7982012788691\\
161	19.7506007980999\\
162	19.7032377526114\\
163	19.6537471504458\\
164	19.6077311781744\\
165	19.5532817128761\\
166	19.4992177680352\\
167	19.4487855349921\\
168	19.3956556822667\\
169	19.3459755929004\\
170	19.3782061966007\\
171	19.4039152131045\\
172	19.4261309985197\\
173	19.4452342614878\\
174	19.4552952939337\\
175	19.4536533920198\\
176	19.4307995043801\\
177	19.3905585982246\\
178	19.3477495218054\\
179	19.2999074399965\\
180	19.2551149898267\\
181	19.2107710315815\\
182	19.1667654800417\\
183	19.1205025116583\\
184	19.0691302772935\\
185	19.0181095151529\\
186	18.970323118386\\
187	18.9201000302155\\
188	18.8702749852511\\
189	18.8239600993865\\
190	18.7748997259444\\
191	18.7262743811007\\
192	18.6780686923684\\
193	18.6301200066336\\
194	18.582561373225\\
195	18.5354007500098\\
196	18.4911683731674\\
197	18.444824925019\\
198	18.398820329332\\
199	18.3531141421492\\
200	0\\
};
\addlegendentry{BP-TOMB/P}

\addplot [color=black, dotted, line width=1.0pt]
  table[row sep=crcr]{%
1	0\\
2	31.6227766016838\\
3	31.6227766016838\\
4	31.6227766016838\\
5	31.4322373504023\\
6	31.2497144077517\\
7	31.0246427092028\\
8	30.7758616499987\\
9	30.4673105054826\\
10	30.1832902979514\\
11	29.8509715144592\\
12	29.5222708208377\\
13	29.2120545729849\\
14	28.9466066815699\\
15	28.559567713722\\
16	28.1871980249094\\
17	27.8358275842873\\
18	27.5228049138138\\
19	27.1483414684088\\
20	27.398425295772\\
21	27.6372394084344\\
22	27.7823780911301\\
23	27.9243676359712\\
24	28.0363374804449\\
25	28.0974832397716\\
26	28.1306010582342\\
27	28.1507002768564\\
28	28.1312728483972\\
29	28.0904102761918\\
30	28.041954706339\\
31	27.9474887071828\\
32	27.8604619805499\\
33	27.7698750481608\\
34	27.6335354355042\\
35	27.5118470899604\\
36	27.4081765254961\\
37	27.2728423418674\\
38	27.1414791170224\\
39	27.0290015700015\\
40	27.0922307943972\\
41	27.1216070419675\\
42	27.1410623841331\\
43	27.1317499083385\\
44	27.1432288179704\\
45	27.1228922172372\\
46	27.0556301296713\\
47	26.9839537130726\\
48	26.9295848034034\\
49	26.8829968259162\\
50	26.8171093453546\\
51	26.7482100359086\\
52	26.6822379483627\\
53	26.6283800708438\\
54	26.5643049040291\\
55	26.4794028965523\\
56	26.4050012516944\\
57	26.3388919188179\\
58	26.2691098350573\\
59	26.2097709080974\\
60	26.146156675102\\
61	26.0640781373898\\
62	25.9490553733671\\
63	25.836092923853\\
64	25.7403927393502\\
65	25.6372546407366\\
66	25.5181758832571\\
67	25.3978748928822\\
68	25.2883673010516\\
69	25.1619518443386\\
70	25.0438603245702\\
71	24.9294249907751\\
72	24.8332698096329\\
73	24.7272833077164\\
74	24.6441780922779\\
75	24.56713581977\\
76	24.4890877061147\\
77	24.4041694514876\\
78	24.3200768391567\\
79	24.2455107224052\\
80	24.170389502033\\
81	24.0798919799484\\
82	23.9873731984706\\
83	23.8995842096601\\
84	23.8139934780866\\
85	23.717588550062\\
86	23.6172059677326\\
87	23.499612932947\\
88	23.3991503270185\\
89	23.2991550072912\\
90	23.1909083108491\\
91	23.0938262933391\\
92	22.9995649094204\\
93	22.9066363724699\\
94	22.8100020391297\\
95	22.7178978326554\\
96	22.6361278256289\\
97	22.5424558197343\\
98	22.4508359868975\\
99	22.3665358467703\\
100	22.2930909399873\\
101	22.2236341055794\\
102	22.13738714136\\
103	22.0477938208286\\
104	21.9551462342921\\
105	21.8771740157345\\
106	21.7957610829215\\
107	21.7194542741853\\
108	21.6528091721387\\
109	21.5750506557003\\
110	21.4977789619318\\
111	21.4184640564463\\
112	21.3592467925751\\
113	21.3011907140768\\
114	21.2290073462681\\
115	21.1689572303609\\
116	21.1103662564802\\
117	21.0476788274278\\
118	20.9859536793305\\
119	20.9210579873016\\
120	20.8615342246152\\
121	20.8066176218967\\
122	20.7371533745595\\
123	20.6605079712919\\
124	20.5968260469931\\
125	20.5446834084114\\
126	20.4853133247908\\
127	20.4240608225812\\
128	20.3644021941812\\
129	20.3008612862068\\
130	20.2416994422856\\
131	20.1797015620643\\
132	20.1223532280619\\
133	20.0773132204275\\
134	20.0254666525946\\
135	19.9805896572305\\
136	19.9213688731069\\
137	19.8591216320619\\
138	19.8052633401046\\
139	19.7484771015442\\
140	19.6954289740279\\
141	19.6432963945531\\
142	19.5916587983657\\
143	19.5442866776799\\
144	19.4937941845119\\
145	19.4438819982425\\
146	19.3906330627852\\
147	19.3378755783936\\
148	19.2794545749951\\
149	19.2176275426644\\
150	19.3326437562154\\
151	19.4345903156165\\
152	19.5315474133084\\
153	19.6301730385979\\
154	19.7238147384678\\
155	19.7932013592313\\
156	19.7998975277317\\
157	19.7905012868547\\
158	19.767090857846\\
159	19.7183789769139\\
160	19.6730799928764\\
161	19.6257025882659\\
162	19.584577231393\\
163	19.5380147705554\\
164	19.4919358813867\\
165	19.4402945425005\\
166	19.3889881629144\\
167	19.3446395309307\\
168	19.2947154283529\\
169	19.2454004426602\\
170	19.2782732980923\\
171	19.3046192052042\\
172	19.3245919019737\\
173	19.3444682947164\\
174	19.3551924415298\\
175	19.3545035876979\\
176	19.3382314969128\\
177	19.2984021017966\\
178	19.258955892719\\
179	19.2112688281468\\
180	19.166795955465\\
181	19.1227245995243\\
182	19.0790442919001\\
183	19.0331228740434\\
184	18.9820427472212\\
185	18.9313075254547\\
186	18.8840251648461\\
187	18.8340460477096\\
188	18.7844420865673\\
189	18.7383401100699\\
190	18.6926178666114\\
191	18.6442097087254\\
192	18.5962078341419\\
193	18.5484352741522\\
194	18.5010464351388\\
195	18.4540682823848\\
196	18.4099622862052\\
197	18.3637684274674\\
198	18.3179091497198\\
199	18.272350411948\\
200	0\\
};
\addlegendentry{BP-LMB}

\addplot [color=red, line width=1.0pt]
  table[row sep=crcr]{%
1	0\\
2	31.6227766016838\\
3	31.6227766016838\\
4	31.6227766016838\\
5	31.4321368475328\\
6	31.3014937781393\\
7	31.0698152349462\\
8	30.8537008138767\\
9	30.6132128427389\\
10	30.2523611160436\\
11	29.8878071291394\\
12	29.5311162293881\\
13	29.196254596605\\
14	28.8097312454472\\
15	28.454486775936\\
16	28.0282864871612\\
17	27.6599489970644\\
18	27.2951638731048\\
19	26.890439089665\\
20	27.0033341628213\\
21	27.2013952491665\\
22	27.3681226194466\\
23	27.4418298016046\\
24	27.562256622132\\
25	27.5923491050208\\
26	27.6036175559069\\
27	27.5361216765322\\
28	27.5202524099841\\
29	27.4959746267948\\
30	27.3869870118291\\
31	27.2394662807942\\
32	27.0930677122953\\
33	26.9397462395431\\
34	26.8171762926011\\
35	26.6961693857903\\
36	26.5493385275675\\
37	26.3924847125367\\
38	26.2699132663375\\
39	26.1214917024014\\
40	26.1660502747659\\
41	26.2051885493821\\
42	26.2266870747234\\
43	26.2285910412262\\
44	26.2103698746585\\
45	26.1867604677447\\
46	26.1431601840748\\
47	26.05783726065\\
48	25.9683149384447\\
49	25.9198081626868\\
50	25.8486959968406\\
51	25.7572327988791\\
52	25.6574158740819\\
53	25.5687190167031\\
54	25.4844510254903\\
55	25.4486868190488\\
56	25.3775172466908\\
57	25.2790970163412\\
58	25.182050328582\\
59	25.072148091584\\
60	24.9778505730056\\
61	24.8772825242344\\
62	24.7735742353738\\
63	24.658272825048\\
64	24.5398775800897\\
65	24.4224427731244\\
66	24.3016409476883\\
67	24.1787712382232\\
68	24.0643660452849\\
69	23.9653485018866\\
70	23.8632225243309\\
71	23.7399263785496\\
72	23.6394768258333\\
73	23.5608325637728\\
74	23.4932166478792\\
75	23.3828608191043\\
76	23.3112634985463\\
77	23.247817964396\\
78	23.1655521448383\\
79	23.0636491445586\\
80	22.9809746998782\\
81	22.8944756087933\\
82	22.8206366857695\\
83	22.741637577225\\
84	22.6401270524021\\
85	22.5559662969721\\
86	22.4590568238924\\
87	22.3548320591046\\
88	22.275230006441\\
89	22.1981295830659\\
90	22.1172570638844\\
91	22.0283418723523\\
92	21.9365684656746\\
93	21.8502969245166\\
94	21.7664461594684\\
95	21.6888089685599\\
96	21.6036944355044\\
97	21.5143598467637\\
98	21.4304857257464\\
99	21.3425117341317\\
100	21.2754965096261\\
101	21.2062314300799\\
102	21.1363263016302\\
103	21.0622405792215\\
104	20.9760114920743\\
105	20.9046545175407\\
106	20.8433733779852\\
107	20.7839167380897\\
108	20.712225892226\\
109	20.6548567699543\\
110	20.5880719703362\\
111	20.5235991090409\\
112	20.4719068244958\\
113	20.4265341735386\\
114	20.3632488591611\\
115	20.3054227363475\\
116	20.2503379036003\\
117	20.1947981518772\\
118	20.1318850276207\\
119	20.0705420918128\\
120	20.0155638225631\\
121	19.9684922022407\\
122	19.9125103050513\\
123	19.8406432193702\\
124	19.7857355757485\\
125	19.7392881621349\\
126	19.682109340321\\
127	19.6262758557594\\
128	19.5757417493646\\
129	19.5318718483569\\
130	19.4887128581644\\
131	19.430759946444\\
132	19.3885717784144\\
133	19.344071339568\\
134	19.2997866464881\\
135	19.263813966446\\
136	19.2083316187701\\
137	19.1569004665662\\
138	19.10995422236\\
139	19.0633979412467\\
140	19.0172719393753\\
141	18.9680122715935\\
142	18.9190397759045\\
143	18.8675702022432\\
144	18.8166815276177\\
145	18.7663644198797\\
146	18.712071917331\\
147	18.6584556386526\\
148	18.6030364421666\\
149	18.5439092313167\\
150	18.6692379624094\\
151	18.7889479807622\\
152	18.9033457402399\\
153	19.0109709462157\\
154	19.1092283952653\\
155	19.1816753359233\\
156	19.1990283297348\\
157	19.1931645746216\\
158	19.1668474385837\\
159	19.1172193619479\\
160	19.0714126487022\\
161	19.030280896526\\
162	18.9946788633809\\
163	18.9483855396567\\
164	18.9057809305824\\
165	18.8608093308273\\
166	18.8125097610279\\
167	18.7673166650074\\
168	18.7195958096787\\
169	18.6724900150332\\
170	18.7131134349917\\
171	18.7436904740655\\
172	18.7706615599131\\
173	18.7941351156886\\
174	18.8082085415392\\
175	18.8103856242388\\
176	18.7965291225646\\
177	18.7583137023879\\
178	18.7174763985864\\
179	18.6713019735297\\
180	18.6282041165535\\
181	18.5854943737283\\
182	18.5431313832448\\
183	18.4960508613813\\
184	18.4463117446228\\
185	18.3969815892061\\
186	18.3479608607682\\
187	18.2993768115966\\
188	18.2511680798913\\
189	18.2033391672283\\
190	18.1558958189595\\
191	18.1088816384594\\
192	18.0622605234191\\
193	18.0158979616345\\
194	17.969899799394\\
195	17.9242784436322\\
196	17.8813950676534\\
197	17.8365841907027\\
198	17.7920952182523\\
199	17.7479030800645\\
200	0\\
};
\addlegendentry{BP-LMB/P (proposed)}

\end{axis}
\end{tikzpicture}%
\caption{Trajectory error of the four filters versus time for TS2.}
\label{fig:TrajMetric1} 
\vspace{-2mm}
\end{figure}

In Fig.\ \ref{fig:TrajMetric1}, we compare BP-LMB/P, Gibbs-LMB, BP-LMB, and BP-TOMB/P for TS2, using instead of the MOSPA metric the trajectory metric proposed in \cite{Gar20TrajMet} with cutoff parameter $c \!=\! 20$, order $p \!=\! 2$, and switching penalty $\gamma \!=\! 2$. This metric can be decomposed into a ``location error'' (the location error of detected objects), a ``false error'' (caused by ``false objects''), a ``missed error'' (caused by ``missed objects''), and a ``switching error.'' Here, false objects are detected objects that do not correspond to any object within the ground truth, whereas missed objects are objects within the ground truth that do not correspond to any detected object. Differently from the OSPA metric, the trajectory metric also takes into account the switching error caused by track switches, i.e., when a detected object is associated with different objects within the ground truth at different times. According to Fig.\ \ref{fig:TrajMetric1}, the trajectory metric performance of BP-LMB/P is slightly better than that of BP-LMB and BP-TOMB/P and significantly better than that of Gibbs-LMB. These results agree with our MOSPA results in Fig.\ \ref{fig:results} (note the different y-axis scales used in the two figures). In addition, they show that BP-LMB/P also succeeds in estimating object trajectories, not just individual object states. 

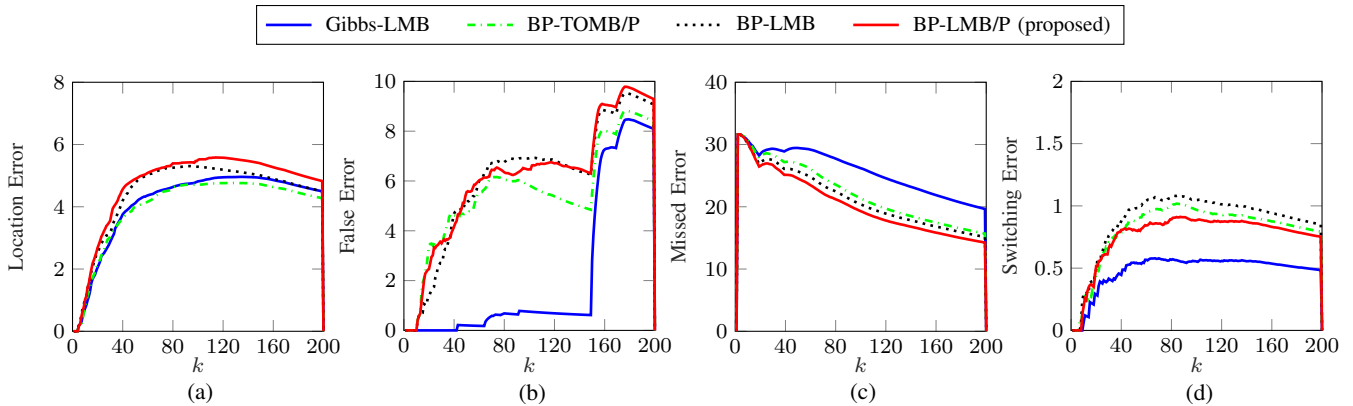
\begin{figure*}[t!]
\centering
\footnotesize
\vspace{-3mm}
\begin{minipage}[H!]{1\textwidth}
\vspace{-1mm}
\hspace{37mm}\ref{Leg:TrajMet2}
\vspace{3mm}
\end{minipage}
\begin{minipage}[H!]{0.235\textwidth}
    \begin{tikzpicture}

\begin{axis}[%
width=1.3in,
height=1.3in,
at={(0.758in,0.481in)},
scale only axis,
clip=false,
xmin=0,
xmax=200,
xtick={  0,  40,  80, 120, 160, 200},
xlabel style={at={(0.5,-2.5mm)},font=\color{white!15!black}},
xlabel={$k$},
ymin=0,
ymax=8,
ylabel style={at={(-5.5mm,0.5)},font=\color{white!15!black}},
ylabel={Location Error},
axis background/.style={fill=white},
legend columns=-1,
legend style={column sep=4mm},
legend to name = {Leg:TrajMet2}
]
\addplot [color=blue, line width=1.0pt]
  table[row sep=crcr]{%
1	0\\
2	0\\
3	0\\
4	0\\
5	0.177968108771983\\
6	0.232260979057966\\
7	0.289347193865575\\
8	0.453571082888164\\
9	0.616659413759542\\
10	0.735530378870607\\
11	0.838812062784099\\
12	0.995679611393333\\
13	1.12091972295484\\
14	1.28976409898425\\
15	1.55420151786575\\
16	1.68022064907285\\
17	1.75575628483062\\
18	1.82720665935112\\
19	1.95224670991553\\
20	2.05418625581951\\
21	2.14663432738688\\
22	2.24930811657908\\
23	2.41743838690057\\
24	2.46348344995454\\
25	2.50554294633024\\
26	2.57039279142814\\
27	2.6418494414566\\
28	2.70982584181841\\
29	2.7802236584033\\
30	2.87207257945913\\
31	2.95671141372391\\
32	3.04755445729461\\
33	3.12412395696316\\
34	3.30949921738827\\
35	3.3704676052991\\
36	3.45724499923821\\
37	3.52219439920459\\
38	3.59189701738558\\
39	3.67293557932602\\
40	3.75866513502846\\
41	3.8043037181201\\
42	3.84742434277136\\
43	3.87950233473952\\
44	3.91392149151021\\
45	3.94981284326592\\
46	3.9918552326319\\
47	4.03480442368982\\
48	4.06160131000282\\
49	4.08848936074682\\
50	4.11808194098522\\
51	4.16221651993874\\
52	4.203388469649\\
53	4.23773425406659\\
54	4.25853481351014\\
55	4.28440037086561\\
56	4.30748884761912\\
57	4.32505170796412\\
58	4.32745283268887\\
59	4.35100236816387\\
60	4.36578355568183\\
61	4.38051372154414\\
62	4.3907956923441\\
63	4.40324210749377\\
64	4.41145227878981\\
65	4.43592302658019\\
66	4.4589561546013\\
67	4.48098500991356\\
68	4.49593876660093\\
69	4.50687585726588\\
70	4.52010927215191\\
71	4.5315572051888\\
72	4.54327449523688\\
73	4.5527013181398\\
74	4.56094385612767\\
75	4.57102432143874\\
76	4.5840869614001\\
77	4.59481947633074\\
78	4.60593640434982\\
79	4.62046405801484\\
80	4.63486815172874\\
81	4.65146938664211\\
82	4.66416086390195\\
83	4.67505341211472\\
84	4.6848202265977\\
85	4.69668579344407\\
86	4.70588699723976\\
87	4.71420965672118\\
88	4.7214671526842\\
89	4.72934359625783\\
90	4.73420620077048\\
91	4.73727451357311\\
92	4.74079985014546\\
93	4.74492888780541\\
94	4.74885542509248\\
95	4.75374056568062\\
96	4.75709852953102\\
97	4.76558697299563\\
98	4.7828608707549\\
99	4.80041434142897\\
100	4.81659283653158\\
101	4.82254771530768\\
102	4.82949207709828\\
103	4.83752924657603\\
104	4.84439718754482\\
105	4.85590522903829\\
106	4.86933034098328\\
107	4.87772788058178\\
108	4.88284350037909\\
109	4.88815096953155\\
110	4.89833250788561\\
111	4.90783364917233\\
112	4.91647238291049\\
113	4.92504377062401\\
114	4.92942034600162\\
115	4.93291495578392\\
116	4.93825518012062\\
117	4.93972478651103\\
118	4.94188535155536\\
119	4.94398552068355\\
120	4.94669599056566\\
121	4.94877047976228\\
122	4.95013315279967\\
123	4.95128050330127\\
124	4.95280250305427\\
125	4.95316626877784\\
126	4.95282228020401\\
127	4.95311582591292\\
128	4.95369622974738\\
129	4.95447224023442\\
130	4.95447063184827\\
131	4.95653671537719\\
132	4.9581976215795\\
133	4.95956968205829\\
134	4.96065804625046\\
135	4.96070715307661\\
136	4.95972219711523\\
137	4.95858752157315\\
138	4.95767699351618\\
139	4.95620802096992\\
140	4.95413247351901\\
141	4.95379049031285\\
142	4.95287199863981\\
143	4.94992685413659\\
144	4.94767685248333\\
145	4.94642446935009\\
146	4.94477231903888\\
147	4.94158724834088\\
148	4.93707072354561\\
149	4.93273965997913\\
150	4.92661597287971\\
151	4.9210084756295\\
152	4.91512393095004\\
153	4.91007099255716\\
154	4.90504864549834\\
155	4.89957009830159\\
156	4.89429121995757\\
157	4.88588438213033\\
158	4.877306653032\\
159	4.86811306256031\\
160	4.85902731530904\\
161	4.84945332978261\\
162	4.84039085643629\\
163	4.83176955730186\\
164	4.82394742539183\\
165	4.81596682193111\\
166	4.80843415202686\\
167	4.80125272966397\\
168	4.79370019055177\\
169	4.7871050176172\\
170	4.77524814249806\\
171	4.76370527264805\\
172	4.75249894196792\\
173	4.74179573502807\\
174	4.73158135906228\\
175	4.72133272102882\\
176	4.71106046755425\\
177	4.70079114049295\\
178	4.69069625893259\\
179	4.68115474049068\\
180	4.67167938207016\\
181	4.66224576885317\\
182	4.65289851152096\\
183	4.64304057377506\\
184	4.63383868794473\\
185	4.62414833461904\\
186	4.61424269683458\\
187	4.60468055816683\\
188	4.59517045308904\\
189	4.58564085365862\\
190	4.57616775760785\\
191	4.56697978899877\\
192	4.55804616823304\\
193	4.54866992979389\\
194	4.53948888074249\\
195	4.5306472681742\\
196	4.52187381008461\\
197	4.51321410717121\\
198	4.50477070276284\\
199	4.49619084221978\\
200	0\\
};
\addlegendentry{\hspace{-3mm}Gibbs-LMB}

\addplot [color=green, dashdotted, line width=1.0pt]
  table[row sep=crcr]{%
1	0\\
2	0\\
3	0\\
4	0\\
5	0.172501125198579\\
6	0.167694658081648\\
7	0.223328079544161\\
8	0.372840910446604\\
9	0.628892616079208\\
10	0.796821670292475\\
11	0.930230339546222\\
12	1.17191632641101\\
13	1.27098284022217\\
14	1.38026240422184\\
15	1.58026619967623\\
16	1.71704438708051\\
17	1.8170326039105\\
18	1.89446480531076\\
19	2.04723355433192\\
20	2.17184360523764\\
21	2.24437855788633\\
22	2.30852868701852\\
23	2.44448525720288\\
24	2.49712501541243\\
25	2.59670872963229\\
26	2.77671212295049\\
27	2.83701146210172\\
28	2.95359123232539\\
29	2.98127692132375\\
30	3.03230040094187\\
31	3.11240618599137\\
32	3.19830840990726\\
33	3.27195616624942\\
34	3.33717096781878\\
35	3.38101661154985\\
36	3.42710722490226\\
37	3.48003490958982\\
38	3.5301536051617\\
39	3.58087822539017\\
40	3.63758967773993\\
41	3.68233587845624\\
42	3.7281219282843\\
43	3.76141312667594\\
44	3.78744741609215\\
45	3.81893795519454\\
46	3.83831491655875\\
47	3.85988086525138\\
48	3.92047553313516\\
49	3.97987404404101\\
50	3.99796726115744\\
51	4.01806636707044\\
52	4.0356238292113\\
53	4.05477610740827\\
54	4.06809933860883\\
55	4.07957790779311\\
56	4.08964171614068\\
57	4.10093098759258\\
58	4.11313581142125\\
59	4.14064500855184\\
60	4.16478887081757\\
61	4.18483868033402\\
62	4.20455158844533\\
63	4.22444428328886\\
64	4.25819959922179\\
65	4.29374032233618\\
66	4.33093894824597\\
67	4.36372871630818\\
68	4.38917094103488\\
69	4.41641076691857\\
70	4.43944447812888\\
71	4.46223676897628\\
72	4.476853143138\\
73	4.49073271479415\\
74	4.49931381316846\\
75	4.50682233463628\\
76	4.51503746358753\\
77	4.52301948515567\\
78	4.53264843816182\\
79	4.58951849758524\\
80	4.6326279684755\\
81	4.64112325335428\\
82	4.64788842354266\\
83	4.65451578347699\\
84	4.66077668274105\\
85	4.66961577402704\\
86	4.67803566409439\\
87	4.6853915408406\\
88	4.69446726117908\\
89	4.70005683682984\\
90	4.70503490961095\\
91	4.70745528599063\\
92	4.7116779385202\\
93	4.71480964841888\\
94	4.71724669409559\\
95	4.71400959367343\\
96	4.71067901330199\\
97	4.70741789232529\\
98	4.70500274822212\\
99	4.70593091701097\\
100	4.70789205124559\\
101	4.70760203142625\\
102	4.71147754033035\\
103	4.71542920069047\\
104	4.71910302935899\\
105	4.72381759192528\\
106	4.72887704703288\\
107	4.73226588279141\\
108	4.73507463340786\\
109	4.73899353941917\\
110	4.74279106203971\\
111	4.74756860778284\\
112	4.75028221342672\\
113	4.75162943966339\\
114	4.75391073597355\\
115	4.75603098213747\\
116	4.75737866490197\\
117	4.75755765418304\\
118	4.75887025963788\\
119	4.75913844585116\\
120	4.76001799063482\\
121	4.76038861816875\\
122	4.76042291412738\\
123	4.7611459705855\\
124	4.76130486750809\\
125	4.76100127504852\\
126	4.76146514900453\\
127	4.7628128759059\\
128	4.76332709419833\\
129	4.76189010263241\\
130	4.76181226921055\\
131	4.76287438313684\\
132	4.76216585770203\\
133	4.76140903961647\\
134	4.76074930280002\\
135	4.76081887874425\\
136	4.76011220858288\\
137	4.75923518530352\\
138	4.75870807470277\\
139	4.75767749319347\\
140	4.75530816096459\\
141	4.75310663002845\\
142	4.75046922807563\\
143	4.7477567034911\\
144	4.74471081434838\\
145	4.74165441725466\\
146	4.73836748156742\\
147	4.73469321465229\\
148	4.73178833344854\\
149	4.72932539053304\\
150	4.72021242136759\\
151	4.71127364595236\\
152	4.70245092317129\\
153	4.69331937831825\\
154	4.68516725383742\\
155	4.67671157049354\\
156	4.66830374603759\\
157	4.66051457024705\\
158	4.65246283685077\\
159	4.64390800676158\\
160	4.6353455743241\\
161	4.62648054555061\\
162	4.61813953444145\\
163	4.61028581168716\\
164	4.60294728070729\\
165	4.59632556063067\\
166	4.58969717200484\\
167	4.58284833164592\\
168	4.57595809802685\\
169	4.5692273513794\\
170	4.55794566704187\\
171	4.54684819461027\\
172	4.53570405802474\\
173	4.52458144631067\\
174	4.51388833741247\\
175	4.50335801033697\\
176	4.49273374755017\\
177	4.4819800050485\\
178	4.47144845730514\\
179	4.46140558815494\\
180	4.45115289947388\\
181	4.44116723436528\\
182	4.43120391391748\\
183	4.42130502019347\\
184	4.41176254637071\\
185	4.40224053211138\\
186	4.3925093672032\\
187	4.38309923866395\\
188	4.37376505211502\\
189	4.36442729082902\\
190	4.35520861657972\\
191	4.34623217352698\\
192	4.33743357788055\\
193	4.32822415221046\\
194	4.31914942899009\\
195	4.31026289974844\\
196	4.30158710310743\\
197	4.29311768834889\\
198	4.28487224379454\\
199	4.27646253854857\\
200	0\\
};
\addlegendentry{\hspace{-3mm}BP-TOMB/P}

\addplot [color=black, dotted, line width=1.0pt]
  table[row sep=crcr]{%
1	0\\
2	0\\
3	0\\
4	0\\
5	0.172953751867792\\
6	0.307059740798232\\
7	0.497723649036468\\
8	0.725836001831886\\
9	0.872288228049042\\
10	1.048998821227\\
11	1.13603590794804\\
12	1.3903652506131\\
13	1.51966687575497\\
14	1.70193791601831\\
15	1.91670303578992\\
16	2.06487575651839\\
17	2.20566678843211\\
18	2.31581131142664\\
19	2.46400530249114\\
20	2.569056801311\\
21	2.62633805345391\\
22	2.73869476401029\\
23	2.82335598727476\\
24	2.89180968429924\\
25	2.96123026703143\\
26	3.03241106348034\\
27	3.07752574573367\\
28	3.08741383139878\\
29	3.10728999161379\\
30	3.16800875309861\\
31	3.27420606309799\\
32	3.41584360830421\\
33	3.52636579176649\\
34	3.60677311126877\\
35	3.6941596444761\\
36	3.79067807503112\\
37	3.88241025027521\\
38	3.98582174224897\\
39	4.0899444808675\\
40	4.20023670498569\\
41	4.26875146578737\\
42	4.34020468825659\\
43	4.40091739754728\\
44	4.44395570513223\\
45	4.54557635507582\\
46	4.59051473147284\\
47	4.63849504234072\\
48	4.68119870671262\\
49	4.72171501529195\\
50	4.76617835185546\\
51	4.8111811863524\\
52	4.8558762595085\\
53	4.88893620762066\\
54	4.91562366850768\\
55	4.93488597255469\\
56	4.95310411197122\\
57	4.96614324276035\\
58	4.99520028204011\\
59	5.01775935653601\\
60	5.04019128030663\\
61	5.04939356435548\\
62	5.05929196690046\\
63	5.06490710697099\\
64	5.08174303838093\\
65	5.09634599776405\\
66	5.1163364950232\\
67	5.14249246604191\\
68	5.15832252071242\\
69	5.16910267536076\\
70	5.1775312636522\\
71	5.18438433458882\\
72	5.18974625910199\\
73	5.1946141960248\\
74	5.23479736324419\\
75	5.23887534571473\\
76	5.24150712923793\\
77	5.24572513949093\\
78	5.25039520450539\\
79	5.2511112664894\\
80	5.25501320438474\\
81	5.25869673789512\\
82	5.26404526466355\\
83	5.26755258872882\\
84	5.2691641865413\\
85	5.27448250832385\\
86	5.27746093406654\\
87	5.28603267740806\\
88	5.29439993561047\\
89	5.2991933896595\\
90	5.30259178606752\\
91	5.30356184388954\\
92	5.30587387671199\\
93	5.3062589195577\\
94	5.3048074995876\\
95	5.29893025872838\\
96	5.29192025162016\\
97	5.28760618239695\\
98	5.28617884632702\\
99	5.28649996232525\\
100	5.2870759024001\\
101	5.2821486808224\\
102	5.27817508297803\\
103	5.27459950894393\\
104	5.27120525172296\\
105	5.2673993386259\\
106	5.26354006383696\\
107	5.258445870542\\
108	5.25170695017827\\
109	5.24660291256658\\
110	5.2420535473866\\
111	5.23890049030493\\
112	5.23295229127886\\
113	5.22661314597502\\
114	5.22255055130325\\
115	5.21765788245807\\
116	5.21207879587145\\
117	5.20628697401045\\
118	5.20143319437748\\
119	5.19538593393593\\
120	5.18982792338022\\
121	5.18424412752039\\
122	5.17779282166568\\
123	5.17241484335129\\
124	5.16649808220672\\
125	5.16066329243244\\
126	5.15520798290903\\
127	5.15096869533947\\
128	5.1462632093154\\
129	5.14017730771072\\
130	5.13511570550835\\
131	5.13048361599814\\
132	5.12425260792623\\
133	5.11738341598818\\
134	5.11166177073739\\
135	5.10616297404139\\
136	5.10034613399251\\
137	5.09451534498605\\
138	5.088946567486\\
139	5.08261610084215\\
140	5.07572367940831\\
141	5.06899833851063\\
142	5.06223117964375\\
143	5.05566579435016\\
144	5.04885958516762\\
145	5.04258341360778\\
146	5.03614127656029\\
147	5.0293418200752\\
148	5.02316687675833\\
149	5.01735660706461\\
150	5.00523363873003\\
151	4.99353682451864\\
152	4.9819780187491\\
153	4.97072912752967\\
154	4.95980990284998\\
155	4.94915054968166\\
156	4.93868823609118\\
157	4.92873008283753\\
158	4.91846801102376\\
159	4.90785458007745\\
160	4.8974911513084\\
161	4.88681372268031\\
162	4.87623027877435\\
163	4.86645118260634\\
164	4.85696754256173\\
165	4.84797585992896\\
166	4.83896900219137\\
167	4.8297946052433\\
168	4.8205985914905\\
169	4.812109495989\\
170	4.79971222707545\\
171	4.78760512797265\\
172	4.77568675628999\\
173	4.76383861742186\\
174	4.75243389647395\\
175	4.74119587241058\\
176	4.72989688140475\\
177	4.71845800220888\\
178	4.70726106858772\\
179	4.69642651319366\\
180	4.68547575731029\\
181	4.67466127199157\\
182	4.66407128509857\\
183	4.65357727777023\\
184	4.64349794572072\\
185	4.63344054266227\\
186	4.62307688102514\\
187	4.61300779093464\\
188	4.60291301936161\\
189	4.59288087125105\\
190	4.58297764095718\\
191	4.5732814450321\\
192	4.56373430950639\\
193	4.55371635308976\\
194	4.54382740746557\\
195	4.53419050135324\\
196	4.52458902762567\\
197	4.5152572289144\\
198	4.5061360851915\\
199	4.49687608947867\\
200	0\\
};
\addlegendentry{\hspace{-3mm}BP-LMB}

\addplot [color=red, line width=1.0pt]
  table[row sep=crcr]{%
1	0\\
2	0\\
3	0\\
4	0\\
5	0.169305267003409\\
6	0.221453776566904\\
7	0.455980012043981\\
8	0.641191892376365\\
9	0.783607760823062\\
10	1.01317657682821\\
11	1.12961874905533\\
12	1.40500622676396\\
13	1.5413103323043\\
14	1.77145515679934\\
15	1.98749768497651\\
16	2.18236987802548\\
17	2.28472130656833\\
18	2.40610531591544\\
19	2.53826556067902\\
20	2.68838877955343\\
21	2.79235364472998\\
22	2.90983415619252\\
23	3.04030520831259\\
24	3.11736580321655\\
25	3.21142680521987\\
26	3.29229237773995\\
27	3.38330319659422\\
28	3.43025252050598\\
29	3.48714226972488\\
30	3.55252430134263\\
31	3.75460144039077\\
32	3.93732077265267\\
33	4.01001929433127\\
34	4.06647200822439\\
35	4.12868917130887\\
36	4.20091947110067\\
37	4.31869700989095\\
38	4.39033768577775\\
39	4.48749539487885\\
40	4.56310555924366\\
41	4.62070851524812\\
42	4.66975412065627\\
43	4.70733606328351\\
44	4.73641099453257\\
45	4.76452837975171\\
46	4.79344653653988\\
47	4.81849589199563\\
48	4.84703320494912\\
49	4.87273947673116\\
50	4.89411207108646\\
51	4.91350581921813\\
52	4.93574468356974\\
53	4.96035995652975\\
54	4.98826991907146\\
55	4.99662534521757\\
56	5.01387553225194\\
57	5.02726476146451\\
58	5.03939834578907\\
59	5.06662892915054\\
60	5.09156673172631\\
61	5.11042189330061\\
62	5.127379631602\\
63	5.14224452347808\\
64	5.16123121347213\\
65	5.17575673962284\\
66	5.18871459709932\\
67	5.20535908168125\\
68	5.21526882038451\\
69	5.22327868698598\\
70	5.23720296546783\\
71	5.25264078136005\\
72	5.25810541679201\\
73	5.2635662574578\\
74	5.27903982822903\\
75	5.29803210621426\\
76	5.30397096503779\\
77	5.30919791037795\\
78	5.3091118060291\\
79	5.35775822681624\\
80	5.3950112100355\\
81	5.39456734682818\\
82	5.39457351583982\\
83	5.39477830709491\\
84	5.3957006346977\\
85	5.39845567201759\\
86	5.40348383026208\\
87	5.41535294411514\\
88	5.41831875287004\\
89	5.4200316152794\\
90	5.42200775620123\\
91	5.42218265072821\\
92	5.42539567201977\\
93	5.42640857651279\\
94	5.42730018127276\\
95	5.42802923913415\\
96	5.42307489518647\\
97	5.4151095301032\\
98	5.46838625244322\\
99	5.46855621774997\\
100	5.50403955191306\\
101	5.52381812794587\\
102	5.52445617673252\\
103	5.52519499522055\\
104	5.52526854891259\\
105	5.52516185030771\\
106	5.52592410238576\\
107	5.54102127077496\\
108	5.55544093804073\\
109	5.56960566578819\\
110	5.56985276094781\\
111	5.57051422731419\\
112	5.57914034220286\\
113	5.57677635951257\\
114	5.58495231653915\\
115	5.582139622899\\
116	5.58053386104223\\
117	5.57906619479889\\
118	5.57872216573408\\
119	5.57759531264713\\
120	5.5757661701694\\
121	5.57421840930703\\
122	5.56916354290702\\
123	5.56429774525192\\
124	5.55905046088695\\
125	5.55314014809881\\
126	5.54822719778422\\
127	5.54376130220266\\
128	5.53869020859348\\
129	5.53224781005293\\
130	5.52660946209854\\
131	5.52205268292489\\
132	5.51498371470285\\
133	5.50886750969901\\
134	5.50311302703768\\
135	5.49833526753698\\
136	5.49301802964567\\
137	5.48728960402967\\
138	5.48186817912227\\
139	5.47630045989698\\
140	5.46952821683726\\
141	5.46290454114693\\
142	5.45566861678528\\
143	5.44832554393148\\
144	5.44081703006661\\
145	5.43375070657273\\
146	5.4262567364752\\
147	5.41815469888698\\
148	5.41040284818037\\
149	5.40331766211229\\
150	5.3903973399521\\
151	5.37768654769006\\
152	5.36520248529243\\
153	5.35289977053788\\
154	5.34109814024901\\
155	5.32892819825985\\
156	5.31684471626709\\
157	5.30521350459613\\
158	5.29357540904484\\
159	5.28175067805064\\
160	5.27011922131233\\
161	5.25829031665753\\
162	5.2466045571701\\
163	5.23602217122992\\
164	5.22597373587308\\
165	5.21638867126385\\
166	5.20668078032539\\
167	5.19688226316972\\
168	5.18705662900294\\
169	5.17793405546985\\
170	5.16445816568219\\
171	5.15118441503764\\
172	5.13793899325636\\
173	5.12483188893582\\
174	5.11208811582712\\
175	5.0995624357189\\
176	5.08700552913493\\
177	5.07429483361536\\
178	5.06183550680734\\
179	5.04974531094655\\
180	5.03755889533892\\
181	5.02553566268049\\
182	5.01356248187765\\
183	5.00170461832171\\
184	4.99010036739015\\
185	4.97862969082615\\
186	4.96698186652024\\
187	4.95559119898502\\
188	4.94425185707579\\
189	4.9329884169503\\
190	4.92184828713284\\
191	4.91096999801739\\
192	4.90024338686704\\
193	4.88926005327491\\
194	4.87836516734091\\
195	4.86762893875697\\
196	4.8571311731943\\
197	4.84688433599681\\
198	4.83681590510174\\
199	4.82665248815146\\
200	0\\
};
\addlegendentry{\hspace{-3mm}BP-LMB/P (proposed)}

\node[right, align=left,scale=1.1]
at (axis cs:29.8mm,-0.7mm) {(a)};
\end{axis}
\end{tikzpicture}%
\end{minipage}
\begin{minipage}[H!]{0.235\textwidth}
    \begin{tikzpicture}

\begin{axis}[%
width=1.3in,
height=1.3in,
at={(0.758in,0.481in)},
scale only axis,
clip=false,
xmin=0,
xmax=200,
xtick={  0,  40,  80, 120, 160, 200},
xlabel style={at={(0.5,-2.5mm)},font=\color{white!15!black}},
xlabel={$k$},
ymin=0,
ymax=10,
ylabel style={at={(-5.5mm,0.5)},font=\color{white!15!black}},
ylabel={False Error},
axis background/.style={fill=white},
]
\addplot [color=blue, line width=1.0pt]
  table[row sep=crcr]{%
1	0\\
2	0\\
3	0\\
4	0\\
5	0\\
6	0\\
7	0\\
8	0\\
9	0\\
10	0\\
11	0\\
12	0\\
13	0\\
14	0\\
15	0\\
16	0\\
17	0\\
18	0\\
19	0\\
20	0\\
21	0\\
22	0\\
23	0\\
24	0\\
25	0\\
26	0\\
27	0\\
28	0\\
29	0\\
30	0\\
31	0\\
32	0\\
33	0\\
34	0\\
35	0\\
36	0\\
37	0\\
38	0\\
39	0\\
40	0\\
41	0\\
42	0\\
43	0.215665546406877\\
44	0.21320071635561\\
45	0.210818510677892\\
46	0.208514414057075\\
47	0.206284249251759\\
48	0.204124145231932\\
49	0.202030508910442\\
50	0.2\\
51	0.198029508595335\\
52	0.196116135138184\\
53	0.194257172471453\\
54	0.192450089729875\\
55	0.190692517849118\\
56	0.188982236504614\\
57	0.187317162316339\\
58	0.185695338177052\\
59	0.184114923579665\\
60	0.182574185835055\\
61	0.181071492085037\\
62	0.179605302026775\\
63	0.17817416127495\\
64	0.176776695296637\\
65	0.350823207722812\\
66	0.420260637914363\\
67	0.472026485948562\\
68	0.514495755427527\\
69	0.55094479996519\\
70	0.583070186551116\\
71	0.578949497555857\\
72	0.607625218510765\\
73	0.633685766504996\\
74	0.629389542280634\\
75	0.625179531537246\\
76	0.621052888871886\\
77	0.617006898643628\\
78	0.640512615220349\\
79	0.662266062261774\\
80	0.682518307093495\\
81	0.678292146910725\\
82	0.674143532171418\\
83	0.670070120062267\\
84	0.666069665681251\\
85	0.662140016838642\\
86	0.658279109191478\\
87	0.654484961686524\\
88	0.650755672288882\\
89	0.647089413975298\\
90	0.64348443097303\\
91	0.639939035226662\\
92	0.78389355923162\\
93	0.779667687156627\\
94	0.775509429842686\\
95	0.771417003229792\\
96	0.767388688472982\\
97	0.763422828908996\\
98	0.759517827193629\\
99	0.755672142598512\\
100	0.751884288456962\\
101	0.748152829749252\\
102	0.744476380818426\\
103	0.740853603208371\\
104	0.737283203616515\\
105	0.733763931954005\\
106	0.730294579506798\\
107	0.72687397719148\\
108	0.723500993900129\\
109	0.720174534928881\\
110	0.716893540485255\\
111	0.713656984269604\\
112	0.710463872126384\\
113	0.707313240761216\\
114	0.704204156519988\\
115	0.701135714226464\\
116	0.698107036075138\\
117	0.695117270576239\\
118	0.69216559155002\\
119	0.689251197167623\\
120	0.686373309035999\\
121	0.683531171324511\\
122	0.680724049930993\\
123	0.677951231685171\\
124	0.675212023587497\\
125	0.672505752081531\\
126	0.669831762358157\\
127	0.667189417689988\\
128	0.664578098794425\\
129	0.661997203223929\\
130	0.659446144782138\\
131	0.656924352964546\\
132	0.654431272422529\\
133	0.651966362449581\\
134	0.649529096488677\\
135	0.647118961659739\\
136	0.644735458306255\\
137	0.642378099560123\\
138	0.640046410923884\\
139	0.637739929869497\\
140	0.635458205452925\\
141	0.633200797943763\\
142	0.630967278469253\\
143	0.628757228672001\\
144	0.626570240380802\\
145	0.624405915293962\\
146	0.622263864674583\\
147	0.620143709057253\\
148	0.618045077965668\\
149	0.615967609640689\\
150	2.81031404580361\\
151	3.80191786063655\\
152	4.55257601435441\\
153	5.17516209422818\\
154	5.71037515214032\\
155	6.17162107031322\\
156	6.59450377398377\\
157	6.89744935336012\\
158	7.08403707532535\\
159	7.21053543915913\\
160	7.26557502800765\\
161	7.29410350982189\\
162	7.30383475425301\\
163	7.31309067059798\\
164	7.33057453287202\\
165	7.3481513443028\\
166	7.34865721568595\\
167	7.3421134078437\\
168	7.33552155073292\\
169	7.32115459631183\\
170	7.54516358309673\\
171	7.75181491325537\\
172	7.92933522908126\\
173	8.07987559525502\\
174	8.22584262992695\\
175	8.35396999560116\\
176	8.44516583074638\\
177	8.46748068829449\\
178	8.46876509494694\\
179	8.47672910876164\\
180	8.46578163535442\\
181	8.45486939231207\\
182	8.44399325014789\\
183	8.43315401162512\\
184	8.41664768991305\\
185	8.3938691699603\\
186	8.37127459554457\\
187	8.34886150419852\\
188	8.32662747935988\\
189	8.30457014927707\\
190	8.28268718594643\\
191	8.26097630408023\\
192	8.23943526010412\\
193	8.21806185118333\\
194	8.19685391427643\\
195	8.17580932521584\\
196	8.15492599781419\\
197	8.1342018829956\\
198	8.11363496795122\\
199	8.09322327531806\\
200	0\\
};

\addplot [color=green, dashdotted, line width=1.0pt]
  table[row sep=crcr]{%
1	0\\
2	0\\
3	0\\
4	0\\
5	0\\
6	0\\
7	0\\
8	0\\
9	0\\
10	0\\
11	0.426401432711221\\
12	0.577350269189626\\
13	0.679366220486757\\
14	1.51185789203691\\
15	1.91194169593806\\
16	2.07313218497099\\
17	2.35423081134452\\
18	2.53191803757505\\
19	2.97888511249567\\
20	3.42532290148367\\
21	3.47060194997799\\
22	3.48663886594017\\
23	3.41000008155464\\
24	3.45777572107699\\
25	3.56771036176239\\
26	3.57274357201643\\
27	3.57020699491604\\
28	3.5629130376544\\
29	3.5524854183585\\
30	3.49277552853431\\
31	3.5411889953633\\
32	3.81487806540577\\
33	3.92456909296003\\
34	4.17394696413588\\
35	4.22133564209963\\
36	4.54188570358994\\
37	4.65875230511795\\
38	4.65120196033634\\
39	4.59118393669419\\
40	4.62458216207336\\
41	4.61497365609328\\
42	4.55970251667446\\
43	4.50637082801937\\
44	4.5431783838959\\
45	4.6886322229333\\
46	4.68661237811075\\
47	4.68051228899993\\
48	4.67156206338149\\
49	4.69521638261393\\
50	4.72159396930532\\
51	4.70721043294029\\
52	4.76682739235705\\
53	4.76750107239269\\
54	4.7642242991489\\
55	4.81874928723294\\
56	5.01189544615791\\
57	5.21735012974248\\
58	5.36931197497732\\
59	5.48393755141083\\
60	5.53032332151177\\
61	5.54357319920757\\
62	5.55289441065358\\
63	5.60900036487025\\
64	5.65935417017808\\
65	5.70486083258796\\
66	5.79995492287094\\
67	5.93559289832587\\
68	6.03749356443505\\
69	6.10587610431493\\
70	6.13903095674548\\
71	6.14940316505633\\
72	6.15872772651974\\
73	6.15038032954541\\
74	6.15833069170074\\
75	6.14785551837281\\
76	6.13726038985312\\
77	6.12656777595249\\
78	6.10079958918043\\
79	6.07696668877117\\
80	6.066691245392\\
81	6.04335303048827\\
82	6.03400153295494\\
83	6.02416659854275\\
84	6.00115027397831\\
85	5.97844014123986\\
86	5.97760273860902\\
87	5.995479292717\\
88	5.99382922532656\\
89	5.99152547690781\\
90	6.0168845518859\\
91	6.03894098039338\\
92	6.02157368165054\\
93	6.00423848260636\\
94	5.98695100551634\\
95	5.95535737598429\\
96	5.92425869148639\\
97	5.8936421625434\\
98	5.88293918780767\\
99	5.85315196254641\\
100	5.82381267020964\\
101	5.79491019567473\\
102	5.7664338061645\\
103	5.73837313450173\\
104	5.71071816325151\\
105	5.68345920969618\\
106	5.65658691159173\\
107	5.63009221365812\\
108	5.60396635475915\\
109	5.58871455584873\\
110	5.56325330928222\\
111	5.55576495582384\\
112	5.5309068223839\\
113	5.50637939854704\\
114	5.4821754158875\\
115	5.45828782767763\\
116	5.46273345096279\\
117	5.43933833938622\\
118	5.42622646338994\\
119	5.40337909259373\\
120	5.38081791225027\\
121	5.3585369970315\\
122	5.33653059194952\\
123	5.31479310611185\\
124	5.2933191067542\\
125	5.27210331353641\\
126	5.25114059308781\\
127	5.23042595378928\\
128	5.20995454077997\\
129	5.18972163117735\\
130	5.16972262949988\\
131	5.14995306328215\\
132	5.13040857887321\\
133	5.11108493740884\\
134	5.09197801094958\\
135	5.08763312313121\\
136	5.06889407926333\\
137	5.05036058364583\\
138	5.03202890579152\\
139	5.0138954093165\\
140	4.99595654891001\\
141	4.97820886742257\\
142	4.96064899306727\\
143	4.94327363672873\\
144	4.9260795893751\\
145	4.90906371956844\\
146	4.89222297106892\\
147	4.87555436052898\\
148	4.85905497527314\\
149	4.84272197116009\\
150	5.51702363179062\\
151	6.07228431682763\\
152	6.57019441952791\\
153	7.0136515601243\\
154	7.40414712829627\\
155	7.71074730363546\\
156	7.87145894861522\\
157	7.96885573276587\\
158	8.03476003642937\\
159	8.03323106375648\\
160	8.01573067531565\\
161	8.00584699479997\\
162	7.99668577262046\\
163	7.98051589570944\\
164	7.97119621395419\\
165	7.95516537346695\\
166	7.93921625887243\\
167	7.92334982819534\\
168	7.89973323370187\\
169	7.87632656526413\\
170	8.07003080723748\\
171	8.24254019840288\\
172	8.40173928561969\\
173	8.54876627542361\\
174	8.67069475461516\\
175	8.76325458014934\\
176	8.80873602703733\\
177	8.81625685310629\\
178	8.81696213814562\\
179	8.80458898004529\\
180	8.79227123183038\\
181	8.78000920025333\\
182	8.76780314944623\\
183	8.74979086866296\\
184	8.72598186978579\\
185	8.70236617866372\\
186	8.68539702132705\\
187	8.6621429045741\\
188	8.63907457359277\\
189	8.62269562017001\\
190	8.5999743800967\\
191	8.57743181346018\\
192	8.55506559074089\\
193	8.53287342471945\\
194	8.51085306949422\\
195	8.48900231952653\\
196	8.46731900871284\\
197	8.44580100948278\\
198	8.42444623192226\\
199	8.40325262292088\\
200	0\\
};

\addplot [color=black, dotted, line width=1.0pt]
  table[row sep=crcr]{%
1	0\\
2	0\\
3	0\\
4	0\\
5	0\\
6	0\\
7	0\\
8	0\\
9	0\\
10	0\\
11	0.426401432711221\\
12	0.577350269189626\\
13	0.679366220486757\\
14	0.755928946018454\\
15	0.730296743340222\\
16	1.06066017177982\\
17	1.02899151085505\\
18	1\\
19	1.2977713690461\\
20	1.33956231322022\\
21	1.50097142641462\\
22	1.62146834766232\\
23	1.71870939315768\\
24	1.75066893989044\\
25	2.10703388540065\\
26	2.23543204843217\\
27	2.5155281732588\\
28	2.62675763940543\\
29	2.70715576391663\\
30	2.70150636473639\\
31	2.84351759535704\\
32	3.19518160137329\\
33	3.27047778795412\\
34	3.25779197038985\\
35	3.48377347986975\\
36	3.68355655278409\\
37	3.76847470961434\\
38	4.22528917655785\\
39	4.44995968201106\\
40	4.6920315769868\\
41	4.77556656248884\\
42	4.82368705686\\
43	4.79052511001963\\
44	4.81915487715059\\
45	4.85882410146943\\
46	4.80572060426065\\
47	4.79480671539566\\
48	4.99447163364693\\
49	5.16528040102681\\
50	5.13279329943721\\
51	5.18792065174424\\
52	5.19642043348332\\
53	5.25454156761588\\
54	5.27234649007724\\
55	5.38611250138048\\
56	5.39291502791698\\
57	5.53752280580146\\
58	5.65273714971163\\
59	5.87853109706879\\
60	5.99527627378211\\
61	6.09484298363484\\
62	6.11355583630788\\
63	6.17215538725207\\
64	6.29870594414637\\
65	6.37882540427034\\
66	6.41660738347178\\
67	6.57424772239983\\
68	6.70842833499355\\
69	6.69461759083877\\
70	6.72192990721874\\
71	6.72242071188047\\
72	6.76330749564988\\
73	6.7372003199303\\
74	6.75412046323011\\
75	6.79141723429076\\
76	6.76408287643813\\
77	6.73985698888803\\
78	6.71339343913593\\
79	6.71728565937279\\
80	6.76236662680624\\
81	6.79296382888707\\
82	6.83925174038531\\
83	6.86418918984967\\
84	6.8868842819372\\
85	6.90760146916886\\
86	6.89667696508634\\
87	6.88875438967936\\
88	6.87702982717521\\
89	6.86766051801549\\
90	6.85805739967843\\
91	6.88373637004074\\
92	6.90787994137366\\
93	6.91441738313521\\
94	6.92020682095102\\
95	6.9253109941013\\
96	6.91422044525134\\
97	6.90305960031065\\
98	6.90365925052368\\
99	6.90310806374043\\
100	6.91683034587565\\
101	6.94041663564337\\
102	6.93712103632783\\
103	6.93067542632091\\
104	6.92391264481922\\
105	6.92771128540988\\
106	6.9058840044183\\
107	6.88794991216593\\
108	6.88789495432567\\
109	6.86991024917818\\
110	6.85196387831484\\
111	6.85111721344359\\
112	6.86610636013602\\
113	6.86402118464029\\
114	6.84924829912921\\
115	6.8464778073847\\
116	6.86246058609587\\
117	6.85881832553942\\
118	6.86506512591993\\
119	6.84953163880496\\
120	6.83083386361082\\
121	6.81553640499875\\
122	6.78754640417157\\
123	6.75989844239359\\
124	6.73258560971854\\
125	6.70560119006877\\
126	6.67893865429746\\
127	6.66343274062234\\
128	6.647993654993\\
129	6.62217610637383\\
130	6.60706605021229\\
131	6.58179993070344\\
132	6.55682147273543\\
133	6.53212525890819\\
134	6.50770601358688\\
135	6.49363213255881\\
136	6.47961613302217\\
137	6.45592458703076\\
138	6.43249102663026\\
139	6.40931080339611\\
140	6.38637938532338\\
141	6.37430280419865\\
142	6.36222749258261\\
143	6.35015952118809\\
144	6.33810436073005\\
145	6.3260669456369\\
146	6.31405172976977\\
147	6.29253871414579\\
148	6.2712441058193\\
149	6.25016423405504\\
150	6.78524962782931\\
151	7.23479642142166\\
152	7.63294469348554\\
153	8.00355553657127\\
154	8.33762619738876\\
155	8.60783728301779\\
156	8.73065665909588\\
157	8.8136721505642\\
158	8.85867502506936\\
159	8.84511076985994\\
160	8.83160282652424\\
161	8.81708951778592\\
162	8.81024682206288\\
163	8.79619763708277\\
164	8.78192406135794\\
165	8.76765878479189\\
166	8.75340654300363\\
167	8.73948093987299\\
168	8.71343175841366\\
169	8.68761412608476\\
170	8.85766552410198\\
171	9.009389236948\\
172	9.14415553164427\\
173	9.27613591507705\\
174	9.38565532193339\\
175	9.46974598518536\\
176	9.5207822148888\\
177	9.52284871795811\\
178	9.5253374380229\\
179	9.50984098912588\\
180	9.49444282653455\\
181	9.47914220163443\\
182	9.46393836180914\\
183	9.44311025971799\\
184	9.41741467396682\\
185	9.39192771336712\\
186	9.3730529256605\\
187	9.34795769207134\\
188	9.32306295361969\\
189	9.30467339624014\\
190	9.2863990583918\\
191	9.26205721034262\\
192	9.23790578146385\\
193	9.21394230197697\\
194	9.19016434671906\\
195	9.16656953411186\\
196	9.1431555251598\\
197	9.1199200224761\\
198	9.09686076933586\\
199	9.07397554875547\\
200	0\\
};

\addplot [color=red, line width=1.0pt]
  table[row sep=crcr]{%
1	0\\
2	0\\
3	0\\
4	0\\
5	0\\
6	0\\
7	0\\
8	0\\
9	0\\
10	0\\
11	0.426401432711221\\
12	0.577350269189626\\
13	0.679366220486757\\
14	1.13389341902768\\
15	1.54679332426795\\
16	1.99767619622864\\
17	2.28102774977509\\
18	2.35483170074866\\
19	2.42641349559491\\
20	2.46548429570847\\
21	2.69784547461212\\
22	2.95240381410683\\
23	3.01098870081307\\
24	3.33264184779148\\
25	3.30896476997885\\
26	3.40049179396979\\
27	3.42342993602588\\
28	3.4964457158273\\
29	3.55367511472821\\
30	3.49394522858291\\
31	3.58591922852533\\
32	3.65544882260508\\
33	3.59963716618484\\
34	3.65532353562874\\
35	3.64964169848817\\
36	3.6919563393631\\
37	3.72724307122056\\
38	3.93538599933707\\
39	4.04114689648724\\
40	4.17521409598328\\
41	4.3070959619377\\
42	4.39174636661277\\
43	4.65665838924824\\
44	4.66819257067048\\
45	4.76458916345972\\
46	4.89279490220036\\
47	5.12164693020823\\
48	5.12201776313657\\
49	5.39281024229248\\
50	5.5586808672826\\
51	5.5896209966273\\
52	5.56926193929184\\
53	5.57190928534948\\
54	5.60116702501491\\
55	5.72582460463505\\
56	5.84906278422662\\
57	5.85898672064658\\
58	5.91988159320641\\
59	5.96798262462842\\
60	6.02731307780609\\
61	6.0554552684507\\
62	6.0699475394472\\
63	6.11651128446828\\
64	6.13885365626521\\
65	6.13834269597581\\
66	6.13702846611173\\
67	6.15852521801311\\
68	6.17621310723997\\
69	6.36275335664945\\
70	6.44962362389308\\
71	6.44323713229694\\
72	6.43651966816393\\
73	6.50314521501606\\
74	6.53943599602637\\
75	6.51990144527468\\
76	6.4894563744586\\
77	6.45954030954096\\
78	6.43013891100851\\
79	6.3893122194382\\
80	6.34925344071995\\
81	6.32171674001725\\
82	6.31385385595963\\
83	6.31690269198975\\
84	6.29026798512326\\
85	6.27485106628647\\
86	6.24888790556675\\
87	6.2409608575773\\
88	6.26100737598171\\
89	6.27975015631663\\
90	6.31375954743982\\
91	6.3450963041423\\
92	6.37629234333327\\
93	6.40989641633941\\
94	6.46055000090225\\
95	6.52151433956273\\
96	6.50122181745316\\
97	6.49098152925895\\
98	6.47033284198362\\
99	6.45957686088882\\
100	6.46867082108183\\
101	6.46417051080446\\
102	6.62084498138496\\
103	6.63563488415189\\
104	6.62485320708168\\
105	6.63742405823564\\
106	6.67551639737855\\
107	6.67752846413693\\
108	6.66598038519779\\
109	6.69314538267757\\
110	6.69034408925446\\
111	6.68744207976061\\
112	6.68750487815934\\
113	6.70073605068372\\
114	6.71340588256364\\
115	6.72554475145485\\
116	6.74057423351615\\
117	6.75160370038523\\
118	6.74235010498901\\
119	6.72499717918091\\
120	6.70775178979429\\
121	6.71567425563232\\
122	6.71561935657568\\
123	6.68826437787941\\
124	6.66908712815269\\
125	6.6423572124592\\
126	6.61594614479157\\
127	6.59867727165392\\
128	6.59571453875831\\
129	6.60007832236375\\
130	6.59067189773949\\
131	6.57293307843232\\
132	6.55537085567674\\
133	6.53798253476684\\
134	6.52181583891821\\
135	6.53687066862863\\
136	6.52798154246465\\
137	6.51912682258994\\
138	6.51030759799634\\
139	6.50089253951077\\
140	6.48446616131325\\
141	6.46819504694285\\
142	6.45207690335961\\
143	6.42947761534785\\
144	6.40711414718862\\
145	6.38498242597158\\
146	6.36307847659301\\
147	6.34139841875659\\
148	6.31993846408582\\
149	6.29869491334326\\
150	6.93264363581084\\
151	7.44169619962564\\
152	7.87785342887993\\
153	8.25448692623898\\
154	8.58863769582869\\
155	8.85081702030739\\
156	8.98241558015859\\
157	9.06231258929114\\
158	9.09446059184328\\
159	9.07698479112546\\
160	9.06157662109859\\
161	9.05421004485424\\
162	9.05033763206465\\
163	9.03753975153209\\
164	9.03205116706811\\
165	9.02654984410451\\
166	9.01367310316283\\
167	8.9990330160269\\
168	8.97916679493589\\
169	8.9594420620099\\
170	9.13216272818238\\
171	9.28039139508834\\
172	9.41800784318569\\
173	9.54489465462548\\
174	9.65074326211777\\
175	9.73253675653515\\
176	9.78318051833225\\
177	9.78321419850383\\
178	9.77729682555878\\
179	9.76005662223947\\
180	9.74294074380326\\
181	9.72594782977061\\
182	9.70907653571083\\
183	9.68797716631116\\
184	9.66161527481719\\
185	9.63546741828079\\
186	9.60953071600387\\
187	9.58380234127762\\
188	9.55827952008848\\
189	9.53295952986183\\
190	9.50783969824206\\
191	9.4829174019079\\
192	9.4581900654216\\
193	9.43365516011108\\
194	9.40931020298372\\
195	9.38515275567085\\
196	9.36118042340196\\
197	9.33739085400744\\
198	9.31378173694926\\
199	9.29035080237827\\
200	0\\
};

\node[right, align=left,scale=1.1]
at (axis cs:29.8mm,-0.9mm) {(b)};
\end{axis}
\end{tikzpicture}%
\end{minipage}
\begin{minipage}[H!]{0.235\textwidth}
    \begin{tikzpicture}

\begin{axis}[%
width=1.3in,
height=1.3in,
at={(0.758in,0.481in)},
scale only axis,
clip=false,
xmin=0,
xmax=200,
xtick={  0,  40,  80, 120, 160, 200},
xlabel style={at={(0.5,-2.5mm)},font=\color{white!15!black}},
xlabel={$k$},
ymin=0,
ymax=40,
ylabel style={at={(-5.5mm,0.5)},font=\color{white!15!black}},
ylabel={Missed Error},
axis background/.style={fill=white},
]
\addplot [color=blue, line width=1.0pt]
  table[row sep=crcr]{%
1	0\\
2	31.6227766016838\\
3	31.6227766016838\\
4	31.6227766016838\\
5	31.4297581359336\\
6	31.2971920334269\\
7	31.2018080739052\\
8	31.0492274269555\\
9	30.8574344926076\\
10	30.6321200397036\\
11	30.3842076935793\\
12	30.1200270317134\\
13	29.9136404116849\\
14	29.7135419979462\\
15	29.374365271599\\
16	29.0575184005644\\
17	28.7743183841335\\
18	28.5171038874164\\
19	28.2265022026866\\
20	28.4805373782954\\
21	28.7250125604022\\
22	28.885214419771\\
23	28.9874970663582\\
24	29.0660515517964\\
25	29.1499353921155\\
26	29.1877869242437\\
27	29.2355212690672\\
28	29.2807213487846\\
29	29.3223415049892\\
30	29.3129130378074\\
31	29.2961146576598\\
32	29.2570948101016\\
33	29.2224953452031\\
34	29.1570963838533\\
35	29.1059674470175\\
36	29.0471090619375\\
37	29.002389087174\\
38	28.9423658300182\\
39	28.8850706177625\\
40	28.9974491806486\\
41	29.0966492003818\\
42	29.1884872063681\\
43	29.283916736441\\
44	29.3305479315524\\
45	29.3877419308819\\
46	29.4221337379104\\
47	29.435819988656\\
48	29.4457526167101\\
49	29.4491016401532\\
50	29.4395120147395\\
51	29.4176949790374\\
52	29.3913145811402\\
53	29.3714042521857\\
54	29.3585502668042\\
55	29.3335500521832\\
56	29.314521966417\\
57	29.284306429676\\
58	29.2719350726256\\
59	29.2189633912023\\
60	29.1674666133611\\
61	29.121752424755\\
62	29.072214561002\\
63	29.013894375162\\
64	28.9428158444663\\
65	28.8689589403269\\
66	28.791574215263\\
67	28.7111648219711\\
68	28.6386717032254\\
69	28.5634741413422\\
70	28.4844793599998\\
71	28.4074251906102\\
72	28.3322390053888\\
73	28.2588055476649\\
74	28.1829691748298\\
75	28.1043049317972\\
76	28.0228476984587\\
77	27.9568554054403\\
78	27.8925136743696\\
79	27.8117846257352\\
80	27.7337745914564\\
81	27.6482035911097\\
82	27.5687435223978\\
83	27.4861785358998\\
84	27.3963047103096\\
85	27.3081844306046\\
86	27.2259648438048\\
87	27.1499619536367\\
88	27.0754318436763\\
89	26.9982385351173\\
90	26.9178776266674\\
91	26.8479578015387\\
92	26.7833119844852\\
93	26.7069297461591\\
94	26.636537377249\\
95	26.5629745731892\\
96	26.4942002851118\\
97	26.4147402796944\\
98	26.3409351785956\\
99	26.2641205861873\\
100	26.189041426438\\
101	26.1271815449123\\
102	26.057614514454\\
103	25.9751359167837\\
104	25.8898090410408\\
105	25.802618312059\\
106	25.7126529609366\\
107	25.6240419066564\\
108	25.5367573025736\\
109	25.4548218029542\\
110	25.3706819760966\\
111	25.2837578691223\\
112	25.198071600169\\
113	25.1176938422639\\
114	25.0378725604806\\
115	24.9591627018263\\
116	24.881432080146\\
117	24.80146676495\\
118	24.7257246576592\\
119	24.6542770446275\\
120	24.5807929918796\\
121	24.5113178206436\\
122	24.4333746704954\\
123	24.3594406920626\\
124	24.2831789926887\\
125	24.2170833815521\\
126	24.148927225629\\
127	24.0784002552733\\
128	24.002871694541\\
129	23.9319763999044\\
130	23.8684542916138\\
131	23.7991836319464\\
132	23.7307377898942\\
133	23.6663620238117\\
134	23.602315067056\\
135	23.53647440776\\
136	23.4644625257748\\
137	23.393270801653\\
138	23.3264590489202\\
139	23.2604055492295\\
140	23.1984824856425\\
141	23.1374615752479\\
142	23.0744311131976\\
143	23.0085648422661\\
144	22.9407327188962\\
145	22.8771412390091\\
146	22.8108073349169\\
147	22.7416998352963\\
148	22.6733123977025\\
149	22.6056327916547\\
150	22.5354291248431\\
151	22.4659360210909\\
152	22.3971416974794\\
153	22.3263991630561\\
154	22.2563525759694\\
155	22.1903172529118\\
156	22.1249285411318\\
157	22.0601760517219\\
158	21.996049627162\\
159	21.9325393347447\\
160	21.8696354602287\\
161	21.8073285017088\\
162	21.7456091636952\\
163	21.6844683513929\\
164	21.6238971651735\\
165	21.5638868952318\\
166	21.5012735147314\\
167	21.4392282906827\\
168	21.3777427267315\\
169	21.3168085030302\\
170	21.2540195345245\\
171	21.1917821548372\\
172	21.1300883349791\\
173	21.068930208633\\
174	21.0083000679413\\
175	20.9481903594247\\
176	20.8885936800298\\
177	20.8295027732994\\
178	20.7709105256608\\
179	20.7128099628289\\
180	20.6551942463191\\
181	20.6011892826375\\
182	20.5476331123191\\
183	20.4914151325164\\
184	20.4356560764196\\
185	20.3803497339381\\
186	20.3254900119961\\
187	20.2710709317123\\
188	20.2170866256635\\
189	20.1635313352269\\
190	20.1103994079996\\
191	20.0576852952936\\
192	20.0053835497007\\
193	19.9534888227289\\
194	19.9019958625044\\
195	19.8508995115395\\
196	19.8001947045624\\
197	19.7498764664082\\
198	19.6999399099686\\
199	19.6503802341978\\
200	0\\
};

\addplot [color=green, dashdotted, line width=1.0pt]
  table[row sep=crcr]{%
1	0\\
2	31.6227766016838\\
3	31.6227766016838\\
4	31.6227766016838\\
5	31.4297581359336\\
6	31.4073476323766\\
7	31.2971936198161\\
8	31.1695900833541\\
9	30.9249358725346\\
10	30.6559639259871\\
11	30.3406578583245\\
12	30.0179900167249\\
13	29.7412183812702\\
14	29.4489007223978\\
15	29.1469060956315\\
16	28.7729763724026\\
17	28.4142141164622\\
18	28.1524608286108\\
19	27.820303701681\\
20	28.0283676260067\\
21	28.2184312839662\\
22	28.3805811992347\\
23	28.4979866815451\\
24	28.5542070678707\\
25	28.5847061416461\\
26	28.5558087780427\\
27	28.513128312814\\
28	28.4387475352284\\
29	28.4037724150282\\
30	28.2992005367875\\
31	28.2279655024781\\
32	28.1045385367206\\
33	27.9871209417138\\
34	27.8552645206545\\
35	27.7415562157393\\
36	27.6108859510155\\
37	27.4670191411108\\
38	27.3070418024945\\
39	27.1654820586801\\
40	27.2003749277568\\
41	27.2233726261819\\
42	27.2383533603777\\
43	27.2517544560163\\
44	27.2365014641544\\
45	27.21707763926\\
46	27.1693222575944\\
47	27.1088267970492\\
48	27.0549572012769\\
49	26.9965286595613\\
50	26.9334041929892\\
51	26.8724676183963\\
52	26.8079359915985\\
53	26.738871278106\\
54	26.6593642730985\\
55	26.5806048039496\\
56	26.4942791480145\\
57	26.3819842340146\\
58	26.2799028559988\\
59	26.1413531944559\\
60	26.0151570621784\\
61	25.8972588223648\\
62	25.7549728901572\\
63	25.6119696228805\\
64	25.4713465402553\\
65	25.3243230820308\\
66	25.169063659874\\
67	25.0305342917067\\
68	24.8870502046794\\
69	24.7471350490944\\
70	24.6054178172657\\
71	24.4671378732666\\
72	24.3376467047835\\
73	24.2411644563384\\
74	24.1572541560366\\
75	24.0524247264999\\
76	23.9615113068029\\
77	23.862498283212\\
78	23.7520671747493\\
79	23.6428752863528\\
80	23.5490921412454\\
81	23.4445945803993\\
82	23.3424221593121\\
83	23.2316293405408\\
84	23.113434352269\\
85	22.99263693535\\
86	22.8746582849824\\
87	22.7485868418993\\
88	22.6240753086582\\
89	22.5158219739545\\
90	22.4000034057525\\
91	22.2863195727834\\
92	22.1752251902153\\
93	22.0659576628925\\
94	21.9584678955642\\
95	21.8570022014718\\
96	21.7566834372545\\
97	21.6531282851533\\
98	21.5511906983227\\
99	21.4466526247231\\
100	21.354182977506\\
101	21.2664904080963\\
102	21.1708790273527\\
103	21.072312922192\\
104	20.9707590158183\\
105	20.8752344487708\\
106	20.7805625979833\\
107	20.6966284886446\\
108	20.6147925823871\\
109	20.5289309915054\\
110	20.4485578320431\\
111	20.3611272317233\\
112	20.2878223632663\\
113	20.2195840748162\\
114	20.1393389332937\\
115	20.0648860456156\\
116	19.9869951423959\\
117	19.9138089051168\\
118	19.8415828941826\\
119	19.774594163765\\
120	19.705276842195\\
121	19.6439788176833\\
122	19.5723588802708\\
123	19.4926340728066\\
124	19.4265910594472\\
125	19.3728633445536\\
126	19.3073634725358\\
127	19.2395561156149\\
128	19.1779017095583\\
129	19.1161612449827\\
130	19.0551538773367\\
131	18.9906694493422\\
132	18.9353774494133\\
133	18.8854990695807\\
134	18.835638504243\\
135	18.7860371224955\\
136	18.7211470175315\\
137	18.6609187257934\\
138	18.6049836071208\\
139	18.550086789733\\
140	18.4996810271708\\
141	18.4459695556374\\
142	18.392841715489\\
143	18.3403645950599\\
144	18.2883723300596\\
145	18.2405646671921\\
146	18.1888882939508\\
147	18.1379744737848\\
148	18.0809441362825\\
149	18.0201676815092\\
150	17.9600000087577\\
151	17.9004310220752\\
152	17.8449975837244\\
153	17.7983297727766\\
154	17.7478998352658\\
155	17.6979601742789\\
156	17.6450284270766\\
157	17.5926065172352\\
158	17.5406863314172\\
159	17.4927079875838\\
160	17.4484340914228\\
161	17.4012920936009\\
162	17.3545946573749\\
163	17.308408784015\\
164	17.2626529745085\\
165	17.2102624045994\\
166	17.1583459593007\\
167	17.1106152898736\\
168	17.0633139749184\\
169	17.0197551058824\\
170	16.9696231704585\\
171	16.9199316342908\\
172	16.8706740868778\\
173	16.8218442475988\\
174	16.77343596235\\
175	16.7254432002865\\
176	16.6778600506663\\
177	16.6306807197926\\
178	16.5838995280504\\
179	16.5375109070346\\
180	16.4947623984844\\
181	16.4523714876847\\
182	16.4103332320221\\
183	16.3686427838447\\
184	16.3241021767021\\
185	16.279923199413\\
186	16.2361009847986\\
187	16.1926307568993\\
188	16.1495078287886\\
189	16.1067276004503\\
190	16.0642855567176\\
191	16.022177265271\\
192	15.9803983746944\\
193	15.938944612586\\
194	15.8978117837235\\
195	15.8569957682803\\
196	15.819420807061\\
197	15.7792189102863\\
198	15.7393219591825\\
199	15.6997261179406\\
200	0\\
};

\addplot [color=black, dotted, line width=1.0pt]
  table[row sep=crcr]{%
1	0\\
2	31.6227766016838\\
3	31.6227766016838\\
4	31.6227766016838\\
5	31.4297581359336\\
6	31.2440407242881\\
7	31.0145609940098\\
8	30.7573989606657\\
9	30.4365869844208\\
10	30.1401067055095\\
11	29.7674842950675\\
12	29.3978122766423\\
13	29.0510009348255\\
14	28.7533840228481\\
15	28.3541982994655\\
16	27.9468457327229\\
17	27.5831835721317\\
18	27.2599844190919\\
19	26.8529283175967\\
20	27.0849752131315\\
21	27.2915516882323\\
22	27.3998559404731\\
23	27.5095534591001\\
24	27.6090447793019\\
25	27.6305778292552\\
26	27.6392903248907\\
27	27.6371348794988\\
28	27.6009145415934\\
29	27.5448846368134\\
30	27.4846708560301\\
31	27.3592072829147\\
32	27.2231369907581\\
33	27.1010553318118\\
34	26.9493495838894\\
35	26.8029030074438\\
36	26.6534047183832\\
37	26.4789790044357\\
38	26.2709168436371\\
39	26.0913964172174\\
40	26.0939993960813\\
41	26.0905328335043\\
42	26.0886382402709\\
43	26.0742421667876\\
44	26.0659270301315\\
45	26.0159608365204\\
46	25.9533707207965\\
47	25.8770552430092\\
48	25.801337715225\\
49	25.743059642188\\
50	25.6707563766467\\
51	25.5873512863353\\
52	25.5091623675327\\
53	25.4378017808936\\
54	25.36492977931\\
55	25.2629382109462\\
56	25.1801180252516\\
57	25.0857168851771\\
58	24.9861328976977\\
59	24.8628812991673\\
60	24.7650201518388\\
61	24.6500323240814\\
62	24.521798007657\\
63	24.3901576590225\\
64	24.2548376520973\\
65	24.1198517796075\\
66	23.9801925949893\\
67	23.8202253693133\\
68	23.6784862822899\\
69	23.5455705007707\\
70	23.4140403663313\\
71	23.2929890185862\\
72	23.1805217421362\\
73	23.0783183736695\\
74	22.9771595501362\\
75	22.8851529852259\\
76	22.8109373835973\\
77	22.7284914836827\\
78	22.6460089864245\\
79	22.5628237977721\\
80	22.4671374054424\\
81	22.3606333703105\\
82	22.2469607336442\\
83	22.1436657588557\\
84	22.043136723745\\
85	21.9305809531907\\
86	21.8250283981095\\
87	21.6992346794786\\
88	21.5920124352759\\
89	21.4857705565866\\
90	21.3707717078387\\
91	21.257690321702\\
92	21.1474895857831\\
93	21.0443426929513\\
94	20.9373215240288\\
95	20.8363025357524\\
96	20.7525837495937\\
97	20.6546519172746\\
98	20.5582506176418\\
99	20.4690228575014\\
100	20.3867564062713\\
101	20.3053715070188\\
102	20.2148458593366\\
103	20.1213246879275\\
104	20.0243538840571\\
105	19.9385552178482\\
106	19.8579373209747\\
107	19.7835593121889\\
108	19.7160178170505\\
109	19.6401337453004\\
110	19.5642209278065\\
111	19.4810957825019\\
112	19.4168724417002\\
113	19.3581243290882\\
114	19.2872385845693\\
115	19.2259217548829\\
116	19.1565105181134\\
117	19.0919137448287\\
118	19.0237512104763\\
119	18.9608510292478\\
120	18.9033702845547\\
121	18.8509197997307\\
122	18.7868386080041\\
123	18.7145405840815\\
124	18.6565622972323\\
125	18.6107483055239\\
126	18.5562749830851\\
127	18.4960123839679\\
128	18.4378849678947\\
129	18.379730809584\\
130	18.3222654475287\\
131	18.2649977660937\\
132	18.2132008063725\\
133	18.1759087149283\\
134	18.1299751468202\\
135	18.0874708573285\\
136	18.0293192038871\\
137	17.9718155371849\\
138	17.9235286526156\\
139	17.8719502457301\\
140	17.8247757385185\\
141	17.7743059996852\\
142	17.7243825116639\\
143	17.6793371296394\\
144	17.630450563002\\
145	17.5820821816097\\
146	17.5296214203888\\
147	17.481737343195\\
148	17.4269036111748\\
149	17.3683256181685\\
150	17.3103343857613\\
151	17.2529201832004\\
152	17.199690053787\\
153	17.1511510816655\\
154	17.1068873326509\\
155	17.0592897608122\\
156	17.0086157310351\\
157	16.9584290659557\\
158	16.9122413552161\\
159	16.866496665424\\
160	16.8244399695953\\
161	16.7798716835888\\
162	16.7391773599716\\
163	16.69512040233\\
164	16.6514731031079\\
165	16.6009374081028\\
166	16.550859051446\\
167	16.5086002666587\\
168	16.4667243906315\\
169	16.4252257042715\\
170	16.3768449638198\\
171	16.3288892387186\\
172	16.2813523423964\\
173	16.2342282136254\\
174	16.1875109132752\\
175	16.1411946211687\\
176	16.0952736330364\\
177	16.0497423575653\\
178	16.0045953135391\\
179	15.9598271270667\\
180	15.918934518971\\
181	15.8783832732852\\
182	15.838168675924\\
183	15.7982861033439\\
184	15.7552974900455\\
185	15.7126579058\\
186	15.6703626530229\\
187	15.6284071221704\\
188	15.5867867896295\\
189	15.5454972156691\\
190	15.504534042451\\
191	15.4638929920975\\
192	15.4235698648153\\
193	15.3835605370723\\
194	15.3438609598263\\
195	15.3044671568039\\
196	15.2683084437359\\
197	15.2295070889034\\
198	15.1910000561336\\
199	15.1527836432482\\
200	0\\
};

\addplot [color=red, line width=1.0pt]
  table[row sep=crcr]{%
1	0\\
2	31.6227766016838\\
3	31.6227766016838\\
4	31.6227766016838\\
5	31.4297581359336\\
6	31.2971920334269\\
7	31.0607379895908\\
8	30.8385166947158\\
9	30.5888634584038\\
10	30.2151541990751\\
11	29.8092290951038\\
12	29.4082009435504\\
13	29.0341413364086\\
14	28.5804566304385\\
15	28.1619538767558\\
16	27.6732674492889\\
17	27.279205314035\\
18	26.8858337371885\\
19	26.4556881337902\\
20	26.5528104712521\\
21	26.7041116244014\\
22	26.810288843028\\
23	26.8552891058086\\
24	26.9395781579985\\
25	26.9611492815872\\
26	26.949771018379\\
27	26.8693550548643\\
28	26.8359598303479\\
29	26.7936498962683\\
30	26.6887293102613\\
31	26.4954736045981\\
32	26.3037692631634\\
33	26.1517766763899\\
34	26.0088905213478\\
35	25.87617571753\\
36	25.7050754670134\\
37	25.5112948924989\\
38	25.3230103025791\\
39	25.1245109978696\\
40	25.1239316996974\\
41	25.1249861572123\\
42	25.119184589189\\
43	25.0919585183076\\
44	25.0672004617152\\
45	25.0242374291546\\
46	24.9494971137546\\
47	24.8449661994821\\
48	24.7454426709997\\
49	24.6491951620324\\
50	24.5395102041862\\
51	24.4339289389343\\
52	24.3341769033651\\
53	24.2380669237238\\
54	24.1386494581022\\
55	24.0651391177801\\
56	23.9520457093368\\
57	23.8404392390318\\
58	23.7154888875081\\
59	23.5800814781562\\
60	23.4564899459211\\
61	23.3356358990684\\
62	23.2156532463486\\
63	23.0727140196129\\
64	22.9332960849074\\
65	22.8031940493367\\
66	22.6703698579425\\
67	22.5274536968399\\
68	22.3953750803058\\
69	22.2725471049965\\
70	22.1457802717262\\
71	22.0096780120177\\
72	21.9000024865432\\
73	21.8004530907354\\
74	21.7167574015413\\
75	21.5982661149281\\
76	21.5307266001052\\
77	21.4713272729713\\
78	21.3923742312652\\
79	21.2861013703466\\
80	21.2022372045311\\
81	21.1174459102717\\
82	21.0411239829295\\
83	20.9539209352159\\
84	20.8518531623636\\
85	20.7633387349296\\
86	20.665364599092\\
87	20.5516907081589\\
88	20.4562970806641\\
89	20.3636280298611\\
90	20.2607483620219\\
91	20.1543792399029\\
92	20.0445452928327\\
93	19.9416789133716\\
94	19.8404732100749\\
95	19.7408841783073\\
96	19.6577827276267\\
97	19.565677495969\\
98	19.4661367187027\\
99	19.3725188392227\\
100	19.2862040423824\\
101	19.2061526587702\\
102	19.1166930455088\\
103	19.0284808751104\\
104	18.9367768190649\\
105	18.8513467431539\\
106	18.7664862930759\\
107	18.6932463868612\\
108	18.6121047628428\\
109	18.5320892938005\\
110	18.4573514234919\\
111	18.384945738509\\
112	18.3223758713685\\
113	18.2655826788641\\
114	18.1852940146754\\
115	18.1155090248639\\
116	18.0470231365736\\
117	17.9790845608971\\
118	17.9120273740667\\
119	17.8505920998136\\
120	17.7954158422125\\
121	17.7402283569016\\
122	17.67740521465\\
123	17.6100185889224\\
124	17.5571752617365\\
125	17.5176029071381\\
126	17.4653750174732\\
127	17.4107025332114\\
128	17.3576942461653\\
129	17.3088151355785\\
130	17.2650261326055\\
131	17.2082205794899\\
132	17.1705515176187\\
133	17.1294635179813\\
134	17.0881205784533\\
135	17.0468431228256\\
136	16.9888117942419\\
137	16.9357295545124\\
138	16.8875940466437\\
139	16.8400255682342\\
140	16.7972206284723\\
141	16.7506713406836\\
142	16.7046254813699\\
143	16.6591767551394\\
144	16.6141099273327\\
145	16.5695203875077\\
146	16.5205156291256\\
147	16.4722594023623\\
148	16.421334182623\\
149	16.3661362644863\\
150	16.3114912496135\\
151	16.2573899687367\\
152	16.2076858813436\\
153	16.1628138342863\\
154	16.1183596492126\\
155	16.0705680785716\\
156	16.0232384217832\\
157	15.9763632927795\\
158	15.9299354696966\\
159	15.8839478902123\\
160	15.8418374230561\\
161	15.8007261074883\\
162	15.7634154991557\\
163	15.7190756629451\\
164	15.675143417684\\
165	15.6275707938079\\
166	15.5804286930546\\
167	15.5377834185325\\
168	15.4955204119405\\
169	15.4536339995935\\
170	15.4081150965944\\
171	15.3629960682632\\
172	15.3182710939787\\
173	15.2739344710495\\
174	15.2299806116591\\
175	15.186404039908\\
176	15.1431993889478\\
177	15.1003613982051\\
178	15.0578849106909\\
179	15.0157648703928\\
180	14.9774983098226\\
181	14.9395513186669\\
182	14.901919482608\\
183	14.8611480806701\\
184	14.8207095075343\\
185	14.7805992594083\\
186	14.7408129173635\\
187	14.7013461452894\\
188	14.6621946879091\\
189	14.6233543688517\\
190	14.5848210887816\\
191	14.5465908235803\\
192	14.5086596225819\\
193	14.4710236068579\\
194	14.4336789675512\\
195	14.3966219642568\\
196	14.3629486607984\\
197	14.3264480969356\\
198	14.2902244028184\\
199	14.2542740957955\\
200	0\\
};

\node[right, align=left,scale=1.1]
at (axis cs:29.8mm,-3.5mm) {(c)};
\end{axis}
\end{tikzpicture}%
\end{minipage}
\begin{minipage}[H!]{0.235\textwidth}
    \begin{tikzpicture}

\begin{axis}[%
width=1.3in,
height=1.3in,
at={(0.758in,0.481in)},
scale only axis,
clip=false,
xmin=0,
xmax=200,
xtick={  0,  40,  80, 120, 160, 200},
xlabel style={at={(0.5,-2.5mm)},font=\color{white!15!black}},
xlabel={$k$},
ymin=0,
ymax=2,
ylabel style={at={(-5.5mm,0.5)},font=\color{white!15!black}},
ylabel={Switching Error},
axis background/.style={fill=white},
]
\addplot [color=blue, line width=1.0pt]
  table[row sep=crcr]{%
1	0\\
2	0\\
3	0\\
4	0\\
5	0\\
6	0\\
7	0\\
8	0\\
9	0\\
10	0.0632455532033676\\
11	0.120604537831105\\
12	0.115470053837925\\
13	0.110940039245046\\
14	0.10690449676497\\
15	0.227949008182319\\
16	0.220710678118655\\
17	0.214120809078803\\
18	0.208088022903976\\
19	0.294304302322867\\
20	0.286852350953347\\
21	0.279939230220829\\
22	0.358783271813407\\
23	0.392599844953278\\
24	0.384333659290053\\
25	0.376568542494924\\
26	0.40174942103339\\
27	0.394239423861396\\
28	0.387135428952991\\
29	0.395785630752608\\
30	0.415864013669141\\
31	0.40910154650555\\
32	0.402658599806889\\
33	0.396510779247043\\
34	0.415745356271773\\
35	0.409763096814749\\
36	0.428433555739967\\
37	0.445033629361153\\
38	0.457109942418308\\
39	0.451211502495665\\
40	0.445535660171872\\
41	0.488042163643086\\
42	0.488783475127634\\
43	0.483066512667927\\
44	0.522236113314894\\
45	0.525876965610205\\
46	0.52803089441443\\
47	0.528609495590417\\
48	0.542766507597299\\
49	0.542806997839606\\
50	0.544930260306316\\
51	0.539561358335939\\
52	0.543163301969486\\
53	0.538014718454987\\
54	0.537981627505143\\
55	0.539434730286676\\
56	0.534596652913775\\
57	0.54540432024939\\
58	0.542084262955417\\
59	0.567054374331547\\
60	0.567025735244171\\
61	0.562358777401507\\
62	0.570607239009579\\
63	0.576332079329054\\
64	0.57771345737253\\
65	0.577277896132275\\
66	0.578141967102267\\
67	0.579025978381024\\
68	0.5747526655247\\
69	0.575298013927982\\
70	0.578771724027799\\
71	0.574681413240215\\
72	0.570676615925095\\
73	0.570770610622226\\
74	0.566900934745753\\
75	0.567641379461586\\
76	0.56389453072943\\
77	0.560220911619077\\
78	0.561451218690907\\
79	0.570756540884174\\
80	0.567178094693405\\
81	0.57145742666964\\
82	0.567962241425397\\
83	0.564530413986051\\
84	0.564378261190192\\
85	0.56104856687269\\
86	0.557777118769294\\
87	0.554562238281748\\
88	0.551402313766099\\
89	0.551294523939193\\
90	0.548223221357012\\
91	0.545202684754253\\
92	0.542231530837425\\
93	0.549708177553007\\
94	0.55230374950165\\
95	0.552909798613968\\
96	0.553524738095925\\
97	0.558592255632417\\
98	0.555734987505456\\
99	0.556183025087671\\
100	0.553395122694189\\
101	0.550648727961538\\
102	0.558439306925506\\
103	0.567733369695035\\
104	0.564997289337641\\
105	0.562300389503212\\
106	0.562640047634947\\
107	0.563799381153961\\
108	0.564021321425776\\
109	0.561428106211906\\
110	0.562353824969729\\
111	0.562614544698499\\
112	0.560097241043903\\
113	0.557613427292903\\
114	0.561025816705863\\
115	0.558581248141761\\
116	0.558607020744677\\
117	0.559215203461523\\
118	0.556840605883432\\
119	0.5544960034451\\
120	0.558131227900829\\
121	0.55582011558062\\
122	0.562329509933662\\
123	0.560038952511068\\
124	0.559987201917905\\
125	0.557742755202925\\
126	0.562276766410303\\
127	0.562180107757438\\
128	0.562708283341438\\
129	0.560523000199258\\
130	0.558362980603235\\
131	0.561604149072815\\
132	0.559472816340132\\
133	0.557365566606352\\
134	0.558672833158125\\
135	0.559393155929481\\
136	0.563408083874551\\
137	0.561348083983042\\
138	0.559310516156097\\
139	0.557294976209281\\
140	0.555301070080947\\
141	0.556061662373728\\
142	0.557166202212441\\
143	0.557681177177261\\
144	0.555741410683784\\
145	0.553821745497951\\
146	0.554095559573574\\
147	0.554540978885227\\
148	0.552664354285398\\
149	0.550806653720656\\
150	0.551488039343273\\
151	0.549658886578505\\
152	0.547847814483393\\
153	0.546054527136765\\
154	0.544278735353919\\
155	0.542520156490733\\
156	0.540778514254687\\
157	0.541407524914955\\
158	0.539691490408297\\
159	0.537991670468254\\
160	0.536307811344278\\
161	0.534639664810757\\
162	0.532986988013306\\
163	0.531349543320252\\
164	0.529727098179108\\
165	0.528119424977839\\
166	0.528497981763022\\
167	0.52691327660115\\
168	0.525342741731866\\
169	0.523786167223827\\
170	0.522243347474198\\
171	0.52071408109458\\
172	0.519198170800583\\
173	0.517695423304924\\
174	0.516205649213898\\
175	0.514728662927119\\
176	0.51326428254039\\
177	0.511812329751599\\
178	0.510372629769527\\
179	0.508945011225458\\
180	0.507529306087505\\
181	0.506125349577523\\
182	0.504732980090558\\
183	0.505364289768453\\
184	0.503989146294939\\
185	0.502625167755287\\
186	0.501272203880569\\
187	0.499930107218157\\
188	0.498598733064222\\
189	0.4972779393982\\
190	0.495967586819153\\
191	0.494667538483974\\
192	0.493377660047357\\
193	0.492097819603481\\
194	0.490827887629354\\
195	0.489567736929749\\
196	0.488317242583696\\
197	0.487076281892463\\
198	0.485844734328988\\
199	0.484622481488709\\
200	0\\
};

\addplot [color=green, dashdotted, line width=1.0pt]
  table[row sep=crcr]{%
1	0\\
2	0\\
3	0\\
4	0\\
5	0\\
6	0\\
7	0.0755928946018454\\
8	0.0707106781186548\\
9	0.133333333333333\\
10	0.189736659610103\\
11	0.266187093288902\\
12	0.25485473884966\\
13	0.244856512922842\\
14	0.311542534351145\\
15	0.279588786131751\\
16	0.270710678118655\\
17	0.322904861235281\\
18	0.31380711874577\\
19	0.469964586001589\\
20	0.427557571080863\\
21	0.46089704285471\\
22	0.479387809644513\\
23	0.541082038917243\\
24	0.583490042446325\\
25	0.61787309798751\\
26	0.625328467061444\\
27	0.636185984273884\\
28	0.673022952325332\\
29	0.670084656963861\\
30	0.706777119677705\\
31	0.706701093722158\\
32	0.738279942243993\\
33	0.74316957502644\\
34	0.749446693015901\\
35	0.760799403233311\\
36	0.749403454336594\\
37	0.766339066575071\\
38	0.775781613884187\\
39	0.776818461733327\\
40	0.784818391780212\\
41	0.793589408069329\\
42	0.803821484691388\\
43	0.806914964920742\\
44	0.844402093127388\\
45	0.850661776081623\\
46	0.875375760215812\\
47	0.880731961464815\\
48	0.88418236430578\\
49	0.884652174231483\\
50	0.880350877304313\\
51	0.878682083048822\\
52	0.883921172038522\\
53	0.888882344716852\\
54	0.902026619730959\\
55	0.906090780467766\\
56	0.908165692082716\\
57	0.930903555972511\\
58	0.930323107273832\\
59	0.939151705630216\\
60	0.943103757892951\\
61	0.945154594845437\\
62	0.956218667010966\\
63	0.95791376280744\\
64	0.96108094888961\\
65	0.9634530014075\\
66	0.968541900804251\\
67	0.973047502368279\\
68	0.974294796855997\\
69	0.975056154053042\\
70	0.971447707699986\\
71	0.974062380462031\\
72	0.967274406610503\\
73	0.96407856511773\\
74	0.965604890386626\\
75	0.981342115381306\\
76	0.983438502568766\\
77	0.985429482680069\\
78	0.995245103958418\\
79	0.999735664235012\\
80	1.00088060861539\\
81	1.00904740884063\\
82	1.01224531902724\\
83	1.01285397486471\\
84	1.01792212261851\\
85	1.01701722089993\\
86	1.01528567585353\\
87	1.01582145877031\\
88	1.01223237713798\\
89	1.00652961414107\\
90	1.00949459698051\\
91	1.00589629391732\\
92	1.00243989059677\\
93	0.997035862600056\\
94	0.991718300084367\\
95	0.986484921601036\\
96	0.983587870849945\\
97	0.980477165905804\\
98	0.975461904546365\\
99	0.972405505947971\\
100	0.967531262191538\\
101	0.964875686074151\\
102	0.964044871564687\\
103	0.961167831204922\\
104	0.958405641377065\\
105	0.953830886658196\\
106	0.951244547758959\\
107	0.9467890452894\\
108	0.946665357987565\\
109	0.946714412802571\\
110	0.942401340654841\\
111	0.942249070349821\\
112	0.938033169693358\\
113	0.933873356869719\\
114	0.93742692023094\\
115	0.935571592302763\\
116	0.935176740050182\\
117	0.935247172510428\\
118	0.931275828997209\\
119	0.927354650196029\\
120	0.928446951997876\\
121	0.926439171273617\\
122	0.931247252776001\\
123	0.930846219861376\\
124	0.927085209653023\\
125	0.925641565294075\\
126	0.928811285966747\\
127	0.928663882622493\\
128	0.929111757971328\\
129	0.927213958495891\\
130	0.92364086636703\\
131	0.923681084604309\\
132	0.921679484681847\\
133	0.919804522721843\\
134	0.918013389638722\\
135	0.919478749524361\\
136	0.923223754603722\\
137	0.919848153704915\\
138	0.91650931091353\\
139	0.916116142645901\\
140	0.912838435741899\\
141	0.911009131617195\\
142	0.907795685513635\\
143	0.90650987701396\\
144	0.904945959617852\\
145	0.901820057498012\\
146	0.902104645155451\\
147	0.90043527182182\\
148	0.901808120384124\\
149	0.900311445041733\\
150	0.897305388419116\\
151	0.894329242945459\\
152	0.891382515847452\\
153	0.888464725642918\\
154	0.888721475361334\\
155	0.887237783255986\\
156	0.886007123365284\\
157	0.883180937142929\\
158	0.880381624436668\\
159	0.877608762039055\\
160	0.874861936014863\\
161	0.87504331111507\\
162	0.872338379415779\\
163	0.870991924222275\\
164	0.868332400687918\\
165	0.868642832977949\\
166	0.866022486636784\\
167	0.863425711715729\\
168	0.860852156925019\\
169	0.858301478260948\\
170	0.855773338812716\\
171	0.853267408575501\\
172	0.850783364269512\\
173	0.848320889164807\\
174	0.845879672911657\\
175	0.843459411376243\\
176	0.841059806481507\\
177	0.838680566052937\\
178	0.836321403669145\\
179	0.833982038517024\\
180	0.83166219525135\\
181	0.829361603858648\\
182	0.827079999525183\\
183	0.82481712250892\\
184	0.823827860252077\\
185	0.821598281440684\\
186	0.819386707357985\\
187	0.817192896975809\\
188	0.815016613759208\\
189	0.812857625559336\\
190	0.810715704509432\\
191	0.808590626923797\\
192	0.806482173199674\\
193	0.804390127721923\\
194	0.802314278770408\\
195	0.800254418429996\\
196	0.798210342503085\\
197	0.797377664002272\\
198	0.795361535203194\\
199	0.793360622510802\\
200	0\\
};

\addplot [color=black, dotted, line width=1.0pt]
  table[row sep=crcr]{%
1	0\\
2	0\\
3	0\\
4	0\\
5	0\\
6	0\\
7	0\\
8	0.0707106781186548\\
9	0.242597228917646\\
10	0.293393492430875\\
11	0.304717736587641\\
12	0.291745021223163\\
13	0.320906422442275\\
14	0.309233166251744\\
15	0.385614132104298\\
16	0.391188975860174\\
17	0.394926422054958\\
18	0.383799494834305\\
19	0.445032378303067\\
20	0.492699380080898\\
21	0.563781904032421\\
22	0.591134753766419\\
23	0.591395923834576\\
24	0.586956433765994\\
25	0.629267453145575\\
26	0.671895132805156\\
27	0.697825255744278\\
28	0.736947602756076\\
29	0.732445119302639\\
30	0.761274345285907\\
31	0.774498226140922\\
32	0.791826895390314\\
33	0.800518245898731\\
34	0.82017460995201\\
35	0.817431257148493\\
36	0.813867021664977\\
37	0.834578957647066\\
38	0.866434677568957\\
39	0.855254406876314\\
40	0.877541887918724\\
41	0.903236530229325\\
42	0.897713789926233\\
43	0.915209667690439\\
44	0.937036564925602\\
45	0.944745498791182\\
46	0.969245149517394\\
47	0.971133133898755\\
48	0.977009916170589\\
49	0.969875337158715\\
50	0.972371012815965\\
51	0.979206414458417\\
52	0.981724092258052\\
53	0.982020709856992\\
54	0.991128304755893\\
55	1.00496916293015\\
56	1.00212994834693\\
57	1.01781415107601\\
58	1.02506872829152\\
59	1.04406696281095\\
60	1.04079322799874\\
61	1.04956014504035\\
62	1.04786096221842\\
63	1.05350512900875\\
64	1.04936985429702\\
65	1.05866369870865\\
66	1.06271672819454\\
67	1.07026863082142\\
68	1.06816816355794\\
69	1.06319873459693\\
70	1.05895843364181\\
71	1.05403418689981\\
72	1.04668891144024\\
73	1.04437575861433\\
74	1.04449054342158\\
75	1.04615125370421\\
76	1.04251808598035\\
77	1.05528880960235\\
78	1.0593054046798\\
79	1.06363179947445\\
80	1.07047558546077\\
81	1.07891879206554\\
82	1.08132256511031\\
83	1.08011548929822\\
84	1.08016617884697\\
85	1.08258351045735\\
86	1.07829098698116\\
87	1.0818031677624\\
88	1.07563899697156\\
89	1.0760136435423\\
90	1.07452311162644\\
91	1.0707653979385\\
92	1.06708090441054\\
93	1.06132840479805\\
94	1.06275656171775\\
95	1.05714830852464\\
96	1.05162791409006\\
97	1.05468510608117\\
98	1.04929026197583\\
99	1.04991213479892\\
100	1.0467576994605\\
101	1.0474137447824\\
102	1.04971727754956\\
103	1.05029827172699\\
104	1.04710653510732\\
105	1.0421083846835\\
106	1.03891176562336\\
107	1.03585657464676\\
108	1.03648440884892\\
109	1.03684766571049\\
110	1.03537163564678\\
111	1.03432455527107\\
112	1.02969668169838\\
113	1.02513037679641\\
114	1.02584884314457\\
115	1.02137889228171\\
116	1.0217994675752\\
117	1.0210502921951\\
118	1.01821792804384\\
119	1.01393067562076\\
120	1.0129287197432\\
121	1.01227406535013\\
122	1.01420223421072\\
123	1.01329139324387\\
124	1.0091972698616\\
125	1.00688994104835\\
126	1.00927498581377\\
127	1.01011531730785\\
128	1.01083681696643\\
129	1.00862163742213\\
130	1.00473483438099\\
131	1.00397035318618\\
132	1.00151133443242\\
133	0.997739150277587\\
134	0.998630812831989\\
135	0.997824668891768\\
136	1.00157345893352\\
137	0.999206781624617\\
138	0.995579884786815\\
139	0.994596532600196\\
140	0.991038036280957\\
141	0.988765971665131\\
142	0.98527825013885\\
143	0.981827177363351\\
144	0.979801038346031\\
145	0.976416568687665\\
146	0.975926146230836\\
147	0.972601004890494\\
148	0.973431097163403\\
149	0.971693671638049\\
150	0.968449276364758\\
151	0.965237163780215\\
152	0.962056802041012\\
153	0.958907671490103\\
154	0.955789264300131\\
155	0.954030613410311\\
156	0.952372365277438\\
157	0.949334487153981\\
158	0.94632549547332\\
159	0.94334493532819\\
160	0.940392361778071\\
161	0.940190710193301\\
162	0.937284395016582\\
163	0.935620744343682\\
164	0.932763880439879\\
165	0.932521386342712\\
166	0.929708344077246\\
167	0.926920606635033\\
168	0.924157796893043\\
169	0.921419545550106\\
170	0.918705490919561\\
171	0.91601527872858\\
172	0.913348561923911\\
173	0.910705000483818\\
174	0.908084261235966\\
175	0.905486017681043\\
176	0.902909949821894\\
177	0.900355743997981\\
178	0.897823092724961\\
179	0.895311694539197\\
180	0.892821253847033\\
181	0.890351480778653\\
182	0.887902091046357\\
183	0.885472805807114\\
184	0.884267236814566\\
185	0.88187408699537\\
186	0.879500262805302\\
187	0.877145505533378\\
188	0.874809561291481\\
189	0.872492180899379\\
190	0.870193119773073\\
191	0.867912137816368\\
192	0.865648999315551\\
193	0.863403472837076\\
194	0.861175331128153\\
195	0.858964351020146\\
196	0.85677031333468\\
197	0.855806024974346\\
198	0.853642163041326\\
199	0.851494632184991\\
200	0\\
};

\addplot [color=red, line width=1.0pt]
  table[row sep=crcr]{%
1	0\\
2	0\\
3	0\\
4	0\\
5	0\\
6	0\\
7	0\\
8	0\\
9	0.133333333333333\\
10	0.189736659610103\\
11	0.254761701334258\\
12	0.243915758875542\\
13	0.312793134971518\\
14	0.345696299258819\\
15	0.364224235719708\\
16	0.370478297741519\\
17	0.359416742020544\\
18	0.349290288820636\\
19	0.431740495786173\\
20	0.516792442852889\\
21	0.516032044171937\\
22	0.556873768273152\\
23	0.557888018996064\\
24	0.571764797497878\\
25	0.602208324812808\\
26	0.602980448477793\\
27	0.61149646455974\\
28	0.610605135254524\\
29	0.609936488558249\\
30	0.662195853754738\\
31	0.685051609840586\\
32	0.674262721248655\\
33	0.698783560553373\\
34	0.720238707777918\\
35	0.739929105157995\\
36	0.762560353745952\\
37	0.784099982711628\\
38	0.789291078369366\\
39	0.790451288370576\\
40	0.807003812325661\\
41	0.806991246966505\\
42	0.8144827683097\\
43	0.818665634972863\\
44	0.80930914900355\\
45	0.81277472739066\\
46	0.817608983238237\\
47	0.815090403507063\\
48	0.812220773722662\\
49	0.803890084028889\\
50	0.809514304519488\\
51	0.805095777437355\\
52	0.800731227580643\\
53	0.798532906767054\\
54	0.807210910641114\\
55	0.799838967087467\\
56	0.815507514940941\\
57	0.822829684490946\\
58	0.834354875390064\\
59	0.845286387931635\\
60	0.842402728482648\\
61	0.844099528026213\\
62	0.853848284109714\\
63	0.865308236371816\\
64	0.858521400320896\\
65	0.855268169302123\\
66	0.855728634295048\\
67	0.859818071534631\\
68	0.859958644429388\\
69	0.857420558925619\\
70	0.858102988736465\\
71	0.858896433732124\\
72	0.852911019809682\\
73	0.847049018938013\\
74	0.8508896250852\\
75	0.866483758324213\\
76	0.86076433107881\\
77	0.864660976569452\\
78	0.877804471558973\\
79	0.892500910356383\\
80	0.892418302155382\\
81	0.903431510449932\\
82	0.905693534580674\\
83	0.905547671085454\\
84	0.910473143434695\\
85	0.907772109556997\\
86	0.907368027325457\\
87	0.912069570762008\\
88	0.909286772042799\\
89	0.909284242406244\\
90	0.908844041840905\\
91	0.909181990877612\\
92	0.903638001816255\\
93	0.898766602437082\\
94	0.893973146379234\\
95	0.889255577056838\\
96	0.884611913061054\\
97	0.886376178186507\\
98	0.8838831775533\\
99	0.883350602983003\\
100	0.884811315486989\\
101	0.888629966263967\\
102	0.890833470272198\\
103	0.888674878671993\\
104	0.886334015675622\\
105	0.884711898963102\\
106	0.882623711154012\\
107	0.880240162518543\\
108	0.880170488352507\\
109	0.876123706051041\\
110	0.87380476683563\\
111	0.871499745089852\\
112	0.871649252884614\\
113	0.872041659801928\\
114	0.879292774554961\\
115	0.876984223081281\\
116	0.874905462132059\\
117	0.878857217817635\\
118	0.875125323176717\\
119	0.871440568608018\\
120	0.874880103635452\\
121	0.873316250887606\\
122	0.875274220338861\\
123	0.875213903172452\\
124	0.873164129845525\\
125	0.871936602908597\\
126	0.877301032033895\\
127	0.877184251294442\\
128	0.877790757254709\\
129	0.876280805889099\\
130	0.872903988681479\\
131	0.876737775044179\\
132	0.874795400169066\\
133	0.874516630528692\\
134	0.8730687050294\\
135	0.875245648717708\\
136	0.87916260327025\\
137	0.875948104012624\\
138	0.872768608569892\\
139	0.872689933891116\\
140	0.869567598536227\\
141	0.867770363117991\\
142	0.864709435191619\\
143	0.863574542218054\\
144	0.86248275890637\\
145	0.859503534947484\\
146	0.860233711331024\\
147	0.858821362030151\\
148	0.859031340106742\\
149	0.857833873698174\\
150	0.854969645756457\\
151	0.852133917726543\\
152	0.849326220084649\\
153	0.846546094065393\\
154	0.845421184354286\\
155	0.844120780722922\\
156	0.841410915685757\\
157	0.838726982481905\\
158	0.836068570144858\\
159	0.833435276768858\\
160	0.83082670925366\\
161	0.829480902681435\\
162	0.826916813385375\\
163	0.827174741896956\\
164	0.824649011598062\\
165	0.823845221797669\\
166	0.821360011846391\\
167	0.818897157690948\\
168	0.816456326158277\\
169	0.814037190985616\\
170	0.811639432637317\\
171	0.809262738127549\\
172	0.806906800848694\\
173	0.804571320405194\\
174	0.802256002452661\\
175	0.799960558542055\\
176	0.797684705968732\\
177	0.795428167626195\\
178	0.793190671864365\\
179	0.790971952352218\\
180	0.788771747944621\\
181	0.786589802553224\\
182	0.784425865021254\\
183	0.783576407256831\\
184	0.781444222604582\\
185	0.779329349382801\\
186	0.777231554596815\\
187	0.775150609618677\\
188	0.773086290082501\\
189	0.77103837578285\\
190	0.769006650576069\\
191	0.766990902284463\\
192	0.76499092260322\\
193	0.763006507009999\\
194	0.76103745467707\\
195	0.759083568385953\\
196	0.757144654444439\\
197	0.75631342137303\\
198	0.754401121419322\\
199	0.752503253956299\\
200	0\\
};

\node[right, align=left,scale=1.1]
at (axis cs:29.8mm,-0.18mm) {(d)};
\end{axis}
\end{tikzpicture}

\end{minipage}
\vspace{0mm}
\caption{Individual components of the trajectory metric of the four filters versus time for TS2: (a) location error, (b) false error, (c) missed error, and (d) switching error.}
\label{fig:TrajMetric2} 
\vspace{-2mm}
\end{figure*}

The four error components of the trajectory metric for TS2---i.e., location error, false error, missed error, and switching error---are shown individually in Fig.\ \ref{fig:TrajMetric2}. Whereas for each error component the results of BP-LMB/P, BP-LMB, and BP-TOMB/P are quite similar, those of Gibbs-LMB are partly very different. This can be explained by the fact that Gibbs-LMB ignores valuable association information and thus detects some of the objects only with a delay or not at all. As a consequence, the number of missed objects is rather large, which leads to a significantly higher missed error (Fig.\ \ref{fig:TrajMetric2}(c)). Furthermore, the smaller number of detected objects (compared to the other three filters) in turn implies a smaller number of false objects (Fig.\ \ref{fig:TrajMetric2}(b)) and also lower location and switching errors (Figs.\ \ref{fig:TrajMetric2}(a) and \ref{fig:TrajMetric2}(d)).

\begin{figure}[t!]
\vspace*{.3mm}
\centering
\footnotesize
\hspace*{-2mm}
    
\begin{tikzpicture}

\begin{axis}[%
width=2.4in,
height=1.3in,
at={(0.758in,0.481in)},
scale only axis,
clip=false,
xmin=1,
xmax=200,
xtick={  0,  40,  80, 120, 160, 200},
xlabel style={at={(0.5,-2.5mm)},font=\color{white!15!black}},
xlabel={$k$},
ymin=0,
ymax=23,
ytick={ 0,  5, 10, 15, 20},
ylabel style={at={(-5.5mm,0.5)},font=\color{white!15!black}},
ylabel={$\text{MOSPA}^{(2)}$ error},
axis background/.style={fill=white},
legend style={row sep=-0.7mm, legend cell align=left, align=left, fill=none, draw=none}
]

\addplot [color=blue, line width=1.0pt]
  table[row sep=crcr]{%
1	20\\
2	20\\
3	20\\
4	20\\
5	20\\
6	19.9562654156201\\
7	19.9050316681298\\
8	19.8591400989384\\
9	19.7649538219442\\
10	19.6615716917879\\
11	19.5026482995054\\
12	19.2849720884814\\
13	19.0293800447019\\
14	18.7130646506029\\
15	18.3417258523512\\
16	17.9547895453893\\
17	17.5793500464697\\
18	17.1487315775402\\
19	16.722249541471\\
20	18.1322849276054\\
21	17.8924009024459\\
22	17.6694541595633\\
23	17.4618871985971\\
24	17.267678182389\\
25	17.0681173469296\\
26	16.8359864811195\\
27	16.5816063654631\\
28	16.3435130133577\\
29	16.1345457385884\\
30	15.9158002973585\\
31	15.6910520202102\\
32	15.4723826154053\\
33	15.2386149772418\\
34	15.019583691821\\
35	14.7993423055216\\
36	14.6220118062092\\
37	14.4409020637477\\
38	14.2650475448216\\
39	14.0790093958184\\
40	15.9264111529595\\
41	15.8142960117499\\
42	15.6958286139292\\
43	15.5740404486566\\
44	15.4632711088819\\
45	15.3517066934707\\
46	15.2243928957334\\
47	15.117849505212\\
48	15.0143841711689\\
49	14.8982382932969\\
50	14.7939205063693\\
51	14.6543895189115\\
52	14.5047792503482\\
53	14.3648522615121\\
54	14.211744949063\\
55	14.0489614053618\\
56	13.894337160759\\
57	13.7399845452636\\
58	13.582615840389\\
59	13.446811554544\\
60	14.9684244605483\\
61	14.8738204851776\\
62	14.7920053238337\\
63	14.7045212382455\\
64	14.6183054393574\\
65	14.5458104004924\\
66	14.4659718698028\\
67	14.3898857193088\\
68	14.3166743524468\\
69	14.2340111443633\\
70	14.1622005383717\\
71	14.0920530931458\\
72	14.0138523756744\\
73	13.9318520048055\\
74	13.8558647566338\\
75	13.7858795629995\\
76	13.7287202812613\\
77	13.6567539755455\\
78	13.5869216423989\\
79	13.5220504564481\\
80	13.4468446548652\\
81	13.3676623162168\\
82	13.2871936126196\\
83	13.2043943247413\\
84	13.1235621675719\\
85	13.0308230829711\\
86	12.9374791352529\\
87	12.8599687878228\\
88	12.7773556946554\\
89	12.6929988958688\\
90	12.6073735887565\\
91	12.5350795476706\\
92	12.4730292617518\\
93	12.4107774659154\\
94	12.3562338904535\\
95	12.3181000433421\\
96	12.2801540367679\\
97	12.2330421560226\\
98	12.1959631319513\\
99	12.1563644835227\\
100	12.1123898805089\\
101	12.0609881601302\\
102	12.0021311050618\\
103	11.9529218381317\\
104	11.8953449706756\\
105	11.8162958714252\\
106	11.7406981989377\\
107	11.6780322490925\\
108	11.6244077208125\\
109	11.5691697578202\\
110	11.5187113665373\\
111	11.469219570183\\
112	11.4243669453964\\
113	11.3771152997614\\
114	11.3298879605987\\
115	11.2928261448738\\
116	11.2533621880214\\
117	11.1980947137878\\
118	11.1336144932237\\
119	11.090579593946\\
120	11.0476716927279\\
121	11.0121063781576\\
122	10.9772604817786\\
123	10.9419837765195\\
124	10.9049209484697\\
125	10.8824462261639\\
126	10.8527529853838\\
127	10.8342373341833\\
128	10.8085862194085\\
129	10.770081294382\\
130	10.7342355884032\\
131	10.6971896919466\\
132	10.6618796096278\\
133	10.6344337369194\\
134	10.6070007911967\\
135	10.5703734708236\\
136	10.5305380447243\\
137	10.4765037631457\\
138	10.4280095601026\\
139	10.385896956958\\
140	10.3434534870241\\
141	10.4036051753685\\
142	10.4536538744744\\
143	10.4974355102806\\
144	10.5304194728567\\
145	10.5490279599752\\
146	10.5684876784175\\
147	10.5913265372377\\
148	10.6088160049071\\
149	10.6154553667862\\
150	9.59298432536439\\
151	9.7061139515167\\
152	9.82426033078877\\
153	9.92832091915097\\
154	10.0165441374583\\
155	10.0932442676314\\
156	10.1480873800829\\
157	10.1963915993938\\
158	10.2523971846134\\
159	10.3022653377088\\
160	9.14660460921156\\
161	9.44483214344344\\
162	9.74229434084438\\
163	10.0350929750263\\
164	10.3078655672502\\
165	10.5472914457171\\
166	10.7886921469205\\
167	11.0377111824972\\
168	11.2820528040376\\
169	11.5618642839734\\
170	9.36532793215353\\
171	9.78165707771286\\
172	10.1696083689461\\
173	10.5684520633117\\
174	10.9361604114712\\
175	11.2942022635935\\
176	11.6543565516937\\
177	11.9807471025767\\
178	12.3065008447855\\
179	12.6937516318913\\
180	10.8043403183865\\
181	11.0845313146803\\
182	11.3210455150202\\
183	11.6089818610867\\
184	11.8380229164203\\
185	12.0790650460992\\
186	12.215295951831\\
187	12.4794369311387\\
188	12.7453301338184\\
189	13.11174238328\\
190	11.0575649335958\\
191	11.1887532628152\\
192	11.3345836944705\\
193	11.4964428987704\\
194	11.5574149435751\\
195	11.706892948784\\
196	11.935125126426\\
197	12.2097490377968\\
198	12.5449155058133\\
199	12.8467916327852\\
200	10.6172717621102\\
};
\addlegendentry{Gibbs-LMB}

\addplot [color=green, dashdotted, line width=1.0pt]
  table[row sep=crcr]{%
1	20\\
2	20\\
3	20\\
4	20\\
5	20\\
6	19.9894889124392\\
7	19.9525485675807\\
8	19.9174244022578\\
9	19.8423838909327\\
10	19.7448436623447\\
11	19.5921194537833\\
12	19.4028706931524\\
13	19.1932711498245\\
14	18.9216377493461\\
15	18.5759093514997\\
16	18.1493062610067\\
17	17.6907574469312\\
18	17.139332200398\\
19	16.5795807116115\\
20	17.9738113871327\\
21	17.6546389044987\\
22	17.3536306309927\\
23	17.0269937359083\\
24	16.704103998524\\
25	16.4091889205878\\
26	16.0781759104122\\
27	15.7666202775879\\
28	15.4974006746412\\
29	15.2352178746993\\
30	14.9627653258765\\
31	14.6566943502151\\
32	14.318414910937\\
33	13.9925423991887\\
34	13.6492171935146\\
35	13.2936415911547\\
36	12.9791343411553\\
37	12.6351453894508\\
38	12.3309693712466\\
39	12.0455276368323\\
40	14.1585892822911\\
41	13.8967817392209\\
42	13.6438511837165\\
43	13.3685168364374\\
44	13.1035309056296\\
45	12.8143850943357\\
46	12.5223558922646\\
47	12.2338487779018\\
48	11.9853217370632\\
49	11.6858263466953\\
50	11.3683462006229\\
51	11.1083530237349\\
52	10.720806210887\\
53	10.42995088558\\
54	10.1512374381814\\
55	9.91066817459799\\
56	9.65252187205576\\
57	9.47619833208011\\
58	9.2844206777778\\
59	9.0624651552403\\
60	10.8875467563411\\
61	10.6381101140631\\
62	10.4238780327484\\
63	10.1853502155353\\
64	9.96554505199475\\
65	9.73609303603023\\
66	9.54309574911996\\
67	9.35371295091889\\
68	9.1410314378373\\
69	8.95749043825715\\
70	8.74771475017806\\
71	8.61306004937832\\
72	8.46463301670899\\
73	8.30043522089236\\
74	8.1739469230056\\
75	7.9300651864489\\
76	7.73178491912828\\
77	7.58802716271234\\
78	7.44865208154819\\
79	7.34109657292862\\
80	7.13263191758889\\
81	6.95897955074509\\
82	6.78237847132028\\
83	6.49824169534255\\
84	6.30634528083352\\
85	6.27872338667497\\
86	6.10426276888405\\
87	5.98008435770707\\
88	5.86506313928917\\
89	5.63038702981499\\
90	5.53956619720509\\
91	5.43420899465767\\
92	5.35078736964977\\
93	5.24453894256713\\
94	5.00519035907004\\
95	4.90574807682247\\
96	4.81710439904008\\
97	4.65110487796497\\
98	4.52724817305235\\
99	4.38944359097251\\
100	4.38248735902931\\
101	4.37859643537464\\
102	4.39904786930977\\
103	4.38921203662066\\
104	4.33693023054032\\
105	4.30044864626139\\
106	4.20642559149179\\
107	4.17378818340502\\
108	4.22035229086675\\
109	4.18649123193718\\
110	4.13658456663718\\
111	4.04640754861011\\
112	3.96622128347844\\
113	3.83553228297151\\
114	3.68579290191483\\
115	3.62999510369794\\
116	3.49152980883896\\
117	3.38405471542684\\
118	3.33716268986877\\
119	3.26657048798396\\
120	3.18697651995379\\
121	3.11564876179996\\
122	3.05827190588291\\
123	2.99420227996432\\
124	2.96428080912187\\
125	3.02086374139541\\
126	3.01967381806519\\
127	2.99877024905232\\
128	2.98931578506109\\
129	2.8950791462127\\
130	2.81279102724883\\
131	2.80924063456573\\
132	2.7703194607071\\
133	2.66293078936756\\
134	2.61431862261547\\
135	2.55175244883693\\
136	2.52074257095588\\
137	2.53594264778449\\
138	2.56428976513067\\
139	2.54269826985928\\
140	2.56974159717883\\
141	2.72580534417354\\
142	2.87270959785244\\
143	2.99616300386335\\
144	3.10945707814727\\
145	3.24996536716948\\
146	3.43607677124638\\
147	3.54905960174654\\
148	3.694960998275\\
149	3.87046855372812\\
150	3.759816494386\\
151	3.88754276898147\\
152	3.98423705056995\\
153	4.03847654669336\\
154	4.05765036782258\\
155	4.06950021231694\\
156	4.12993503725863\\
157	4.1544184688252\\
158	4.24416436494477\\
159	4.30922524066533\\
160	4.32596546390416\\
161	4.59159876074731\\
162	4.85921635977202\\
163	5.01504209154174\\
164	5.00207309227445\\
165	5.13689427015662\\
166	5.21450318098171\\
167	5.46193844477659\\
168	5.68417244911553\\
169	6.09908495702023\\
170	6.64187229166179\\
171	6.85553108496848\\
172	7.01410315858615\\
173	7.08475586043365\\
174	7.1249552733468\\
175	6.79602208001009\\
176	6.86585989927167\\
177	6.98867281412823\\
178	7.11553492858393\\
179	7.61144611682159\\
180	8.40180317933302\\
181	8.67956371476687\\
182	8.83735393809006\\
183	8.68701676223821\\
184	8.55489727490168\\
185	8.15753505599517\\
186	8.04758270303239\\
187	7.97276675412605\\
188	8.21738340294108\\
189	8.65925639013981\\
190	9.57782179417512\\
191	9.69776747546424\\
192	9.73362130524314\\
193	9.63706048417267\\
194	9.07425143179672\\
195	8.54490837401998\\
196	8.48647687083458\\
197	8.69457712338345\\
198	8.86969965464553\\
199	9.23274504084103\\
200	9.8705255910235\\
};
\addlegendentry{BP-TOMB/P}

\addplot [color=black, dotted, line width=1.0pt]
  table[row sep=crcr]{%
1	20\\
2	20\\
3	20\\
4	20\\
5	20\\
6	19.9458261167666\\
7	19.8404464429306\\
8	19.7476894405098\\
9	19.6035052123038\\
10	19.4182387789827\\
11	19.1739428821837\\
12	18.8758261612056\\
13	18.4960254532172\\
14	18.033819267858\\
15	17.4780421275478\\
16	16.9186599037665\\
17	16.393366995684\\
18	15.7526916915134\\
19	15.1783886161979\\
20	17.2485930970039\\
21	16.9251815411989\\
22	16.603156133582\\
23	16.3025047112462\\
24	16.0343191109004\\
25	15.7668019202392\\
26	15.485377254624\\
27	15.1749096672417\\
28	14.9218998117001\\
29	14.6859119009489\\
30	14.4272306285016\\
31	14.1214177064539\\
32	13.850851304914\\
33	13.5303979812408\\
34	13.2532865796551\\
35	12.9971330400523\\
36	12.6244068305967\\
37	12.2931333840046\\
38	11.9262799022058\\
39	11.6372688144198\\
40	13.8087819246284\\
41	13.5442051747636\\
42	13.2798061733718\\
43	13.0123109776935\\
44	12.736159921112\\
45	12.4541898553881\\
46	12.1910584028162\\
47	11.948050861006\\
48	11.6497113914366\\
49	11.3328339196005\\
50	11.0681101402207\\
51	10.8340369759584\\
52	10.6014657362928\\
53	10.3243328426812\\
54	10.0103008195089\\
55	9.83169287585324\\
56	9.67985873057825\\
57	9.40471902308876\\
58	9.23719539447218\\
59	8.9617697159785\\
60	10.7760572231652\\
61	10.5457812686317\\
62	10.3134199439192\\
63	10.145545620382\\
64	9.94651112043662\\
65	9.71844140933769\\
66	9.48707247206156\\
67	9.29784166167159\\
68	9.1410441932814\\
69	8.98566091697651\\
70	8.77392624024239\\
71	8.6497907030044\\
72	8.4065050435014\\
73	8.25933074196035\\
74	8.15510891411857\\
75	7.99819868147198\\
76	7.87684031966434\\
77	7.72128103746795\\
78	7.47310817795305\\
79	7.39901492054374\\
80	7.16380849511283\\
81	7.02247848974572\\
82	6.94222181352683\\
83	6.75679995126984\\
84	6.59935860501375\\
85	6.344723936106\\
86	6.18935332611885\\
87	6.22873042558131\\
88	6.10281607061811\\
89	5.88017761139745\\
90	5.77129818797949\\
91	5.72913817042177\\
92	5.59110275277672\\
93	5.42838502413682\\
94	5.36115367414024\\
95	5.28622386094019\\
96	5.22570746967956\\
97	5.13642144940104\\
98	4.9929332359777\\
99	4.85597205914513\\
100	4.73344532229435\\
101	4.60472299130743\\
102	4.56649935033512\\
103	4.49908219005818\\
104	4.43020341321252\\
105	4.39784865390796\\
106	4.32957009355885\\
107	4.1816610130734\\
108	4.03329612482578\\
109	3.99203419727221\\
110	3.94712711616053\\
111	3.86924729320606\\
112	3.84231444091424\\
113	3.75403765682224\\
114	3.68481174789014\\
115	3.72826593929841\\
116	3.6207176205501\\
117	3.50357536318449\\
118	3.40135174924793\\
119	3.27020068321598\\
120	3.20951456050901\\
121	3.22301518577593\\
122	3.11684579379525\\
123	3.0816394406235\\
124	3.02248528731289\\
125	3.00062522870019\\
126	2.92832072638461\\
127	2.86651078564181\\
128	2.81527847090142\\
129	2.78759132280026\\
130	2.6827807217469\\
131	2.59369374081212\\
132	2.58857198621596\\
133	2.49223072519181\\
134	2.44779123106449\\
135	2.38227311262744\\
136	2.37478274428455\\
137	2.36064366730046\\
138	2.34675943741851\\
139	2.35662892521301\\
140	2.38010393526421\\
141	2.56447445753196\\
142	2.67647736854416\\
143	2.86387824853735\\
144	2.99755371893144\\
145	3.1222761495851\\
146	3.25247052623683\\
147	3.40022128680114\\
148	3.57373539104477\\
149	3.77664674949372\\
150	3.74225534243369\\
151	3.81616245561166\\
152	3.97400333816493\\
153	4.06200543288884\\
154	4.09295659367329\\
155	4.03618101978036\\
156	4.10438884506911\\
157	4.14604578528076\\
158	4.19889935278829\\
159	4.36571812547565\\
160	4.441818044886\\
161	4.73551463138179\\
162	5.00849764864221\\
163	5.17411710333326\\
164	5.24274169109225\\
165	5.29011951957926\\
166	5.35628696523881\\
167	5.54067591441389\\
168	5.75505132689162\\
169	6.14655566795253\\
170	6.75387820281077\\
171	7.01088513457112\\
172	7.1037804580771\\
173	7.22123849115137\\
174	7.25454140614843\\
175	6.76898401165803\\
176	6.88433096806507\\
177	6.94221997462124\\
178	7.20624702570698\\
179	7.58951111010258\\
180	8.34346657141506\\
181	8.64966103971092\\
182	8.93811073768516\\
183	8.73308922597766\\
184	8.50386384925593\\
185	8.2508795736735\\
186	8.03497706412448\\
187	7.97949757690349\\
188	8.14769550289738\\
189	8.65242205086932\\
190	9.47560998183965\\
191	9.63502803774806\\
192	9.74079256989115\\
193	9.59332992426819\\
194	9.12171804854474\\
195	8.47003338759539\\
196	8.39346846844775\\
197	8.46020290907962\\
198	8.67490385996205\\
199	9.25889916122143\\
200	10.0507925153009\\
};
\addlegendentry{BP-LMB}

\addplot [color=red, line width=1.0pt]
  table[row sep=crcr]{%
1	20\\
2	20\\
3	20\\
4	20\\
5	19.9864471980083\\
6	19.9338781547158\\
7	19.8491597570673\\
8	19.7703166876452\\
9	19.617760509513\\
10	19.4397269362375\\
11	19.175330124769\\
12	18.8455858034009\\
13	18.4565177130696\\
14	17.9535096622048\\
15	17.4087174781096\\
16	16.8113614613948\\
17	16.2024755187898\\
18	15.5191641196592\\
19	14.8948708214607\\
20	17.0599483522091\\
21	16.7000879122358\\
22	16.3845747140783\\
23	16.060491476204\\
24	15.765764270866\\
25	15.4791490512382\\
26	15.1727354546865\\
27	14.868114848279\\
28	14.5875699551719\\
29	14.316439553225\\
30	14.0057389096896\\
31	13.6660048816266\\
32	13.342223524851\\
33	12.9746834133613\\
34	12.6418110074255\\
35	12.3471093148922\\
36	12.0450169589973\\
37	11.7062457533431\\
38	11.4040100678388\\
39	11.1498197836948\\
40	13.3092074595219\\
41	13.0332970460092\\
42	12.7546741047026\\
43	12.4724951840235\\
44	12.2162508579147\\
45	11.9806196399649\\
46	11.7046228795812\\
47	11.4683855242329\\
48	11.313751655866\\
49	11.1419073796547\\
50	10.8118797654878\\
51	10.5307991217148\\
52	10.2711630689475\\
53	10.0534017964173\\
54	9.89297781426316\\
55	9.69522604555133\\
56	9.48619307638263\\
57	9.28543210479315\\
58	9.04980583088699\\
59	8.76805789910403\\
60	10.4890866350837\\
61	10.2789003150396\\
62	10.0479073399022\\
63	9.81198385412423\\
64	9.56062621725008\\
65	9.33094428730411\\
66	9.15036099096518\\
67	8.91633017947397\\
68	8.78059879098221\\
69	8.64541611895081\\
70	8.44310662578825\\
71	8.29930827324261\\
72	8.15802164853605\\
73	8.05160339958298\\
74	7.9043628963365\\
75	7.75716774701893\\
76	7.56581107149155\\
77	7.35228177921936\\
78	7.26694067116206\\
79	7.12604281168437\\
80	6.97408792756806\\
81	6.79675674468243\\
82	6.54983430358956\\
83	6.45307345378942\\
84	6.42388134721266\\
85	6.32381328063007\\
86	6.23901818985391\\
87	6.14198687541707\\
88	6.05816156359936\\
89	5.9930846643821\\
90	5.82332947484823\\
91	5.67218168938313\\
92	5.63006108389825\\
93	5.54847807013702\\
94	5.37046488437246\\
95	5.24420440396167\\
96	5.17555076641697\\
97	5.08399952456571\\
98	4.92053514904414\\
99	4.89672619817686\\
100	4.91399202139436\\
101	4.82958852220644\\
102	4.86331713965246\\
103	4.715259704139\\
104	4.80422614375341\\
105	4.71363452187877\\
106	4.72083303839401\\
107	4.66582284201464\\
108	4.69891660713158\\
109	4.66390364944876\\
110	4.57160217123264\\
111	4.45548445129443\\
112	4.30256775579486\\
113	4.17013165487421\\
114	4.09477019698372\\
115	3.950190781445\\
116	3.9674714835379\\
117	3.88560675624825\\
118	3.77777411764263\\
119	3.73066152193427\\
120	3.68132413013835\\
121	3.66479668221199\\
122	3.61206852945616\\
123	3.53497223480753\\
124	3.5897893262778\\
125	3.48761715517086\\
126	3.41414826393234\\
127	3.37146103545143\\
128	3.29431873433254\\
129	3.20402555218984\\
130	3.12115018108781\\
131	3.06622075550158\\
132	3.02287827292933\\
133	2.96005192969245\\
134	2.88674153194768\\
135	2.83369672734137\\
136	2.81635287762309\\
137	2.82632512522736\\
138	2.82924227394557\\
139	2.85143710001297\\
140	2.86488245788265\\
141	3.01047440250786\\
142	3.08645408247988\\
143	3.1988463241743\\
144	3.33166474625432\\
145	3.45833687411496\\
146	3.55113459492121\\
147	3.72160844019398\\
148	3.84515695406301\\
149	3.98475232795375\\
150	3.90364739560439\\
151	4.09585807850759\\
152	4.1573047660002\\
153	4.35810376191956\\
154	4.34872074646419\\
155	4.37432693964934\\
156	4.49092199226301\\
157	4.37569941593237\\
158	4.49494464172373\\
159	4.64139449021407\\
160	4.65450848980813\\
161	4.92578860029758\\
162	5.10727938395737\\
163	5.26199012021719\\
164	5.28273845420523\\
165	5.30621047648809\\
166	5.4382278259799\\
167	5.59756496889885\\
168	5.85905313280163\\
169	6.20670236023547\\
170	6.77830941128941\\
171	7.01060382299537\\
172	7.00756427375549\\
173	7.10825220823551\\
174	7.13441386474079\\
175	6.8075152477621\\
176	6.89670578651499\\
177	6.96925803233042\\
178	7.24215584137413\\
179	7.70107387788725\\
180	8.45443250724285\\
181	8.67846266845697\\
182	8.91137926567316\\
183	8.76432900221638\\
184	8.5526106835713\\
185	8.1081661545493\\
186	8.10785703074502\\
187	7.99047927684413\\
188	8.34711193583811\\
189	8.71258653750093\\
190	9.67612595984666\\
191	9.83014417313118\\
192	9.85332841522654\\
193	9.79850462159539\\
194	9.41633871388309\\
195	8.73261611900699\\
196	8.71514203835317\\
197	8.86148684712534\\
198	8.86223094874793\\
199	9.40760547208425\\
200	10.1745956729861\\
};
\addlegendentry{BP-LMB/P (proposed)}

\end{axis}

\end{tikzpicture}%
\vspace{-.5mm}
\caption{\rd{$\text{MOSPA}^{(2)}$ error of the four filters versus time for TS2.}} 
\label{fig:OSPA2} 
\end{figure}
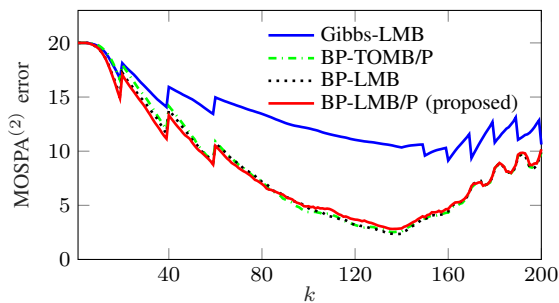

It can also be seen that for all filters, the missed error shown in Fig.\ \ref{fig:TrajMetric2}(c) is much higher than the other error components (note the widely different y-axis scale used in Fig.\ \ref{fig:TrajMetric2}(c) compared to the other parts of Fig.\ \ref{fig:TrajMetric2}). Thus, the missed error dominates the overall trajectory metric, which explains why Fig.\ \ref{fig:TrajMetric2}(c) is similar to Fig.\ \ref{fig:TrajMetric1}. Furthermore, the high missed error of Gibbs-LMB (compared to the other three filters) is not compensated by the fact that the other error components are lower. 
The other three filters, i.e., BP-LMB/P, BP-LMB, and BP-TOMB/P, exhibit a similar performance, with BP-LMB/P performing best. The latter fact can be attributed to the proposed transfer scheme between the Poisson part and the LMB part. Indeed, these simulation results suggest that our transfer scheme, with an appropriate choice of the thresholds  $\gamma_{\text{tr}}$, $\gamma_{\text{leg}}$, and $\gamma_{\text{C}}$, can result in performance advantages compared to both BP-LMB (using a pruning of Bernoulli components) and BP-TOMB/P (using a recycling of Bernoulli components). These advantages come in addition to the lower filter runtimes obtained with BP-LMB/P, as reported presently.  

\rd{Another trajectory metric that is closely related to the OSPA metric is the $\text{OSPA}^{(2)}$ metric proposed in \cite{Beard17OSPA2}.
In Fig.\ \ref{fig:OSPA2}, we compare BP-LMB/P, Gibbs-LMB, BP-LMB, and BP-TOMB/P for TS2, using the $\text{OSPA}^{(2)}$ metric with cutoff parameter 
$c \!=\! 20$, order $p \!=\! 2$, and window length $L \!=\! 10$. The results are seen to be similar to those for the MOSPA error shown in Fig. \ref{fig:results}(b). We note that for window length $L \!=\! 1$, the $\text{OSPA}^{(2)}$ metric would simplify to the OSPA metric.}

Table \ref{fig:results_2} lists the average runtime per time ($k$) step required by MATLAB implementations of the various filters on an Intel quad core i7-6600U CPU. Also shown is the average number of Bernoulli components per time step employed by each filter. Again, these numbers were obtained by averaging over 1,000 Monte Carlo runs. One can see that BP-LMB/P achieves the lowest runtimes of all filters; furthermore, it employs the lowest numbers of Bernoulli components of all filters except Gibbs-LMB. We note that, as is demonstrated by Fig.\ \ref{fig:results}, this low complexity of BP-LMB/P does not come at the cost of a poorer MOSPA performance. Also, while Gibbs-LMB employs fewer Bernoulli components (especially for TS2), its MOSPA performance for TS2 is significantly poorer.

We can conclude from the results in Figs.\ \ref{fig:results}\rd{--\ref{fig:OSPA2}} and Table \ref{fig:results_2} that BP-LMB/P offers a superior performance-complexity compromise relative to the other filters. It has a significantly better performance than Gibbs-LMB (especially for TS2) and also a lower runtime. When compared to BP-LMB and BP-TOMB/P, the runtime of BP-LMB/P is much lower while its performance is almost identical. 
The low runtime of BP-LMB/P is a direct consequence of the fact that objects of unlikely existence are modeled by the Poisson RFS. The performance advantage of BP-LMB/P over Gibbs-LMB is mainly due to the fact that BP-LMB/P takes into account more association information. Gibbs-LMB ignores relevant association information, which allows it to employ fewer Bernoulli components but also results in a poorer performance. For challenging scenarios with a high number of (closely spaced) objects and/or a low detection probability and/or strong clutter, 
the number of samples used by the Gibbs sampler must be increased significantly to obtain an acceptable MOSPA performance, and this entails a higher complexity.

\begin{table}
\centering
\vspace*{2.5mm}
{\small
\hspace*{-.5mm}	\begin{tabular}{|lrrrr|}	
	\hline
	\rule[0mm]{0mm}{2.8mm}\!\!Filter \!\!\!\!&\!\!\!\! RT-TS1\!\!\!\!& RT-TS2\!\!\!\!& NBC-TS1\!\!\!\!& NBC-TS2\!\\[.12mm] \hline
	\rule{0mm}{3.2mm}\!\!BP-LMB/P (proposed)\hspace{0mm} \!\!\!\!&\!\!\!\!  1.33{\ist}s\!&  5.05{\ist}s\!\rmv & \!\!15.21\,\, & 162.82\,\,\,\, \\[.1mm]
	\!\!Gibbs-LMB  \!\!\!\!&\!\!\!\!  5.12{\ist}s\!& 7.94{\ist}\text{s}\!\rmv & \!\!9.69\,\, & 34.23\,\,\,\, \\[.1mm]
	\!\!BP-LMB \!\!\!\!&\!\!\!\! 5.55{\ist}s\!& 21.68{\ist}\text{s}\!\rmv & \!\!34.15\,\, & 861.96\,\,\,\, \\[.1mm]
	\!\!BP-TOMB/P  \!\!\!\!&\!\!\!\!  10.66{\ist}s\!&  16.09{\ist}\text{s}\!\rmv & \!\!63.33\,\, & 521.93\,\,\,\, \\[.1mm]
	\hline	
	\end{tabular}
}
\vspace{.5mm}
\caption{Measured complexity of the four filters for TS1 and TS2. RT-TS1 and RT-TS2 designate the average runtime per time step, and NBC-TS1 and NBC-TS2 designate the average number of Bernoulli components per time step.}
\label{fig:results_2} 
\vspace{-2mm}
\end{table}

\section{Conclusion}
\label{sec:Con}

We proposed an efficient multiobject tracking algorithm that maintains track continuity. Low complexity is achieved by a combination of a labeled multi-Bernoulli random finite set (RFS) and a Poisson RFS as well as complexity-reducing approximations in the update step. Objects of unlikely existence are tracked in an efficient manner by the Poisson RFS, and a new labeled Bernoulli component is created and maintained only if the existence of an object is sufficiently likely. Our simulation results showed that the proposed algorithm offers an attractive accuracy-complexity compromise. The complexity is significantly smaller than that of other RFS-based algorithms with comparable performance, especially in scenarios with many objects and strong clutter. Interesting directions of future research include extensions of our algorithm to multiple-detection measurement models and multi-sensor scenarios \cite{Mey21ExtObj,Mey18Proc,Beard16LMBExTarg,Gra12ExtPHD,Sau17MB}.

\appendices
\renewcommand*\thesubsectiondis{\thesection.\arabic{subsection}}
\renewcommand*\thesubsection{\thesection.\arabic{subsection}}

\section{} 
\label{sec:app}

\vspace{-1.2mm}

In Table \ref{tab:PartAlg}, we present an algorithm for constructing the partitionings \eqref{eq:upda_Ldecom} and \eqref{eq:upda_Zdecom2}. This algorithm is further explained in the following. In Step 1, the sets $\mathcal{M}_k(\la) \!\subseteq\rmv \mathcal{M}_k$ comprise the indices of all those measurements whose association with the object with state $(\stR_k,\la)$ is plausible. (Note that the $\mathcal{M}_k(\la)$ for different $\la \!\in\! \mathbb{L}_{k-1}^{*}$ are not necessarily disjoint.) Then, after an initialization step in Step 2, we perform the iterative procedure constituted by Step 3, which generates the label subsets $\mathbb{L}_{k-1}^{(c)} \!\subseteq \mathbb{L}_{k-1}^{*}$, $c \rmv\in\rmv 
\{1,\ldots,C\}$ and the  corresponding measurement index subsets $\mathcal{M}_k^{(c)} \!\subseteq\rmv \mathcal{M}_k$, $c \rmv\in\rmv \{1,\ldots,C\}$.
 
  \begin{table}[t!]
\vspace{1.5mm}
      \small    
      {\hrule height .5pt} 
\vspace{.9mm}
			\caption{{\ist}Algorithm for constructing the partitionings \eqref{eq:upda_Ldecom} and \eqref{eq:upda_Zdecom2}}
	  \label{tab:alg-part}
      \vspace{-1.5mm}

      {\hrule height .5pt} 
  
  
      \vspace{1.5mm}

      \textbf{Input:}\, Label set $\mathbb{L}_{k-1}^{*} \!=\rmv \{\la^{(1)},\ldots,\la^{(|\mathbb{L}_{k-1}^{*}|)}\}$; measurement index set $\mathcal{M}_k$;
			association weight $\assw_k^{(\la,m)}$; threshold $\gamma_{\text{C}}$\ist.

      \vspace{1.5mm}

      \textbf{Output:}\, Number of subsets $C$, label subsets $\mathbb{L}_{k-1}^{(c)}$, $c \rmv\in\rmv 
      \{1,\ldots,C\}$;  
      measurement index subsets $\mathcal{M}_{k}^{(c)}\rmv$, $c \rmv\in\rmv \{1,\ldots,C\}$ and $\mathcal{M}^{\text{res}}_{k}$.
 
      \vspace{1.5mm}

      \textbf{Operations:}\,

      \vspace{.4mm}
			
			\begin{enumerate}[leftmargin=5mm]
			
			\item[1)]
			For each $\la \!\in\! \mathbb{L}_{k-1}^{*}$, determine $\mathcal{M}_k(\la) \rmv\subseteq\rmv \mathcal{M}_k$  
			as the subset of all measurement indices $m \!\in\! \mathcal{M}_k$ for which $\assw_k^{(\la,m)} \!\geq\rmv \gamma_{\text{C}}$.
			
			\vspace{0.8mm}
			\item[2)]
			Initialization: Set $C \!=\! 1$, $\mathbb{L}_{k-1}^{(1)} \!=\! \{\la^{(1)}\}$,
			and $\mathcal{M}_k^{(1)} \!=\rmv \mathcal{M}_k(\la^{(1)})$.

		\vspace{1.2mm}
			\item[3)]
			Iteration: For $j = 2,\ldots,|\mathbb{L}_{k-1}^{*}|$, do the following:

			\begin{enumerate}[leftmargin=7mm]		
					
			\vspace{.7mm}
			\item[3.1)] Determine 
			$\mathcal{C}' \subseteq \{1,\ldots,C\}$ 
			as the set of all $c \rmv\in\rmv\{1,\ldots,C\}$ for which $\mathcal{M}_k^{(c)} \rmv\cap \mathcal{M}_k(\la^{(j)})  \neq \emptyset$.
			
			\vspace{0mm}
			\item[3.2)] If $\mathcal{C}' \!=\rmv \emptyset$, then increment $C$ by one and set $\mathbb{L}_{k-1}^{(C)} \rmv=\rmv \{\la^{(j)}\}$ and 
		$\mathcal{M}_k^{(C)} \rmv=\rmv \mathcal{M}_k(\la^{(j)})$; else do the following:
		
\begin{itemize}[leftmargin=3mm]	

\vspace{.2mm}
\item Select an arbitrary $c \rmv\in\rmv \mathcal{C'}$ and set 
$\mathbb{L}_{k-1}^{(c)} \rmv=\rmv \{\la^{(j)}\} \cup \bigcup_{c' \rmv\in \mathcal{C}'} \rmv\mathbb{L}_{k-1}^{(c')}$ and 
$\mathcal{M}_{k}^{(c)} \!=\rmv \mathcal{M}_k(\la^{(j)}) \cup \ist \bigcup_{c' \rmv\in \mathcal{C}'} \rmv\mathcal{M}_{k}^{(c')}\rmv$.

\vspace{1.5mm}
\item Set $\mathcal{C}'' \!= (\{1,\ldots,C\} \!\setminus\rmv \mathcal{C}') \ist\cup\ist \{c\}$ and 
$C \rmv=\rmv |\mathcal{C}''|$.

\vspace{1mm}
\item Perform a reindexing whereby the indices contained in $\mathcal{C''}$ are replaced by the new indices $1,2,\ldots,C$.
\end{itemize}			
									
			\end{enumerate}

\vspace{.5mm}
\item[4)] Set $\mathcal{M}_k^{\text{res}} = \mathcal{M}_k \!\setminus \bigcup_{c \in \{1,\ldots,C\}} \mathcal{M}_k^{(c)}\rmv$.

			\end{enumerate}

\vspace{1.5mm}
{\hrule height .5pt}
\label{tab:PartAlg}
\vspace{-2.5mm}
  \end{table}

The generation of these subsets is done such that for each $c \rmv\in\rmv \{1,\ldots,C\}$, the association of an object state $(\stR_k,\la)$, $\la \rmv\in\rmv \mathbb{L}_{k-1}^{(c)}$ with a measurement index $m$ is plausible for $m \rmv\in\rmv \mathcal{M}_k^{(c)}$ and implausible for $m \rmv\in\rmv \mathcal{M}_k^{(c')}$ with $c' \!\neq\rmv c$. This is achieved by doing the following for each $\la^{(j)} \!\in\rmv \mathbb{L}_{k-1}^{*}$: In Step 3.1, we determine the subset $\mathcal{C}'$ of those indices $c \rmv\in\rmv \{1,\ldots,C\}$ for which the measurement index subsets $\mathcal{M}_k^{(c)} \!\subseteq\rmv \mathcal{M}_k$ have some elements in common with $\mathcal{M}_k(\la^{(j)})$, i.e., with the measurement indices corresponding to object state $(\stR_k,\la^{(j)})$; this expresses the fact that the association between object state $(\stR_k,\la^{(j)})$ and some measurement indices from $\bigcup_{c'\in\mathcal{C}'}\mathcal{M}_k^{(c')}$ is plausible. If none of the $\mathcal{M}_k^{(c)}$ has an element in common with $\mathcal{M}_k(\la^{(j)})$, i.e., if the association between object state $(\stR_k,\la^{(j)})$ with \emph{any} measurement index $m \rmv\in\rmv \bigcup_{c \in \{1,\ldots,C\}} \rmv \mathcal{M}_k^{(c)}$ is implausible, then $\mathcal{C}'$ is empty. In that case, $C$ is incremented by $1$, and a new label subset and a new measurement index subset are created as $\mathbb{L}_{k-1}^{(C)} \!=\! \{\la^{(j)}\}$ and $\mathcal{M}_k^{(C)} \!=\rmv \mathcal{M}_k(\la^{(j)})$, respectively (see Step 3.2). Otherwise, i.e., if $|\mathcal{C}'| \!\ge\rmv 1$, we merge all the label subsets $\mathbb{L}_{k-1}^{(c')}$ with $c' \!\in\rmv \mathcal{C}'$ as well as the considered label $\la^{(j)}$ into one common label subset $\mathbb{L}_{k-1}^{(c)}$, and we merge all the corresponding measurement index subsets $\mathcal{M}_{k}^{(c')}\rmv$, $c' \!\in\rmv \mathcal{C}'$ as well as $\mathcal{M}_k(\la^{(j)})$ into one common measurement index subset $\mathcal{M}_k^{(c)}$ (see Step 3.2, first bullet item). Here, the index $c$ is picked arbitrarily from $\mathcal{C'}\rmv$. Next, we perform a reindexing such that the index values in $\mathcal{C}'' \!\triangleq (\{1,\ldots,C\} \!\setminus \mathcal{C}') \ist\cup\ist \{c\}$ become $1,2,\ldots,|\mathcal{C}''|$. Furthermore, we update $C$ as $C = |\mathcal{C}''|$, so that the new set of subset indices is given by $\{1,\ldots,C\}$ (see Step 3.2, second and third bullet items). Subsequently, Steps 3.1 and 3.2 are repeated for the next $\la^{(j)} \!\in\rmv \mathbb{L}_{k-1}^{*}$ (if available).  

The final number $C$ of subsets $\mathbb{L}_{k-1}^{(c)}$, $c \rmv\in\rmv\{1,\ldots,C\}$ is determined by this iterative procedure. Finally, in Step 4, the measurement indices $m \rmv\in\rmv \mathcal{M}_k$ that are not part of any subset $\mathcal{M}_k^{(c)}$ are collected in $\mathcal{M}^{\text{res}}_{k}$. We note that a larger threshold $\gamma_{\text{C}}$ used in the definition of the sets $\mathcal{M}_k(\la)$ in Step 1 tends to result in smaller subsets $\mathcal{M}_k(\la)$, $\mathbb{L}_{k-1}^{(c)}$, and $\mathcal{M}_k^{(c)}\rmv$, a larger residual set $\mathcal{M}^{\text{res}}_{k}\rmv$,
a larger number $C$ of subsets $\mathbb{L}_{k-1}^{(c)}$ and $\mathcal{M}_k^{(c)}\rmv$, and a higher probability of $\mathcal{C}'$ being empty.

\section{} 
\label{sec:appB}

We will derive the approximation of the posterior pgfl $G_{\lRFSstR_k,\RFSstR_k}[\tilde{h},h|\RFSme_{1:k}]$ 
given by \eqref{eq:up_post_pgfl_6} and subsequent equations.

\vspace{-1mm}

\subsection{Pruning and Clustering}
\label{sec:first_appr:prun-1}

Our approximation is based on the partitioning of the label set $\mathbb{L}_{k-1}^{(*)}$ in \eqref{eq:upda_Ldecom} and the partitioning of the measurement index set $\mathcal{M}_k$ in \eqref{eq:upda_Zdecom2}. As described in Section \ref{sec:first_appr:part}, only the associations between objects with labeled state $(\stR_k,\la)$, $\la \rmv\in\rmv \mathbb{L}_{k-1}^{(c)}$ 
\vspace{-.5mm} 
and measurements with index $m \rmv\in\rmv \mathcal{M}_k^{(c)}$ are plausible. Thus, by pruning all the association hypotheses $\assv_k \!\in\! \mathcal{A}_k$ that associate some $\la \!\in\! \mathbb{L}_{k-1}^{(c)}$ with some $m \!\in\! \mathcal{M}_k \!\setminus\rmv \mathcal{M}_k^{(c)}\rmv$, we obtain a more efficient representation of the \emph{relevant} association information.
Let $\mathcal{A}^{\text{rem}}_k \!\subseteq\! \mathcal{A}_k$ denote the set of the remaining (nonpruned) $\assv_k$. 
Note that our pruning does not include missed detections (described by $\ass_k^{(\la)} = 0$), i.e., all $\assv_k$ with $\ass_k^{(\la)} \rmv=\rmv 0$, $\la \rmv\in\rmv\mathbb{L}_{k-1}^{(c)}$ are part of $\mathcal{A}_k^{\text{rem}}$. Therefore, each $\assv_k \!\in\! \mathcal{A}^{\text{rem}}_{k}$ associates each object label $\la \in \mathbb{L}_{k-1}^{(c)}$ with some measurement index $m \!\in\! \{0\} \cup \mathcal{M}_k^{(c)}\rmv$. 
(Note, also, that an $\assv_k \!\in\! \mathcal{A}^{\text{rem}}_{k}$ does not associate any object label with any $m \!\in\! \mathcal{M}^{\text{res}}_k$.)   

The pruning yields the following approximation of $G'_{\lRFSstR_k,\RFSstR_k}[\tilde{h},h]$ in 
\vspace{-1mm}
\eqref{eq:upd_LMB2}:
\begin{align}
G'_{\lRFSstR_k,\RFSstR_k}[\tilde{h},h] 
 & \approx\rmv \sum_{\assv_k \in \mathcal{A}^{\text{rem}}_{k}} \!\!\rmv\rd{\wei_{\assv_k}} \ist 
  L_{\mathbb{L}_{k-1}^{*}}\big[\tilde{h};\ex_k^{(\cdot,\ass_k^{(\cdot)})}\!,\sd_k^{(\cdot,\ass_k^{(\cdot)})}\big]\nn\\[-2mm]
&\hspace*{26mm} \times M_{\mathcal{M}_{\assv_k}}\big[h;\bar{\ex}_k^{(\cdot)}\!,\bar{\sd}_k^{(\cdot)}\big] .
\label{eq:joint_post_32}
\end{align}	
Using the fact that the Bernoulli component factors in $M_{\mathcal{M}_{\assv_k}}\big[h;\bar{\ex}_k^{(\cdot)}\!,\bar{\sd}_k^{(\cdot)}\big]$ 
with $m \rmv\in\rmv \mathcal{M}^{\text{res}}_k \rmv\subseteq\rmv \mathcal{M}_{\assv_k}$
appear in each one\linebreak 
of the 
summation terms in \eqref{eq:joint_post_32}, we obtain 
further
\begin{align}	
&\hspace*{-2mm}G'_{\lRFSstR_k,\RFSstR_k}[\tilde{h},h] \nn\\[-.5mm]
&\hspace*{-2mm}\approx \ist \rd{M_{\mathcal{M}^{\text{res}}_k}\big[h;\bar{\ex}_k^{(\cdot)}\!,\bar{\sd}_k^{(\cdot)}\big]} \rmv\sum_{\assv_k \in \mathcal{A}^{\text{rem}}_{k}} \!\!\rmv\rd{\wei_{\assv_k} \ist L_{\mathbb{L}_{k-1}^{*}}\big[\tilde{h};\ex_k^{(\cdot,\ass_k^{(\cdot)})}\!,\sd_k^{(\cdot,\ass_k^{(\cdot)})}\big]} \nn\\[-0.5mm]
 &\hspace{35mm}\times \rmv \rd{M_{\mathcal{M}_{\assv_k} \!\rmv\setminus\ist \mathcal{M}^{\text{res}}_k}\big[h;\bar{\ex}_k^{(\cdot)}\!,\bar{\sd}_k^{(\cdot)}\big]} . 
\label{eq:joint_post_3} \\[-5mm]
\nn
\end{align}
Here, the $\wei_{\assv_k}$ are given by expression \eqref{eq:upd_LMB_2}.

As a consequence of the pruning, all objects with labels $\la \!\in\rmv \mathbb{L}_{k-1}^{(c)}$, i.e., corresponding to cluster $c$,
are now associated only with measurements of the same cluster $c$, $m \rmv\in\rmv \{0\} \cup \mathcal{M}_k^{(c)}\rmv$, and not with any other measurements $m \rmv\in\rmv \mathcal{M}_k \!\setminus\rmv \mathcal{M}_k^{(c)}\rmv$. This implies that each entry $\ass_k^{(\la)}$ of $\assv_k \!\in\! \mathcal{A}^{\text{rem}}_{k}$ associates labels $\la \!\in\rmv \mathbb{L}_{k-1}^{(c)}$ of cluster $c$ only with measurements $m \!\in\! \{0\} \cup \mathcal{M}_k^{(c)}\rmv$ of cluster $c$.   
Therefore, $\assvR_k$ (of dimension $|\mathbb{L}_{k-1}^{*}|$) can be split into $C$ subvectors $\assvR_k^{\prime(c)}\rmv\in \big(\{0\} \cup \mathcal{M}_k^{(c)}\big)^{|\mathbb{L}_{k-1}^{(c)}|}\rmv$, $c \!\in\! \mathcal{C}$ of lower dimensions $|\mathbb{L}_{k-1}^{(c)}|$. Here, the entry $\assR_k^{\prime(c,\la)}$ of $\assvR_k^{\prime(c)}\rmv\rmv$, with $\la\rmv \in \mathbb{L}_{k-1}^{(c)}$, is defined similarly to $\assR_k^{(\la)}$ in Section \ref{sec:upda_SM1_det}
as $\assR_{k}^{\prime(c,\la)} \!\triangleq m \rmv\in\rmv \mathcal{M}_k^{(c)}$ if the labeled object with state $(\stR_k,\la)$ generates measurement $\me_k^{(m)}$ and $\assR_{k}^{\prime(c,\la)} \!\triangleq\rmv 0$ if it does not generate a measurement. The admissible association vectors $\assv_k^{\prime(c)}$ (where admissibility was defined in Section \ref{sec:upda_SM1_det}) are collected in the association alphabet 
$\mathcal{A}_k^{\prime(c)}\rmv$. We can now factor the weights as
\be
\wei_{\assv_k} = \prod_{c \in \mathcal{C}} \wei_{\assv_k^{\prime(c)}} ,
\label{eq:w_prod}
\vspace{-2.5mm}
\ee
where (cf.\ \eqref{eq:upd_LMB_2})
\vspace{-.5mm}
\[
\label{eq:upd_LMB1_approx}
\wei_{\assv_k^{\prime(c)}} \rmv\propto \ist\Bigg( \prod_{\la \in \mathbb{L}_{k-1}^{(c)}} \!\!\rmv \assw_k^{(\la,\ass_k^{\prime(c,\la)})} \Bigg) 
   \!\prod_{m \in \mathcal{M}_{\!\assv_k^{\prime(c)}}} \!\!\rmv \assw_k^{(m)}.
\vspace{-1mm}
\]
Here, $\mathcal{M}_{\rmv\assv_k^{\prime(c)}} \!\subseteq\rmv \mathcal{M}_k^{(c)}$ comprises all measurement indices 
\vspace{-.8mm}
$m \rmv\in \mathcal{M}_k^{(c)}$ that are not associated with any labeled object via $\assv_k^{\prime(c)} \!\rmv\in\! \mathcal{A}_k^{\prime(c)}$ and, thus, originate from an unlabeled object 
\vspace{-.3mm}
or from clutter. In particular, $\mathcal{M}_{\rmv\assv_k^{\prime(c)}} \!=\rmv \emptyset$ indicates that
\vspace{-1mm}
all $m \rmv\in\rmv \mathcal{M}_k^{(c)}$ are associated with an object with label $\la \!\in\rmv \mathbb{L}_{k-1}^{(c)}$. 
Furthermore, we have 
\begin{align}
\hspace{-.6mm}\rd{L_{\mathbb{L}_{k-1}^{*}}\big[\tilde{h};\ex_k^{(\cdot,\ass_k^{(\cdot)})}\!,\sd_k^{(\cdot,\ass_k^{(\cdot)})}\big]} 
  &= \prod_{c \in \mathcal{C}} \rd{L_{\mathbb{L}_{k-1}^{(c)}}\big[\tilde{h};\ex_k^{(\cdot,\ass_k^{\prime(c,\cdot)})} \!, \sd_k^{(\cdot,\ass_k^{\prime(c,\cdot)})}\big]} , \rmv\nn\\[-3mm]
  \label{eq:G-LMB_prod}\\[-.5mm]
\hspace{-.6mm}\rd{M_{\mathcal{M}_{\assv_k} \!\rmv\setminus\ist \mathcal{M}^{\text{res}}_k}\big[h;\bar{\ex}_k^{(\cdot)}\!,\bar{\sd}_k^{(\cdot)}\big]} 
  &= \prod_{c \in \mathcal{C}} \rd{M_{\mathcal{M}_{\assv_k^{\prime(c)}}}\big[h;\bar{\ex}_k^{(\cdot)}\!,\bar{\sd}_k^{(\cdot)}\big]} . 
  \label{eq:G-MB_prod} 
\end{align}
Using the factorizations \eqref{eq:w_prod}--\eqref{eq:G-MB_prod}
 as well as the identity $\sum_{\assv_k \in \mathcal{A}^{\text{rem}}_{k}} = \sum_{\assv_k^{\prime(1)} \rmv\in \mathcal{A}^{\prime(1)}_k} \cdots \sum_{\assv_k^{\prime(C)}\rmv \in \mathcal{A}^{\prime(C)}_k}$, we can rewrite the approximation \eqref{eq:joint_post_3} in terms of the $\assv_k^{\prime(c)}$ as
\be
G'_{\lRFSstR_k,\RFSstR_k}[\tilde{h},h] \approx\ist \rd{M_{\mathcal{M}^{\text{res}}_k}\big[h;\bar{\ex}_k^{(\cdot)}\!,\bar{\sd}_k^{(\cdot)}\big]}\ist \prod_{c\in\mathcal{C}} \rmv G^{(c)}[\tilde{h},h] \ist, 
\label{eq:approx_post1}
\vspace{-2mm}
\ee
where 
\begin{align}
G^{(c)}[\tilde{h},h] &\triangleq\! \sum_{\assv_k^{\prime(c)}\rmv \in \mathcal{A}_{k}^{\prime(c)}} \!\!\rmv \wei_{\assv_k^{\prime(c)}} \ist 
  \rd{L_{\mathbb{L}_{k-1}^{(c)}}\big[\tilde{h};\ex_k^{(\cdot,\ass_k^{\prime(c,\cdot)})} \!, \sd_k^{(\cdot,\ass_k^{\prime(c,\cdot)})}\big]} \nn \\[-2.5mm]
& \hspace{23mm}\times \rd{M_{\mathcal{M}_{\assv_k^{\prime(c)}}}\big[h;\bar{\ex}_k^{(\cdot)}\!,\bar{\sd}_k^{(\cdot)}\big]} .
\label{eq:cluster_pgfl} \\[-6mm]
\nn
\end{align}
\rd{We note that $G^{(c)}[\tilde{h},h]$ and $M_{\mathcal{M}^{\text{res}}_k}\big[h;\bar{\ex}_k^{(\cdot)}\!,\bar{\sd}_k^{(\cdot)}\big]$ represent \emph{clustered} objects
and \emph{nonclustered} objects, respectively, which, in both cases, may be likely or unlikely to exist.}

So far, we approximated $G'_{\lRFSstR_k,\RFSstR_k}[\tilde{h},h]$ in \eqref{eq:upd_LMB2} by expression \eqref{eq:approx_post1}, 
which is the product of the $C$ LMB--MB mixture pgfls $G^{(c)}[\tilde{h},h]$ in \eqref{eq:cluster_pgfl} and the MB pgfl 
\rd{$M_{\mathcal{M}^{\text{res}}_k}\big[h;\bar{\ex}_k^{(\cdot)}\!,\bar{\sd}_k^{(\cdot)}\big]$}. 
As visualized in Fig.~\ref{fig:PGFLs}, this is the first step in a series of pgfl approximations or modifications that are used in the development of the proposed LMB/P filter.
Next, we will develop approximations of $G^{(c)}[\tilde{h},h]$ and \rd{$M_{\mathcal{M}^{\text{res}}_k}\big[h;\bar{\ex}_k^{(\cdot)}\!,\bar{\sd}_k^{(\cdot)}\big]$}.

\vspace{-1mm}

\subsection{Approximation of the pgfl of Clustered Objects}
\label{sec:approx3}

\vspace{.5mm}

We will approximate \rd{the pgfl representing clustered objects,} $G^{(c)}[\tilde{h},h]$, by an LMBM pgfl. To this end, we recall from Section
\ref{sec:first_appr:prun-1} that the MB pgfl 
 \vspace{-1.2mm}
$\rd{M_{\mathcal{M}_{\assv_k^{\prime(c)}}}\big[h;\bar{\ex}_k^{(\cdot)}\!,\bar{\sd}_k^{(\cdot)}\big]}$ involved in $G^{(c)}[\tilde{h},h]$ in \eqref{eq:cluster_pgfl} 
corresponds to measurements $m \!\in\! \mathcal{M}_{k}^{(c)}$ that originate from an unlabeled object or from clutter. 
We want to transfer unlabeled objects that are likely to exist, or, more specifically, (unlabeled) Bernoulli components $\rd{B\big[h;\bar{\ex}_k^{(m)}\!,\bar{\sd}_k^{(m)}\big]}$, 
$m \!\in\! \mathcal{M}_{\assv_k^{\prime(c)}}$ with $\bar{\ex}_k^{(m)} \!\geq\rmv \gamma_{\text{tr}}$\ist,
\vspace{-1mm}
to the labeled RFS part. 
Here, $\bar{\ex}_k^{(m)}$ is given by \eqref{eq:undet3_2}.
This transfer is motivated by the fact that the labeled RFS part guarantees track continuity and, after further modifications that are
described in Section \ref{sec:second_appr_CO}, achieves a higher tracking accuracy than the unlabeled RFS part. The transfer is accomplished by formally replacing the measurement index $m$ arising in $\rd{B\big[h;\bar{\ex}_k^{(m)}\!,\bar{\sd}_k^{(m)}\big]}$ by the label $\la \rmv=\rmv (k,m)$. 
Let $\mathbb{L}_k^{(c)\text{tr}}$ collect the labels of the transferred Bernoulli components, i.e., all $l \!=\! (k,m)$ with $m \!\in\! \mathcal{M}_k^{(c)}$ such that $\bar{\ex}^{(m)} \!\geq\rmv \gamma_{\text{tr}}$ (see also Fig.~\ref{fig:Labels}). We note that a higher threshold $\gamma_{\text{tr}}$ tends to imply a smaller number of transferred Bernoulli components, $|\mathbb{L}_k^{(c)\text{tr}}|$. Furthermore, since the other Bernoulli components $\rd{B\big[h;\bar{\ex}_k^{(m)}\!,\bar{\sd}_k^{(m)}\big]}$ (with $\bar{\ex}_k^{(m)} \!<\rmv \gamma_{\text{tr}}$) are unlikely, we prune them. This is done by setting $h \!=\! 1$ because $\rd{B\big[1;\bar{\ex}_k^{(m)}\!,\bar{\sd}_k^{(m)}\big]} \rmv=\! 1$.

With these modifications, \rd{and by introducing the association vector $\assvR_k^{(c)}$ as in Section \ref{sec:Stage1_approx}}, $G^{(c)}[\tilde{h},h]$ in \eqref{eq:cluster_pgfl} is replaced by $G^{(c)}[\tilde{h}]$ as defined in \eqref{eq:cluster_pgfl2_2} 
(see Fig.~\ref{fig:PGFLs}). 
Accordingly, Eq.\ \eqref{eq:approx_post1} 
\vspace{.5mm}
becomes
\begin{equation}
\label{eq:approx_post2}
G'_{\lRFSstR_k,\RFSstR_k}[\tilde{h},h] \ist\approx \rd{M_{\mathcal{M}^{\text{res}}_k}\big[h;\bar{\ex}_k^{(\cdot)}\!,\bar{\sd}_k^{(\cdot)}\big]}\ist \prod_{c\in\mathcal{C}} \rmv G^{(c)}[\tilde{h}] \ist.
\vspace*{-2.5mm}
\end{equation}

\subsection{\rd{Approximation of the pgfl of Nonclustered Objects}}
\label{sec:first_appr_part2}

\vspace{.5mm}

Next, we consider \rd{$M_{\mathcal{M}^{\text{res}}_k}\big[h;\bar{\ex}_k^{(\cdot)}\!,\bar{\sd}_k^{(\cdot)}\big]$} in \eqref{eq:approx_post2}. 
Similarly to the measurements $m \!\in\! \mathcal{M}_{\assv_k^{\prime(c)}}$ involved 
\vspace*{-1mm}
in $\rd{M_{\mathcal{M}_{\assv_k^{\prime(c)}}}\big[h;\bar{\ex}_k^{(\cdot)}\!,\bar{\sd}_k^{(\cdot)}\big]}$ in \eqref{eq:cluster_pgfl}, 
\vspace*{-.7mm}
the measurements $m \!\in\! \mathcal{M}^{\text{res}}_k$ 
involved in $\rd{M_{\mathcal{M}^{\text{res}}_k}\big[h;\bar{\ex}_k^{(\cdot)}\!,\bar{\sd}_k^{(\cdot)}\big]}$ 
originate from an unlabeled object or from clutter. As in Section \ref{sec:approx3}, we transfer objects that are likely to exist to the labeled RFS part, 
and thus we formally replace the measurement index $m$ in each Bernoulli component $\rd{B\big[h;\bar{\ex}_k^{(m)}\!,\bar{\sd}_k^{(m)}\big]}$, 
$m \!\in\! \mathcal{M}^{\text{res}}_k$ of $\rd{M_{\mathcal{M}^{\text{res}}_k}\big[h;\bar{\ex}_k^{(\cdot)}\!,\bar{\sd}_k^{(\cdot)}\big]}$ 
with $\bar{\ex}_k^{(m)} \!\geq\rmv \gamma_{\text{tr}}$ by the label $\la \rmv=\rmv (k,m)$. 
These labels are collected in $\mathbb{L}_k^{\text{res,tr}}\rmv$ (see Fig.~\ref{fig:Labels}), and the corresponding 
\linebreak 
measurement indices are collected in $\mathcal{M}^{\text{res},\text{tr}}_k \!\subseteq\rmv \mathcal{M}^{\text{res}}_k$ (see Fig.~\ref{fig:Meas}). 
The remaining measurement indices are collected in $\mathcal{M}'_k \rmv=\rmv \mathcal{M}^{\text{res}}_k \setminus \mathcal{M}^{\text{res},\text{tr}}_k$ (again see Fig.~\ref{fig:Meas}). As before, a higher threshold $\gamma_{\text{tr}}$ tends to imply a smaller number of transferred Bernoulli components, $|\mathbb{L}_k^{\text{res,tr}}|$. 

Using these modifications, $\rd{M_{\mathcal{M}^{\text{res}}_k}\big[h;\bar{\ex}_k^{(\cdot)}\!,\bar{\sd}_k^{(\cdot)}\big]}$ is approximated according to (see Fig.~\ref{fig:PGFLs})
\begin{equation}
\rd{M_{\mathcal{M}^{\text{res}}_k}\big[h;\bar{\ex}_k^{(\cdot)}\!,\bar{\sd}_k^{(\cdot)}\big]} 
  \approx \rd{L_{\mathbb{L}_k^{\text{res,tr}}}\big[\tilde{h};\bar{\ex}_k^{(\cdot)}\rmv,\bar{\sd}_k^{(\cdot)}\big] \ist M_{\mathcal{M}'_k}\big[h;\bar{\ex}_k^{(\cdot)}\!,\bar{\sd}_k^{(\cdot)}\big]}. 
\vspace{1mm}
\label{eq:approx_pgfl-MB}
\end{equation}
Inserting \eqref{eq:approx_pgfl-MB} into \eqref{eq:approx_post2} yields $\rmv G'_{\lRFSstR_k,\RFSstR_k}[\tilde{h},h] 
\!\approx\!\rmv L_{\mathbb{L}_k^{\text{res,tr}}}\big[\tilde{h};\bar{\ex}_k^{(\cdot)}\rmv,\bar{\sd}_k^{(\cdot)}\big]$ $\times M_{\mathcal{M}'_k}\big[h;\bar{\ex}_k^{(\cdot)}\!,\bar{\sd}_k^{(\cdot)}\big] \prod_{c\in\mathcal{C}} G^{(c)}[\tilde{h}]$. Finally, inserting this latter approximation into \eqref{eq:up_post_pgfl_3} and grouping terms, we obtain \eqref{eq:up_post_pgfl_6}, \eqref{eq:final_appr_stage1_1}, and \eqref{eq:final_appr_stage1_2} 
(again see 
\vspace{4mm}
Fig.~\ref{fig:PGFLs}).

\end{document}